\documentclass[ 
	fontsize=12pt,          % Schriftgröße
	numbers=noenddot,    	% keine Punkt nach Gliederungsziffer z.B. 1.2.2 (alternativ: enddot)
    parskip=half,        	% Absatzumbruch als halbe Zeile (alternativ: parindent, parskip)
    listof=totoc,        	% Abbildungs- und Tabellenverzeichnis unnummeriert im Inhaltsverzeichnis
    bibliography=totoc,  	% Literaturverzeichnis unnummeriert im Inhaltsverzeichnis
	headsepline=true,       % Linie zwischen Kopf und Textteil einer Seite
	footsepline=false, 		% Linie zwischen Textteil und Fußzeile
    DIV=12,                	% Textbereichsausdehnung (6-15)
    toc=sectionentrywithdots
]{scrartcl}
\usepackage{amsthm}
\usepackage{amsmath, amsfonts, amssymb, bbm, bm}
\usepackage{prodint}
\usepackage{placeins}
\usepackage{array}
\usepackage[numbers]{natbib}
\usepackage{etoolbox}
\usepackage{pdfpages}

\AtBeginEnvironment{tabular}{\scriptsize}

\usepackage[english]{babel}

\usepackage[a4paper, left=2cm, right=2cm, top=3cm, bottom=3cm]{geometry}

\usepackage[hidelinks]{hyperref} 

\usepackage{todonotes} % Paket fuer Notizen

\usepackage{threeparttable}
\usepackage{mathabx}
\usepackage{floatrow, float}
\usepackage[morefloats = 100]{morefloats} %damit man alle Tabellen sieht.

\theoremstyle{plain}
\newtheorem{theorem}{Theorem}
\newtheorem{example}{Example}

\newtheorem{assumption}{Assumption}

\newtheorem{proposition}{Proposition}

\usepackage{apptools}
\AtAppendix{
\setcounter{table}{0}
\setcounter{equation}{0}
\setcounter{lemma}{0}
\setcounter{theorem}{0}
\setcounter{figure}{0}

}

\newcommand{\R}{\mathbb{R}}
\newcommand{\T}{\mathcal{T}}
\newcommand{\E}{\mathbb{E}}
\renewcommand{\P}{P}

\newcommand{\mbf}{\mathbf}
\newcommand{\bs}{\boldsymbol}
\newcommand{\ji}{, j\in\{1,\ldots,n_i\}, i\in\{1,\ldots,k\}}
\newcommand{\im}{, i\in\{1,\ldots,k\}, m \in\{1,\ldots,M\}}
\newcommand{\ijmnew}{, i,j\in\{1,\ldots,k\}, m \in\{1,\ldots,M\}}
\renewcommand{\i}{, i\in\{1,\ldots,k\}}
\newcommand{\ii}{i\in\{1,\ldots,k\}}
\newcommand{\iimm}{i\in\{1,\ldots,k\}, m\in\{1,\ldots,M\}}

\newcommand{\iim}{i\in\{1,\ldots,k\}, m,\Tilde{m}\in\{1,\ldots,M\}}

\newcommand{\m}{, m\in\{1,\ldots,M\}}
\newcommand{\mm}{, m\in\{1,2,3\}}
\newcommand{\mtilde}{, m\in\{1,\ldots,\Tilde{M}\}}
\renewcommand{\l}{, \ell\in\{1,\ldots,L\}}
\renewcommand{\ll}{\ell\in\{1,\ldots,L\}}
\newcommand{\jjii}{j\in\{1,\ldots,n_i\},i\in\{1,\ldots,k\}}

\renewcommand{\citet}{\cite}





\renewcommand{\title}[1]{ \noindent{\centering \Large \textbf{ #1 } \\} }
\newcommand{\inst}[1]{\textsuperscript{#1}}
\newcommand{\institute}[1]{{\centering \footnotesize{#1}} \vspace{2ex}}

\newenvironment{keywords}{%
    \noindent\textbf{Keywords: }%
}

\begin{document}

\title { Multiple tests for restricted mean time lost with competing risks data }
\begin{center}
		Merle Munko\inst{1,*}\let\thefootnote\relax\footnote{*Corresponding author. Email address: \url{merle.munko@ovgu.de}},
        Dennis Dobler\inst{2,3},
        and
        Marc Ditzhaus\inst{1}
\end{center}

	% Institutes of all authors
	% Include city and country of each institute, do not include the full address.
	\institute{
		\inst{1} Otto-von-Guericke University Magdeburg; Magdeburg (Germany)\newline
        \inst{2} TU Dortmund University; Dortmund (Germany)\newline
        \inst{3} Research Center Trustworthy Data Science and Security, University Alliance Ruhr; Dortmund (Germany)
	}

\hrule

%%%%%%%%%%%%%%%%%%%%%%%%%%%%%%%%%%%%%%%%%%%%%%%%%%%%%%%%%%%%%%%%%%%%%%%%%%%%%%%%%%%%%%%%%%%%%%%%%%%%%

\begin{abstract}
Easy-to-interpret effect estimands are highly desirable in survival analysis. In the competing risks framework, one good candidate is the restricted mean time lost (RMTL). It is defined as the area under the cumulative incidence function up to a prespecified time point and, thus, it summarizes the cumulative incidence function into a meaningful estimand. 
While existing RMTL-based tests are limited to two-sample comparisons and mostly to two event types, we aim to develop general contrast tests for factorial designs and an arbitrary number of event types based on a Wald-type test statistic. Furthermore, we avoid the often-made, rather restrictive continuity assumption on the event time distribution. This allows for ties in the data, which often occur in practical applications, e.g., when event times are measured in whole days. In addition, we develop more reliable tests for RMTL comparisons that are based on a permutation approach to improve the small sample performance.
In a second step, multiple tests for RMTL comparisons are developed to test several null hypotheses simultaneously. Here, we incorporate the asymptotically exact dependence structure between the local test statistics to gain more power. The small sample performance of the proposed testing procedures is analyzed in simulations and finally illustrated by analyzing a real data example about leukemia patients who underwent bone marrow transplantation.
\end{abstract}

\begin{keywords}%3-8 keywords in alphabetical order, nouns, singular
    Competing risks;
    Factorial design; 
    Multiple testing; %Nonproportional hazards; 
    Permutation; 
    Restricted mean time lost; Survival analysis.
\end{keywords}

\section{Introduction} \label{sec:Intro}
% Warum ist das alles mega toll?
%RMST ist super
The restricted mean survival time is becoming increasingly important as an alternative effect measure to the popular hazard ratio, especially in situations where the proportional hazards assumption is violated \citep{alternative}. It is defined as the area under the survival curve up to a prespecified end point $\tau$ and, thus, it offers a straightforward interpretation as the expected duration of time alive before $\tau$. By integrating across the distribution function rather than the survival curve, we derive the restricted mean time lost (RMTL), which can be interpreted as expected time lost before $\tau$.
Naturally, it equals $\tau$ minus the restricted mean survival time. % for a single event.
In the context of competing-risks frameworks, where multiple events like death from various causes occur, the RMTL for a specific event can be defined simply as the area under the corresponding sub-distribution function.

% Was gibt es schon? Literaturrecherche!
For the restricted mean survival time, asymptotic tests for two sample comparisons and resampling-based tests have already been developed \citep{RMST,Perm} as well as multiple contrast tests in more complex factorial designs \citep{munko2023rmstbased}.
In competing-risks settings, the RMTL has also been studied in several papers \citep{andersen2013, lyu2020use, wen2023, wu2022implementation, zhao2018}. 
%However, these works are limited to one- and two-sample cases.
%\citet{andersen2013} studied the RMTL for one event and one sample. %one event, one sample
%For the two-sample case, inference for the difference of RMTLs of one event is developed in \citet{lyu2020use, wu2022implementation}.
%In \citet{conner2021}, the RMTL difference of two samples is adjusted for covariates.
% Was ist neu?
% discontinuous subdistribution functions
However, the considered settings are limited to one- and two-sample cases and mostly allow for only two different event types such that there is a lack of suitable RMTL-based tests for more complex factorial designs, more general hypotheses and more event types to, e.g., compare the RMTLs of various event types across several groups. Additionally, all proposed methods seem to require existing sub-distribution hazards, i.e., in particular continuous sub-distribution functions. This assumption is often not justified in practice, e.g., when the event times are measured in whole days or weeks.
Hence, our first aim is to develop flexible Wald-type tests that are applicable (i) for general RMTL contrasts in factorial designs and (ii) without a continuity assumption on the event times. 
Moreover, for the restricted mean survival time in the classical survival setup, resampling procedures have proven to be useful in ensuring an accurate type I error control for finite samples \citep{RMST, Perm, munko2023rmstbased}. 
Hence, to improve the small sample performance of the constructed Wald-type test, a studentized permutation approach is applied and its asymptotic validity is shown. 

Furthermore, if a global test indicates a significant difference between the RMTLs across several groups, it would also be of interest which concrete RMTL differences cause the significant result. In order to answer such questions, multiple tests need to be performed simultaneously. 
Recently, a maximum joint test for testing the equalities of two RMTLs in the two-sample case jointly was studied in \citet{wen2023}. 
However, the two-sample case and the two considered hypotheses are rather restrictive. 
Hence, there is still a lack of multiple testing procedures based on RMTLs for multiple contrast hypotheses addressing (i) and (ii).
Thus, we secondly aim to develop powerful multiple tests for RMTL contrasts by taking the asymptotically exact dependence structure of the local Wald-type test statistics into account.

The remainder of this paper is organized as follows. 
Section~\ref{sec:Inference} is divided into three subsections. The general factorial competing risks setup is presented in Section~\ref{ssec:Setup} including the formal definition of the RMTL and the global testing problem.
The Wald-type test statistic is investigated in Section~\ref{ssec:Wald}. In Section~\ref{ssec:Perm}, the studentized permutation approach is introduced and its asymptotic validity is proven. Multiple tests for several RMTL contrast hypotheses are developed in Section~\ref{sec:Multiple}. In Section~\ref{sec:Simu}, the finite sample performance of our proposed methods is analyzed in extensive simulations. Additionally, we illustrate our methods by analyzing a data example %about blood and marrow transplantation 
 about leukemia patients who underwent bone marrow transplantation
in Section~\ref{sec:Data}. Finally, an outlook is given in Section~\ref{sec:Discussion}.
The technical proofs of all stated theorems can be found in the Supplementary Material.
%%Further material such as technical details, extended proofs, code, or additional simulations, ﬁgures and examples may appear online, and should be brieﬂy mentioned as Supplementary Material where appropriate. Please submit any such content as a PDF ﬁle along with your paper, entitled ‘Supplementary material for Title-of-paper’. The Supplementary Material should be produced using the same latex template as the main paper

\section{Global tests}\label{sec:Inference}
% Biometrika will hier keinen Text zwischen haben!!!
\subsection{Factorial competing risks setup  }\label{ssec:Setup}
In the following, we are interpreting a factorial competing risks design as a $k$-sample setup with $M$ competing events, where $k, M \in\mathbb N$. 
% Zitieren wie man Indexe aufsplitten kann.
We assume that there are independent survival and right-censoring times $T_{ij} \sim S_i, C_{ij} \sim G_i\ji,$ respectively, and random variables $D_{ij}\ji$ indicating the event types and taking values in $\{1,\ldots,M\}$.
Here, \begin{align*}
    S_i:[0,\infty) \to [0,1], \: S_i(t) := \P(T_{i1} > t) \quad\text{ and }\quad G_i:[0,\infty)\to [0,1], \: G_i(t) := \P(C_{i1} > t)
\end{align*} denote the survival functions of the survival and censoring times, respectively, and $n_i\geq 2$ denotes the sample size of group $i$ for all $\ii$. We do not suppose the continuity of the survival functions and, thus, we explicitly allow for ties in the data.
Additionally, we assume that $(T_{ij}, C_{ij}, D_{ij})\ji$ are mutually independent and that the censoring time $C_{ij}$ is independent of the survival time and event type $(T_{ij}, D_{ij})$ for all $\jjii$.
 Due to right-censoring, we can only observe the right-censored event times $X_{ij} := \min\{T_{ij},C_{ij}\}$ and the event indicator $\delta_{ij} := D_{ij}\mathbbm{1}\{X_{ij} = T_{ij}\}\ji$, where here and throughout $\mathbbm{1}$ denotes the indicator function.
Furthermore, let
\begin{align*}
    &F_{im}:[0,\infty) \to [0,1], \: F_{im}(t) := \P (T_{i1} \leq t, D_{i1} = m) \\ \text{ and }\quad &A_{im}:[0,\infty)\to [0,\infty], \: A_{im}(t) := \int_{[0,t]} \frac{1}{S_{i-}} \;\mathrm{d}F_{im}
\end{align*}
 denote the cumulative incidence function and the cause-specific cumulative hazard functions, respectively, 
for all $\iimm$, where here and throughout $S_-$ denotes the left-continuous version of a c\`adl\`ag function $S$. The sum of all cause-specific hazard functions of group $i$ is denoted by $A_i := \sum_{m=1}^M A_{im}\i$ in the following.

In order to introduce suitable estimators for these quantities, we firstly define the number of individuals at risk just before time $t\geq 0$ by $Y_i(t) := \sum_{j=1}^{n_i} \mathbbm{1}\{X_{ij} \geq t\}$ and the number of individuals with an event of type $m$ before or at time $t\geq 0$ by $N_{im}(t) := \sum_{j=1}^{n_i} \mathbbm{1}\{X_{ij} \leq t, \delta_{ij} = m\}$ for all $\iimm.$ Then, we set 
\begin{align*}
   \widehat{A}_{im}(t) := \int_{[0,t]} \frac{1}{Y_i} \;\mathrm{d}N_{im} , \quad \widehat{A}_{i} := \sum_{m=1}^M\widehat{A}_{im} \quad\text{ and }\quad \widehat{S}_i(t) := \prodi\limits_{x \leq t} \left\{ 1 - \mathrm{d} \widehat{A}_{i} (x) \right\}
\end{align*}
for all $t\geq 0; \iimm$, where here and throughout $\prodi$ denotes the product integral as in \citet{GillJohansen}. %\Marcinline{referenz für produtintegration, da es nicht allen bekannt ist?}. 
These estimators are the cause-specific and all-cause Nelson--Aalen estimators and the Kaplan--Meier estimator, respectively.
Thus, we obtain the Aalen--Johansen estimator at $t$ for $F_{im}(t)$ as $\widehat{F}_{im}(t) := \int_{[0,t]} \widehat{S}_{i-}\;\mathrm{d}\widehat{A}_{im}\im$ for all $t\geq 0$.

The restricted mean time lost (RMTL) 
%of group $i$ and event of type $m$ 
due to the event type $m$ in group $i$
is defined as the area under the corresponding cumulative incidence function up to a prespecified time point $\tau > 0$, that is,
\begin{align*}
    \mu_{im} := \int_0^{\tau} F_{im}(t) \;\mathrm{d}t\quad\im.
\end{align*} 
Of note, in the case of only one event type, i.e., $M=1$, the RMTL equals $\tau$ minus the more popular restricted mean survival time.

By replacing $F_{im}$ through the corresponding Aalen--Johansen estimator, we obtain a natural estimator for the RMTL, that is,
\begin{align*}
    \widehat\mu_{im} := \int_0^{\tau} \widehat F_{im}(t) \;\mathrm{d}t\quad\im.
\end{align*}
% Hypothese formulieren
Let $\bs{\mu} := (\mu_{11}, \ldots, \mu_{1M}, \mu_{21},\ldots, \mu_{kM})^{\top}$ denote the vector of the RMTLs and its estimator by $\widehat{\bs{\mu}} := (\widehat\mu_{11}, \ldots, \widehat\mu_{1M}, \widehat\mu_{21},\ldots, \widehat\mu_{kM})^{\top}$.
Moreover, let $r\in\mathbb N$, $\mbf{c} \in \R^r$ and $\mbf{H} \in \R^{r\times kM}\setminus \{\mbf0_{r\times kM}\}$ satisfying $\mbf{H}(\mbf{1}_{k}\otimes \mbf{e}_m) = \mbf{0}_r\m$, where here and throughout $\mbf0_{r\times kM}\in\R^{r\times kM}$ denotes the matrix of zeros, $\mbf{1}_{k}\in\R^{k}$ denotes the vector of ones, $\mbf{e}_m\in\R^M$ denotes the $m$th unit vector, $\mbf{0}_r\in\R^{r}$ denotes the vector of zeros and $\otimes$ denotes the Kronecker product.  %^{r\times kM}$. % a contrast matrix with $\mathrm{rank}(\mbf{H}) > 0$.
This means that $\mbf{H}$ has the contrast property in terms of the different groups and not in terms of the different event types.
In this paper, we consider the testing problem
\begin{align}\label{eq:Hypos}
    \mathcal{H}_0: \mbf{H}\bs{\mu} = \mbf{c} \quad \text{vs.} \quad \mathcal{H}_1: \mbf{H}\bs{\mu} \neq \mbf{c}.
\end{align}
This testing problem is very general and covers various types of hypotheses and factorial designs as illustrated in the following examples.

\begin{example}[Two-sample case]\label{twosample}
%{Der two-Sample case für Marc ;) Marc: Gene mit KOnfidenzintervallen, dann auch mit permutieren.}
    The simplest but perhaps most relevant case in practice is the two-sample case, i.e., $k=2$. 
    The null hypothesis of equal RMTLs of all event types, i.e., $\mathcal{H}_0: \mu_{1m} = \mu_{2m}\m$, can be realized by choosing $\mbf{c}:=\mbf{0}_r$ and $\mbf{H} := [-1, 1] \otimes \mbf{I}_M$, where here and throughout $\mbf{I}_M \in \R^{M\times M}$ denotes the unit matrix.
    If not the RMTLs of all $M$ event types but only the first $\Tilde{M} < M$ %\in\{1,...,M\}$ 
    event types are of interest, we may choose $\mbf{H} := [-1, 1] \otimes [\mbf{I}_{\Tilde{M}},\mbf{0}_{\Tilde{M}\times (M-\Tilde{M})}]$ instead. This yields the null hypothesis $\mathcal{H}_0: \mu_{1m} = \mu_{2m}\mtilde$.
\end{example}

\begin{example}[One-way design]\label{ex1}
    In many applications, the hypothesis of equal RMTLs across the groups is of interest, that is $$\mathcal{H}_0: \mu_{1m} = \ldots = \mu_{km} \quad\m.$$ 
    %is of interest. 
    This hypothesis can be formulated with $\mbf{c}:=\mbf{0}_r$ and various hypothesis matrices $\mbf{H}$. For example, $\mbf{H}$ may be chosen as the Kronecker product of the Dunnett-type \citep{dunnett_1955} contrast matrix $[-\mbf{1}_{k-1}, \mbf{I}_{k-1}] \in \R^{(k-1)\times k}$ and the unit matrix $\mbf{I}_M \in \R^{M\times M}$. Another possibility is the Kronecker product of the Tukey-type \citep{tukey} contrast matrix $$\begin{bmatrix}
-1 & 1 & 0 & 0 & \cdots & \cdots & 0 \\
-1 & 0 & 1 & 0 &\cdots & \cdots & 0 \\
\vdots  & \vdots &\vdots & \vdots & \ddots & \vdots & \vdots  \\
-1 & 0 & 0 & 0& \cdots & \cdots & 1\\
0 & -1 & 1 & 0& \cdots & \cdots & 0 \\
0 & -1 & 0 & 1& \cdots & \cdots & 0 \\
\vdots  & \vdots & \vdots  & \vdots & \ddots & \vdots & \vdots \\
0 & 0 & 0 & 0 & \cdots & -1 & 1
\end{bmatrix} \in \R^{k(k-1)/2 \times k}$$ and the unit matrix $\mbf{I}_M$.
\end{example}

\begin{example}[Factorial 2-by-2 design]\label{ex2}
    In a factorial 2-by-2 design with factors A and B, we have $k = 4$ groups (group $1$: A=1, B=1; group $2$: A=1, B=2; group $3$: A=2, B=1; group $4$: A=2, B=2). The hypothesis of no main effect of factor A can be formulated by using $\mbf{H}_A = [\mbf{I}_M, \mbf{I}_M, -\mbf{I}_M, -\mbf{I}_M]$. Analogously, the hypothesis of no main effect of factor B uses $\mbf{H}_B = [\mbf{I}_M, -\mbf{I}_M, \mbf{I}_M, -\mbf{I}_M]$ and the hypothesis of no interaction effect between A and B uses $\mbf{H}_{AB} = [\mbf{I}_M, -\mbf{I}_M, -\mbf{I}_M, \mbf{I}_M]$. % If all three effects should be tested simultaneously, we can put all matrices together in one hypothesis matrix $[(\mbf{H}^A)^{\top}, (\mbf{H}^B)^{\top}, (\mbf{H}^{AB})^{\top}]^{\top}$
\end{example}

More general factorial designs can be incorporated easily by splitting up the indices similarly as in Example~\ref{ex2}, see \citet{paulyETAL2015} for details.

\subsection{The Wald-type test statistic and its asymptotic behaviour}
\label{ssec:Wald}

In this section, a suitable test statistic for the testing problem (\ref{eq:Hypos}) is constructed and its asymptotic behaviour is studied.

For technical reasons, we need the following assumptions throughout this paper.
\begin{assumption}\label{Assumption}
    In the following, we assume $S_{i-}(\tau) > 0$, $G_{i-}(\tau) > 0$ %for all $i\in\{1,\ldots,k\},$    and no group size vanishes asymptotically, i.e. 
    and $n_i / n \to \kappa_i \in (0,1)$ as $n\to\infty$ for all $\ii$, where here and throughout $n := \sum_{i=1}^k n_i$ denotes the total sample size.
\end{assumption}

By applying empirical process theory and the delta method, we proved the asymptotic normality of the vector of RMTL estimators $\widehat{\bs{\mu}}$:
\begin{theorem}\label{AsymptoticNormality}
Under Assumption~\ref{Assumption}, we have     
        $${n}^{1/2} (\widehat{\bs{\mu}}-\bs{\mu}) \xrightarrow{d} \mbf{Z} \sim \mathcal{N}_{kM}(\mbf{0}_{kM}, \bs{\Sigma})$$
     as $n\to\infty$, where here and throughout $\xrightarrow{d}$ denotes the convergence in distribution and $\mbf{0}_{kM}\in\R^{kM}$ denotes the vector of zeros. The definition of the covariance matrix $\bs{\Sigma}$ is given in (S3) in the Supplementary Material.
\end{theorem}

In the following, we also need that the limit distribution is not degenerated. In Lemma~S2 in the Supplementary Material, we show that the following assumption together with Assumption~\ref{Assumption} suffices for the positive definiteness of $\bs{\Sigma}$:

\begin{assumption}\label{Assumption2}
    We assume that $F_{im-}(\tau) > 0\im$.
\end{assumption}

As shown in (S2), the entries of $\bs{\Sigma}$ depend on the unknown functions $F_{im}, A_i$ and $\sigma_{im\Tilde{m}}$ for all $\iim$, where $\sigma_{im\Tilde{m}}$ denotes the asymptotic covariance function of %the limit variables for 
the cause-specific cumulative hazard functions $\widehat A_{im}, \widehat A_{i\Tilde{m}}$. Thus, the plug-in estimator
$$\widehat{\bs{\Sigma}} := \bigoplus_{i=1}^k \left(\frac{n}{n_i}\widehat{\bs{\Sigma}}_i\right)$$ for $\bs{\Sigma}$ can be obtained by replacing $F_{im}, A_i$ and $\sigma_{im\Tilde{m}}$ in (S2) by $\widehat{F}_{im}, \widehat{A}_i$ and $\widehat{\sigma}_{im\Tilde{m}}$, respectively, for all $\iim,$ 
with
\begin{align*}
    &\widehat{\sigma}_{im{m}}(t) := n_i\int_{[0,t]} \frac{1 - \Delta \widehat{A}_{im}}{Y_i} \;\mathrm{d}\widehat{A}_{im}
   \quad\text{and}\quad \widehat{\sigma}_{im\Tilde{m}}(t) := - n_i\int_{[0,t]} \frac{\Delta \widehat{A}_{im}}{Y_i} \;\mathrm{d}\widehat{A}_{i\Tilde{m}}
\end{align*}
for all $t\geq 0, m \neq \Tilde{m}$. Here, $\bigoplus$ denotes the direct sum and $\Delta A := A - A_-$ denotes the increment of a c\`adl\`ag function $A$. 
Then, the Wald-type test statistic can be defined by
\begin{align*}
    W_n(\mbf{H},\mbf{c}) := n (\mbf{H}\widehat{\bs{\mu}}-\mbf{c})^{\top} \left(\mbf{H} \widehat{\bs{\Sigma}} \mbf{H}^{\top}  \right)^+ (\mbf{H}\widehat{\bs{\mu}}-\mbf{c}),
\end{align*} where here and throughout $\mbf{A}^+$ denotes the Moore--Penrose inverse of a matrix $\mbf{A}$.
Since the Wald-type test statistic is a quadratic form of the vector $(\mbf{H}\widehat{\bs{\mu}}-\mbf{c})$, we would reject the null hypothesis in (\ref{eq:Hypos}) for large values of $ W_n(\mbf{H},\mbf{c})$.
The following theorem provides the asymptotic distribution of the Wald-type test statistic.

\begin{theorem}\label{AsymptoticWald}
    Under Assumptions~\ref{Assumption} and \ref{Assumption2} and the null hypothesis in (\ref{eq:Hypos}), we have, as $n\to\infty$,
        $$W_n(\mbf{H},\mbf{c}) \xrightarrow{d} \chi^2_{\mathrm{rank}(\mbf{H})}.$$
\end{theorem}

Thus, an asymptotic level-$\alpha$ test for (\ref{eq:Hypos}) is
%$$
$\varphi = \mathbbm{1}\{ W_n(\mbf{H},\mbf{c}) > \chi^2_{\mathrm{rank}(\mbf{H}), 1-\alpha} \}, $ 
%$$
where $\chi^2_{\mathrm{rank}(\mbf{H}), 1-\alpha}$ denotes the $(1-\alpha)$-quantile of the $\chi^2_{\mathrm{rank}(\mbf{H})}$ distribution for $\alpha\in(0,1)$.

Due to the direct connection between tests and confidence regions, we can obtain a confidence region with level $1-\alpha$ for $\mbf{H}\bs{\mu}$ from Theorem~\ref{AsymptoticWald}, that is
$ \{\bs{\xi} \in \R^{r} \mid  W_n(\mbf{H},\bs{\xi}) \leq \chi^2_{\mathrm{rank}(\mbf{H}), 1-\alpha}\} .$

\subsection{Studentized permutation test}\label{ssec:Perm}
We showed in the previous section that the proposed test for the RMTLs is asymptotically valid but, however, this does not guarantee a good small sample performance of the test in terms of type I error control. As we will see in Section~\ref{ssec:SimuResults}, the asymptotic test has in fact an increased type I error in simulations.
For the restricted mean survival time, permutation methods solved this problem \citep{RMST,Perm,  munko2023rmstbased}.
Permutation tests are known to control the type I error exactly under exchangeable data \citep{hemerik2018exact, lehmann1986testing}, which means $F_{im} = F_{jm}, G_i = G_j\ijmnew$ in our case.
However, since our null hypothesis in \eqref{eq:Hypos} may hold even when the data is not exchangeable, we aim to use a studentized permutation approach in this section that preserves the finite exact control of the type I error under exchangeability but is also asymptotically valid under non-exchangeable data as the studentized permutation tests in \citet{RMST, munko2023rmstbased}.

Therefore, let $(\mbf{X},\bs{\delta}) := (X_{ij},\delta_{ij})_{\jjii}$ denote the pooled sample and $(X_{ij}^{\pi},\delta_{ij}^{\pi})_{\jjii}$ denote the permuted data. In detail, the data are permuted as pairs $(X_{ij},\delta_{ij})$ by shuffling the groups of the original data randomly. Furthermore, we denote the statistics $\widehat{\bs{\mu}},\widehat{\bs{\Sigma}}$ based on the permuted data $(X_{ij}^{\pi},\delta_{ij}^{\pi})_{\jjii}$ with a $\pi$ in the superscript, i.e., $\widehat{\bs{\mu}}^{\pi},\widehat{\bs{\Sigma}}^{\pi}$. Therewith, we define the permutation counterpart of the Wald-type test statistic by
\begin{align*}
    W_n^{\pi}(\mbf{H}) := n (\mbf{H}\widehat{\bs{\mu}}^{\pi})^{\top} \left(\mbf{H} \widehat{\bs{\Sigma}}^{\pi} \mbf{H}^{\top}  \right)^+ (\mbf{H}\widehat{\bs{\mu}}^{\pi}).
\end{align*}
The permutation approach mimics the correct distribution asymptotically, which is shown in the following theorem.

\begin{theorem}\label{Permutation}
    Under Assumptions~\ref{Assumption} and \ref{Assumption2}, we have under both hypotheses $\mathcal{H}_0$ and $\mathcal{H}_1$, as $n\to\infty$,
        $$W_n^{\pi}(\mbf{H}) \xrightarrow{d^*} \chi^2_{\mathrm{rank}(\mbf{H})},$$
        where here and throughout $\xrightarrow{d^*}$ denote conditional convergence in distribution in probability given the data $(\mbf{X},\bs{\delta})$.
        Mathematically, this means, as $n\to\infty$,
        $$ \sup\limits_{z\in\R} \left|\P\left(W_n^{\pi}(\mbf{H}) \leq z  \mid (\mbf{X},\bs{\delta})\right) - \chi^2_{\mathrm{rank}(\mbf{H})}((-\infty,z])\right| \xrightarrow{\P} 0.$$
\end{theorem}

By using this result, we can construct a permutation test for \eqref{eq:Hypos}.
In practice, usually a Monte Carlo method is applied to approximate the resulting critical value, which is the $(1-\alpha)$-quantile of the conditional distribution of $W_n^{\pi}(\mbf{H})$ given the data $(\mbf{X},\bs{\delta})$. 
  Therefore, the quantile is approximated by the empirical $(1-\alpha)$-quantile ${q}_{1-\alpha}^{\pi}$ of $B_n$ conditional independent random variables distributed as $W_n^{\pi}(\mbf{H})$ given $(\mbf{X},\bs{\delta})$. Here and throughout, $(B_n)_{n\in\mathbb{N}}$ denotes a sequence of natural numbers with $B_n \to\infty$ as $n\to\infty$.
Hence, we receive the permutation test 
   $ \varphi^{\pi} = \mathbbm{1}\left\{W_n(\mbf{H}, \mbf{c}) > q_{1-\alpha}^{\pi}  \right\}.$ % where $q_{1-\alpha}^{\pi}$ denotes the $(1-\alpha)$-quantile of the conditional distribution of $W_n^{\pi}(\mbf{H})$ given the data $(\mbf{X},\bs{\delta})$. 
   %Lemma~1 in \cite{janssenPauls2003} ensures the asymptotic validity of the studentized permutation test. 
   %Among the asymptotic validity, permutation tests are also known to be finitely exact under exchangeability, which means here that the cumulative incidence functions as well as the censoring distributions coincide across all groups.
  By the following proposition, the permutation test is asymptotically valid.

  \begin{proposition}\label{PermProp}
       Under Assumptions~\ref{Assumption}, \ref{Assumption2} and the null hypothesis in \eqref{eq:Hypos}, we have 
       \begin{align*}
           \lim\limits_{n\to\infty} \E\left( \varphi^{\pi} \right) = \lim\limits_{n\to\infty} \P\left( W_n(\mbf{H}, \mbf{c}) > {q}_{1-\alpha}^{\pi} \right) = \alpha.
       \end{align*}
       %where here and throughout $\E^*, \P^*$ denote the outer expectation and probability, respectively.
  \end{proposition}
%This is a direct consequence of Lemma~1 in \cite{janssenPauls2003} together with Theorem~\ref{AsymptoticWald} and \ref{Permutation}.%stimmt nicht wegen Monte Carlo
  
The corresponding confidence region with level $1-\alpha$ for $\mbf{H}\bs{\mu}$ is
$ \left\{\bs{\xi} \in \R^{r} \mid  W_n(\mbf{H},\bs{\xi}) \leq q^{\pi}_{1-\alpha}\right\} .$

\begin{example}
    The confidence interval with level $1-\alpha$ for $\mu_{11} - \mu_{21} = \mbf{H}\bs{\mu}$ with $\mbf{H} = (\Tilde{\mbf{e}}_{1} - \Tilde{\mbf{e}}_{2})^{\top} \otimes \mbf{e}_1^{\top}$, where $\Tilde{\mbf{e}}_{1},\Tilde{\mbf{e}}_{2}\in\R^k$ denote the first and second unit vector in $\R^k$, based on the permutation test is
    $$ \left[\widehat\mu_{11} - \widehat\mu_{21} - \left( \left({\widehat{\Sigma}_{111}/n_1 + \widehat{\Sigma}_{211}/n_2}\right) q^{\pi}_{1-\alpha}\right)^{1/2} , \widehat\mu_{11} - \widehat\mu_{21} + \left( \left({\widehat{\Sigma}_{111}/n_1 + \widehat{\Sigma}_{211}/n_2}\right) q^{\pi}_{1-\alpha}\right)^{1/2} \right], $$
    where $\widehat{\Sigma}_{i11}$ denotes the top-left entry of $\widehat{\bs{\Sigma}}_i\i.$
%    \Dennisinline{Meinst du hier dasselbe $q^\pi$ wie in Theorem 3? Denn dort war es ja für die Wald-type stat. Dh wir sind mit diesem CI konservativ? Ggf Definition von $q^\pi$ anpassen. Oder wollen wir sagen: man bekommt ein Konfidenzellipsoid für die Differenz, dh insbesondere CIs für die einzelnen Differenzen (wenngleich konservativ).}
    Analogously, the confidence interval with level $1-\alpha$ for $\mu_{11} - \mu_{21}$  based on the asymptotic test in Section~\ref{ssec:Wald} is
    $$ \left[\widehat\mu_{11} - \widehat\mu_{21} - \left( \left({\widehat{\Sigma}_{111}/n_1 + \widehat{\Sigma}_{211}/n_2}\right) \chi^2_{1,1-\alpha}\right)^{1/2} , \widehat\mu_{11} - \widehat\mu_{21} + \left( \left({\widehat{\Sigma}_{111}/n_1 + \widehat{\Sigma}_{211}/n_2}\right) \chi^2_{1,1-\alpha}\right)^{1/2} \right]. $$
\end{example}

%\Merleinline{Hier sind die Konfidenzintervall-Beispiele für Marc}
%andere Resampling Techniken:
% parametric Bootstrap?

\section{Multiple tests}\label{sec:Multiple}
\subsection{Multiple testing problem} 
In many applications, not only the global test decisions for (\ref{eq:Hypos}) are of interest but a more in-depth analysis of local hypotheses. Thereby, conclusions on which specific hypotheses cause a rejection of the global hypothesis can be drawn.

Formally, we split up the hypothesis matrix $\mbf{H} = [\mbf{H}_1^{\top},\ldots,\mbf{H}_L^{\top}]^{\top}$ into $L$ matrices with $\mathrm{rank}(\mbf{H}_{\ell}) > 0\l$ and the vector $\mbf{c} = (\mbf{c}_1^{\top},\ldots,\mbf{c}_L^{\top})^{\top}$ into $L$ vectors of lengths corresponding to the number of rows of the matrices $\mbf{H}_1,\ldots,\mbf{H}_L$, respectively. 
This covers but is not restricted to the case that $\mbf{H}_{1},\ldots,\mbf{H}_L$ can be chosen to be the $r$ rows of $\mbf{H}$. 
Then, the multiple testing problem is
\begin{align}\label{eq:MultipleHypos}
    \mathcal{H}_{0,\ell}: \mbf{H}_{\ell}\bs{\mu} = \mbf{c}_{\ell} \quad \text{vs.} \quad \mathcal{H}_{1,\ell}: \mbf{H}_{\ell}\bs{\mu} \neq \mbf{c}_{\ell} \quad\l.
\end{align}

As we will see in the following examples, this formulation covers the most interesting cases for multiple hypotheses about the RMTLs in practice.

\begin{example}[Two-sample case, continued]
If it is also of interest which event type differences in Example~\ref{twosample} cause the significant result, multiple tests need to be performed for $\Tilde{M} > 1$.
In our notation, this can be realized by choosing the matrices $\mbf{H}_{m} \in \R^{1\times kM}\mtilde$ as the rows of the hypothesis matrix $\mbf{H}$ given in Example~\ref{twosample}, i.e., $\mbf{H}_{m}:=[-\mbf{e}_{m}^{\top},\mbf{e}_{m}^{\top}]\mtilde$. Hence, we receive the multiple hypotheses $\mathcal{H}_{0,m}: \mu_{1m} = \mu_{2m}\mtilde$. For $\Tilde{M}=2$, we receive the hypotheses of \citet{wen2023} as special case.\end{example}

\begin{example}[One-way design, continued]\label{ex3}
Now, the choice of the hypothesis matrix leading to the hypothesis of equal RMTLs across the groups in Example~\ref{ex1} becomes important and depends on the question of interest.
E.g., if all RMTLs should be compared to the RMTLs of the first group (many-to-one), the Dunnett-type contrast matrix is the hypothesis matrix to go with. However, the Tukey-type contrast matrix should be used if the RMTLs of all pairs of groups should be compared.

The second choice is how to split up the hypothesis matrix $\mbf{H}$. This, again, depends on the question of interest.
If it is only of interest which groups have different RMTLs but it does not matter for which event types the RMTLs exhibit differences,
it is enough to consider $\mbf{h}_{\ell} \otimes \mbf{I}_M \l,$ as the hypothesis matrices, where $\mbf{h}_{\ell}$ denotes the $\ell$th row of the Dunnett- and Tukey-type contrast matrix, respectively. However, if also the event types that cause a rejection of equal RMTLs should be detected, each row of the (global) hypothesis matrix $\mbf{H}$ corresponds to a hypothesis matrix of the multiple tests, i.e., $\mbf{H}_{\ell} \in \R^{1\times kM}\l$.
\end{example}

\begin{example}[Factorial 2-by-2 design, continued]\label{ex4}
In Example~\ref{ex2}, main effects $A$ and $B$, and an interaction effect could be tested simultaneously with the help of $\mbf{H}_1 = \mbf{H}_A, \mbf{H}_2 = \mbf{H}_B$ and $\mbf{H}_3 = \mbf{H}_{AB}$, respectively.
However, the resulting tests cannot determine which event type(s) caused a significant difference between the groups.
For this, all $M$ rows of each of the three hypotheses matrix must be considered as separate hypothesis matrices which results in $3M$ multiple hypotheses.
\iffalse
In Example~\ref{ex2}, there are also different possibilities how to construct multiple hypotheses. If the question of interest is which effect (main effect of factor A, main effect of factor B and/or interaction effect) is significant, it makes sense to consider $L=3$ hypotheses with $\mbf{H}_1 = \mbf{H}_A, \mbf{H}_2 = \mbf{H}_B$ and $\mbf{H}_3 = \mbf{H}_{AB}$. However, these tests cannot identify the events to which the effect is significant. If this is the question of interest, all $M$ rows of all three hypotheses matrices must be considered as separate hypothesis matrices resulting in $3M$ multiple hypotheses.
\fi
\end{example}

The local test statistics $W_n(\mbf{H}_{\ell},\mbf{c}_{\ell})\l$ can be used to derive (local) test decisions for $\mathcal{H}_{0,\ell}\l$, respectively.
As we already developed global tests in Section~\ref{sec:Inference}, a simple application of the  Bonferroni-correction can solve the multiple testing problem. However, the Bonferroni-correction is known to lead to conservative decisions and low power.
Hence, we aim to incorporate the asymptotic exact dependence structure of the local test statistics as described in \citet{munko2023rmstbased} for constructing powerful multiple tests. The multivariate limit distribution of the local test statistics is given by the following theorem. 

\begin{theorem}\label{AsymptoticMutiple}
    Let $\T\subset \{1,\ldots,L\}$ denote the indices of the true null hypotheses in \eqref{eq:MultipleHypos} and let $\mbf{Z}$ be as in Theorem~\ref{AsymptoticNormality} . Under Assumptions~\ref{Assumption} and \ref{Assumption2}, we have, as $n \to \infty$,
    \begin{align*}
        \left( W_n(\mbf{H}_{\ell},\mbf{c}_{\ell}) \right)_{\ell\in\T} \xrightarrow{d} \left( (\mbf{H}_{\ell}\mbf{Z})^{\top}  (\mbf{H}_{\ell}\bs{\Sigma}\mbf{H}_{\ell}^{\top})^+ (\mbf{H}_{\ell}\mbf{Z}) \right)_{\ell\in\T} =: (W_{\ell})_{\ell\in\T}
    \end{align*} 
    %as $n\to\infty$, where $\mbf{Z} \sim \mathcal{N}_{kM}(\mbf{0}_{kM}, \bs{\Sigma})$.
\end{theorem}

\subsection{Asymptotic multiple tests}\label{ssec:MultipleAsy}
Motivated by Theorem~\ref{AsymptoticMutiple}, asymptotic multiple tests are given by
\begin{align*}
    \varphi_{\ell} = \mathbbm{1}\left\{ W_n(\mbf{H}_{\ell},\mbf{c}_{\ell}) > \chi^2_{{\mathrm{rank}(\mbf{H}_{\ell})}, 1-\beta_n(\alpha)} \right\}\quad\l,
\end{align*}
where $\beta_n(\alpha)$ denotes the local level for each test and can be derived from the multivariate limit distribution in Theorem~\ref{AsymptoticMutiple}. In practice, this local level can be approximated by a Monte Carlo method as $\beta_n(\alpha) = \max\left\{\beta\in\{0,1/B_n,\ldots,(B_n-1)/B_n\} \mid \mathrm{FWER}_n(\beta) \leq \alpha\right\}$ with approximated family-wise error rate %$\beta_n(\alpha) = \mathrm{FWER}_n^{-1}(\alpha)$ with
\begin{align*}
    %\mathrm{FWER}_n(\beta) = \P\left(\exists\; \ell\in\{1,\ldots,L\}:\: (\mbf{H}_{\ell}\widehat{\bs{\Sigma}}^{1/2}\mbf{Y})^{\top}  (\mbf{H}_{\ell}\widehat{\bs{\Sigma}}\mbf{H}_{\ell}^{\top})^+ (\mbf{H}_{\ell}\widehat{\bs{\Sigma}}^{1/2}\mbf{Y}) > \chi^2_{\textcolor{blue}{\mathrm{rank}(\mbf{H}_{\ell}\widehat{\bs{\Sigma}}\mbf{H}_{\ell}^{\top})}, 1-\beta} \mid \widehat{\bs{\Sigma}}\right)
    \mathrm{FWER}_n(\beta) = \frac{1}{B_n}\sum_{b=1}^{B_n} \max\limits_{\ll}\mathbbm{1}\left\{%\exists \ell\in\{1,\ldots,L\}: 
    (\mbf{H}_{\ell}\widehat{\bs{\Sigma}}^{1/2}\mbf{Y}^{(b)})^{\top}  (\mbf{H}_{\ell}\widehat{\bs{\Sigma}}\mbf{H}_{\ell}^{\top})^+ (\mbf{H}_{\ell}\widehat{\bs{\Sigma}}^{1/2}\mbf{Y}^{(b)}) > \chi^2_{{\mathrm{rank}(\mbf{H}_{\ell}\widehat{\bs{\Sigma}}\mbf{H}_{\ell}^{\top})}, 1-\beta}\right\}
\end{align*} for $\beta\in [0,1)$ and $\mbf{Y}^{(1)},\ldots,\mbf{Y}^{(B_n)}\sim\mathcal{N}_{kM}(\mbf{0}_{kM}, \mbf{I}_{kM})$ i.i.d.~and independent of $\widehat{\bs{\Sigma}}$. Here, $(B_n)_{n\in\mathbb{N}}$ is a sequence of natural numbers with $B_n \to \infty$ as $n\to\infty$.
%\Merleinline{Man könnte hier auch $\mathrm{rank}(\mbf{H}_{\ell})$ nehmen bzw oben $\mathrm{rank}(\mbf{H}_{\ell}\widehat{\bs{\Sigma}}\mbf{H}_{\ell}^{\top})$ bei $\varphi$. (asymptotisch ist das äquivalent.) Ich habe mich hier aber für die Version entschieden, sodass es konsistent zu dem globalen Test ist. } 

\begin{proposition}\label{AsymptotikProposition}
    Let $\T\subset \{1,\ldots,L\}$ denote the indices of the true null hypotheses in \eqref{eq:MultipleHypos}. Under Assumptions~\ref{Assumption} and \ref{Assumption2}, we have
    \begin{align*}
     \lim\limits_{n\to\infty}\E\left( \max_{\ell\in\T} \varphi_{\ell} \right)  = \lim\limits_{n\to\infty} \P\left( \exists\; \ell\in\T:\: W_n(\mbf{H}_{\ell},\mbf{c}_{\ell}) > \chi^2_{\mathrm{rank}(\mbf{H}_{\ell}), 1-\beta_n(\alpha)} \right)  \leq \alpha.
    \end{align*} If $\T = \{1,\ldots,L\}$, the inequality is an equality.
\end{proposition}

%\Merleinline{Wie im RMST Papier kann man hier auch noch Paragraphen zu Adjusted p-values, simultaneous confidence regions, non-inf. \& equivalence tests, stepwise extension machen oder kurz drauf verweisen.}\Marcinline{Fände ich gut, wenn man hier bisschen was dazu schreibt, insbesondere sim. conf., da dies so auxh im antrag steht. bei den anderen dingen würde mir auch eine Verweisung reichen, müssen mal sehen, wie groß das Papier so schon wird.}

%\paragraph{Simultaneous confidence regions and intervals}
%\begin{remark}[Confidence regions]
   By Proposition~\ref{AsymptotikProposition}, simultaneous confidence regions for $\mbf{H}_{\ell}\bs{\mu} \l$ with asymptotic global confidence level $1-\alpha$ of the following form are immediate:
   %
   %can directly be derived as 
   %$$
   $$ \bigtimes_{\ell=1}^L \left\{ \bs\xi \mid   W_n(\mbf{H}_{\ell},\bs\xi) \leq \chi^2_{\mathrm{rank}(\mbf{H}_{\ell}), 1-\beta_n(\alpha)} \right\} \subset \R^{r}.$$ %$ \l $.
   %$$ 
   For row vectors $\mbf{H}_{\ell} \in \R^{1\times k}$, the confidence region simplifies to a confidence interval, that is $$\left[ \mbf{H}_{\ell}\widehat{\bs{\mu}} - \left(\frac{{\mbf{H}_{\ell}\widehat{\bs\Sigma}\mbf{H}_{\ell}^{\prime}}}{{n}}{\chi^2_{\mathrm{rank}(\mbf{H}_{\ell}), 1-\beta_n(\alpha)}}\right)^{1/2}, \mbf{H}_{\ell}\widehat{\bs{\mu}} + \left(\frac{{\mbf{H}_{\ell}\widehat{\bs\Sigma}\mbf{H}_{\ell}^{\prime}}}{{n}}{\chi^2_{\mathrm{rank}(\mbf{H}_{\ell}), 1-\beta_n(\alpha)}}\right)^{1/2}\right].$$

Moreover, we can derive adjusted p-values, non-inferiority and equivalence tests and a more powerful stepwise extension of the proposed multiple testing procedure analogously as in Section~3 of \citet{munko2023rmstbased}.
%\end{remark}

%\subsection{Resampling multiple tests} \label{ssec:Resampling}
%???
%\Merleinline{Wir wollten jetzt keine weiteren Methoden mehr außer Asymptotic/Asymptotic Bonferroni/Permutation Bonferroni aufnehmen, oder? (zB Bootstrap oder p-Wert Permutation)}

\section{Simulation study}\label{sec:Simu}
%\Merleinline{code implementing the method should be made available, either as part of the Supplementary Material or via a public repository, e.g., on GitHub; details of how to reproduce those results should also be available}
%code implementing the method should be made available, either as partof the Supplementary Material or via a public repository, e.g., on GitHub. If simulation resultsare central to evaluation of such methods, then details of how to reproduce those results shouldalso be available.
\subsection{Simulation setup}\label{ssec:SimuSetup}
For the simulation study, we used the computing environment R, version~4.2.1 \citep{R}. 
The simulation setup is based on \citet{munko2023rmstbased} and adapted for competing risks data. 
We considered $k=4$ groups with equal survival time distributions for the first three groups while the distribution of the fourth group may differ, using the same survival and censoring distributions as in \citet{munko2023rmstbased} with the same censoring rates stated there. More details on the settings can be found in Section~B.1 in the Supplementary Material.
An illustration of the survival curves of the event times can be found in \citet{RMST} and of the censoring times in \citet{munko2023rmstbased}.

Beyond these {continuous} settings, we also added corresponding {discrete} settings since the proposed methods also work theoretically under the existence of ties. Therefore, we generated the event times as in the continuous case but round them up to obtain integers. Of course, the rounding typically results in altered values of the restricted mean survival times $\int_{[0,\tau]} S_i(t)\mathrm{d}t\i$ and RMTLs. However, it is still possible to obtain a specific restricted mean survival time difference $\delta$ as in \citet{munko2023rmstbased} by adjusting the parameters $\lambda_{\delta,1},\ldots,\lambda_{\delta,9}$ adequately. 

%The censoring times are generated as in \citet{munko2023rmstbased}.
As in the data example in Section~\ref{sec:Data} below, we are considering $M=3$ event types. The causes $D_{ij}\in \{1,2,3\}$ were drawn independently of the survival and censoring times with probabilities $p_1 = 33\%$, $p_2 = 25\%$, and $p_3 = 42\%$, respectively, across all $\jjii$. This results in a direct connection between the survival function and the cumulative incidence functions, that is, $F_{im} = p_m (1 - S_i)$ for all $\iimm$. Hence, the RMTLs $\mu_{1m},\mu_{2m},\mu_{3m},\mu_{4m}$ of the event type $m$ coincide whenever the restricted mean survival times $\int_{[0,\tau]} S_i(t)\mathrm{d}t\i$ coincide. Furthermore, a restricted mean survival time difference of $\delta = \int_{[0,\tau]} S_1(t)\mathrm{d}t - \int_{[0,\tau]} S_4(t)\mathrm{d}t$ results in an RMTL difference of $\mu_{4m} - \mu_{1m} = p_m \delta\m$.

Motivated by the data example in Section~\ref{sec:Data}, the hypothesis matrix in Example~\ref{ex4} is considered for testing on the two main effects and an interaction effect simultaneously in a 2-by-2 design  ({2x2}). 
This results in the local null hypotheses
\begin{align}\begin{split}\label{eq:SimuHypos}
    &\mathcal{H}_{0,m}^A: \mu_{1m} + \mu_{2m} = \mu_{3m} + \mu_{4m},\quad\mathcal{H}_{0,m}^B: \mu_{1m} + \mu_{3m} = \mu_{2m} + \mu_{4m},  \\\text{and } \quad &\mathcal{H}_{0,m}^{AB}: \mu_{1m} + \mu_{4m} = \mu_{2m} + \mu_{3m} \quad \m.
\end{split}
\end{align}
Moreover, {Dunnett}- and {Tukey}-type contrast matrices are used for many-to-one and all-pairs comparisons of the RMTLs, respectively, as in Example~\ref{ex3}. The block matrices $\mbf{H}_{\ell}\l$, for the local hypotheses are always chosen to be the rows of the global hypothesis matrices. The global hypothesis matrices all lead to the same global null hypothesis, that is, all RMTLs are equal across the groups for each respective event type. However, the local hypotheses differ for the different matrices.
 
The restricted mean survival time difference is chosen as $\delta = 0$ for simulating under the null and as $\delta = 1.5$ for simulating under the alternative hypothesis.
Since the survival settings \textit{exp early}, \textit{exp late} and \textit{exp prop} defined in the Supplementary Material result in the same survival functions under the null hypothesis, the results for these scenarios are only included once in the figures and tables, respectively. The same holds for the settings \textit{Weib late} and \textit{Weib prop}.

%by the following three scenarios:
%\begin{itemize}
%    \item Equally Weibull distributed censoring times (\textit{equal}): $C_{11}, C_{21}, C_{31}, C_{41} \sim Weib(3,10)$,
%    \item unequally Weibull distributed censoring times with high censoring rates (\textit{unequal, high}): $C_{11} \sim Weib(0.5,15), C_{21}\sim Weib(0.5,10), C_{31}\sim Weib(1,8)$ and $C_{41} \sim Weib(1,10)$,
%    \item unequally Weibull distributed censoring times with low censoring rates (\textit{unequal, low}): $C_{11} \sim Weib(1,20), C_{21}\sim Weib(3,10), C_{31}\sim Weib(1,15)$ and $C_{41} \sim Weib(3,20)$.
%\end{itemize}

%The resulting censoring rates of the different groups are presented in Table~\ref{tab:cens} in the Supplementary Material. The censoring rates ranged from 20\% up to 60\% in groups 1--3 and from 1\% up to 57\% in group 4. 

{Balanced} and {unbalanced} designs with sample sizes $\mbf{n}=(n_1,n_2,n_3,n_4)=K \cdot (60,60,60,60)$ and $\mbf{n}=K \cdot (128,44,52,16)$ are considered, where the factor $K\in\{1,5,25\}$ resulted in {small, medium}, and {large} samples, respectively. 

In total, $N_{sim} = 5000$ simulation runs with $B = 1999$ resampling iterations were conducted. We set the level of significance to $\alpha = 5\%$ and the terminal time point to $\tau = 10.$

We included the following methods in our simulation study:
multiple asymptotic Wald-type tests as in Section~\ref{ssec:MultipleAsy} (\textit{asymptotic}),
global asymptotic Wald-type tests as in Section~\ref{ssec:Wald} adjusted with the Bonferroni-correction (\textit{asymptotic\_bonf}), and
global studentized permutation tests as in Section~\ref{ssec:Perm} 
adjusted with the Bonferroni-correction (\textit{permutation\_bonf}).
In our first simulations, we also compared a pooled bootstrap, wild bootstrap, and groupwise bootstrap method similar to that in \citet{munko2023rmstbased} as well as a random p-value permutation approach similar as the prepivoting method in \citet{CHUNG201676}. However, the results of these methods were not as convincing in terms of type I error control and/or power. Moreover, the runtime of the random p-value permutation approach was quite high since for each permutation sample, the calculations need to be done for several groupwise bootstrap samples. Therefore, in this paper we only focus on the three methods mentioned above.

\subsection{Simulation results}\label{ssec:SimuResults}
\begin{figure}[bt]
    \centering
    \includegraphics[width=\textwidth]{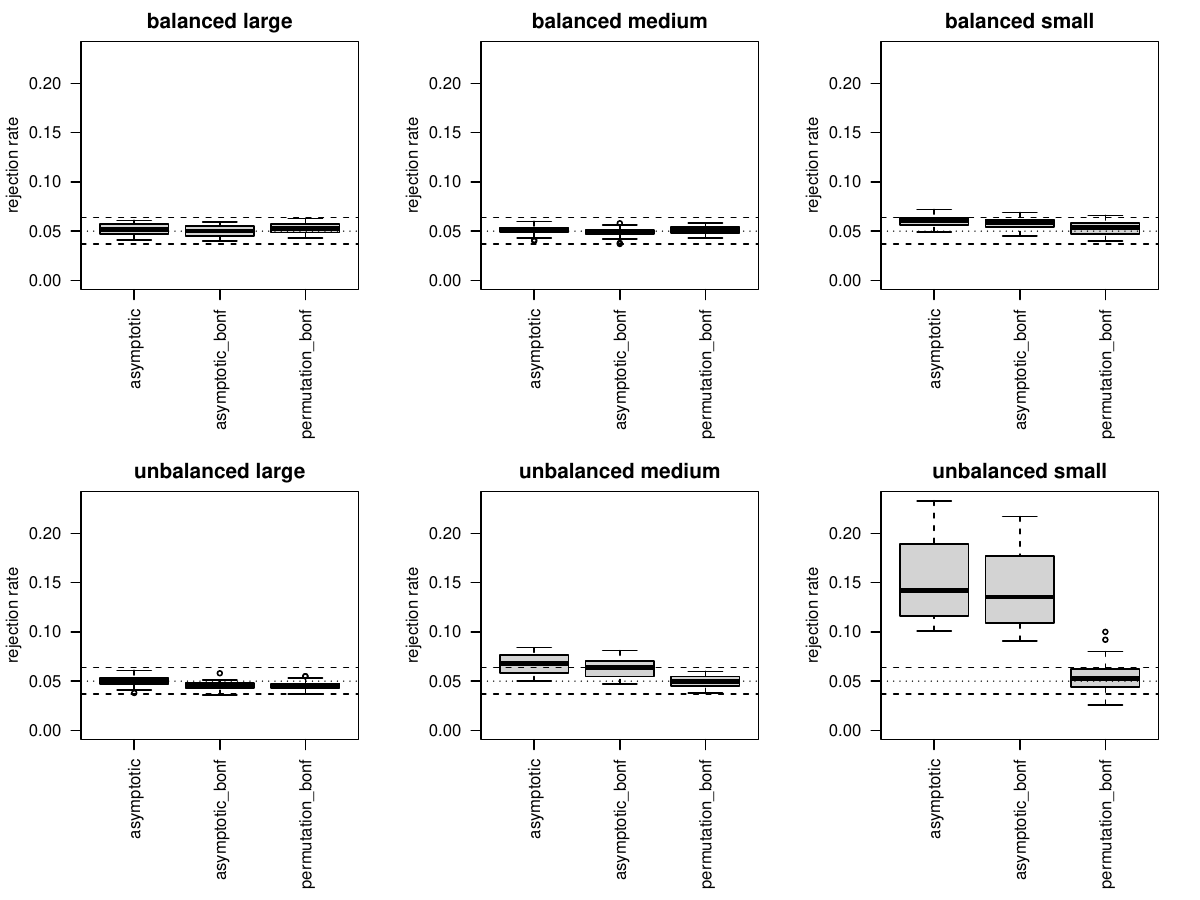}
    \caption{Empirical family wise error rates for the 2-by-2 design across all scenarios under the global null hypothesis. The dotted line represents the desired global level of 0.05 and the dashed lines represent the borders of the binomial interval [0.037, 0.064].}
    \label{fig:22}
\end{figure}
\begin{figure}[tbh]
    \centering
    \includegraphics[width=\textwidth]{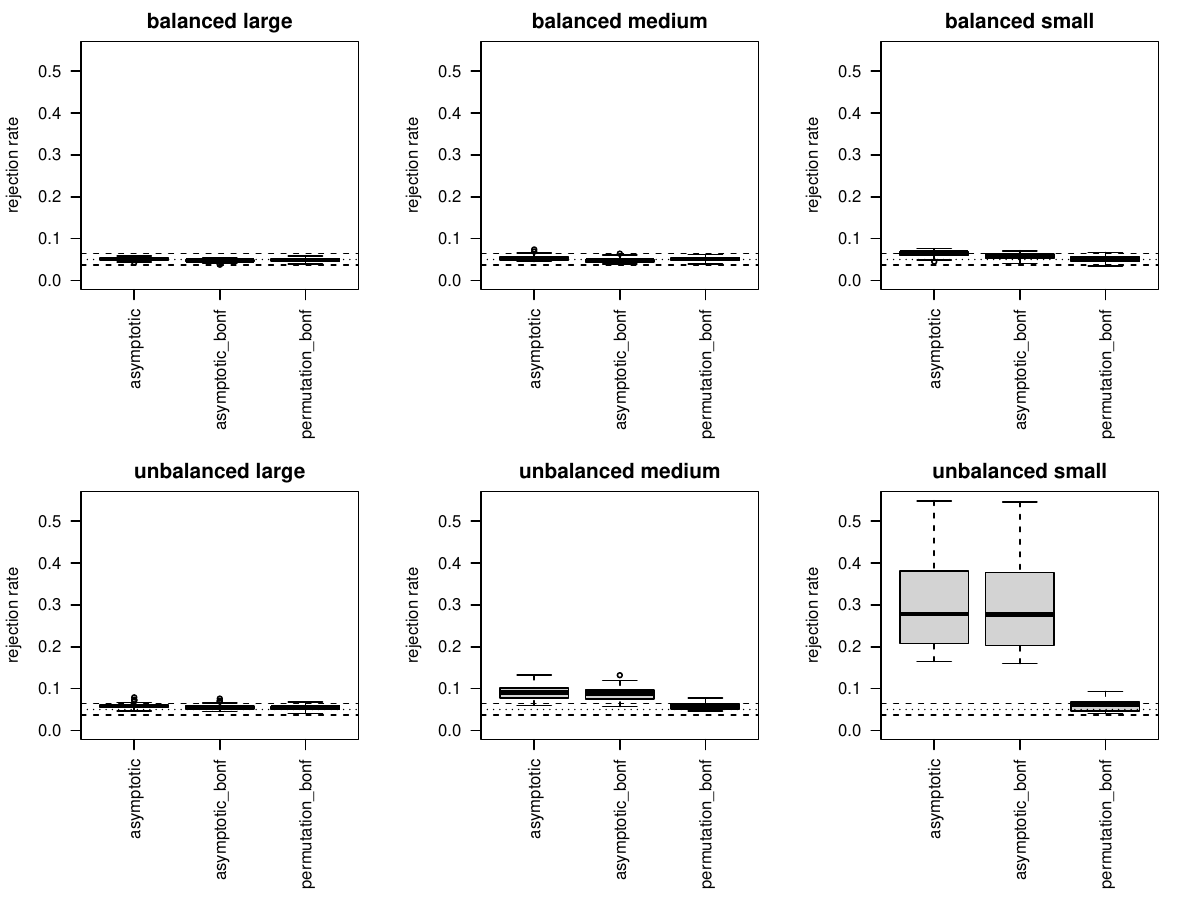}
    \caption{Empirical family wise error rates for the Dunnett-type contrast hypotheses across all scenarios under the global null hypothesis. The dotted line represents the desired global level of 0.05 and the dashed lines represent the borders of the binomial interval [0.037, 0.064].}
    \label{fig:dunn}
\end{figure}
\begin{figure}[tbh]
    \centering
    \includegraphics[width=\textwidth]{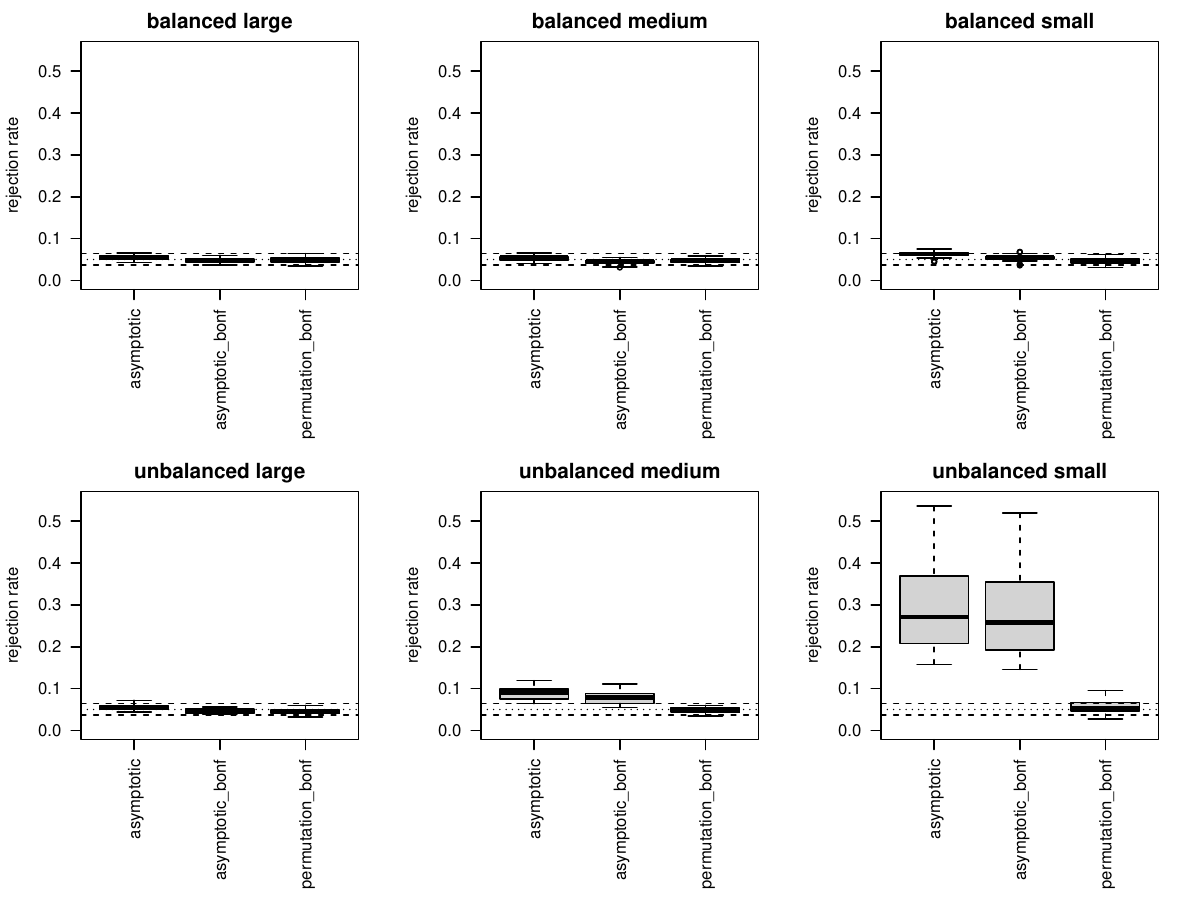}
    \caption{Empirical family wise error rates for the Tukey-type contrast hypotheses across all scenarios under the global null hypothesis. The dotted line represents the desired global level of 0.05 and the dashed lines represent the borders of the binomial interval [0.037, 0.064].}
    \label{fig:tuk}
\end{figure}
%\Dennisinline{My suggestion: The boxplots in Figure ... summarize the rejection levels under all null hypotheses (perhaps split according to A, B, AB) and the boxplots in ... alternative. Detailed results can be found in the Supplementary Material.}
The boxplots in Figures~\ref{fig:22}, \ref{fig:dunn} and \ref{fig:tuk} summarize the rejection levels under all null hypotheses. 
It is observable that the multiple asymptotic Wald-type tests of Section~\ref{ssec:MultipleAsy} as well as the Bonferroni-corrected global asymptotic Wald-type tests of Section~\ref{ssec:Wald} can not control the type I error in unbalanced designs with smaller sample sizes as they perform to liberal in all scenarios. For Dunnett- and Tukey-type contrast matrices, the empirical family wise error rates are exceeding even 50\% in some scenarios. 
The Bonferroni-corrected permutation tests also have a slight liberal behaviour in some of these scenarios but not nearly as dramatic.
The highest empirical family wise error rates for the Bonferroni-corrected permutation tests are only up to 10\% and reached under the non-exchangeable survival distribution settings \textit{Weib scale} and \textit{Weib shape} for unbalanced small sample sizes.
A possible reason for the liberal behaviour of the tests for unbalanced small sample sizes could be that it is more likely to observe no event of a specific type in at least one of the samples. 
In this case, the permutation approach may still use the information of the events of the same type that occur in other groups through the randomization across groups. 
However, the asymptotic approach can not benefit from observations in other samples and, thus, probably underestimates the variance systematically.
Even for unbalanced designs and medium sample sizes with 80--640 observations in the groups, the liberality of the asymptotic approaches is still notable.
In balanced designs, this issue is only slightly present, even for small sample sizes with 60 observations per group.
For large sample sizes, all methods seem to perform quite well under the global null hypothesis in terms of family wise error rate control, which underlines the asymptotic validity of the proposed tests. 

Figures~S1, S2 and S3 in the Supplementary Material visualize the empirical rejection rates of the global null hypothesis under the alternative hypothesis, i.e., the empirical (global) powers. 
Here, we observe that all methods have a comparable power in all scenarios with balanced designs or large sample sizes. 
Moreover, the power naturally increases for larger sample sizes.
In unbalanced designs with small to medium sample sizes, the multiple and Bonferroni-corrected asymptotic tests have usually a higher power than the Bonferroni-corrected permutation tests. 
Here, the multiple asymptotic tests that take the multivariate distribution of the test statistics into account (Section~\ref{ssec:MultipleAsy}) are slightly more powerful than the asmptotic tests with Bonferroni-correction.
However, both asymptotic testing procedures performed too liberal in unbalanced designs with small to medium sample sizes and, thus, we do not recommend their application in these scenarios.
Furthermore, it is observable that only under a few scenarios, all methods can detect the alternative in unbalanced designs with small sample sizes as the most rejection rates are similar as under the null hypothesis.
A possible explanation for this may be the small sample size of 16 in group four, which was sampled with different RMTLs under the alternative.

% ggf noch was über die lokale Power schreiben?

\section{Data example about blood and marrow transplantation}\label{sec:Data}
\begin{figure}[tb]
    \centering
    \includegraphics[width=\linewidth]{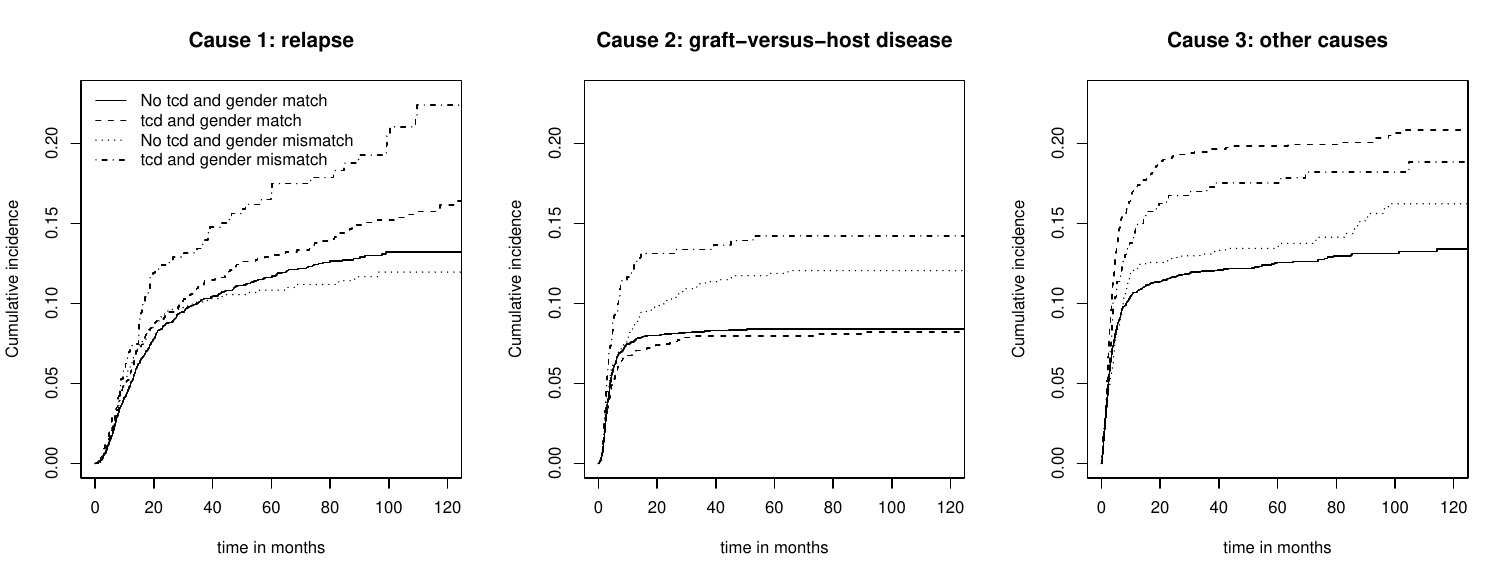}
    \caption{Aalen-Johansen estimators of the cumulative incidence functions for the data example for the different causes and groups}
    \label{fig:RPlot}
\end{figure}

For illustrating the proposed methods, we analyze the data set \texttt{ebmt2} in the R package \texttt{mstate} \citep{mstate2, mstate1, mstate3} from the European Society for Blood and Marrow Transplantation. The data consists of $8966$  leukemia patients who underwent bone marrow transplantation. 
%transplanted at the European Society for Blood and Marrow Transplantation.
An initial statistical analysis \citep{fiocco05} focused on reduced rank models for proportional cause-specific hazard models.
First of all, the data set contains \texttt{time}, which is the time in months from transplantation to death or the last follow-up, and \texttt{status}, which indicates the survival status;
for simplicity, we aggregate the status levels into the following $M=3$ causes of the death, next to censoring: relapse (1), graft-versus-host disease (2), and all other causes (3).
We included the following factor variables in our analysis within a factorial 2-by-2 design; see Example~\ref{ex2}:
($A$) \texttt{match}: yes/no, according to whether the donor's and the recipient's genders matched;
 ($B$) \texttt{tcd}: yes/no, depending on whether a T-cell depletion took place. 
Because it was unknown for $2856$ patients whether a T-cell depletion took place or not, we assumed the missingness to have been completely at random, and the incomplete records were removed from our further analysis. Hence, $n = 6110$ patients remained. 
Thereof, 1296 (3313) patients with donor-recipient gender match did (not) receive a T-cell depletion.
For those without a match, the numbers were 424 and 1077, respectively.
An illustration of the resulting Aalen-Johansen estimators of the cumulative incidence functions can be found in Figure~\ref{fig:RPlot}

\iffalse
In our analysis, we include the variables  the factors \texttt{match}, which contains the donor-recipient gender match (yes/no), and \texttt{tcd}, which indicates whether a T-cell depletion took place (yes/no). This is an example for the factorial 2-by-2 design, see Example~\ref{ex2}.
For simplicity, we summarize the status into the following $M=3$ causes for the death: censored (0), relapse (1), graft versus host disease (2) and all other causes (3).
Moreover, it is unknown whether a T-cell depletion took place for $2856$ patients and, thus, they are omitted in our data analysis. Hence, $n = 6110$ patients remain. 
Thereof, 3313 patients with donor-recipient gender match did not receive a T-cell depletion and 1296 did receive a T-cell depletion. Furthermore, 1077 patients with donor-recipient gender mismatch did not receive a T-cell depletion and 424 did receive a T-cell depletion.
\fi

Here, the question of interest is whether the donor-recipient gender match and/or the T-cell depletion have a main or interaction effect on any event type-specific RMTL. If there is a significant effect, we are also interested in which specific main/interaction effects are present and which of the event types are affected. Hence, the nine hypothesis matrices for the multiple testing problem are, for the event type $m\in \{1,2,3\}$ and the effect ($\ell=1$ for main $A$, $\ell=2$ for main $B$, and $\ell=3$ for interaction),
$ \mbf{h}_\ell \otimes \mbf{e}_m^{\top}\mm$, where $\mbf{h}_1 = [1, 1, -1, -1]$, $\mbf{h}_2 = [1, -1, 1, -1]$, and $\mbf{h}_3 = [1, -1, -1, 1]$.
The vector of no RMTL differences, i.e., $\mbf{c} = \mbf{0}_9$, is tested under the global null hypothesis.
%
%
\iffalse
$[1, -1, 1, -1] \otimes \mbf{e}_m\m$ for the main effect of the T-cell depletion on event of type $m$, $[1, 1, -1, -1] \otimes \mbf{e}_m\m$ for the main effect of factor match on event of type $m$, and $[1, -1, -1, 1] \otimes \mbf{e}_m\m$ for the interaction effect between the two factors on event of type $m$, where $\mbf{e}_m \in \R^{1\times 3}$ denotes the transposed $m$th unit vector, and $\mbf{c} = \mbf{0}_9$.
\fi
%
%
This results in the hypotheses \eqref{eq:SimuHypos} as in the simulation study.
We used the terminal time point ten years, i.e., $\tau = 120$ months.

The resulting method-specific and adjusted p-values based on $B = 4999$ resampling iterations are presented in Table~\ref{tab:pValues}. 
Comparing the adjusted p-values with the global level of significance allows for testing all nine local hypotheses simultaneously.
Due to the large sample sizes, all methods should yield reliable results in terms of type I error control regarding the simulation results of Section~\ref{sec:Simu}.
For $\alpha = 5\%$, all methods indicate that $\mathcal{H}_{0,1}^B, \mathcal{H}_{0,3}^B,$ and $\mathcal{H}_{0,2}^A$ can be rejected simultaneously. Thus, there is a significant main effect of the T-cell depletion on the RMTL for relapse and other causes and a significant main effect of the donor-recipient gender match on the RMTL for the graft versus host disease. %Thus, graft versus host disease can be affected by taking the gender match into account and relapse and other causes can be affected by a T-cell depletion.
\begin{table}[tb]
    \centering
    \caption{Adjusted p-values in \% for the data example}
    {\begin{tabular}{crrrrrrrrr}
    \phantom{huhu} Method \phantom{huhu} & $\mathcal{H}_{0,1}^A$ & $\mathcal{H}_{0,2}^A$ & $\mathcal{H}_{0,3}^A$ & $\mathcal{H}_{0,1}^B$ & $\mathcal{H}_{0,2}^B$ & $\mathcal{H}_{0,3}^B$ & $\mathcal{H}_{0,1}^{AB}$ & $\mathcal{H}_{0,2}^{AB}$ & $\mathcal{H}_{0,3}^{AB}$ \\[0.15cm]
    asymptotic & 66.3 & $<$ 0.1 & 100.0 & 0.8 & 95.0 & $<$ 0.1 & 29.5 & 77.3 & 61.4 \\
    asymptotic\_bonf & 100.0 & $<$ 0.1 & 100.0 & 0.8 & 100.0 & $<$ 0.1 & 39.6 & 100.0 & 100.0 \\
    %\ldots & & & & & & & & & \\
    permutation\_bonf & 100.0 & 0.2 & 100.0 & 0.7 & 100.0 & $<$ 0.1 & 39.2 & 100.0 & 100.0 \\
\end{tabular}}
    \label{tab:pValues}
\end{table}

As a word of caution, no clinical conclusions should be drawn from this analysis since the data were simplified. % and do not reflect any real life situation.
For example, aspects that would be relevant for causal interpretations were not taken into consideration.
Instead, the present real data analysis was meant to illustrate the potential of our new statistical techniques.

\section{Discussion}\label{sec:Discussion}
%In the present paper, we firstly developed Wald-type tests for general contrast hypotheses of the RMTLs in general factorial competing-risks designs. Thereby, we did not assume continuity of the cumulative incidence functions and, thus, we allow for ties in the data. Secondly, we constructed asymptotically valid multiple tests to infer several RMTL contrast hypotheses simultaneously. For the global and multiple testing problem, we introduced asymptotic as well as resampling based tests. However, the simulation results indicated that the asymptotic tests can not control the type I error well for small samples since they performed way too liberal in different scenarios. ??? %(mehr zu den Simu Ergebnissen)
%Finally, we illustrated the methods by an application to a data set about blood and marrow transplantation.
%%%Biometrika: This is the concluding part of the paper. It is only needed if it contains new material such as open questions and future research areas. It should not repeat the summary or reiterate the contents of the paper.
%\Marcinline{Hier sollten wir Werbung für Dennis Resamplinverfahren im Einstichprobenfall machen, als Anwendung, wenn man Zwei Risiken miteinander vergleichen will. Merle, du sagtest doch, das diese Verfahren funktionieren, wenn wir (Block-)Kontrastmatirizen für ede Gruppe haben, das hätten wir ja im Fall beim Vergleich zweier (oder mehrerer) Risiken. Die konkrete Ausarbeitung würde hier bestimmt den Rahmen sprengen, ist vielleicht was für eine Masterarbeit?!}

In this paper, we have constructed tests that cover RMTL comparisons for the same event types across different groups.
However, one may also be interested in comparing two or more RMTLs within each group.
A potentially suitable resampling procedure for this problem could be motivated from the randomization approach in \citet{dobler2023nonparametric}. As an adaption of this, the event indicators $\delta_{ij}$ are re-drawn as $\widetilde{\delta}_{ij}$ from $\{1,\ldots,M\}$ with equal probability $1/M$ if $\delta_{ij} \neq 0$, which leads to the randomized data $(X_{ij},\widetilde{\delta}_{ij})\ji.$ 
By the theory of \citet{dobler2023}, the asymptotic validity of this randomization approach can be shown 
if the hypothesis matrix %satisfies $\mbf{H}(\mbf{e}_i^{\top} \otimes \mbf{1}_M) = \mbf{0}_r\i$. Here, $\mbf{e}_i\in\R^{1\times k}$ denotes the transposed $i$th unit vector for all $\ii$, i.e., $\mbf{H}$ 
can be partitioned into a block matrices with one row block and $k$ column blocks for $k$ contrast matrices.
 Moreover, finitely exact tests could be achieved by this randomization approach under the event type exchangeability of the data, i.e., $F_{i1} \equiv \ldots \equiv F_{iM}\i$.
 More analysis on this matter is a point of future research.

 \iffalse
 For example, two (or more) RMTLs of the cumulative incidence functions within the groups could be compared. However, in this paper, the considered hypothesis matrices do not satisfy the given condition as mainly RMTLs between different groups should be compared.
 Moreover, finitely exact tests could be achieved by this randomization approach under exchangeability of the competing risks within each group, i.e. $F_{i1} = \ldots = F_{iM}\i$.
\fi

%\Marcinline{Dann wäre die Frage, was es schon für andere Effektmaße gibt. Falls es ncihts dazu gibt (und wenn nur fürs multiple Testen), dann könnte man das hier auch als Ausblick angeben.}
%\Merleinline{Zu Quantilen gibts schon was für competing risks, sonst habe ich gerade nichts gefunden, auch nichts multiples. }
As an outlook, multiple tests based on other effect estimands than the RMTL could be developed in future research. This includes cumulative incidence quantiles \citep{beyersmann2008, Lee2011, peng2007, SANKARAN2010886}, extensions of the probabilistic index (or relative treatment effect) \citep{doblerpauly2018, doblerpauly2020} in the presence of competing risks, and the area between curves statistic \citep{liu2020, lyu2020}.

\section*{Acknowledgements}
The authors would like to thank Hein Putter and Liesbeth de Wreede for pointing us to the \texttt{ebmt2} data set.
Merle Munko and Marc Ditzhaus gratefully acknowledge support from the \textit{Deutsche Forschungsgemeinschaft}. 
% (grant no. DI 2906/1--2 and GRK 2297 MathCoRe). 
%% Information, such as contract numbers, of no interest to readers, should be excluded.
% Hein und Liesbeth danken.

\section*{Supplementary Material}
The Supplementary Material includes the proofs of the stated theorems and further details on the simulation study.

\bibliographystyle{abbrv}



\section*{Supplementary material}

# Supplementary material to
# Multiple tests for the restricted mean time lost with competing risks data

Merle Munko[1,*], Dennis Dobler[2,3], and Marc Ditzhaus[1]

[1] Otto-von-Guericke University Magdeburg; Magdeburg (Germany)
[2] TU Dortmund University; Dortmund (Germany)
[3] Research Center Trustworthy Data Science and Security, University Alliance Ruhr; Dortmund (Germany)

In this supplement, the theoretical proofs of all stated theorems can be found in Section A. In Section B, details on the simulations are given. This includes an explanation of the simulation settings in Section B.1 as well as additional figures and detailed tables of the simulation results in Section B.2 and B.3, respectively.

# Contents

# List of Figures

---

*Corresponding author. Email address: `merle.munko@ovgu.de`

# List of Tables







# A Proofs

## Proof of Theorem 1

Firstly, let $i \in \{1, ..., k\}$ be arbitrary but fixed. By Theorem 4.1 in [3], we have

$$n_i^{1/2} \left( \hat{A}_{i1} - A_{i1}, ..., \hat{A}_{iM} - A_{iM} \right) \xrightarrow{d} (U_{i1}, ..., U_{iM}) \tag{S1}$$

as $n \to \infty$ on $(D[0,\tau])^M$ equipped with the sup-norm, where $U_{i1}, ..., U_{iM}$ are centered Gaussian-martingales with

$$\mathbb{C}ov(U_{im_1}(t), U_{im_1}(s)) = \int_{[0,\min\{t,s\}]} \frac{1 - \Delta A_{im_1}}{y_i} \, dA_{im_1} =: \sigma_{im_1 m_1}(\min\{t, s\}),$$

$$\mathbb{C}ov(U_{im_1}(t), U_{im_2}(s)) = -\int_{[0,\min\{t,s\}]} \frac{\Delta A_{im_1}}{y_i} \, dA_{im_2} =: \sigma_{im_1 m_2}(\min\{t, s\})$$

and $y_i(t) := P(X_{i1} \geq t)$ for all $t, s \in [0, \tau]; m_1, m_2 \in \{1, \ldots, M\}; m_1 \neq m_2$. By [4], the limit $(U_{i1}, ..., U_{iM})$ is separable.

Let $BV[0, \tau] \subset D[0, \tau]$ denote the set of all càdlàg functions on $[0, \tau]$ of total variation bounded by a fixed and sufficiently large constant. Then, we consider the functional

$$\Phi : (BV[0, \tau])^M \to \mathbb{R}^M,$$

$$(\Lambda_1, ..., \Lambda_M) \mapsto \left( \int_0^\tau \int_{[0,t]} \prod_{0 \leq x < u} \left\{ 1 - d\left( \sum_{\tilde{m}=1}^M \Lambda_{\tilde{m}} \right)(x) \right\} \, d\Lambda_m(u) \, dt \right)_{m \in \{1,...,M\}}.$$

Here and throughout, we define the jump $\Delta \Lambda(0)$ of a function $\Lambda \in BV[0, \tau]$ at $0$ as $\Lambda(0)$. Note that $\Phi(A_{i1}, ..., A_{iM}) = (\mu_{i1}, ..., \mu_{iM})$ and $\Phi(\hat{A}_{i1}, ..., \hat{A}_{iM}) = (\hat{\mu}_{i1}, ..., \hat{\mu}_{iM})$ holds. We aim to apply the delta-method. Therefore, we firstly show that $\Phi$ is Hadamard-differentiable at $(A_{i1}, ..., A_{iM})$ with Hadamard-derivative $\Phi'_{(A_{i1},...,A_{iM})}$ at $(\alpha_1, ..., \alpha_M) \in (D[0, \tau])^M$ given by

$$\left( \int_0^\tau \left( \int_{[0,t]} \prod_{0 \leq x < u} (1 - dA_i(x)) \, d\alpha_m(u) - \int_{[0,t]} \prod_{0 \leq x < u} (1 - dA_i(x)) \int_{[0,u)} \frac{d\sum_{\tilde{m}=1}^M \alpha_{\tilde{m}}}{1 - \Delta A_i} \, dA_{im}(u) \right) dt \right)_{m \in \{1,...,M\}}.$$

Here, the integrals with respect to $\alpha_m$ are defined via integration by parts because $\alpha_m$ need not have finite variation. The same holds for other integrals of this kind below. In order to prove the Hadamard-differentiability, we consider the Wilcoxon functional and the product integral functional, that are

$$\Psi : \tilde{D}[0, \tau] \times BV[0, \tau] \to D[0, \tau], \quad (A, B) \mapsto \int_{[0,\cdot]} A \, dB$$

and $\quad \phi : D[0, \tau] \to \tilde{D}[0, \tau], \quad A \mapsto \prod_{0 \leq x < \cdot} (1 + dA(x)),$

where $\tilde{D}[0, \tau]$ denotes the space of all left-continuous functions on $[0, \tau]$ with existing limits from the right. Then, we can write

$$\Phi(\Lambda_1, ..., \Lambda_M) = \left( \int_0^\tau \Psi\left( \phi\left( -\sum_{\tilde{m}=1}^M \Lambda_{\tilde{m}} \right), \Lambda_m \right)(t) \, dt \right)_{m \in \{1,...,M\}}.$$

for $\Lambda_1, ..., \Lambda_M \in BV[0, \tau]$. Applying the chain rule (Lemma 3.10.3 in [9]) yields

$$\Phi'_{(A_{i1},...,A_{iM})}(\alpha_1, ..., \alpha_M) = \left( \int_0^\tau \Psi'_{(\phi(-A_i), A_{im})} \left( \phi'_{-A_i} \left( -\sum_{\tilde{m}=1}^M \alpha_{\tilde{m}} \right), \alpha_m \right)(t) \, \mathrm{d}t \right)_{m \in \{1,...,M\}}$$

for $\alpha_1, ..., \alpha_M \in D[0, \tau]$. Note here that the final integral functional is linear, so it equals its Hadamard-derivative. Analogously to Lemma 3.10.18 and Lemma 3.10.32 in [9], one can show that

$$\Psi'_{(\phi(-A_i), A_{im})}(\alpha, \beta) = \int_{[0,.]} S_{i-} \, \mathrm{d}\beta + \int_{[0,.]} \alpha \, \mathrm{d}A_{im}, m \in \{1, \ldots, M\}$$

$$\text{and} \quad \phi'_{-A_i}(\beta) = S_{i-}(.) \int_{[0,.)} \frac{1}{1 - \Delta A_i} \, \mathrm{d}\beta$$

holds for all $\alpha \in \tilde{D}[0, \tau], \beta \in D[0, \tau]$. Thus, it follows that

$$\Phi'_{(A_{i1},...,A_{iM})}(\alpha_1, ..., \alpha_M)$$

$$= \left( \int_0^\tau \left( \int_{[0,t]} S_{i-} \, \mathrm{d}\alpha_m - \int_{[0,t]} S_{i-}(s) \int_{[0,s)} \frac{\mathrm{d} \sum_{\tilde{m}=1}^M \alpha_{\tilde{m}}}{1 - \Delta A_i} \, \mathrm{d}A_{im}(s) \right) \mathrm{d}t \right)_{m \in \{1,...,M\}}$$

$$= \left( \int_0^\tau \left( \int_{[0,t]} S_{i-} \, \mathrm{d}\alpha_m - \int_{[0,t)} \frac{\int_{(u,t]} S_{i-} \, \mathrm{d}A_{im}}{1 - \Delta A_i(u)} \, \mathrm{d} \sum_{\tilde{m}=1}^M \alpha_{\tilde{m}}(u) \right) \mathrm{d}t \right)_{m \in \{1,...,M\}}$$

$$= \left( \int_0^\tau \left( \int_{[0,t]} \frac{S_i}{1 - \Delta A_i} \, \mathrm{d}\alpha_m - \int_{[0,t]} \frac{F_{im}(t) - F_{im}(u)}{1 - \Delta A_i(u)} \, \mathrm{d} \sum_{\tilde{m}=1}^M \alpha_{\tilde{m}}(u) \right) \mathrm{d}t \right)_{m \in \{1,...,M\}}$$

$$= \left( \int_0^\tau \left( \int_{[0,t]} \frac{S_i(u) - F_{im}(t) + F_{im}(u)}{1 - \Delta A_i(u)} \, \mathrm{d}\alpha_m(u) - \int_{[0,t]} \frac{F_{im}(t) - F_{im}(u)}{1 - \Delta A_i(u)} \, \mathrm{d} \sum_{\tilde{m} \neq m} \alpha_{\tilde{m}}(u) \right) \mathrm{d}t \right)_{m \in \{1,...,M\}}$$

$$= \left( \int_0^\tau \left( \int_{[0,t]} \frac{1 - \sum_{\tilde{m} \neq m} F_{i\tilde{m}}(u) - F_{im}(t)}{1 - \Delta A_i(u)} \, \mathrm{d}\alpha_m(u) - \int_{[0,t]} \frac{F_{im}(t) - F_{im}(u)}{1 - \Delta A_i(u)} \, \mathrm{d} \sum_{\tilde{m} \neq m} \alpha_{\tilde{m}}(u) \right) \mathrm{d}t \right)_{m \in \{1,...,M\}}$$

$$= \left( \int_{[0,\tau)} \frac{(\tau - u)\left(1 - \sum_{\tilde{m} \neq m} F_{i\tilde{m}}(u)\right) - \int_u^\tau F_{im}(t) \, \mathrm{d}t}{1 - \Delta A_i(u)} \, \mathrm{d}\alpha_m(u) \right.$$

$$\left. + \int_{[0,\tau)} \frac{(\tau - u)F_{im}(u) - \int_u^\tau F_{im}(t) \, \mathrm{d}t}{1 - \Delta A_i(u)} \, \mathrm{d} \sum_{\tilde{m} \neq m} \alpha_{\tilde{m}}(u) \right)_{m \in \{1,...,M\}}$$

for $\alpha_1, ..., \alpha_M \in D[0, \tau]$, where here and throughout, we write $\sum_{\tilde{m} \neq m}$ instead of $\sum_{\tilde{m}=1, \tilde{m} \neq m}^{M}$ for the sake of brevity. Thus, an application of the delta-method (Theorem 3.10.4 in [9]) yields

$$n_i^{1/2}(\hat{\mu}_{im} - \mu_{im})_{m \in \{1,...,M\}} \xrightarrow{d} \Phi'_{(A_{i1},...,A_{iM})}(U_{i1}, ..., U_{iM})$$

$$= \left( \int_{[0,\tau)} \frac{(\tau - u)\left(1 - \sum_{\tilde{m} \neq m} F_{i\tilde{m}}(u)\right) - \int_u^\tau F_{im}(t)\, dt}{1 - \Delta A_i(u)} \, dU_{im}(u) \right.$$

$$\left. + \int_{[0,\tau)} \frac{(\tau - u)F_{im}(u) - \int_u^\tau F_{im}(t)\, dt}{1 - \Delta A_i(u)} \, d \sum_{\tilde{m} \neq m} U_{i\tilde{m}}(u) \right)_{m \in \{1,...,M\}}$$

as $n \to \infty$, where the limit is a centered $M$-dimensional random vector which follows a multivariate normal distribution. Its covariance matrix $\Sigma_i := [\Sigma_{im_1 m_2}]_{m_1, m_2 \in \{1,...,M\}}$ has the following entries: $\Sigma_{im_1 m_2}$ given by

$$\int_{[0,\tau)} \frac{\left\{(\tau - u)\left(1 - \sum_{\tilde{m} \neq m_1} F_{i\tilde{m}}(u)\right) - \int_u^\tau F_{im_1}(t)\, dt\right\}\left\{(\tau - u)\left(1 - \sum_{\tilde{m} \neq m_2} F_{i\tilde{m}}(u)\right) - \int_u^\tau F_{im_2}(t)\, dt\right\}}{(1 - \Delta A_i(u))^2} \, d\sigma_{im_1 m_2}(u)$$

$$+ \int_{[0,\tau)} \frac{\left\{(\tau - u)\left(1 - \sum_{\tilde{m} \neq m_1} F_{i\tilde{m}}(u)\right) - \int_u^\tau F_{im_1}(t)\, dt\right\}\left\{(\tau - u)F_{im_2}(u) - \int_u^\tau F_{im_2}(t)\, dt\right\}}{(1 - \Delta A_i(u))^2} \, d \sum_{\tilde{m} \neq m_2} \sigma_{im_1 \tilde{m}}(u)$$

$$+ \int_{[0,\tau)} \frac{\left\{(\tau - u)\left(1 - \sum_{\tilde{m} \neq m_2} F_{i\tilde{m}}(u)\right) - \int_u^\tau F_{im_2}(t)\, dt\right\}\left\{(\tau - u)F_{im_1}(u) - \int_u^\tau F_{im_1}(t)\, dt\right\}}{(1 - \Delta A_i(u))^2} \, d \sum_{\tilde{m} \neq m_1} \sigma_{im_2 \tilde{m}}(u)$$

$$+ \int_{[0,\tau)} \frac{\left\{(\tau - u)F_{im_1}(u) - \int_u^\tau F_{im_1}(t)\, dt\right\}\left\{(\tau - u)F_{im_2}(u) - \int_u^\tau F_{im_2}(t)\, dt\right\}}{(1 - \Delta A_i(u))^2} \, d \sum_{\tilde{m} \neq m_1} \sum_{\check{m} \neq m_2} \sigma_{i\tilde{m}\check{m}}(u)$$

(S2)

for all $m_1, m_2 \in \{1, \ldots, M\}$. Since $i \in \{1, ..., k\}$ was arbitrary and the groups are independent, the statement of the theorem follows with

$$\Sigma := \bigoplus_{i=1}^{k} \left(\kappa_i^{-1} \Sigma_i\right) \tag{S3}$$

by Assumption 1.

## Proof of Theorem 2

First of all, we investigate the covariance matrix estimator $\hat{\Sigma}$.

**Lemma S1.** *Under Assumption 1, we have $\hat{\Sigma} \xrightarrow{P} \Sigma$ as $n \to \infty$, where here and throughout $\xrightarrow{P}$ denotes convergence in outer probability.*

*Proof.* By (S1), Slutsky's lemma, and the continuous mapping theorem, we have
$$\sup_{t \in [0,\tau]} \left| \widehat{A}_{im}(t) - A_{im}(t) \right| \xrightarrow{P} 0, i \in \{1, \ldots, k\}, m \in \{1, \ldots, M\}$$
as $n \to \infty$. Moreover, it holds
$$\sup_{t \in [0,\tau]} |\widehat{\sigma}_{im_1 m_2}(t) - \sigma_{im_1 m_2}(t)| \xrightarrow{P} 0, i \in \{1, \ldots, k\}, m_1, m_2 \in \{1, \ldots, M\}$$
as $n \to \infty$ by [3]. Note that $\Delta A_i(u) \leq 1 - S_{i-}(\tau) < 1, i \in \{1, \ldots, k\}$ holds for all $u \in [0, \tau)$. Hence, the covariance estimator $\widehat{\Sigma}_i$ is a continuous functional of $\widehat{A}_{im_1}$ and $\widehat{\sigma}_{im_1 m_2}, m_1, m_2 \in \{1, \ldots, M\}$ for all $i \in \{1, \ldots, k\}$ and, thus, the consistency $\widehat{\Sigma} \xrightarrow{P} \Sigma$ as $n \to \infty$ follows by Assumption 1. $\square$

Furthermore, we need that $\Sigma$ is positive definite. The following lemma ensures this under Assumption 2.

**Lemma S2.** *Under Assumptions 1 and 2, $\Sigma$ is positive definite.*

*Proof.* Since $\Sigma$ is positive definite whenever $\Sigma_i$ is positive definite for all $i \in \{1, \ldots, k\}$, we fix $i \in \{1, ..., k\}$. Now, let $\mathbf{a} = (a_1, ..., a_M)^\top \in \mathbb{R}^M \setminus \{\mathbf{0}_M\}$ be arbitrary. We aim to show $\mathbf{a}^\top \Sigma_i \mathbf{a} > 0$. By the proof of Theorem 1, it holds that

$$\mathbf{a}^\top \Sigma_i \mathbf{a} = \mathbb{V}ar\left( \sum_{m=1}^M a_m \left( \int_{[0,\tau)} f_{im} \, dU_{im} + \int_{[0,\tau)} g_{im} \, d\sum_{\widetilde{m} \neq m} U_{i\widetilde{m}} \right) \right) = \mathbb{V}ar\left( \sum_{m=1}^M \int_{[0,\tau)} h_{im} \, dU_{im} \right)$$

with
$$f_{im}(u) := \frac{(\tau - u)\left(1 - \sum_{\widetilde{m} \neq m} F_{i\widetilde{m}}(u)\right) - \int_u^\tau F_{im}(t) \, dt}{1 - \Delta A_i(u)},$$
$$g_{im}(u) := \frac{(\tau - u) F_{im}(u) - \int_u^\tau F_{im}(t) \, dt}{1 - \Delta A_i(u)}$$
and $\quad h_{im}(u) := a_m f_{im}(u) + \sum_{\widetilde{m} \neq m} a_{\widetilde{m}} g_{i\widetilde{m}}(u)$

for all $u \in [0, \tau)$. We can calculate this variance further as

$$\mathbf{a}^\top \Sigma_i \mathbf{a} = \sum_{m=1}^M \mathbb{E}\left( \left( \int_{[0,\tau)} h_{im} \, dU_{im} \right)^2 \right) + \sum_{m=1}^M \sum_{\widetilde{m} \neq m} \mathbb{E}\left( \int_{[0,\tau)} h_{im} \, dU_{im} \int_{[0,\tau)} h_{i\widetilde{m}} \, dU_{i\widetilde{m}} \right)$$

$$= \sum_{m=1}^M \int_{[0,\tau)} h_{im}^2 \frac{1 - \Delta A_{im}}{y_i} \, dA_{im} - \sum_{m=1}^M \sum_{\widetilde{m} \neq m} \int_{[0,\tau)} h_{im} h_{i\widetilde{m}} \frac{\Delta A_{im}}{y_i} \, dA_{i\widetilde{m}}$$

$$= \sum_{m=1}^M \int_{[0,\tau)} \frac{h_{im}^2}{y_i} \, dA_{im} - \sum_{m=1}^M \sum_{\widetilde{m}=1}^M \int_{[0,\tau)} h_{im} h_{i\widetilde{m}} \frac{\Delta A_{im}}{y_i} \, dA_{i\widetilde{m}}$$

$$= \sum_{m=1}^M \int_{[0,\tau)} \frac{h_{im}^2}{y_i} \, dA_{im}^c + \sum_{x \in \mathcal{D}_i} \frac{\sum_{m=1}^M h_{im}^2(x) \Delta A_{im}(x) - \left(\sum_{m=1}^M h_{im}(x) \Delta A_{im}(x)\right)^2}{y_i(x)} \quad (S4)$$

where $\mathcal{D}_i = \{x \in [0, \tau) : \Delta A_i(x) > 0\}$ is the set of discontinuity time points and

$$A_{im}^c(x) := A_{im}(x) - \sum_{y \leqslant x, y \in \mathcal{D}_i} \Delta A_{im}(y), m \in \{1, \ldots, M\}$$

denotes the continuous part of $A_{im}$ at $x \in [0, \tau)$. The Cauchy-Schwarz inequality yields

$$\left(\sum_{m=1}^{M} h_{im}(x) \Delta A_{im}(x)\right)^2 \leqslant \left(\sum_{m=1}^{M} h_{im}^2(x) \Delta A_{im}(x)\right) \left(\sum_{m=1}^{M} \Delta A_{im}(x)\right)$$

and, thus,

$$\sum_{m=1}^{M} h_{im}^2(x) \Delta A_{im}(x) - \left(\sum_{m=1}^{M} h_{im}(x) \Delta A_{im}(x)\right)^2 \geqslant \sum_{m=1}^{M} h_{im}^2(x) \Delta A_{im}(x) \left(1 - \Delta A_i(x)\right) \geqslant 0$$

for all $x \in \mathcal{D}_i$. Let $m^* \in \{1, ..., M\}$ be the index with $|a_{m^*}| = \max\{|a_1|, ..., |a_M|\} > 0$. Then, it holds

$$|h_{im^*}(u)| = \left|a_{m^*} f_{im^*}(u) + \sum_{\tilde{m} \neq m^*} a_{\tilde{m}} g_{i\tilde{m}}(u)\right| \geqslant |a_{m^*}| f_{im^*}(u) - \left|\sum_{\tilde{m} \neq m^*} a_{\tilde{m}} g_{i\tilde{m}}(u)\right|$$

$$\geqslant |a_{m^*}| \left(f_{im^*}(u) + \sum_{\tilde{m} \neq m^*} g_{i\tilde{m}}(u)\right) = |a_{m^*}| \frac{\int_u^{\tau} S_i(t) \, dt}{1 - \Delta A_i(u)} \geqslant |a_{m^*}| (\tau - u) \frac{S_{i-}(\tau)}{1 - \Delta A_i(u)} > 0$$

for all $u \in [0, \tau)$ since $|f_{im^*}(u)| = f_{im^*}(u)$, $|g_{i\tilde{m}}(u)| = -g_{i\tilde{m}}(u)$ and $1 - \Delta A_i(u) \geqslant S_{i-}(\tau) > 0$ for all $u \in [0, \tau); \tilde{m} \in \{1, \ldots, M\}$. Note that $A_{im^*-}(\tau) > 0$ due to Assumption 2. Hence, it follows that at least one of the summands in (S4) is strictly positive and, thus, $\mathbf{a}^\top \boldsymbol{\Sigma}_i \mathbf{a} > 0$. □

For $\boldsymbol{\Sigma}$ positive definite, it holds that $\operatorname{rank}\left(\mathbf{H}\boldsymbol{\Sigma}\mathbf{H}^\top\right) = \operatorname{rank}(\mathbf{H})$. Furthermore, the consistency of the covariance matrix estimator provides

$$P\left(\operatorname{rank}\left(\mathbf{H}\hat{\boldsymbol{\Sigma}}\mathbf{H}^\top\right) \neq \operatorname{rank}(\mathbf{H})\right) \to 0 \tag{S5}$$

as $n \to \infty$. Note that we use the regular probability $P$ instead of the outer probability since $\hat{\boldsymbol{\Sigma}}$ is measurable. Hence, it follows that $\left(\mathbf{H}\hat{\boldsymbol{\Sigma}}\mathbf{H}^\top\right)^+ \xrightarrow{P} \left(\mathbf{H}\boldsymbol{\Sigma}\mathbf{H}^\top\right)^+$ as $n \to \infty$. Theorem 1, Slutsky's lemma and Theorem 9.2.2 in [8] yield

$$W_n(\mathbf{H}, \mathbf{c}) = n(\mathbf{H}\hat{\boldsymbol{\mu}} - \mathbf{c})^\top \left(\mathbf{H}\hat{\boldsymbol{\Sigma}}\mathbf{H}^\top\right)^+ (\mathbf{H}\hat{\boldsymbol{\mu}} - \mathbf{c})$$

$$= \left(\mathbf{H}\left(n^{1/2}(\hat{\boldsymbol{\mu}} - \boldsymbol{\mu})\right)\right)^\top \left(\mathbf{H}\hat{\boldsymbol{\Sigma}}\mathbf{H}^\top\right)^+ \mathbf{H}\left(n^{1/2}(\hat{\boldsymbol{\mu}} - \boldsymbol{\mu})\right) \xrightarrow{d} \chi^2_{\operatorname{rank}(\mathbf{H})}$$

as $n \to \infty$ under the null hypothesis $\mathcal{H}_0$ in (1).

## Proof of Theorem 3

In the following, we denote all permutation counterparts of the counting processes and estimators with a $\pi$ in the superscript, e.g., $Y_i^\pi, N_{im}^\pi, \widehat{A}_{im}^\pi, i \in \{1,\ldots,k\}, m \in \{1,\ldots,M\}$, and all counterparts for the pooled sample with a subscript $\bullet$ instead of $i$, e.g., $Y_\bullet := \sum_{i=1}^k Y_i, N_{\bullet m} := \sum_{i=1}^k N_{im}, \widehat{A}_{\bullet m}(t) := \int_{[0,t]} Y_\bullet^{-1}\, dN_{\bullet m}, m \in \{1,\ldots,M\}$ for all $t \geq 0$. Furthermore, set $y_\bullet := \sum_{i=1}^k \kappa_i y_i, F_{\bullet m} := \sum_{i=1}^k \kappa_i F_{im}, A_{\bullet m}(t) := \int_{[0,t]} y_\bullet^{-1}\, dF_{\bullet m}, A_\bullet := \sum_{m=1}^M A_{\bullet m}$,

$$S_\bullet(t) := \prod_{x \leq t}\{1 - dA_\bullet(x)\}, \qquad \sigma_{\bullet m\tilde{m}}(t) := \int_{[0,t]}(\mathbb{1}\{m = \tilde{m}\} - \Delta A_{\bullet m}) y_\bullet^{-1}\, dA_{\bullet\tilde{m}}$$

for all $t \geq 0; m, \tilde{m} \in \{1,\ldots,M\}$ and $\boldsymbol{\mu}_\bullet := \mathbf{1}_k \otimes \left(\int_0^\tau F_{\bullet m}(t)\, dt\right)_{m \in \{1,\ldots,M\}}$.

**Lemma S3.** *Under Assumption 1, we have*

$$n^{1/2}(\widehat{\boldsymbol{\mu}}^\pi - \boldsymbol{\mu}_\bullet) \xrightarrow{d^*} \mathbf{Z}^\pi \sim \mathcal{N}_{kM}(\mathbf{0}_{kM}, \boldsymbol{\Sigma}^\pi)$$

*as $n \to \infty$. The definition of the covariance matrix $\boldsymbol{\Sigma}^\pi$ is given at the end of the proof.*

*Proof of Lemma S3.* Let us consider the class

$$\mathcal{F} := \{(x,d) \mapsto \mathbb{1}\{x \geq t\}, (x,d) \mapsto \mathbb{1}\{x \leq t, d = m\} \mid t \in [0,\tau], m \in \{1,...,M\}\}$$

with finite envelope function $F \equiv 1$. It can be shown that $\mathcal{F}$ is $P^{(X_{i1},D_{i1})}$-Donsker for all $i = 1,\ldots,k$, e.g., by applying Theorem 2.6.8 in [9]. By Theorem 1 in [7], it follows that

$$n^{1/2}\left(n_i^{-1}(Y_i^\pi, N_{i1}^\pi, \ldots, N_{iM}^\pi) - n^{-1}(Y_\bullet, N_{\bullet 1}, \ldots, N_{\bullet M})\right)_{i \in \{1,\ldots,k\}} \xrightarrow{d^*} (\overline{G}_i, G_{i1}, \ldots, G_{iM})_{i \in \{1,\ldots,k\}} \quad (S6)$$

on $(D[0,\tau])^{k(M+1)}$ as $n \to \infty$, where $(\overline{G}_i, G_{i1}, \ldots, G_{iM})_{i \in \{1,\ldots,k\}}$ denotes a tight centered Gaussian process with covariance structure

$$\mathbb{E}\left(\overline{G}_i(t)\overline{G}_j(s)\right) = \left(\kappa_i^{-1}\mathbb{1}\{i = j\} - 1\right)\left(y_\bullet(\max\{t,s\}) - y_\bullet(t)y_\bullet(s)\right),$$
$$\mathbb{E}\left(\overline{G}_i(t)G_{jm_1}(s)\right) = \left(\kappa_i^{-1}\mathbb{1}\{i = j\} - 1\right)\left((F_{\bullet m_1}(s) - F_{\bullet m_1-}(t))\mathbb{1}\{s \geq t\} - y_\bullet(t)F_{\bullet m_1}(s)\right),$$
$$\mathbb{E}\left(G_{im_1}(t)G_{jm_2}(s)\right) = \left(\kappa_i^{-1}\mathbb{1}\{i = j\} - 1\right)\left(F_{\bullet m_1}(\min\{t,s\})\mathbb{1}\{m_1 = m_2\} - F_{\bullet m_1}(t)F_{\bullet m_2}(s)\right)$$

at $(t,s) \in [0,\tau]^2$ for all $i,j \in \{1,\ldots,k\}, m_1, m_2 \in \{1,\ldots,M\}$. By the conditional delta-method (Theorem 4 in [7]) and the uniform Hadamard-differentiability of the Wilcoxon functional (Example 1 in [7]), we get

$$n^{1/2}\left(\widehat{A}_{im}^\pi - \widehat{A}_{\bullet m}\right)_{i \in \{1,\ldots,k\}, m \in \{1,\ldots,M\}} \xrightarrow{d^*} \left(\int_{[0,\cdot]} y_\bullet^{-1}\, dG_{im} - \int_{[0,\cdot]} \overline{G}_i y_\bullet^{-2}\, dF_{\bullet m}\right)_{i \in \{1,\ldots,k\}, m \in \{1,\ldots,M\}}$$
$$=: (U_{im}^\pi)_{i \in \{1,\ldots,k\}, m \in \{1,\ldots,M\}}$$

on $(D[0,\tau])^{kM}$ as $n \to \infty$. The limit variable $(U_{im}^\pi)_{i \in \{1,\ldots,k\}, m \in \{1,\ldots,M\}}$ is a separable centered Gaussian process and the covariance structure can be calculated similarly as in [3] as

$$\mathbb{E}\left(U_{im_1}^\pi(t)U_{jm_2}^\pi(s)\right) = \left(\kappa_i^{-1}\mathbb{1}\{i = j\} - 1\right)\int_{[0,\min\{t,s\}]}(\mathbb{1}\{m_1 = m_2\} - \Delta A_{\bullet m_1})y_\bullet^{-1}\, dA_{\bullet m_2}$$
$$=: \sigma_{ijm_1m_2}^\pi(\min\{t,s\})$$

for all $i,j \in \{1,\ldots,k\}, m_1, m_2 \in \{1,\ldots,M\}$ and $t,s \in [0,\tau]$. Furthermore, we aim to apply the conditional delta-method with the function

$$(D[0,\tau])^{kM} \ni (\Lambda_{im})_{i\in\{1,\ldots,k\}, m\in\{1,\ldots,M\}} \mapsto (\Phi(\Lambda_{11},\ldots,\Lambda_{1M}),\ldots,\Phi(\Lambda_{k1},\ldots,\Lambda_{kM})),$$

where $\Phi$ is defined as in the proof of Theorem 2. To show the uniform Hadamard-differentiability of $\Phi$ at $(A_{\bullet 1},\ldots,A_{\bullet M})$, we remind that $\Phi$ is a composition of linear functionals, the Wilcoxon functional $\Psi$ and the product integral functional $\phi$. Hence, the chain rule together with the examples in Section 4 of [7] implies that it remains to show that $-\sum_{m=1}^{M} A_{\bullet m} : [0,\tau] \to \mathbb{R}$ is a càdlàg function of bounded variation with jumps contained in $(-1,\infty)$, which follows by Assumption 1. Thus, we receive

$$n^{1/2}(\hat{\boldsymbol{\mu}}^\pi - \boldsymbol{\mu}_\bullet) \xrightarrow{d^*} \left(\Phi'_{A_{\bullet 1},\ldots,A_{\bullet M}}(U_{11}^\pi,\ldots,U_{1M}^\pi),\ldots,\Phi'_{A_{\bullet 1},\ldots,A_{\bullet M}}(U_{k1}^\pi,\ldots,U_{kM}^\pi)\right) =: \mathbf{Z}^\pi$$

as $n \to \infty$. The limit variable $\mathbf{Z}^\pi$ is a centered normal variable with covariance matrix $\boldsymbol{\Sigma}^\pi$, where the entries $\Sigma^\pi_{im_1, jm_2}$ are given by

$$\int_{[0,\tau)} \frac{\left\{(\tau-u)\left(1 - \sum_{\tilde{m} \neq m_1} F_{\bullet \tilde{m}}(u)\right) - \int_u^\tau F_{\bullet m_1}(t)\,\mathrm{d}t\right\}\left\{(\tau-u)\left(1 - \sum_{\tilde{m} \neq m_2} F_{\bullet \tilde{m}}(u)\right) - \int_u^\tau F_{\bullet m_2}(t)\,\mathrm{d}t\right\}}{(1 - \Delta A_\bullet(u))^2}\,\mathrm{d}\sigma^\pi_{ijm_1 m_2}(u)$$

$$+ \int_{[0,\tau)} \frac{\left\{(\tau-u)\left(1 - \sum_{\tilde{m} \neq m_1} F_{\bullet \tilde{m}}(u)\right) - \int_u^\tau F_{\bullet m_1}(t)\,\mathrm{d}t\right\}\left\{(\tau-u)F_{\bullet m_2}(u) - \int_u^\tau F_{\bullet m_2}(t)\,\mathrm{d}t\right\}}{(1 - \Delta A_\bullet(u))^2}\,\mathrm{d}\sum_{\tilde{m} \neq m_2} \sigma^\pi_{ijm_1 \tilde{m}}(u)$$

$$+ \int_{[0,\tau)} \frac{\left\{(\tau-u)\left(1 - \sum_{\tilde{m} \neq m_2} F_{\bullet \tilde{m}}(u)\right) - \int_u^\tau F_{\bullet m_2}(t)\,\mathrm{d}t\right\}\left\{(\tau-u)F_{\bullet m_1}(u) - \int_u^\tau F_{\bullet m_1}(t)\,\mathrm{d}t\right\}}{(1 - \Delta A_\bullet(u))^2}\,\mathrm{d}\sum_{\tilde{m} \neq m_1} \sigma^\pi_{ijm_2 \tilde{m}}(u)$$

$$+ \int_{[0,\tau)} \frac{\left\{(\tau-u)F_{\bullet m_1}(u) - \int_u^\tau F_{\bullet m_1}(t)\,\mathrm{d}t\right\}\left\{(\tau-u)F_{\bullet m_2}(u) - \int_u^\tau F_{\bullet m_2}(t)\,\mathrm{d}t\right\}}{(1 - \Delta A_\bullet(u))^2}\,\mathrm{d}\sum_{\tilde{m} \neq m_1}\sum_{\check{m} \neq m_2} \sigma^\pi_{ij\tilde{m}\check{m}}(u)$$

for all $i,j \in \{1,\ldots,k\}, m_1, m_2 \in \{1,\ldots,M\}$. $\square$

Now, we turn to the permutation counterpart of the covariance matrix estimator.

**Lemma S4.** *Under Assumption 1, we have*

$$\hat{\boldsymbol{\Sigma}}^\pi \xrightarrow{P} \widetilde{\boldsymbol{\Sigma}}^\pi := \bigoplus_{i=1}^{k} \kappa_i^{-1} (\widetilde{\Sigma}^\pi_{m_1 m_2})_{m_1, m_2 \in \{1,\ldots,M\}}$$

*as $n \to \infty$, where $\widetilde{\Sigma}^\pi_{m_1 m_2}$ is defined as in (S7) for all $m_1, m_2 \in \{1,\ldots,M\}$.*

*Proof of Lemma S4.* Due to the definition of $\hat{\boldsymbol{\Sigma}}^\pi$, it remains to show $\hat{\Sigma}^\pi_{im_1, im_2} \xrightarrow{P} \widetilde{\Sigma}^\pi_{m_1 m_2}, i \in \{1,\ldots,k\}, m_1, m_2 \in \{1,\ldots,M\}$. Therefore, let $i \in \{1,\ldots,k\}$ be arbitrary but fixed. Then, (S6) implies

$$n^{1/2}\left(n_i^{-1}(Y_i^\pi, N_{i1}^\pi, \ldots, N_{iM}^\pi) - n^{-1}(Y_\bullet, N_{\bullet 1}, \ldots, N_{\bullet M})\right) \xrightarrow{d} (\overline{G}_i, G_{i1}, \ldots, G_{iM})$$

on $(D[0,\tau])^{M+1}$ as $n \to \infty$ unconditionally by Lemma 1 of [7]. Hence, Slutsky's lemma provides

$$\sup_{t\in[0,\tau]} \left|n_i^{-1}Y_i^\pi(t) - n^{-1}Y_\bullet(t)\right| \xrightarrow{P} 0 \quad \text{and} \quad \sup_{t\in[0,\tau]} \left|n_i^{-1}N_{im}^\pi(t) - n^{-1}N_{\bullet m}(t)\right| \xrightarrow{P} 0, m \in \{1,\ldots,M\}$$

as $n \to \infty$. By the definitions of $Y_\bullet, N_{\bullet 1}, \ldots, N_{\bullet M}$, we further get

$$\sup_{t\in[0,\tau]} \left|n^{-1}Y_\bullet(t) - y_\bullet(t)\right| = \sup_{t\in[0,\tau]} \left|\sum_{i=1}^{k}\left(\frac{n_i}{n}n_i^{-1}Y_i(t) - \kappa_i y_i(t)\right)\right| \xrightarrow{P} 0 \quad \text{and}$$

$$\sup_{t\in[0,\tau]} \left|n^{-1}N_{\bullet m}(t) - F_{\bullet m}(t)\right| = \sup_{t\in[0,\tau]} \left|\sum_{i=1}^{k}\left(\frac{n_i}{n}n_i^{-1}N_{im}(t) - \kappa_i F_{im}(t)\right)\right| \xrightarrow{P} 0, m \in \{1,\ldots,M\}$$

as $n \to \infty$ since $\mathcal{F}$ in the proof of Lemma S3 is $P^{(X_{i1},D_{i1})}$-Donsker. Thus, it follows

$$\sup_{t\in[0,\tau]} \left|n_i^{-1}Y_i^\pi(t) - y_\bullet(t)\right| \xrightarrow{P} 0 \quad \text{and} \quad \sup_{t\in[0,\tau]} \left|n_i^{-1}N_{im}^\pi(t) - F_{\bullet m}(t)\right| \xrightarrow{P} 0, m \in \{1,\ldots,M\}$$

as $n \to \infty$. Since $\hat{A}_{im}^\pi = \int_{[0,\cdot]} n_i(Y_i^\pi)^{-1} \, \mathrm{d}(n_i^{-1}N_{im}^\pi), \hat{F}_{im}^\pi$ and $\hat{\sigma}_{im\tilde{m}}^\pi$ are depending continuously on $n_i^{-1}(Y_i^\pi, N_{i1}^\pi, \ldots, N_{iM}^\pi)$, we can conclude

$$\sup_{t\in[0,\tau]} \left|\hat{A}_{im}^\pi(t) - A_{\bullet m}(t)\right| \xrightarrow{P} 0, \sup_{t\in[0,\tau]} \left|\hat{F}_{im}^\pi(t) - F_{\bullet m}(t)\right| \xrightarrow{P} 0 \text{ and } \sup_{t\in[0,\tau]} |\hat{\sigma}_{im\tilde{m}}^\pi(t) - \sigma_{\bullet m\tilde{m}}(t)| \xrightarrow{P} 0$$

as $n \to \infty$ for all $m, \tilde{m} \in \{1, \ldots, M\}$. Hence, we get

$$\hat{\Sigma}_{im_1,im_2}^\pi \xrightarrow{P} \tilde{\Sigma}_{m_1m_2}^\pi :=$$

$$\int_{[0,\tau)} \frac{\left\{(\tau-u)\left(1 - \sum_{\tilde{m}\neq m_1} F_{\bullet\tilde{m}}(u)\right) - \int_u^\tau F_{\bullet m_1}(t)\,\mathrm{d}t\right\}\left\{(\tau-u)\left(1 - \sum_{\tilde{m}\neq m_2} F_{\bullet\tilde{m}}(u)\right) - \int_u^\tau F_{\bullet m_2}(t)\,\mathrm{d}t\right\}}{(1-\Delta A_\bullet(u))^2} \,\mathrm{d}\sigma_{\bullet m_1 m_2}(u)$$

$$+ \int_{[0,\tau)} \frac{\left\{(\tau-u)\left(1 - \sum_{\tilde{m}\neq m_1} F_{\bullet\tilde{m}}(u)\right) - \int_u^\tau F_{\bullet m_1}(t)\,\mathrm{d}t\right\}\left\{(\tau-u)F_{\bullet m_2}(u) - \int_u^\tau F_{\bullet m_2}(t)\,\mathrm{d}t\right\}}{(1-\Delta A_\bullet(u))^2} \,\mathrm{d}\sum_{\tilde{m}\neq m_2}\sigma_{\bullet m_1\tilde{m}}(u)$$

$$+ \int_{[0,\tau)} \frac{\left\{(\tau-u)\left(1 - \sum_{\tilde{m}\neq m_2} F_{\bullet\tilde{m}}(u)\right) - \int_u^\tau F_{\bullet m_2}(t)\,\mathrm{d}t\right\}\left\{(\tau-u)F_{\bullet m_1}(u) - \int_u^\tau F_{\bullet m_1}(t)\,\mathrm{d}t\right\}}{(1-\Delta A_\bullet(u))^2} \,\mathrm{d}\sum_{\tilde{m}\neq m_1}\sigma_{\bullet m_2\tilde{m}}(u)$$

$$+ \int_{[0,\tau)} \frac{\left\{(\tau-u)F_{\bullet m_1}(u) - \int_u^\tau F_{\bullet m_1}(t)\,\mathrm{d}t\right\}\left\{(\tau-u)F_{\bullet m_2}(u) - \int_u^\tau F_{\bullet m_2}(t)\,\mathrm{d}t\right\}}{(1-\Delta A_\bullet(u))^2} \,\mathrm{d}\sum_{\tilde{m}\neq m_1}\sum_{\breve{m}\neq m_2}\sigma_{\bullet\tilde{m}\breve{m}}(u)$$

(S7)

as $n \to \infty$. □

As in the proof of Theorem 2, we need the positive definiteness of $\tilde{\Sigma}^\pi$, which is given by the following lemma.

**Lemma S5.** *Under Assumptions 1 and 2, $\widetilde{\Sigma}^\pi$ is positive definite.*

*Proof of Lemma S5.* The proof is similar to the proof of Lemma S2. Firstly note that $\widetilde{\Sigma}^\pi$ is positive definite if $(\widetilde{\Sigma}^\pi_{m_1 m_2})_{m_1, m_2 \in \{1,\ldots,M\}}$ is positive definite. Let $\mathbf{a} = (a_1, \ldots, a_M)^\top \in \mathbb{R}^M \setminus \{\mathbf{0}_M\}$ be arbitrary. With

$$f_m(u) := \frac{(\tau - u)(1 - \sum_{\tilde{m} \neq m} F_{\bullet \tilde{m}}(u)) - \int_u^\tau F_{\bullet m}(t) \, \mathrm{d}t}{1 - \Delta A_\bullet(u)},$$

$$g_m(u) := \frac{(\tau - u) F_{\bullet m}(u) - \int_u^\tau F_{\bullet m}(t) \, \mathrm{d}t}{1 - \Delta A_\bullet(u)}$$

$$\text{and} \quad h_m(u) := a_m f_m(u) + \sum_{\tilde{m} \neq m} a_{\tilde{m}} g_{\tilde{m}}(u)$$

for all $u \in [0, \tau)$, we get

$$\mathbf{a}^\top (\widetilde{\Sigma}^\pi_{m_1 m_2})_{m_1, m_2 \in \{1,\ldots,M\}} \mathbf{a} = \sum_{m=1}^M \int_{[0,\tau)} \frac{h_m^2}{y_\bullet} \, \mathrm{d} A^c_{\bullet m} \tag{S8}$$

$$+ \sum_{x \in \mathcal{D}} \frac{\sum_{m=1}^M h_m^2(x) \Delta A_{\bullet m}(x) - \left( \sum_{m=1}^M h_m(x) \Delta A_{\bullet m}(x) \right)^2}{y_\bullet(x)} \tag{S9}$$

analogously to the proof of Lemma S2, where $\mathcal{D} := \{x \in [0, \tau) : \Delta A_\bullet(x) > 0\}$ and

$$A^c_{\bullet m}(x) := A_{\bullet m}(x) - \sum_{y \leq x, y \in \mathcal{D}} \Delta A_{\bullet m}(y), m \in \{1, \ldots, M\}$$

for all $x \in [0, \tau)$. The Cauchy-Schwarz inequality implies

$$\sum_{m=1}^M h_m^2(x) \Delta A_{\bullet m}(x) - \left( \sum_{m=1}^M h_m(x) \Delta A_{\bullet m}(x) \right)^2 \geq \sum_{m=1}^M h_m^2(x) \Delta A_{\bullet m}(x) (1 - \Delta A_\bullet(x)) \geq 0$$

for all $x \in \mathcal{D}$. For $m^* \in \{1, \ldots, M\}$ with $|a_{m^*}| = \max\{|a_1|, \ldots, |a_M|\} > 0$, it holds

$$|h_{m^*}(u)| = \left| a_{m^*} f_{m^*}(u) + \sum_{\tilde{m} \neq m^*} a_{\tilde{m}} g_{\tilde{m}}(u) \right| \geq |a_{m^*}| f_{m^*}(u) - \left| \sum_{\tilde{m} \neq m^*} a_{\tilde{m}} g_{\tilde{m}}(u) \right|$$

$$\geq |a_{m^*}| \left( f_{im^*}(u) + \sum_{\tilde{m} \neq m^*} g_{i\tilde{m}}(u) \right) = |a_{m^*}| \frac{\sum_{i=1}^k \kappa_i \int_u^\tau S_i(t) \, \mathrm{d}t}{1 - \Delta A_\bullet(u)}$$

$$\geq |a_{m^*}| \frac{(\tau - u) \sum_{i=1}^k \kappa_i S_i(\tau)}{1 - \Delta A_\bullet(u)} > 0$$

for all $u \in [0, \tau)$ since $|f_{m^*}(u)| = f_{m^*}(u)$, $|g_{\tilde{m}}(u)| = -g_{\tilde{m}}(u)$ and

$$\Delta A_\bullet(u) = \frac{\sum_{i=1}^k \kappa_i y_i(u) \Delta A_i(u)}{\sum_{i=1}^k \kappa_i y_i(u)} < 1$$

for all $u \in [0, \tau); \tilde{m} \in \{1, \ldots, M\}$. Note that $A_{\bullet m^*-}(\tau) > 0$ due to Assumption 2. Thus, at least one of the summands (S8) and (S9) is strictly positive. □

Note that $\sigma^\pi_{ijm_1m_2} = (\kappa_i^{-1}\mathbb{1}\{i = j\}-1)\sigma_{\bullet m_1m_2}$ for all $i,j \in \{1,\ldots,k\}, m_1, m_2 \in \{1,\ldots,M\}$. Hence, it holds $\widetilde{\boldsymbol{\Sigma}}^\pi = \boldsymbol{\Sigma}^\pi + (\mathbf{1}_k\mathbf{1}_k^\top) \otimes (\widetilde{\Sigma}^\pi_{m_1m_2})_{m_1,m_2\in\{1,\ldots,M\}}$ and, thus, $\mathbf{H}\widehat{\boldsymbol{\Sigma}}^\pi\mathbf{H}^\top \xrightarrow{P} \mathbf{H}\widetilde{\boldsymbol{\Sigma}}^\pi\mathbf{H}^\top = \mathbf{H}\boldsymbol{\Sigma}^\pi\mathbf{H}^\top$ as $n \to \infty$ due to the contrast property of the hypothesis matrix $\mathbf{H}$. The positive definiteness of $\widetilde{\boldsymbol{\Sigma}}^\pi$ ensures

$$\mathrm{rank}\left(\mathbf{H}\boldsymbol{\Sigma}^\pi\mathbf{H}^\top\right) = \mathrm{rank}\left(\mathbf{H}\widetilde{\boldsymbol{\Sigma}}^\pi\mathbf{H}^\top\right) = \mathrm{rank}(\mathbf{H}).$$

As in the proof of Theorem 2, we get $(\mathbf{H}\widehat{\boldsymbol{\Sigma}}^\pi\mathbf{H}^\top)^+ \xrightarrow{P} (\mathbf{H}\boldsymbol{\Sigma}^\pi\mathbf{H}^\top)^+$ as $n \to \infty$. Combining this with Lemma S3, Slutsky's lemma and Theorem 9.2.2 in [8] yields

$$\begin{aligned} W_n^\pi(\mathbf{H}) &= n(\mathbf{H}\widehat{\boldsymbol{\mu}}^\pi)^\top \left(\mathbf{H}\widehat{\boldsymbol{\Sigma}}^\pi\mathbf{H}^\top\right)^+ (\mathbf{H}\widehat{\boldsymbol{\mu}}^\pi) \\ &= \left(\mathbf{H}(n^{1/2}(\widehat{\boldsymbol{\mu}}^\pi - \boldsymbol{\mu}_\bullet))\right)^\top \left(\mathbf{H}\widehat{\boldsymbol{\Sigma}}^\pi\mathbf{H}^\top\right)^+ \left(\mathbf{H}(n^{1/2}(\widehat{\boldsymbol{\mu}}^\pi - \boldsymbol{\mu}_\bullet))\right) \\ &\xrightarrow{d^*} (\mathbf{H}\mathbf{Z}^\pi)^\top(\mathbf{H}\boldsymbol{\Sigma}^\pi\mathbf{H}^\top)^+(\mathbf{H}\mathbf{Z}^\pi) \sim \chi^2_{\mathrm{rank}(\mathbf{H})} \end{aligned}$$

as $n \to \infty$.

## Proof of Proposition 1

First of all, we aim to show $q^\pi_{1-\alpha} \xrightarrow{P} \chi^2_{\mathrm{rank}(\mathbf{H}),1-\alpha}$ as $n \to \infty$. Therefore, we firstly state an in-probability version of Lemma S4 in the supplement of [5].

**Lemma S6.** *Let $L \in \mathbb{N}$, $F : \mathbb{R}^L \to [0,1]$ be a continuous cumulative distribution function and $(F_n)_{n\in\mathbb{N}}$ denote a sequence of maps taking values in the space of all cumulative distribution functions with*

$$F_n(\mathbf{t}) \xrightarrow{P} F(\mathbf{t}) \text{ as } n \to \infty \text{ for all } \mathbf{t} \in \mathbb{R}^L. \tag{S10}$$

*Then, we have $\sup_{\mathbf{t}\in\mathbb{R}^L} |F_n(\mathbf{t}) - F(\mathbf{t})| \xrightarrow{P} 0$ as $n \to \infty$.*

*Proof of Lemma S6.* To show this, we proceed similarly as in the proof of Lemma S4 in the supplement of [5]. Let $f_\mathbf{t} : \mathbb{R}^L \to \mathbb{R}, f_\mathbf{t}(x_1,\ldots,x_L) = \mathbb{1}\{x_1 \leqslant t_1,\ldots,x_L \leqslant t_L\}$ for all $\mathbf{t} = (t_1,\ldots,t_L)^\top \in (\mathbb{R} \cup \{-\infty,\infty\})^L$, $\mathbf{F} := \{f_\mathbf{t} \mid \mathbf{t} \in (\mathbb{R} \cup \{-\infty,\infty\})^L\}$ and $\varepsilon > 0$ be arbitrary. Then, the bracketing number of $\mathbf{F}$ is bounded by $(m+1)^L$ for some $m \in \mathbb{N}$ as shown in the proof of Lemma S4 in the supplement of [5]. As in the proof of Proposition 2.1 in [1], it holds

$$\sup_{\mathbf{w}\in\mathbb{R}^L} |F_n(\mathbf{w}) - F(\mathbf{w})| \leqslant 3\max\{|F_n(\mathbf{w}) - F(\mathbf{w})|\} + 2\varepsilon,$$

where the maximum is taken over $2(m+1)^L$ different values $\mathbf{w} \in \mathbb{R}^L$. Hence, we get

$$P^*\left(\sup_{\mathbf{w}\in\mathbb{R}^L} |F_n(\mathbf{w}) - F(\mathbf{w})| > 3\varepsilon\right) \leqslant P^*\left(\max\{|F_n(\mathbf{w}) - F(\mathbf{w})|\} > \varepsilon\right),$$

where $P^*$ denotes the outer probability. The latter tends to 0 by assumption since the maximum is taken over a finite number of values $\mathbf{w} \in \mathbb{R}^L$. □

The following lemma can be used to show condition (S10) for a consistent resampling scheme. Here, $\mathbf{X}_n$ represents the randomness of the data while $\mathbf{Y}_n^{(1)}, \mathbf{Y}_n^{(2)}, \ldots$ can be interpreted as the randomness of the resampling method.

**Lemma S7.** *Let $\mathbf{X}_n : \Omega \to \chi_{1n}, \mathbf{Y}_n^{(1)} : \Omega_1 \to \chi_{2n}, \mathbf{Y}_n^{(2)} : \Omega_2 \to \chi_{2n}, \ldots$ denote independent maps defined on probability spaces $\Omega, \Omega_1, \Omega_2, \ldots$ for $n \in \mathbb{N}$, where independence is understood in terms of a product probability space, cf. Section 2.3.1 in [10]. Furthermore, assume that $\mathbf{Y}_n^{(1)}, \mathbf{Y}_n^{(2)}, \ldots$ are identically distributed and $\mathbf{W}_n : \chi_{1n} \times \chi_{2n} \to \mathbb{R}^L$ is a function for $L, n \in \mathbb{N}$. Suppose that $F_n$ is the empirical distribution function of $\mathbf{W}_n^{(b)} := \mathbf{W}_n(\mathbf{X}_n, \mathbf{Y}_n^{(b)}), b \in \{1, \ldots, B_n\}$ for $B_n, n \in \mathbb{N}$ satisfying $\mathbf{W}_n^{(1)} \xrightarrow{d*} \mathbf{W} \sim F$ as $n \to \infty$ conditionally on $\mathbf{X}_n$, where $F : \mathbb{R}^L \to [0, 1]$ denotes a continuous cumulative distribution function. Then, (S10) is satisfied if $B_n \to \infty$ as $n \to \infty$.*

*Proof of Lemma S7.* Let $\mathbf{t} \in \mathbb{R}^L$ be arbitrary. Approximate $f : \mathbb{R}^L \to \{0, 1\}, \mathbf{w} \mapsto \mathbb{1}\{\mathbf{w} \leqslant \mathbf{t}\}$ through sequences of Lipschitz functions $(g_m)_{m \in \mathbb{N}}, (h_m)_{m \in \mathbb{N}}$ with $1 \geqslant g_m \geqslant f \geqslant h_m \geqslant 0$ and $\mathbb{E}\left[g_m(\mathbf{W}) - h_m(\mathbf{W})\right] \leqslant m^{-1}$ for all $m \in \mathbb{N}$, where $\mathbf{x} \leqslant \mathbf{t}$ means that all components of $\mathbf{x}$ are less than or equal to the corresponding values in the vector $\mathbf{t}$. Let $\mathbb{E}_2^*$ denote the outer expectation with respect to the product space $\times_{i=1}^{\infty} \Omega_i$, cf. Section 1.2 in [10]. Then, it holds

$$\left|\mathbb{E}_2\left[f(\mathbf{W}_n^{(1)})^*\right] - \mathbb{E}\left[f(\mathbf{W})\right]\right|$$
$$\leqslant \mathbb{E}_2\left[f(\mathbf{W}_n^{(1)})^*\right] - \mathbb{E}_2^*\left[f(\mathbf{W}_n^{(1)})\right] + \left|\mathbb{E}_2^*\left[f(\mathbf{W}_n^{(1)})\right] - \mathbb{E}\left[f(\mathbf{W})\right]\right|$$
$$\leqslant \mathbb{E}_2\left[g_m(\mathbf{W}_n^{(1)})^*\right] - \mathbb{E}_2^*\left[h_m(\mathbf{W}_n^{(1)})\right]$$
$$\quad + \max\left\{\mathbb{E}_2^*\left[g_m(\mathbf{W}_n^{(1)})\right] - \mathbb{E}\left[g_m(\mathbf{W})\right], \mathbb{E}\left[h_m(\mathbf{W})\right] - \mathbb{E}_2^*\left[h_m(\mathbf{W}_n^{(1)})\right]\right\}$$
$$\quad + \mathbb{E}\left[g_m(\mathbf{W})\right] - \mathbb{E}\left[h_m(\mathbf{W})\right]$$
$$\leqslant \mathbb{E}_2\left[g_m(\mathbf{W}_n^{(1)})^* - g_m(\mathbf{W}_n^{(1)})_*\right] + \mathbb{E}_2^*\left[g_m(\mathbf{W}_n^{(1)})\right] - \mathbb{E}_2^*\left[h_m(\mathbf{W}_n^{(1)})\right]$$
$$\quad + \max\left\{\mathbb{E}_2^*\left[g_m(\mathbf{W}_n^{(1)})\right] - \mathbb{E}\left[g_m(\mathbf{W})\right], \mathbb{E}\left[h_m(\mathbf{W})\right] - \mathbb{E}_2^*\left[h_m(\mathbf{W}_n^{(1)})\right]\right\} + m^{-1}$$
$$\xrightarrow{P} 0 + \mathbb{E}\left[g_m(\mathbf{W})\right] - \mathbb{E}\left[h_m(\mathbf{W})\right] + 0 + m^{-1} \leqslant 2m^{-1},$$

where here and throughout $A^*$ and $A_*$ denote the minimal measurable majorant and maximal measurable minorant, respectively, for a map $A$ with respect to all probability spaces jointly. By choosing $m$ sufficiently large, it follows

$$\mathbb{E}_2\left[f(\mathbf{W}_n^{(1)})^*\right] \xrightarrow{P} \mathbb{E}\left[f(\mathbf{W})\right] = F(\mathbf{t})$$

as $n \to \infty$. Analogously, one can show $\mathbb{E}_2\left[f(\mathbf{W}_n^{(1)})_*\right] \xrightarrow{P} F(\mathbf{t})$ as $n \to \infty$. Since $B_n \to \infty$ as

$n \to \infty$, we have

$$\mathbb{E}_2\left[\left(|F_n(\mathbf{t}) - F(\mathbf{t})|^2\right)^*\right]$$
$$\leqslant \mathbb{E}_2\left[\left(F_n^2(\mathbf{t})\right)^*\right] + 2\mathbb{E}_2\left[(-F_n(\mathbf{t}))^*\right]F(\mathbf{t}) + F^2(\mathbf{t})$$
$$\leqslant \frac{1}{B_n^2}\sum_{b_1,b_2=1}^{B_n}\mathbb{E}_2\left[\left(f(\mathbf{W}_n^{(b_1)})f(\mathbf{W}_n^{(b_2)})\right)^*\right] + \frac{2}{B_n}\sum_{b=1}^{B_n}\mathbb{E}_2\left[\left(-f(\mathbf{W}_n^{(b)})\right)^*\right]F(\mathbf{t}) + F^2(\mathbf{t})$$
$$\leqslant \frac{1}{B_n} + \frac{1}{B_n^2}\sum_{b_1,b_2=1, b_1\neq b_2}^{B_n}\mathbb{E}_2\left[f(\mathbf{W}_n^{(b_1)})^*\right]\mathbb{E}_2\left[f(\mathbf{W}_n^{(b_2)})^*\right] - 2\mathbb{E}_2\left[f(\mathbf{W}_n^{(1)})_*\right]F(\mathbf{t}) + F^2(\mathbf{t})$$
$$\leqslant \frac{1}{B_n} + \frac{B_n-1}{B_n}\mathbb{E}_2\left[f(\mathbf{W}_n^{(1)})^*\right]^2 - 2\mathbb{E}_2\left[f(\mathbf{W}_n^{(1)})_*\right]F(\mathbf{t}) + F^2(\mathbf{t})$$
$$\xrightarrow{P} F^2(\mathbf{t}) - 2F(\mathbf{t})F(\mathbf{t}) + F^2(\mathbf{t}) = 0$$

as $n \to \infty$. Thus, $\mathbb{E}^*\left[|F_n(\mathbf{t}) - F(\mathbf{t})|^2\right] \to 0$ as $n \to \infty$ for all $\mathbf{t} \in \mathbb{R}^L$ by the dominated convergence theorem (dominated by 1). Hence, (S10) in Lemma S8 follows. □

We apply both lemmas with $F(.) = \chi^2_{\text{rank}(\mathbf{H})}((-\infty, .])$ and $F_n$ being the empirical distribution function of the permutation counterparts of the Wald-type test statistic. Hence, we receive

$$\sup_{t\in\mathbb{R}}|F_n(t) - F(t)| \xrightarrow{P} 0$$

as $n \to \infty$ by Theorem 3.

Moreover, the subsequence criterion (Lemma 1.9.2 (ii) in [10]) provides that every subsequence has a further subsequence such that $\sup_{t\in\mathbb{R}}|F_n(t) - F(t)| \to 0$ holds outer almost surely along the latter subsequence. By proceeding as in the proof of Lemma S3 in the supplement of [5], we get $q^\pi_{1-\alpha} \to \chi^2_{\text{rank}(\mathbf{H}),1-\alpha}$ outer almost surely along the subsequence. Applying the subsequence criterion again yields $q^\pi_{1-\alpha} \xrightarrow{P} \chi^2_{\text{rank}(\mathbf{H}),1-\alpha}$ as $n \to \infty$.

By Slutsky's lemma and Theorem 2, it follows $(W_n(\mathbf{H},\mathbf{c}), q^\pi_{1-\alpha}) \xrightarrow{d} (W, \chi^2_{\text{rank}(\mathbf{H}),1-\alpha})$ as $n \to \infty$ for $W \sim \chi^2_{\text{rank}(\mathbf{H})}$. By approximating the function $f : \mathbb{R}^2 \to \{0,1\}, f(w,q) := \mathbb{1}\{w > q\}$ with sequences of Lipschitz functions $(g_m)_{m\in\mathbb{N}}, (h_m)_{m\in\mathbb{N}}$ with $1 \geqslant g_m \geqslant f \geqslant h_m \geqslant 0$ and

$$\mathbb{E}\left[g_m(W, \chi^2_{\text{rank}(\mathbf{H}),1-\alpha}) - h_m(W, \chi^2_{\text{rank}(\mathbf{H}),1-\alpha})\right] \leqslant m^{-1}$$

for all $m \in \mathbb{N}$, one can follow similarly as in the proof of Lemma S7 that

$$\mathbb{E}(\varphi^\pi) = \mathbb{E}\left[f(W_n(\mathbf{H},\mathbf{c}), q^\pi_{1-\alpha})\right] \to \mathbb{E}\left[f(W, \chi^2_{\text{rank}(\mathbf{H}),1-\alpha})\right] = P(W > \chi^2_{\text{rank}(\mathbf{H}),1-\alpha}) = \alpha$$

holds as $n \to \infty$. Here, we can use the regular expectation $\mathbb{E}$ instead of the outer expectation due to the measurability of $W_n(\mathbf{H},\mathbf{c})$ and $q^\pi_{1-\alpha}$.

## Proof of Theorem 4

By Slutsky's lemma, we combine Theorem 1 and Lemma S1 for Theorem 4. Since the map

$$\mathbb{R}^{kM} \times \mathbb{R}^{kM \times kM} \ni (\mathbf{m}, \mathbf{S}) \mapsto \left((\mathbf{H}_\ell \mathbf{m})'(\mathbf{H}_\ell \mathbf{S} \mathbf{H}_\ell')^+ \mathbf{H}_\ell \mathbf{m}\right)_{\ell \in \mathcal{T}} \in \mathbb{R}^{\mathcal{T}}$$

is continuous on $\mathbb{R}^{kM} \times \{\mathbf{\Sigma}\}$ due to Lemma S2, the statement follows by the continuous mapping theorem.

## Proof of Proposition 2

The following lemma is a version of Lemma S3 in [6] that helps to prove the convergence of the critical values.

**Lemma S8.** *Let $L \in \mathbb{N}$ and $\alpha \in (0,1)$. Moreover, let $F_n$ denote a map taking values in the space of all cumulative distribution functions on $\mathbb{R}^L$ and $F_{n,\ell}, \ell \in \{1, \ldots, L\}$ denote maps taking values in the space of all cumulative distribution functions on $\mathbb{R}$ for all $n \in \mathbb{N}$. Additionally, let $F : \mathbb{R}^L \to [0,1]$ denote a continuous cumulative distribution function with marginal distribution functions $F_1, \ldots, F_L : \mathbb{R} \to [0,1]$, where $F_\ell$ is strictly increasing on $\left[F_\ell^{-1}(a), F_\ell^{-1}(b)\right], \ell \in \{1, \ldots, L\}$ with*

$$0 < a < 1 - \alpha \leq 1 - \alpha/L < b < 1.$$

*If* (S10) *and*

$$F_{n,\ell}(t) \xrightarrow{P} F_\ell(t) \text{ as } n \to \infty \text{ for all } t \in \mathbb{R}; \ell \in \{1, \ldots, L\} \tag{S11}$$

*hold, we have*

$$F_{n,\ell}^{-1}\left(1 - \beta_n(\alpha)\right) \xrightarrow{P} F_\ell^{-1}\left(1 - \mathrm{FWER}^{-1}(\alpha)\right)$$

*as $n \to \infty$ for all $\ell \in \{1, \ldots, L\}$, where*

$$\mathrm{FWER} : \mathbb{R} \to [0,1], \quad \mathrm{FWER}(\beta) := 1 - F\left((F_\ell^{-1}(1-\beta))_{\ell \in \{1,\ldots,L\}}\right)$$

*and $\beta_n(\alpha) \in [0,1]$ satisfying $\beta_n(\alpha) \in \left[\mathrm{FWER}_n^{-1}(\alpha) - \varepsilon_n, \mathrm{FWER}_n^{-1}(\alpha + \varepsilon_n)\right]$ for all $n \in \mathbb{N}$ with*

$$\mathrm{FWER}_n(\beta) := 1 - F_n\left((F_{n,\ell}^{-1}(1-\beta))_{\ell \in \{1,\ldots,L\}}\right) \quad \text{for all } \beta \in [0,1]$$

*and some null sequence $(\varepsilon_n)_{n \in \mathbb{N}} \subset [0, \infty)$.*

The statement can be proved analogously to Lemma S3 in the supplement of [6]. However, we want to emphasize that all convergences stated there hold in outer probability. That is because

$$\sup_{\mathbf{t} \in \mathbb{R}^L} |F_n(\mathbf{t}) - F(\mathbf{t})| \xrightarrow{P} 0 \quad \text{as } n \to \infty$$

holds by Lemma S6 and there is an outer probability analogue of the subsequence criterion in Lemma 1.9.2 (ii) in [10].

**Remark 1.** *If $F_{n,1}, \ldots, F_{n,L}$ are the marginal cumulative distribution functions of $F_n$ in Lemma S8, (S11) is a direct consequence of (S10).*

In order to prove Proposition 2, let $F_n$ denote the empirical distribution function of

$$\left( (\mathbf{H}_\ell \hat{\boldsymbol{\Sigma}}^{1/2} \mathbf{Y}^{(b)})^\top (\mathbf{H}_\ell \hat{\boldsymbol{\Sigma}} \mathbf{H}_\ell^\top)^+ (\mathbf{H}_\ell \hat{\boldsymbol{\Sigma}}^{1/2} \mathbf{Y}^{(b)}) \right)_{\ell \{1,\ldots,L\}}, b \in \{1, \ldots, B_n\},$$

$F_{n,\ell}(.) = \chi^2_{\text{rank}(\mathbf{H}_\ell \hat{\boldsymbol{\Sigma}} \mathbf{H}_\ell^\top)}((-\infty, .])$ and $F$ being the distribution function of $(W_\ell)_{\ell \in \{1,\ldots,L\}}$. It should be noted that

$$\beta_n(\alpha) := \max \{\beta \in \{0, 1/B_n, ..., (B_n - 1)/B_n\} \mid \text{FWER}_n(\beta) \leqslant \alpha\}$$
$$= \begin{cases} \text{FWER}_n^{-1}(\alpha + 1/B_n) - 1/B_n & \text{if } \alpha B_n \in \mathbb{N} \\ \text{FWER}_n^{-1}(\alpha) - 1/B_n & \text{if } \alpha B_n \notin \mathbb{N} \end{cases}$$

suffices the conditions on $\beta_n(\alpha)$ in Lemma S8 with $\varepsilon_n = 1/B_n$ for all $n \in \mathbb{N}$. Due to Lemma S1, (S5) and Lemma S7, (S10) and (S11) hold. Thus, Lemma S8 provides

$$\chi^2_{\text{rank}(\mathbf{H}_\ell), 1-\beta_n(\alpha)} \xrightarrow{P} \chi^2_{\text{rank}(\mathbf{H}_\ell), 1-\text{FWER}^{-1}(\alpha)}$$

as $n \to \infty$ by (S5). Furthermore, it follows

$$\left( W_n(\mathbf{H}_\ell, \mathbf{c}_\ell), \chi^2_{\text{rank}(\mathbf{H}_\ell), 1-\beta_n(\alpha)} \right)_{\ell \in \mathcal{T}} \xrightarrow{d} \left( W_\ell, \chi^2_{\text{rank}(\mathbf{H}_\ell), 1-\text{FWER}^{-1}(\alpha)} \right)_{\ell \in \mathcal{T}}$$

as $n \to \infty$ under $\mathcal{H}_{0,\ell}, \ell \in \mathcal{T}$, by Theorem 4 and Slutsky's lemma. Similarly as in the proof of Proposition 1, we approximate the function $f : (w_\ell, q_\ell)_{\ell \in \mathcal{T}} \mapsto \max_{\ell \in \mathcal{T}} \mathbb{1}\{w_\ell > q_\ell\}$ through Lipschitz functions to obtain

$$\mathbb{E}\left( \max_{\ell \in \mathcal{T}} \varphi_\ell \right) = \mathbb{E}\left( f\left( \left( W_n(\mathbf{H}_\ell, \mathbf{c}_\ell), \chi^2_{\text{rank}(\mathbf{H}_\ell), 1-\beta_n(\alpha)} \right)_{\ell \in \mathcal{T}} \right) \right)$$
$$\to \mathbb{E}\left( f\left( \left( W_\ell, \chi^2_{\text{rank}(\mathbf{H}_\ell), 1-\text{FWER}^{-1}(\alpha)} \right)_{\ell \in \mathcal{T}} \right) \right)$$
$$\leqslant 1 - F\left( \left( \chi^2_{\text{rank}(\mathbf{H}_\ell), 1-\text{FWER}^{-1}(\alpha)} \right)_{\ell \in \{1,\ldots,L\}} \right)$$
$$= \text{FWER}(\text{FWER}^{-1}(\alpha)) = \alpha$$

as $n \to \infty$ holds under $\mathcal{H}_{0,\ell}, \ell \in \mathcal{T}$. Here, we can use the regular expectation $\mathbb{E}$ instead of the outer expectation due to the measurability of $W_n(\mathbf{H}_\ell, \mathbf{c}_\ell)$ and $\chi^2_{\text{rank}(\mathbf{H}_\ell), 1-\beta_n(\alpha)}$. The inequality is an equality if $\mathcal{T} = \{1, ..., L\}$.

# B  Details on the Simulation Study

## B.1  Simulation Settings

For the survival and censoring distributions, the same settings as in [5] are considered. Hence, the survival distributions are given by the following:

- Exponential distributions and early departures (*exp early*): $T_{11}, T_{21}, T_{31} \sim Exp(0.2)$ and $T_{41}$ with piece-wise constant hazard function $t \mapsto \lambda_{\delta,1} \cdot \mathbb{1}\{t \leqslant 2\} + 0.2 \cdot \mathbb{1}\{t > 2\}$,

- exponential distributions and late departures (*exp late*): $T_{11}, T_{21}, T_{31} \sim Exp(0.2)$ and $T_{41}$ with piece-wise constant hazard function $t \mapsto 0.2 \cdot \mathbb{1}\{t \leqslant 2\} + \lambda_{\delta,2} \cdot \mathbb{1}\{t > 2\}$,

- exponential distributions and proportional hazard alternative (*exp prop*): $T_{11}, T_{21}, T_{31} \sim Exp(0.2)$ and $T_{41} \sim Exp(\lambda_{\delta,3})$,

- lognormal distributions with scale alternatives (*logn*): $T_{11}, T_{21}, T_{31} \sim logN(2, 0.25)$ and $T_{41} \sim logN(\lambda_{\delta,4}, 0.25)$,

- exponential distributions and piece-wise exponential distributions (*pwExp*): $T_{11}, T_{21}, T_{31} \sim Exp(0.2)$ and $T_{41}$ with piece-wise constant hazard function $t \mapsto 0.5 \cdot \mathbb{1}\{t \leqslant \lambda_{\delta,5}\} + 0.05 \cdot \mathbb{1}\{t > \lambda_{\delta,5}\}$,

- Weibull distributions and late departures (*Weib late*): $T_{11}, T_{21}, T_{31} \sim Weib(3, 8)$ and $T_{41} \sim Weib(3 \cdot \lambda_{\delta,6}, 8/\lambda_{\delta,6})$,

- Weibull distributions and proportional hazard alternative (*Weib prop*): $T_{11}, T_{21}, T_{31} \sim Weib(3, 8)$ and $T_{41} \sim Weib(3, \lambda_{\delta,7})$,

- Weibull distributions with crossing curves and scale alternatives (*Weib scale*): $T_{11}, T_{21}, T_{31} \sim Weib(3, 8)$ and $T_{41} \sim Weib(1.5, \lambda_{\delta,8})$,

- Weibull distributions with crossing curves and shape alternatives (*Weib shape*): $T_{11}, T_{21}, T_{31} \sim Weib(3, 8)$ and $T_{41} \sim Weib(\lambda_{\delta,9}, 14)$.

The parameters $\lambda_{\delta,1}, \ldots, \lambda_{\delta,9}$ are determined as in [2] such that the restricted mean survival time difference equals $\delta = \int_{[0,\tau]} S_1(t)\mathrm{d}t - \int_{[0,\tau]} S_4(t)\mathrm{d}t$. As explained in Section 4.1, this results in RMTL differences of $\mu_{4m} - \mu_{1m} = p_m \delta$ for all events $m = 1, \ldots, M$.

The following three scenarios are chosen for the censoring times:

- Equally Weibull distributed censoring times (*equal*): $C_{11}, C_{21}, C_{31}, C_{41} \sim Weib(3, 10)$,

- unequally Weibull distributed censoring times with high censoring rates (*unequal, high*): $C_{11} \sim Weib(0.5, 15), C_{21} \sim Weib(0.5, 10), C_{31} \sim Weib(1, 8)$ and $C_{41} \sim Weib(1, 10)$,

- unequally Weibull distributed censoring times with low censoring rates (*unequal, low*): $C_{11} \sim Weib(1, 20), C_{21} \sim Weib(3, 10), C_{31} \sim Weib(1, 15)$ and $C_{41} \sim Weib(3, 20)$.

## B.2 Additional Figures of Simulation Results

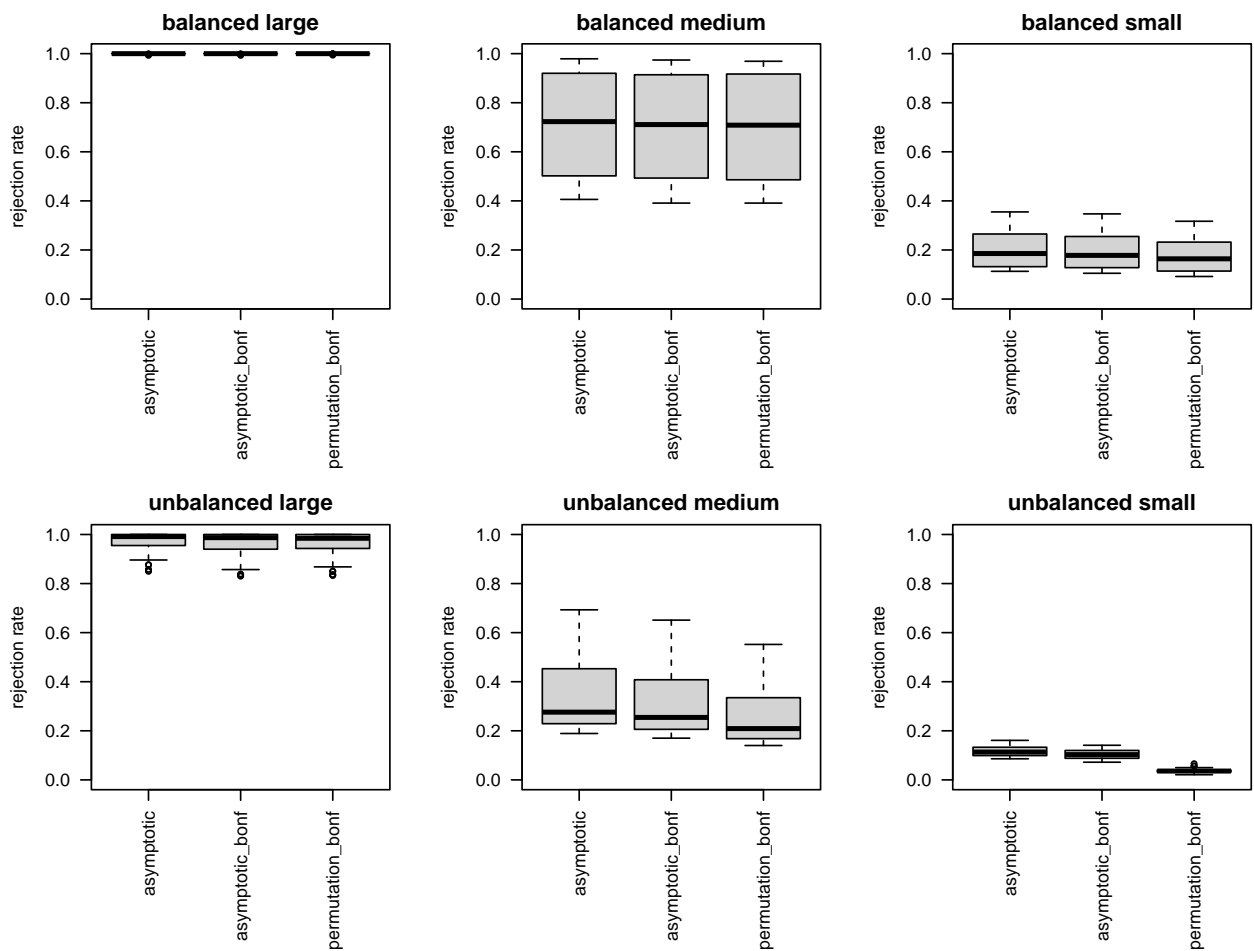

Figure S1: Empirical powers for the 2-by-2 design across all scenarios under the alternative hypothesis.

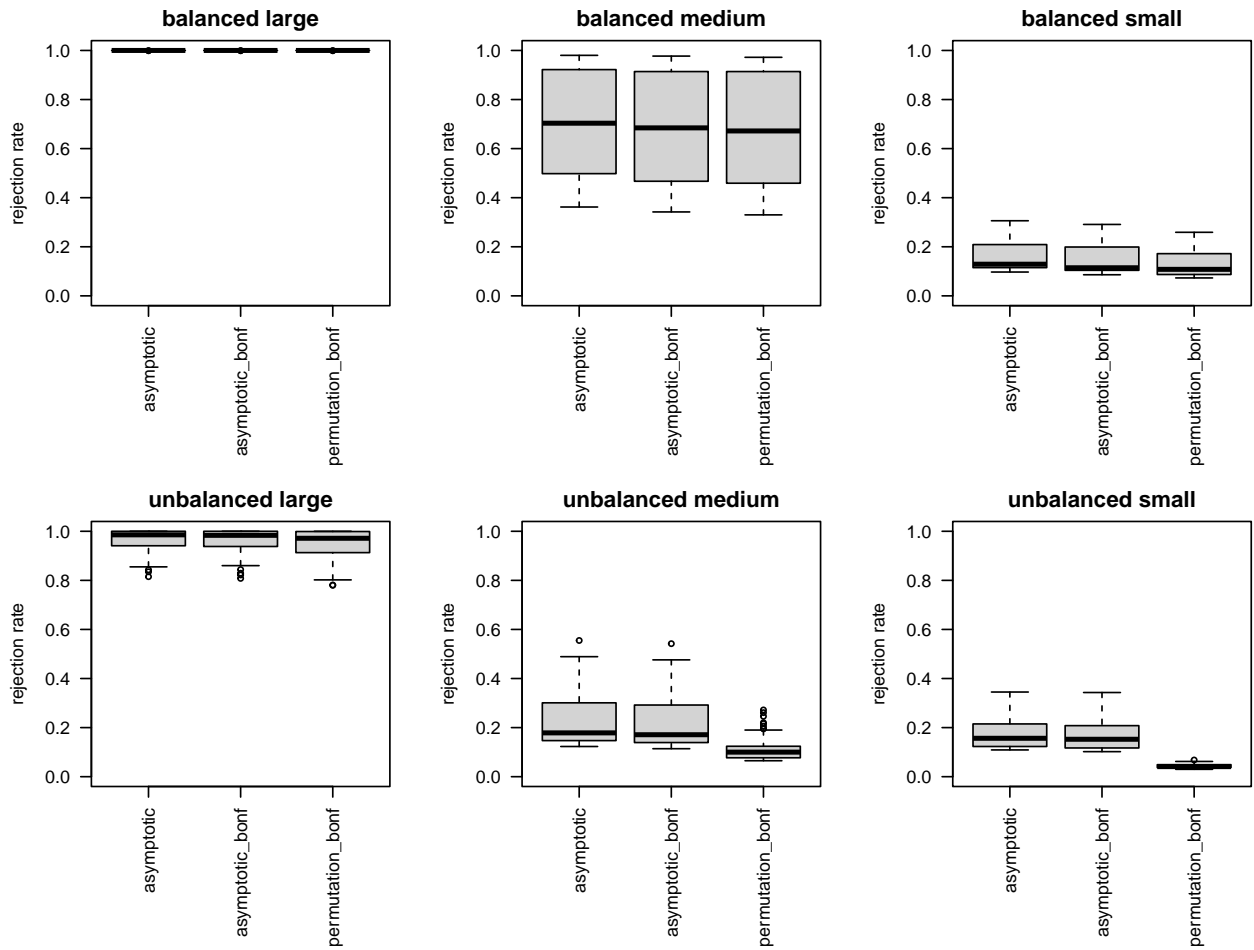

Figure S2: Empirical powers for the Dunnett-type contrast hypotheses across all scenarios under the alternative hypothesis.

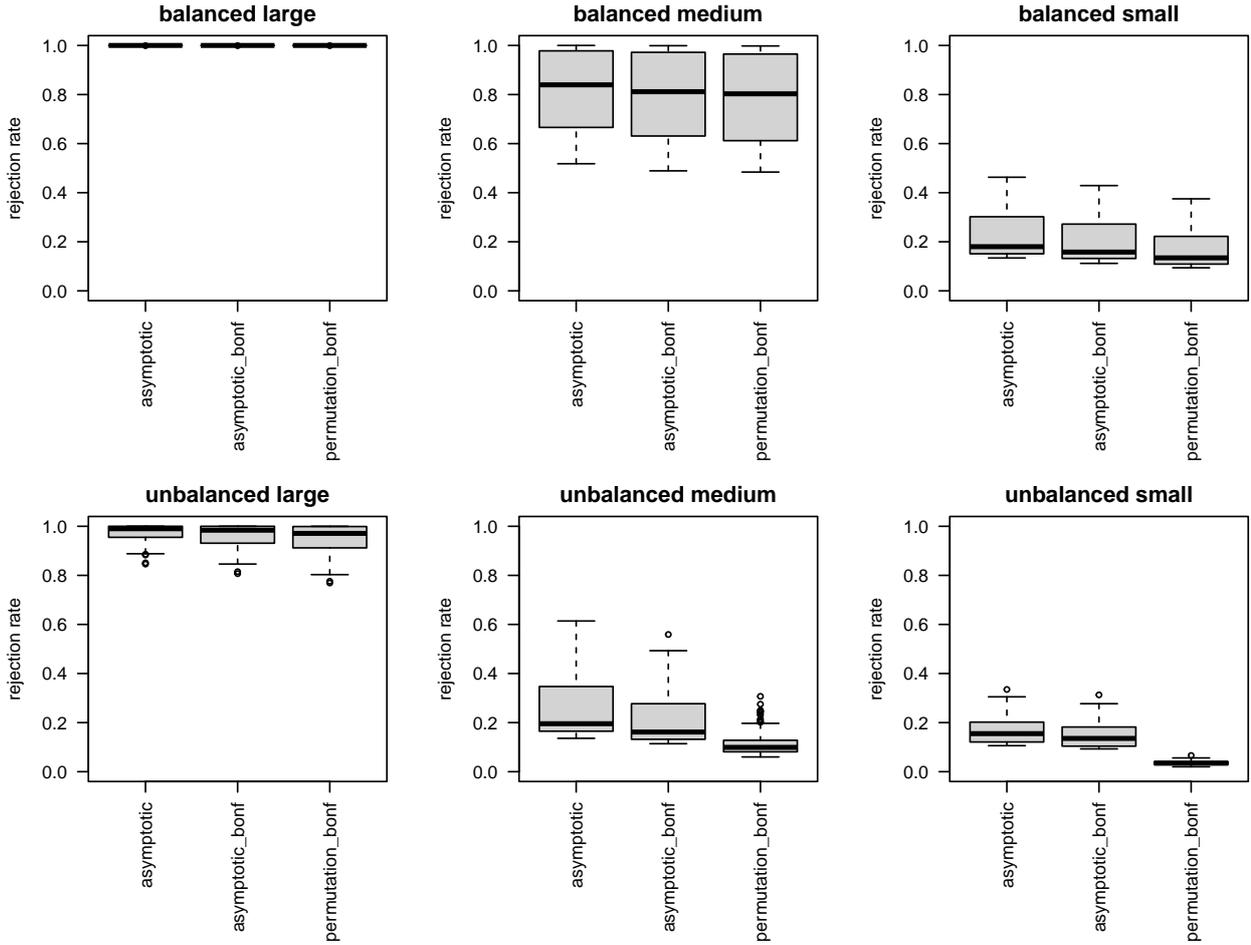

Figure S3: Empirical powers for the Tukey-type contrast hypotheses across all scenarios under the alternative hypothesis.

## B.3 Tables of Simulation Results

Tables S1–S36 contain the global rejection rates for all scenarios under the null and alternative hypothesis. For analyzing whether the false local hypotheses are rejected, Tables S37–S90 contain the rejection rates of all false hypotheses (under the alternative hypothesis). In detail, the following hypotheses are false under the alternative hypothesis:

- $\mathcal{H}_{0,1} = \mathcal{H}_{0,1}^A, \mathcal{H}_{0,2} = \mathcal{H}_{0,2}^A, \mathcal{H}_{0,3} = \mathcal{H}_{0,3}^A, \mathcal{H}_{0,4} = \mathcal{H}_{0,1}^B, \mathcal{H}_{0,5} = \mathcal{H}_{0,2}^B, \mathcal{H}_{0,6} = \mathcal{H}_{0,3}^B, \mathcal{H}_{0,7} = \mathcal{H}_{0,1}^{AB}, \mathcal{H}_{0,8} = \mathcal{H}_{0,2}^{AB}, \mathcal{H}_{0,9} = \mathcal{H}_{0,3}^{AB}$ for the 2-by-2 design,

- $\mathcal{H}_{0,7} : \mu_{11} = \mu_{41}, \mathcal{H}_{0,8} : \mu_{12} = \mu_{42}, \mathcal{H}_{0,9} : \mu_{13} = \mu_{43}$ for the Dunnett-type contrast matrix, and

- $\mathcal{H}_{0,7} : \mu_{11} = \mu_{41}, \mathcal{H}_{0,8} : \mu_{12} = \mu_{42}, \mathcal{H}_{0,9} : \mu_{13} = \mu_{43}, \mathcal{H}_{0,13} : \mu_{21} = \mu_{41}, \mathcal{H}_{0,14} : \mu_{22} = \mu_{42}, \mathcal{H}_{0,15} : \mu_{23} = \mu_{43}, \mathcal{H}_{0,16} : \mu_{31} = \mu_{41}, \mathcal{H}_{0,17} : \mu_{32} = \mu_{42}, \mathcal{H}_{0,18} : \mu_{33} = \mu_{43}$ for the Tukey-type contrast matrix.

Table S1: Rejection rates in percent for the 2-by-2 design with $\delta = 0.0$ and balanced large sample sizes.

| Distribution | Censoring distribution | asymptotic | asymptotic_bonf | permutation_bonf |
|---|---|---|---|---|
| exp early continuous | equal | **5.4** | **5.1** | **4.9** |
| exp early discrete | equal | **6.1** | **5.7** | **5.3** |
| exp early continuous | unequal, high | **5.3** | **5.2** | **5.3** |
| exp early discrete | unequal, high | **5.7** | **5.2** | **5.7** |
| exp early continuous | unequal, low | **6.0** | **5.7** | **5.9** |
| exp early discrete | unequal, low | **6.1** | **5.8** | **6.3** |
| logn continuous | equal | 4.7 | 4.5 | 5.1 |
| logn discrete | equal | 4.8 | 4.6 | 5.6 |
| logn continuous | unequal, high | 4.3 | 4.2 | 4.5 |
| logn discrete | unequal, high | 4.1 | 4.0 | 4.4 |
| logn continuous | unequal, low | 4.5 | 4.4 | 5.5 |
| logn discrete | unequal, low | 4.7 | 4.7 | 5.0 |
| pwExp continuous | equal | **5.8** | **5.5** | **5.9** |
| pwExp discrete | equal | **5.7** | **5.6** | **6.0** |
| pwExp continuous | unequal, high | **6.0** | **5.9** | **6.3** |
| pwExp discrete | unequal, high | **5.3** | **5.2** | **6.2** |
| pwExp continuous | unequal, low | **5.7** | **5.4** | **5.5** |
| pwExp discrete | unequal, low | **5.9** | **5.5** | **5.7** |
| Weib prop continuous | equal | 4.9 | 4.6 | **5.5** |
| Weib prop discrete | equal | **5.1** | 4.7 | 4.7 |
| Weib prop continuous | unequal, high | **5.3** | **5.2** | **5.3** |
| Weib prop discrete | unequal, high | **5.7** | **5.5** | **5.8** |
| Weib prop continuous | unequal, low | 4.4 | 4.4 | **5.1** |
| Weib prop discrete | unequal, low | **5.1** | 4.9 | **5.2** |
| Weib scale continuous | equal | 4.3 | 4.1 | 4.9 |
| Weib scale discrete | equal | 4.7 | 4.6 | 5.0 |
| Weib scale continuous | unequal, high | **5.6** | **5.6** | **5.5** |
| Weib scale discrete | unequal, high | **6.1** | **5.8** | **5.8** |
| Weib scale continuous | unequal, low | 4.6 | 4.5 | 4.8 |
| Weib scale discrete | unequal, low | 4.6 | 4.5 | 4.7 |
| Weib shape continuous | equal | 4.5 | 4.1 | 4.4 |
| Weib shape discrete | equal | 4.9 | 4.7 | 4.3 |
| Weib shape continuous | unequal, high | **6.1** | **5.6** | **6.2** |
| Weib shape discrete | unequal, high | **5.9** | **5.7** | **5.7** |
| Weib shape continuous | unequal, low | 5.0 | 4.4 | 4.8 |
| Weib shape discrete | unequal, low | **5.1** | 4.9 | 4.6 |

All values in the binomial interval $[3.7, 6.4]$ are printed in bold type.

Table S2: Rejection rates in percent for the 2-by-2 design with $\delta = 0.0$ and balanced medium sample sizes.

| Distribution | Censoring distribution | asymptotic | asymptotic_bonf | permutation_bonf |
|---|---|---|---|---|
| exp early continuous | equal | **4.9** | **4.8** | **4.8** |
| exp early discrete | equal | **5.0** | **4.7** | **4.8** |
| exp early continuous | unequal, high | **5.2** | **5.1** | **5.5** |
| exp early discrete | unequal, high | **5.2** | **5.0** | **4.8** |
| exp early continuous | unequal, low | **4.7** | **4.6** | **5.1** |
| exp early discrete | unequal, low | **5.3** | **5.0** | **5.4** |
| logn continuous | equal | **5.5** | **5.4** | **5.8** |
| logn discrete | equal | **4.8** | **4.8** | **5.0** |
| logn continuous | unequal, high | **4.1** | **3.8** | **4.9** |
| logn discrete | unequal, high | **5.0** | **4.9** | **5.4** |
| logn continuous | unequal, low | **4.4** | **4.3** | **4.3** |
| logn discrete | unequal, low | **4.0** | **3.7** | **4.3** |
| pwExp continuous | equal | **5.3** | **4.8** | **4.9** |
| pwExp discrete | equal | **5.0** | **5.0** | **4.4** |
| pwExp continuous | unequal, high | **4.3** | **4.3** | **4.7** |
| pwExp discrete | unequal, high | **4.4** | **4.2** | **4.7** |
| pwExp continuous | unequal, low | **5.3** | **4.9** | **5.4** |
| pwExp discrete | unequal, low | **5.0** | **4.7** | **4.6** |
| Weib prop continuous | equal | **5.0** | **5.0** | **5.1** |
| Weib prop discrete | equal | **5.5** | **4.9** | **5.3** |
| Weib prop continuous | unequal, high | **4.6** | **4.3** | **5.0** |
| Weib prop discrete | unequal, high | **5.2** | **5.2** | **4.6** |
| Weib prop continuous | unequal, low | **5.4** | **5.0** | **5.6** |
| Weib prop discrete | unequal, low | **5.1** | **4.9** | **5.1** |
| Weib scale continuous | equal | **6.0** | **5.5** | **5.5** |
| Weib scale discrete | equal | **5.1** | **4.9** | **5.1** |
| Weib scale continuous | unequal, high | **5.5** | **5.3** | **5.0** |
| Weib scale discrete | unequal, high | **5.7** | **5.6** | **5.8** |
| Weib scale continuous | unequal, low | **5.1** | **5.1** | **5.6** |
| Weib scale discrete | unequal, low | **4.9** | **4.8** | **5.0** |
| Weib shape continuous | equal | **5.8** | **5.8** | **5.6** |
| Weib shape discrete | equal | **5.5** | **5.3** | **5.3** |
| Weib shape continuous | unequal, high | **5.5** | **5.4** | **5.4** |
| Weib shape discrete | unequal, high | **5.1** | **4.9** | **4.9** |
| Weib shape continuous | unequal, low | **5.2** | **5.1** | **5.5** |
| Weib shape discrete | unequal, low | **4.9** | **4.7** | **5.0** |

All values in the binomial interval $[3.7, 6.4]$ are printed in bold type.

Table S3: Rejection rates in percent for the 2-by-2 design with $\delta = 0.0$ and balanced small sample sizes.

| Distribution | Censoring distribution | asymptotic | asymptotic_bonf | permutation_bonf |
|---|---|---|---|---|
| exp early continuous | equal | **5.4** | **5.1** | **4.8** |
| exp early discrete | equal | **5.5** | **5.3** | **4.6** |
| exp early continuous | unequal, high | **5.4** | **5.3** | **4.2** |
| exp early discrete | unequal, high | **6.0** | **6.0** | **4.7** |
| exp early continuous | unequal, low | **6.0** | **5.4** | **5.2** |
| exp early discrete | unequal, low | **6.1** | **6.0** | **4.9** |
| logn continuous | equal | **5.9** | **5.5** | **5.8** |
| logn discrete | equal | **6.4** | **6.1** | **6.2** |
| logn continuous | unequal, high | **6.2** | **6.0** | **4.8** |
| logn discrete | unequal, high | **5.9** | **5.7** | **4.7** |
| logn continuous | unequal, low | **6.0** | **5.9** | **5.5** |
| logn discrete | unequal, low | **6.0** | **5.9** | **5.5** |
| pwExp continuous | equal | **5.7** | **5.4** | **4.3** |
| pwExp discrete | equal | **5.2** | **5.0** | **4.2** |
| pwExp continuous | unequal, high | **5.1** | **4.8** | **4.4** |
| pwExp discrete | unequal, high | **4.9** | **4.5** | **4.0** |
| pwExp continuous | unequal, low | **5.4** | **4.9** | **4.5** |
| pwExp discrete | unequal, low | **5.6** | **5.5** | **4.5** |
| Weib prop continuous | equal | **6.1** | **5.9** | **5.7** |
| Weib prop discrete | equal | **5.8** | **5.6** | **4.7** |
| Weib prop continuous | unequal, high | **6.3** | **6.0** | **5.6** |
| Weib prop discrete | unequal, high | 6.5 | **6.1** | **4.7** |
| Weib prop continuous | unequal, low | 7.2 | 6.8 | **5.9** |
| Weib prop discrete | unequal, low | **5.9** | **5.7** | **5.8** |
| Weib scale continuous | equal | 6.9 | 6.9 | 6.5 |
| Weib scale discrete | equal | **5.5** | **5.4** | 5.7 |
| Weib scale continuous | unequal, high | 7.1 | **6.4** | **5.3** |
| Weib scale discrete | unequal, high | **6.2** | **6.2** | **5.3** |
| Weib scale continuous | unequal, low | **6.4** | **6.3** | 6.6 |
| Weib scale discrete | unequal, low | **6.2** | **5.9** | **6.0** |
| Weib shape continuous | equal | 6.5 | **6.3** | **6.4** |
| Weib shape discrete | equal | **6.3** | **6.0** | **5.7** |
| Weib shape continuous | unequal, high | **6.4** | **6.4** | **6.0** |
| Weib shape discrete | unequal, high | 6.9 | 6.8 | **5.4** |
| Weib shape continuous | unequal, low | 6.5 | **6.4** | 6.5 |
| Weib shape discrete | unequal, low | **6.2** | **6.2** | **5.6** |

All values in the binomial interval $[3.7, 6.4]$ are printed in bold type.

Table S4: Rejection rates in percent for the 2-by-2 design with $\delta = 0.0$ and unbalanced large sample sizes.

| Distribution | Censoring distribution | asymptotic | asymptotic_bonf | permutation_bonf |
|---|---|---|---|---|
| exp early continuous | equal | **4.3** | **4.0** | **4.2** |
| exp early discrete | equal | **5.1** | **4.8** | **4.6** |
| exp early continuous | unequal, high | **5.4** | **4.6** | **4.7** |
| exp early discrete | unequal, high | **5.6** | **5.0** | **4.9** |
| exp early continuous | unequal, low | **5.3** | **4.9** | **5.3** |
| exp early discrete | unequal, low | **5.3** | **5.0** | **5.5** |
| logn continuous | equal | **4.9** | **4.3** | **4.3** |
| logn discrete | equal | **5.0** | **4.8** | **4.6** |
| logn continuous | unequal, high | **5.1** | **4.5** | **5.0** |
| logn discrete | unequal, high | **4.9** | **4.4** | **5.5** |
| logn continuous | unequal, low | **5.1** | **4.8** | **4.6** |
| logn discrete | unequal, low | **5.0** | **4.8** | **4.5** |
| pwExp continuous | equal | **5.7** | **5.1** | **4.7** |
| pwExp discrete | equal | **5.5** | **4.8** | **4.7** |
| pwExp continuous | unequal, high | **3.8** | **3.7** | **4.1** |
| pwExp discrete | unequal, high | **4.1** | 3.6 | **4.4** |
| pwExp continuous | unequal, low | **5.6** | **4.8** | **4.5** |
| pwExp discrete | unequal, low | **5.5** | **4.9** | **4.5** |
| Weib prop continuous | equal | **5.2** | **4.8** | **4.7** |
| Weib prop discrete | equal | **5.5** | **5.1** | **4.7** |
| Weib prop continuous | unequal, high | **4.8** | **4.5** | **4.6** |
| Weib prop discrete | unequal, high | **6.1** | **5.8** | **4.6** |
| Weib prop continuous | unequal, low | **5.0** | **4.5** | **4.4** |
| Weib prop discrete | unequal, low | **4.6** | **4.6** | **4.7** |
| Weib scale continuous | equal | **4.8** | **4.7** | **4.4** |
| Weib scale discrete | equal | **4.8** | **4.3** | **4.7** |
| Weib scale continuous | unequal, high | **5.0** | **4.6** | **4.9** |
| Weib scale discrete | unequal, high | **4.6** | **4.3** | **3.7** |
| Weib scale continuous | unequal, low | **4.3** | **4.1** | **4.0** |
| Weib scale discrete | unequal, low | **4.3** | **4.2** | **4.1** |
| Weib shape continuous | equal | **5.1** | **4.4** | **4.4** |
| Weib shape discrete | equal | **5.0** | **4.4** | **4.3** |
| Weib shape continuous | unequal, high | **5.3** | **4.3** | **4.7** |
| Weib shape discrete | unequal, high | **4.9** | **4.6** | **4.0** |
| Weib shape continuous | unequal, low | **4.4** | **4.2** | **4.2** |
| Weib shape discrete | unequal, low | **4.2** | **4.1** | **4.1** |

All values in the binomial interval $[3.7, 6.4]$ are printed in bold type.

Table S5: Rejection rates in percent for the 2-by-2 design with $\delta = 0.0$ and unbalanced medium sample sizes.

| Distribution | Censoring distribution | asymptotic | asymptotic_bonf | permutation_bonf |
|---|---|---|---|---|
| exp early continuous | equal | 6.6 | **5.9** | **5.4** |
| exp early discrete | equal | **6.2** | **6.0** | **5.4** |
| exp early continuous | unequal, high | **5.5** | **5.4** | **4.7** |
| exp early discrete | unequal, high | **6.0** | **5.6** | **4.5** |
| exp early continuous | unequal, low | **5.9** | **5.3** | **4.4** |
| exp early discrete | unequal, low | **5.7** | **5.5** | **5.0** |
| logn continuous | equal | 8.1 | 7.7 | **5.8** |
| logn discrete | equal | 7.3 | 7.0 | **5.7** |
| logn continuous | unequal, high | 7.8 | 7.6 | **5.4** |
| logn discrete | unequal, high | 7.4 | 6.6 | **4.8** |
| logn continuous | unequal, low | 6.9 | 6.5 | **4.8** |
| logn discrete | unequal, low | **6.3** | **6.0** | **4.0** |
| pwExp continuous | equal | **5.3** | **4.9** | **4.5** |
| pwExp discrete | equal | **5.4** | **5.0** | **4.6** |
| pwExp continuous | unequal, high | **5.4** | **5.2** | **4.4** |
| pwExp discrete | unequal, high | **5.0** | **4.7** | **4.0** |
| pwExp continuous | unequal, low | **5.7** | **4.8** | **4.6** |
| pwExp discrete | unequal, low | **6.0** | **5.5** | **4.9** |
| Weib prop continuous | equal | **5.3** | **4.8** | **4.3** |
| Weib prop discrete | equal | **5.5** | **5.5** | **4.5** |
| Weib prop continuous | unequal, high | 7.5 | 6.7 | **5.0** |
| Weib prop discrete | unequal, high | 7.2 | 6.5 | **4.7** |
| Weib prop continuous | unequal, low | **5.9** | **5.4** | 3.8 |
| Weib prop discrete | unequal, low | **6.2** | **6.1** | **4.1** |
| Weib scale continuous | equal | 7.0 | **6.3** | **5.1** |
| Weib scale discrete | equal | 7.5 | 7.3 | **5.9** |
| Weib scale continuous | unequal, high | 8.4 | 8.1 | **5.5** |
| Weib scale discrete | unequal, high | 8.2 | 7.7 | **5.3** |
| Weib scale continuous | unequal, low | 6.7 | **6.4** | **5.0** |
| Weib scale discrete | unequal, low | 7.5 | 7.0 | **5.8** |
| Weib shape continuous | equal | 7.7 | 6.7 | **5.5** |
| Weib shape discrete | equal | 8.4 | 7.1 | **6.0** |
| Weib shape continuous | unequal, high | 7.8 | 7.3 | **4.2** |
| Weib shape discrete | unequal, high | 7.9 | 7.6 | **5.5** |
| Weib shape continuous | unequal, low | 7.6 | 6.9 | **5.3** |
| Weib shape discrete | unequal, low | 7.8 | 7.6 | **5.9** |

All values in the binomial interval $[3.7, 6.4]$ are printed in bold type.

Table S6: Rejection rates in percent for the 2-by-2 design with $\delta = 0.0$ and unbalanced small sample sizes.

| Distribution | Censoring distribution | asymptotic | asymptotic_bonf | permutation_bonf |
|---|---|---|---|---|
| exp early continuous | equal | 12.4 | 11.0 | **5.5** |
| exp early discrete | equal | 12.7 | 12.2 | **5.7** |
| exp early continuous | unequal, high | 12.4 | 11.2 | **3.8** |
| exp early discrete | unequal, high | 10.9 | 10.2 | **3.9** |
| exp early continuous | unequal, low | 11.1 | 10.5 | **4.4** |
| exp early discrete | unequal, low | 11.5 | 10.8 | **4.3** |
| logn continuous | equal | 15.1 | 14.3 | **6.1** |
| logn discrete | equal | 14.9 | 14.3 | **5.5** |
| logn continuous | unequal, high | 17.2 | 16.0 | **4.9** |
| logn discrete | unequal, high | 16.6 | 15.8 | **5.8** |
| logn continuous | unequal, low | 11.7 | 11.5 | **4.3** |
| logn discrete | unequal, low | 11.7 | 11.4 | **3.9** |
| pwExp continuous | equal | 10.9 | 10.0 | **4.7** |
| pwExp discrete | equal | 11.5 | 10.8 | **5.4** |
| pwExp continuous | unequal, high | 12.2 | 11.4 | 3.5 |
| pwExp discrete | unequal, high | 12.2 | 11.8 | **4.4** |
| pwExp continuous | unequal, low | 10.4 | 9.8 | **5.0** |
| pwExp discrete | unequal, low | 10.7 | 9.9 | **4.7** |
| Weib prop continuous | equal | 13.5 | 11.9 | **5.1** |
| Weib prop discrete | equal | 13.8 | 13.1 | **4.9** |
| Weib prop continuous | unequal, high | 17.8 | 16.8 | **5.1** |
| Weib prop discrete | unequal, high | 16.8 | 15.7 | **4.8** |
| Weib prop continuous | unequal, low | 10.1 | 9.1 | 2.6 |
| Weib prop discrete | unequal, low | 10.5 | 9.7 | 3.4 |
| Weib scale continuous | equal | 19.8 | 18.6 | 8.0 |
| Weib scale discrete | equal | 20.2 | 18.9 | 7.0 |
| Weib scale continuous | unequal, high | 19.2 | 17.8 | **6.4** |
| Weib scale discrete | unequal, high | 19.6 | 18.0 | **6.1** |
| Weib scale continuous | unequal, low | 14.8 | 14.3 | **5.5** |
| Weib scale discrete | unequal, low | 14.6 | 14.0 | **5.6** |
| Weib shape continuous | equal | 23.3 | 21.7 | 10.0 |
| Weib shape discrete | equal | 22.8 | 21.1 | 9.2 |
| Weib shape continuous | unequal, high | 21.6 | 19.8 | 6.7 |
| Weib shape discrete | unequal, high | 21.3 | 19.4 | 6.5 |
| Weib shape continuous | unequal, low | 19.0 | 18.1 | 7.5 |
| Weib shape discrete | unequal, low | 18.8 | 17.6 | 7.1 |

All values in the binomial interval $[3.7, 6.4]$ are printed in bold type.

Table S7: Rejection rates in percent for the Dunnett-type contrast matrix with $\delta = 0.0$ and balanced large sample sizes.

| Distribution | Censoring distribution | asymptotic | asymptotic_bonf | permutation_bonf |
|---|---|---|---|---|
| exp early continuous | equal | **5.7** | **5.1** | **5.1** |
| exp early discrete | equal | **5.3** | **4.8** | **5.0** |
| exp early continuous | unequal, high | **5.5** | **5.1** | **5.1** |
| exp early discrete | unequal, high | **5.8** | **5.4** | **5.3** |
| exp early continuous | unequal, low | **5.5** | **5.0** | **5.5** |
| exp early discrete | unequal, low | **5.6** | **4.7** | **5.0** |
| logn continuous | equal | **4.4** | **4.1** | **3.9** |
| logn discrete | equal | **5.1** | **4.5** | **4.3** |
| logn continuous | unequal, high | **4.9** | **4.5** | **4.6** |
| logn discrete | unequal, high | **4.8** | **4.4** | **4.3** |
| logn continuous | unequal, low | **4.6** | **4.0** | **4.8** |
| logn discrete | unequal, low | **4.9** | **4.3** | **4.9** |
| pwExp continuous | equal | **5.5** | **5.0** | **5.6** |
| pwExp discrete | equal | **5.3** | **4.8** | **5.2** |
| pwExp continuous | unequal, high | **5.1** | **4.6** | **4.6** |
| pwExp discrete | unequal, high | **4.9** | **4.5** | **5.0** |
| pwExp continuous | unequal, low | **4.7** | **4.4** | **5.0** |
| pwExp discrete | unequal, low | **5.2** | **4.8** | **4.9** |
| Weib prop continuous | equal | **5.1** | **4.6** | **4.9** |
| Weib prop discrete | equal | **4.3** | **4.1** | **4.8** |
| Weib prop continuous | unequal, high | **5.3** | **5.2** | **5.3** |
| Weib prop discrete | unequal, high | **5.8** | **5.4** | **5.9** |
| Weib prop continuous | unequal, low | **5.2** | **4.9** | **4.3** |
| Weib prop discrete | unequal, low | **5.3** | **4.7** | **4.8** |
| Weib scale continuous | equal | **4.9** | **4.6** | **4.7** |
| Weib scale discrete | equal | **4.4** | 3.8 | **4.2** |
| Weib scale continuous | unequal, high | **4.8** | **4.7** | **4.6** |
| Weib scale discrete | unequal, high | **5.6** | **5.4** | **5.3** |
| Weib scale continuous | unequal, low | **5.0** | **4.8** | **4.5** |
| Weib scale discrete | unequal, low | **5.2** | **4.5** | **4.7** |
| Weib shape continuous | equal | **5.1** | **4.8** | **4.7** |
| Weib shape discrete | equal | **4.5** | 3.8 | **4.5** |
| Weib shape continuous | unequal, high | **5.1** | **4.9** | **5.2** |
| Weib shape discrete | unequal, high | **5.3** | **4.9** | **5.2** |
| Weib shape continuous | unequal, low | **5.2** | **5.0** | **5.1** |
| Weib shape discrete | unequal, low | **5.3** | **4.6** | **5.0** |

All values in the binomial interval $[3.7, 6.4]$ are printed in bold type.

Table S8: Rejection rates in percent for the Dunnett-type contrast matrix with $\delta = 0.0$ and balanced medium sample sizes.

| Distribution | Censoring distribution | asymptotic | asymptotic_bonf | permutation_bonf |
|---|---|---|---|---|
| exp early continuous | equal | **5.6** | **4.7** | **4.6** |
| exp early discrete | equal | **5.3** | **4.9** | **5.3** |
| exp early continuous | unequal, high | 6.6 | **5.6** | **5.4** |
| exp early discrete | unequal, high | **5.1** | **4.7** | **5.5** |
| exp early continuous | unequal, low | **5.8** | **5.2** | **6.2** |
| exp early discrete | unequal, low | **6.1** | **5.3** | **6.2** |
| logn continuous | equal | **4.5** | **3.9** | **4.0** |
| logn discrete | equal | **5.4** | **4.8** | **5.2** |
| logn continuous | unequal, high | **4.5** | **4.1** | **4.0** |
| logn discrete | unequal, high | **5.1** | **4.8** | **4.7** |
| logn continuous | unequal, low | **4.9** | **4.2** | **4.5** |
| logn discrete | unequal, low | **5.5** | **5.1** | **5.3** |
| pwExp continuous | equal | **6.1** | **5.3** | **5.6** |
| pwExp discrete | equal | 6.5 | **5.6** | **5.7** |
| pwExp continuous | unequal, high | **5.5** | **4.6** | **5.1** |
| pwExp discrete | unequal, high | **6.4** | **5.4** | **4.9** |
| pwExp continuous | unequal, low | 7.4 | **6.1** | **5.9** |
| pwExp discrete | unequal, low | 6.9 | **6.4** | **6.2** |
| Weib prop continuous | equal | **5.0** | **4.4** | **5.1** |
| Weib prop discrete | equal | **4.8** | **4.4** | **5.5** |
| Weib prop continuous | unequal, high | **4.8** | **4.0** | **4.2** |
| Weib prop discrete | unequal, high | **5.0** | **4.2** | **3.9** |
| Weib prop continuous | unequal, low | **5.1** | **4.6** | **5.0** |
| Weib prop discrete | unequal, low | **4.8** | **4.0** | **4.7** |
| Weib scale continuous | equal | **5.3** | **4.7** | **5.0** |
| Weib scale discrete | equal | **4.8** | **4.3** | **5.1** |
| Weib scale continuous | unequal, high | **5.0** | **4.4** | **4.4** |
| Weib scale discrete | unequal, high | **5.6** | **5.0** | **4.8** |
| Weib scale continuous | unequal, low | **5.5** | **4.8** | **4.8** |
| Weib scale discrete | unequal, low | **4.7** | **4.3** | **5.3** |
| Weib shape continuous | equal | **5.4** | **4.7** | **5.1** |
| Weib shape discrete | equal | **4.6** | **4.1** | **5.1** |
| Weib shape continuous | unequal, high | **5.1** | **4.6** | **5.2** |
| Weib shape discrete | unequal, high | **5.2** | **4.9** | **5.0** |
| Weib shape continuous | unequal, low | **5.5** | **4.9** | **5.2** |
| Weib shape discrete | unequal, low | **4.6** | **4.4** | **4.8** |

All values in the binomial interval $[3.7, 6.4]$ are printed in bold type.

Table S9: Rejection rates in percent for the Dunnett-type contrast matrix with $\delta = 0.0$ and balanced small sample sizes.

| Distribution | Censoring distribution | asymptotic | asymptotic_bonf | permutation_bonf |
|---|---|---|---|---|
| exp early continuous | equal | **5.6** | **5.2** | **4.7** |
| exp early discrete | equal | **5.5** | **5.1** | **4.8** |
| exp early continuous | unequal, high | **6.4** | **6.2** | **4.7** |
| exp early discrete | unequal, high | 7.0 | **6.2** | **4.9** |
| exp early continuous | unequal, low | **5.4** | **4.6** | **4.2** |
| exp early discrete | unequal, low | **5.1** | **4.5** | **4.0** |
| logn continuous | equal | **5.5** | **4.8** | **4.0** |
| logn discrete | equal | **4.9** | **4.3** | 3.5 |
| logn continuous | unequal, high | 7.0 | 6.6 | **5.7** |
| logn discrete | unequal, high | 7.3 | 6.7 | **6.1** |
| logn continuous | unequal, low | **5.4** | **4.5** | **4.1** |
| logn discrete | unequal, low | **4.5** | **4.0** | 3.7 |
| pwExp continuous | equal | 6.6 | 5.5 | **4.9** |
| pwExp discrete | equal | **6.2** | **5.2** | **4.8** |
| pwExp continuous | unequal, high | **6.4** | **5.2** | 3.9 |
| pwExp discrete | unequal, high | **6.4** | **5.4** | **4.1** |
| pwExp continuous | unequal, low | 6.7 | **5.8** | **5.4** |
| pwExp discrete | unequal, low | **6.3** | **5.6** | **4.7** |
| Weib prop continuous | equal | 7.3 | 6.6 | 6.5 |
| Weib prop discrete | equal | 6.7 | **5.7** | **5.2** |
| Weib prop continuous | unequal, high | 6.6 | **5.9** | **4.6** |
| Weib prop discrete | unequal, high | **6.1** | **5.7** | **5.2** |
| Weib prop continuous | unequal, low | 6.9 | 6.5 | **5.9** |
| Weib prop discrete | unequal, low | 7.0 | **6.3** | **5.7** |
| Weib scale continuous | equal | 7.6 | 7.1 | 6.7 |
| Weib scale discrete | equal | 6.7 | **6.0** | **5.4** |
| Weib scale continuous | unequal, high | 6.7 | **6.1** | **4.5** |
| Weib scale discrete | unequal, high | **5.9** | **5.4** | **4.6** |
| Weib scale continuous | unequal, low | 7.0 | 6.5 | **6.2** |
| Weib scale discrete | unequal, low | 6.9 | **6.1** | **5.5** |
| Weib shape continuous | equal | 7.5 | 6.9 | 6.5 |
| Weib shape discrete | equal | 6.5 | **6.0** | **5.7** |
| Weib shape continuous | unequal, high | 6.9 | **6.4** | **5.3** |
| Weib shape discrete | unequal, high | 6.7 | **6.2** | **5.5** |
| Weib shape continuous | unequal, low | 6.8 | 6.5 | **6.4** |
| Weib shape discrete | unequal, low | **6.3** | **5.6** | **5.6** |

All values in the binomial interval $[3.7, 6.4]$ are printed in bold type.

Table S10: Rejection rates in percent for the Dunnett-type contrast matrix with $\delta = 0.0$ and unbalanced large sample sizes.

| Distribution | Censoring distribution | asymptotic | asymptotic_bonf | permutation_bonf |
|---|---|---|---|---|
| exp early continuous | equal | **5.5** | **5.4** | **5.3** |
| exp early discrete | equal | **5.7** | **5.5** | **5.6** |
| exp early continuous | unequal, high | **4.8** | **4.7** | **4.4** |
| exp early discrete | unequal, high | **5.6** | **5.2** | **5.6** |
| exp early continuous | unequal, low | **6.2** | **5.6** | **5.5** |
| exp early discrete | unequal, low | **6.3** | **6.0** | **6.2** |
| logn continuous | equal | 7.9 | 7.6 | 6.6 |
| logn discrete | equal | 6.7 | 6.6 | 6.8 |
| logn continuous | unequal, high | 7.3 | 7.0 | **6.2** |
| logn discrete | unequal, high | 7.3 | 7.1 | 5.4 |
| logn continuous | unequal, low | 7.1 | 6.9 | **6.0** |
| logn discrete | unequal, low | 6.6 | 6.5 | **5.6** |
| pwExp continuous | equal | **5.8** | **5.4** | **5.2** |
| pwExp discrete | equal | **5.4** | **5.1** | **5.4** |
| pwExp continuous | unequal, high | **5.1** | **5.0** | **5.2** |
| pwExp discrete | unequal, high | **5.7** | **5.7** | **5.8** |
| pwExp continuous | unequal, low | **6.1** | **5.9** | **5.4** |
| pwExp discrete | unequal, low | **6.0** | **5.8** | **5.6** |
| Weib prop continuous | equal | **6.0** | **5.7** | **5.6** |
| Weib prop discrete | equal | **6.0** | **5.8** | **5.7** |
| Weib prop continuous | unequal, high | **4.9** | **4.5** | **5.1** |
| Weib prop discrete | unequal, high | **5.9** | **5.6** | **5.0** |
| Weib prop continuous | unequal, low | **5.1** | **4.9** | **4.7** |
| Weib prop discrete | unequal, low | **6.0** | **5.5** | **5.6** |
| Weib scale continuous | equal | **5.5** | **5.2** | **4.9** |
| Weib scale discrete | equal | **5.9** | **5.7** | **5.9** |
| Weib scale continuous | unequal, high | **4.7** | **4.5** | **4.1** |
| Weib scale discrete | unequal, high | **5.5** | **5.3** | **4.8** |
| Weib scale continuous | unequal, low | **5.0** | **4.8** | **4.9** |
| Weib scale discrete | unequal, low | **5.8** | **5.5** | **5.0** |
| Weib shape continuous | equal | **5.7** | **5.3** | **5.8** |
| Weib shape discrete | equal | **6.1** | **5.9** | **6.0** |
| Weib shape continuous | unequal, high | **5.3** | **5.1** | **5.1** |
| Weib shape discrete | unequal, high | **6.0** | **5.8** | **5.2** |
| Weib shape continuous | unequal, low | **5.1** | **5.0** | **4.7** |
| Weib shape discrete | unequal, low | **6.1** | **5.8** | **5.5** |

All values in the binomial interval $[3.7, 6.4]$ are printed in bold type.

Table S11: Rejection rates in percent for the Dunnett-type contrast matrix with $\delta = 0.0$ and unbalanced medium sample sizes.

| Distribution | Censoring distribution | asymptotic | asymptotic_bonf | permutation_bonf |
|---|---|---|---|---|
| exp early continuous | equal | 8.0 | 7.8 | **6.1** |
| exp early discrete | equal | 7.7 | 7.2 | **6.2** |
| exp early continuous | unequal, high | 7.9 | 7.6 | **5.3** |
| exp early discrete | unequal, high | 9.0 | 8.5 | 7.1 |
| exp early continuous | unequal, low | 7.5 | 7.4 | **5.9** |
| exp early discrete | unequal, low | 8.0 | 7.6 | **6.1** |
| logn continuous | equal | 10.2 | 9.5 | **5.9** |
| logn discrete | equal | 9.3 | 9.0 | **5.1** |
| logn continuous | unequal, high | 9.4 | 9.1 | **4.6** |
| logn discrete | unequal, high | 10.7 | 10.4 | **4.9** |
| logn continuous | unequal, low | 9.0 | 8.8 | **5.1** |
| logn discrete | unequal, low | 9.2 | 9.0 | **4.7** |
| pwExp continuous | equal | 6.7 | 6.7 | **4.7** |
| pwExp discrete | equal | 6.7 | **6.2** | **5.2** |
| pwExp continuous | unequal, high | **6.0** | 5.7 | **5.0** |
| pwExp discrete | unequal, high | 6.7 | 6.5 | **5.0** |
| pwExp continuous | unequal, low | **6.4** | **6.0** | **5.1** |
| pwExp discrete | unequal, low | **6.4** | **6.0** | **5.2** |
| Weib prop continuous | equal | 8.4 | 7.9 | **5.3** |
| Weib prop discrete | equal | 7.8 | 7.3 | **5.0** |
| Weib prop continuous | unequal, high | 11.6 | 11.1 | 6.9 |
| Weib prop discrete | unequal, high | 11.7 | 11.6 | **6.4** |
| Weib prop continuous | unequal, low | 7.8 | 7.6 | **5.1** |
| Weib prop discrete | unequal, low | 7.9 | 7.8 | **5.8** |
| Weib scale continuous | equal | 9.1 | 8.5 | **5.9** |
| Weib scale discrete | equal | 9.7 | 9.1 | **5.7** |
| Weib scale continuous | unequal, high | 11.7 | 10.9 | 7.8 |
| Weib scale discrete | unequal, high | 12.0 | 11.9 | 6.8 |
| Weib scale continuous | unequal, low | 9.4 | 9.2 | **5.8** |
| Weib scale discrete | unequal, low | 8.9 | 8.8 | 6.6 |
| Weib shape continuous | equal | 10.4 | 10.0 | **6.3** |
| Weib shape discrete | equal | 10.1 | 9.7 | **5.6** |
| Weib shape continuous | unequal, high | 11.1 | 10.6 | 7.3 |
| Weib shape discrete | unequal, high | 13.2 | 13.2 | 6.8 |
| Weib shape continuous | unequal, low | 9.8 | 9.7 | **6.4** |
| Weib shape discrete | unequal, low | 9.4 | 9.2 | **6.3** |

All values in the binomial interval $[3.7, 6.4]$ are printed in bold type.

Table S12: Rejection rates in percent for the Dunnett-type contrast matrix with $\delta = 0.0$ and unbalanced small sample sizes.

| Distribution | Censoring distribution | asymptotic | asymptotic_bonf | permutation_bonf |
|---|---|---|---|---|
| exp early continuous | equal | 17.6 | 17.0 | **6.3** |
| exp early discrete | equal | 18.3 | 18.0 | 6.8 |
| exp early continuous | unequal, high | 21.6 | 21.2 | **6.2** |
| exp early discrete | unequal, high | 22.9 | 22.4 | **5.4** |
| exp early continuous | unequal, low | 16.5 | 16.0 | **4.3** |
| exp early discrete | unequal, low | 17.6 | 17.1 | **4.1** |
| logn continuous | equal | 28.3 | 28.1 | **5.4** |
| logn discrete | equal | **31.7** | **31.3** | **4.8** |
| logn continuous | unequal, high | **43.9** | **43.6** | **6.2** |
| logn discrete | unequal, high | **47.4** | **46.9** | **6.4** |
| logn continuous | unequal, low | 24.0 | 23.4 | **4.1** |
| logn discrete | unequal, low | 26.3 | 25.5 | **4.7** |
| pwExp continuous | equal | 16.7 | 16.4 | **5.3** |
| pwExp discrete | equal | 16.6 | 16.4 | **4.7** |
| pwExp continuous | unequal, high | 20.0 | 19.7 | **5.2** |
| pwExp discrete | unequal, high | 23.3 | 23.0 | **4.7** |
| pwExp continuous | unequal, low | 16.5 | 16.2 | **4.6** |
| pwExp discrete | unequal, low | 16.7 | 16.4 | **4.3** |
| Weib prop continuous | equal | 23.5 | 23.3 | **6.1** |
| Weib prop discrete | equal | 27.4 | 27.2 | **6.0** |
| Weib prop continuous | unequal, high | **38.8** | 38.4 | **5.1** |
| Weib prop discrete | unequal, high | **43.8** | **43.4** | **6.4** |
| Weib prop continuous | unequal, low | 21.5 | 21.0 | **4.1** |
| Weib prop discrete | unequal, low | 23.7 | 23.2 | **4.2** |
| Weib scale continuous | equal | **31.5** | **31.2** | 8.8 |
| Weib scale discrete | equal | **34.2** | **34.2** | 8.1 |
| Weib scale continuous | unequal, high | **45.7** | **45.4** | 6.7 |
| Weib scale discrete | unequal, high | **51.4** | **51.0** | 7.1 |
| Weib scale continuous | unequal, low | 29.0 | 28.7 | **6.1** |
| Weib scale discrete | unequal, low | **32.0** | **31.7** | **5.6** |
| Weib shape continuous | equal | **35.8** | **35.6** | 9.3 |
| Weib shape discrete | equal | **39.4** | **39.2** | 9.2 |
| Weib shape continuous | unequal, high | **47.2** | **47.1** | 6.8 |
| Weib shape discrete | unequal, high | **54.9** | **54.6** | 7.5 |
| Weib shape continuous | unequal, low | **34.5** | **34.0** | 7.7 |
| Weib shape discrete | unequal, low | **37.3** | **37.2** | **6.4** |

All values in the binomial interval $[3.7, 6.4]$ are printed in bold type.

Table S13: Rejection rates in percent for the Tukey-type contrast matrix with $\delta = 0.0$ and balanced large sample sizes.

| Distribution | Censoring distribution | asymptotic | asymptotic_bonf | permutation_bonf |
|---|---|---|---|---|
| exp early continuous | equal | **5.5** | **4.4** | **5.0** |
| exp early discrete | equal | **5.3** | **4.9** | **5.2** |
| exp early continuous | unequal, high | **6.0** | **5.6** | **4.8** |
| exp early discrete | unequal, high | **5.3** | **4.5** | **4.4** |
| exp early continuous | unequal, low | **4.9** | **3.8** | **4.1** |
| exp early discrete | unequal, low | **5.0** | **3.9** | **4.6** |
| logn continuous | equal | **5.2** | **4.6** | **4.9** |
| logn discrete | equal | **5.3** | **4.3** | **4.6** |
| logn continuous | unequal, high | **4.8** | **3.7** | **4.2** |
| logn discrete | unequal, high | **4.5** | **3.8** | **4.3** |
| logn continuous | unequal, low | **5.2** | **4.7** | **4.9** |
| logn discrete | unequal, low | **5.4** | **4.5** | **5.0** |
| pwExp continuous | equal | **5.9** | **5.4** | **5.7** |
| pwExp discrete | equal | **6.1** | **5.2** | **5.9** |
| pwExp continuous | unequal, high | **5.7** | **5.0** | **4.4** |
| pwExp discrete | unequal, high | **5.8** | **5.1** | **4.5** |
| pwExp continuous | unequal, low | **5.7** | **5.1** | **5.4** |
| pwExp discrete | unequal, low | **5.5** | **5.3** | **4.9** |
| Weib prop continuous | equal | **4.3** | **3.9** | 3.4 |
| Weib prop discrete | equal | **5.1** | **4.4** | **4.2** |
| Weib prop continuous | unequal, high | 6.6 | **5.6** | **5.5** |
| Weib prop discrete | unequal, high | 6.5 | **6.0** | **5.9** |
| Weib prop continuous | unequal, low | **5.7** | **5.2** | **4.6** |
| Weib prop discrete | unequal, low | **6.1** | **4.9** | **5.3** |
| Weib scale continuous | equal | **4.3** | **3.8** | **4.5** |
| Weib scale discrete | equal | **4.8** | **4.4** | **4.3** |
| Weib scale continuous | unequal, high | **5.9** | **5.1** | **5.7** |
| Weib scale discrete | unequal, high | **6.4** | **5.8** | **6.4** |
| Weib scale continuous | unequal, low | **6.1** | **5.1** | **5.0** |
| Weib scale discrete | unequal, low | **5.6** | **5.0** | **5.3** |
| Weib shape continuous | equal | **4.3** | **3.8** | **4.4** |
| Weib shape discrete | equal | **5.0** | **4.5** | **4.4** |
| Weib shape continuous | unequal, high | **5.7** | **5.1** | **4.9** |
| Weib shape discrete | unequal, high | **6.3** | **5.6** | **5.5** |
| Weib shape continuous | unequal, low | **5.7** | **4.9** | **4.9** |
| Weib shape discrete | unequal, low | **6.0** | **5.0** | **5.0** |

All values in the binomial interval $[3.7, 6.4]$ are printed in bold type.

Table S14: Rejection rates in percent for the Tukey-type contrast matrix with $\delta = 0.0$ and balanced medium sample sizes.

| Distribution | Censoring distribution | asymptotic | asymptotic_bonf | permutation_bonf |
|---|---|---|---|---|
| exp early continuous | equal | **4.5** | 3.1 | **4.2** |
| exp early discrete | equal | **5.2** | **4.2** | **4.0** |
| exp early continuous | unequal, high | **5.5** | **4.5** | **5.0** |
| exp early discrete | unequal, high | **5.5** | **4.7** | **4.9** |
| exp early continuous | unequal, low | **5.7** | **4.8** | **5.0** |
| exp early discrete | unequal, low | **5.9** | **5.4** | **5.2** |
| logn continuous | equal | **4.9** | **4.5** | **5.2** |
| logn discrete | equal | **5.4** | **4.8** | **5.9** |
| logn continuous | unequal, high | **4.1** | 3.2 | **4.1** |
| logn discrete | unequal, high | **4.3** | **3.9** | **4.5** |
| logn continuous | unequal, low | **4.7** | **4.1** | **4.7** |
| logn discrete | unequal, low | **4.8** | **3.9** | **4.3** |
| pwExp continuous | equal | **5.5** | **4.8** | **4.1** |
| pwExp discrete | equal | **5.9** | **5.3** | **4.3** |
| pwExp continuous | unequal, high | **4.6** | **4.2** | **4.2** |
| pwExp discrete | unequal, high | **4.3** | 3.3 | 3.4 |
| pwExp continuous | unequal, low | **6.3** | **4.9** | **4.9** |
| pwExp discrete | unequal, low | 6.6 | **5.5** | **4.8** |
| Weib prop continuous | equal | **6.0** | **5.3** | **5.4** |
| Weib prop discrete | equal | **5.4** | **4.4** | **4.9** |
| Weib prop continuous | unequal, high | **4.9** | **4.3** | **4.2** |
| Weib prop discrete | unequal, high | **4.3** | 3.6 | **4.1** |
| Weib prop continuous | unequal, low | **5.2** | **4.3** | **4.6** |
| Weib prop discrete | unequal, low | **5.5** | **4.5** | **5.0** |
| Weib scale continuous | equal | **6.2** | **5.5** | **4.8** |
| Weib scale discrete | equal | **5.4** | **4.6** | **5.0** |
| Weib scale continuous | unequal, high | **5.1** | **4.2** | 3.9 |
| Weib scale discrete | unequal, high | **4.1** | 3.5 | **4.5** |
| Weib scale continuous | unequal, low | **5.3** | **4.1** | **5.0** |
| Weib scale discrete | unequal, low | **5.5** | **4.5** | **5.1** |
| Weib shape continuous | equal | **6.1** | **5.4** | **5.0** |
| Weib shape discrete | equal | **5.9** | **5.2** | **5.7** |
| Weib shape continuous | unequal, high | **5.5** | **4.7** | **4.5** |
| Weib shape discrete | unequal, high | **5.1** | **4.3** | **4.6** |
| Weib shape continuous | unequal, low | **5.3** | **4.5** | **5.7** |
| Weib shape discrete | unequal, low | **5.6** | **4.8** | **5.4** |

All values in the binomial interval $[3.7, 6.4]$ are printed in bold type.

Table S15: Rejection rates in percent for the Tukey-type contrast matrix with $\delta = 0.0$ and balanced small sample sizes.

| Distribution | Censoring distribution | asymptotic | asymptotic_bonf | permutation_bonf |
|---|---|---|---|---|
| exp early continuous | equal | **6.3** | **5.3** | **4.2** |
| exp early discrete | equal | **6.1** | **5.5** | **4.4** |
| exp early continuous | unequal, high | 6.9 | **6.0** | **5.0** |
| exp early discrete | unequal, high | 6.9 | **5.7** | **4.5** |
| exp early continuous | unequal, low | **6.3** | **5.9** | **4.2** |
| exp early discrete | unequal, low | **6.4** | **5.7** | **4.7** |
| logn continuous | equal | **6.2** | **5.5** | **5.1** |
| logn discrete | equal | **6.0** | **5.3** | **4.8** |
| logn continuous | unequal, high | 7.5 | 6.8 | **5.4** |
| logn discrete | unequal, high | 7.3 | 6.8 | **6.2** |
| logn continuous | unequal, low | **6.3** | **5.7** | **4.4** |
| logn discrete | unequal, low | **6.1** | **5.3** | **4.2** |
| pwExp continuous | equal | **4.4** | 3.6 | 3.1 |
| pwExp discrete | equal | **4.6** | **3.9** | **3.8** |
| pwExp continuous | unequal, high | **6.3** | **5.3** | **3.8** |
| pwExp discrete | unequal, high | **6.0** | **4.9** | **4.2** |
| pwExp continuous | unequal, low | **5.5** | **5.0** | **4.0** |
| pwExp discrete | unequal, low | **5.4** | **4.6** | **4.1** |
| Weib prop continuous | equal | 6.7 | 5.8 | 5.3 |
| Weib prop discrete | equal | **5.7** | **5.0** | **5.1** |
| Weib prop continuous | unequal, high | 6.8 | 5.7 | 4.4 |
| Weib prop discrete | unequal, high | **6.2** | **5.7** | **4.1** |
| Weib prop continuous | unequal, low | 6.6 | **6.0** | **5.1** |
| Weib prop discrete | unequal, low | 6.9 | **5.8** | **4.9** |
| Weib scale continuous | equal | 7.0 | **6.4** | 5.8 |
| Weib scale discrete | equal | **5.7** | **4.9** | **4.6** |
| Weib scale continuous | unequal, high | 6.5 | 5.5 | 4.3 |
| Weib scale discrete | unequal, high | **6.1** | **5.2** | 3.7 |
| Weib scale continuous | unequal, low | 7.0 | **5.7** | **5.1** |
| Weib scale discrete | unequal, low | **6.0** | **5.2** | **4.8** |
| Weib shape continuous | equal | 6.7 | 5.6 | 5.3 |
| Weib shape discrete | equal | **5.6** | **4.7** | **4.5** |
| Weib shape continuous | unequal, high | **6.4** | **5.2** | **4.2** |
| Weib shape discrete | unequal, high | **6.4** | **5.3** | **4.7** |
| Weib shape continuous | unequal, low | 6.5 | 5.8 | 5.2 |
| Weib shape discrete | unequal, low | **5.8** | **4.9** | **4.8** |

All values in the binomial interval $[3.7, 6.4]$ are printed in bold type.

Table S16: Rejection rates in percent for the Tukey-type contrast matrix with $\delta = 0.0$ and unbalanced large sample sizes.

| Distribution | Censoring distribution | asymptotic | asymptotic_bonf | permutation_bonf |
|---|---|---|---|---|
| exp early continuous | equal | **5.5** | **4.9** | **4.5** |
| exp early discrete | equal | **5.9** | **5.2** | **5.0** |
| exp early continuous | unequal, high | **5.0** | **4.4** | **4.0** |
| exp early discrete | unequal, high | **5.4** | **4.8** | **4.8** |
| exp early continuous | unequal, low | **5.7** | **5.4** | **5.5** |
| exp early discrete | unequal, low | **6.0** | **4.9** | **5.4** |
| logn continuous | equal | 6.8 | **5.5** | **5.3** |
| logn discrete | equal | **6.4** | **5.6** | **4.9** |
| logn continuous | unequal, high | 6.7 | **5.2** | **4.3** |
| logn discrete | unequal, high | **5.7** | **4.7** | **4.7** |
| logn continuous | unequal, low | 7.2 | **5.6** | **6.0** |
| logn discrete | unequal, low | **6.3** | **5.6** | **5.0** |
| pwExp continuous | equal | **5.2** | **4.4** | **4.7** |
| pwExp discrete | equal | **5.0** | **4.4** | **4.6** |
| pwExp continuous | unequal, high | **5.6** | **4.3** | **4.7** |
| pwExp discrete | unequal, high | **5.8** | **5.0** | **5.3** |
| pwExp continuous | unequal, low | **5.5** | **4.0** | **4.3** |
| pwExp discrete | unequal, low | **4.9** | **4.2** | **4.4** |
| Weib prop continuous | equal | **5.5** | **4.2** | **4.1** |
| Weib prop discrete | equal | **6.0** | **5.1** | **5.4** |
| Weib prop continuous | unequal, high | **4.5** | **4.0** | **4.4** |
| Weib prop discrete | unequal, high | **5.2** | **4.2** | **4.4** |
| Weib prop continuous | unequal, low | **5.5** | **4.0** | 3.4 |
| Weib prop discrete | unequal, low | **4.8** | **4.0** | **4.1** |
| Weib scale continuous | equal | **5.2** | **3.8** | 3.6 |
| Weib scale discrete | equal | **6.0** | **5.2** | **4.6** |
| Weib scale continuous | unequal, high | **4.4** | **4.0** | 3.2 |
| Weib scale discrete | unequal, high | **4.7** | **3.8** | 3.6 |
| Weib scale continuous | unequal, low | **5.1** | **4.0** | **4.4** |
| Weib scale discrete | unequal, low | **5.0** | **4.1** | 3.9 |
| Weib shape continuous | equal | **5.4** | **4.1** | 3.7 |
| Weib shape discrete | equal | **6.4** | **5.3** | **5.1** |
| Weib shape continuous | unequal, high | **5.0** | **4.5** | **4.0** |
| Weib shape discrete | unequal, high | **4.9** | **4.0** | **4.3** |
| Weib shape continuous | unequal, low | **4.9** | **4.0** | **4.0** |
| Weib shape discrete | unequal, low | **5.0** | **4.2** | **4.1** |

All values in the binomial interval $[3.7, 6.4]$ are printed in bold type.

Table S17: Rejection rates in percent for the Tukey-type contrast matrix with $\delta = 0.0$ and unbalanced medium sample sizes.

| Distribution | Censoring distribution | asymptotic | asymptotic_bonf | permutation_bonf |
|---|---|---|---|---|
| exp early continuous | equal | 7.0 | 6.6 | **5.1** |
| exp early discrete | equal | 6.8 | **6.0** | **4.5** |
| exp early continuous | unequal, high | 7.6 | 6.9 | **4.8** |
| exp early discrete | unequal, high | 8.7 | 7.8 | **5.3** |
| exp early continuous | unequal, low | 6.9 | **6.1** | **5.1** |
| exp early discrete | unequal, low | 7.1 | **6.3** | **5.9** |
| logn continuous | equal | 9.7 | 8.5 | **5.6** |
| logn discrete | equal | 9.7 | 8.7 | **4.6** |
| logn continuous | unequal, high | 9.1 | 8.3 | **3.7** |
| logn discrete | unequal, high | 10.6 | 9.2 | **4.3** |
| logn continuous | unequal, low | 9.0 | 8.3 | **4.8** |
| logn discrete | unequal, low | 9.7 | 8.5 | **4.4** |
| pwExp continuous | equal | 7.5 | **6.0** | **4.8** |
| pwExp discrete | equal | 8.0 | 6.5 | **4.5** |
| pwExp continuous | unequal, high | **6.4** | **5.5** | 3.4 |
| pwExp discrete | unequal, high | 6.8 | **5.7** | 3.5 |
| pwExp continuous | unequal, low | 7.5 | **6.3** | **4.7** |
| pwExp discrete | unequal, low | 7.4 | **6.2** | **4.9** |
| Weib prop continuous | equal | 7.7 | 6.7 | **3.8** |
| Weib prop discrete | equal | 7.4 | 6.5 | **4.1** |
| Weib prop continuous | unequal, high | 10.6 | 9.3 | **5.4** |
| Weib prop discrete | unequal, high | 11.1 | 9.4 | **5.4** |
| Weib prop continuous | unequal, low | 8.0 | **6.4** | **3.8** |
| Weib prop discrete | unequal, low | 7.9 | 7.2 | **3.8** |
| Weib scale continuous | equal | 9.3 | 7.6 | **4.8** |
| Weib scale discrete | equal | 8.8 | 8.1 | **5.9** |
| Weib scale continuous | unequal, high | 10.7 | 9.7 | **5.8** |
| Weib scale discrete | unequal, high | 11.9 | 10.9 | **6.0** |
| Weib scale continuous | unequal, low | 9.3 | 7.4 | **4.5** |
| Weib scale discrete | unequal, low | 9.8 | 9.0 | **4.5** |
| Weib shape continuous | equal | 9.9 | 8.5 | **5.5** |
| Weib shape discrete | equal | 9.4 | 8.5 | **5.8** |
| Weib shape continuous | unequal, high | 10.5 | 9.3 | **5.5** |
| Weib shape discrete | unequal, high | 11.6 | 11.1 | **5.8** |
| Weib shape continuous | unequal, low | 10.0 | 8.1 | **4.6** |
| Weib shape discrete | unequal, low | 10.7 | 9.2 | **5.1** |

All values in the binomial interval $[3.7, 6.4]$ are printed in bold type.

Table S18: Rejection rates in percent for the Tukey-type contrast matrix with $\delta = 0.0$ and unbalanced small sample sizes.

| Distribution | Censoring distribution | asymptotic | asymptotic_bonf | permutation_bonf |
|---|---|---|---|---|
| exp early continuous | equal | 16.9 | 16.0 | **4.6** |
| exp early discrete | equal | 18.2 | 16.4 | **5.2** |
| exp early continuous | unequal, high | 20.8 | 19.3 | **5.4** |
| exp early discrete | unequal, high | 22.8 | 21.9 | **4.7** |
| exp early continuous | unequal, low | 15.8 | 14.6 | **3.7** |
| exp early discrete | unequal, low | 16.4 | 15.3 | 2.8 |
| logn continuous | equal | 27.9 | 26.5 | **5.3** |
| logn discrete | equal | 30.3 | 29.7 | **5.3** |
| logn continuous | unequal, high | 43.1 | 41.4 | **5.8** |
| logn discrete | unequal, high | 46.3 | 43.9 | **5.2** |
| logn continuous | unequal, low | 23.9 | 21.7 | **3.7** |
| logn discrete | unequal, low | 25.9 | 23.4 | **4.1** |
| pwExp continuous | equal | 16.5 | 14.8 | **4.6** |
| pwExp discrete | equal | 16.4 | 15.3 | **4.8** |
| pwExp continuous | unequal, high | 20.8 | 19.6 | **4.9** |
| pwExp discrete | unequal, high | 22.9 | 22.4 | **4.8** |
| pwExp continuous | unequal, low | 16.3 | 15.1 | **4.1** |
| pwExp discrete | unequal, low | 16.8 | 15.1 | **4.4** |
| Weib prop continuous | equal | 23.0 | 22.3 | **4.7** |
| Weib prop discrete | equal | 26.7 | 25.7 | **4.7** |
| Weib prop continuous | unequal, high | **37.9** | 35.8 | **5.8** |
| Weib prop discrete | unequal, high | **42.4** | 40.8 | **5.6** |
| Weib prop continuous | unequal, low | 21.1 | 19.1 | 2.8 |
| Weib prop discrete | unequal, low | 22.4 | 21.0 | **3.9** |
| Weib scale continuous | equal | **31.1** | 30.4 | 7.9 |
| Weib scale discrete | equal | **34.4** | 33.1 | 7.0 |
| Weib scale continuous | unequal, high | **44.1** | 42.1 | 7.3 |
| Weib scale discrete | unequal, high | **49.8** | 48.1 | 6.6 |
| Weib scale continuous | unequal, low | 27.6 | 25.8 | **5.7** |
| Weib scale discrete | unequal, low | **30.8** | 29.0 | **6.1** |
| Weib shape continuous | equal | **35.9** | 35.1 | 9.5 |
| Weib shape discrete | equal | **39.1** | 38.1 | 8.7 |
| Weib shape continuous | unequal, high | **46.3** | 44.7 | 7.5 |
| Weib shape discrete | unequal, high | **53.6** | 52.0 | 6.8 |
| Weib shape continuous | unequal, low | **33.7** | 32.2 | 7.6 |
| Weib shape discrete | unequal, low | **35.9** | 34.3 | 7.7 |

All values in the binomial interval $[3.7, 6.4]$ are printed in bold type.

Table S19: Rejection rates in percent for the 2-by-2 design with $\delta = 1.5$ and balanced large sample sizes.

| Distribution | Censoring distribution | asymptotic | asymptotic_bonf | permutation_bonf |
|---|---|---|---|---|
| exp early continuous | equal | 100.0 | 100.0 | 100.0 |
| exp early discrete | equal | 100.0 | 100.0 | 100.0 |
| exp early continuous | unequal, high | 99.8 | 99.8 | 99.9 |
| exp early discrete | unequal, high | 99.8 | 99.8 | 99.9 |
| exp early continuous | unequal, low | 100.0 | 100.0 | 100.0 |
| exp early discrete | unequal, low | 100.0 | 100.0 | 100.0 |
| exp late continuous | equal | 100.0 | 100.0 | 100.0 |
| exp late discrete | equal | 100.0 | 100.0 | 100.0 |
| exp late continuous | unequal, high | 100.0 | 100.0 | 100.0 |
| exp late discrete | unequal, high | 99.9 | 99.9 | 99.9 |
| exp late continuous | unequal, low | 100.0 | 100.0 | 100.0 |
| exp late discrete | unequal, low | 100.0 | 100.0 | 100.0 |
| exp prop continuous | equal | 100.0 | 100.0 | 100.0 |
| exp prop discrete | equal | 100.0 | 100.0 | 100.0 |
| exp prop continuous | unequal, high | 100.0 | 100.0 | 99.9 |
| exp prop discrete | unequal, high | 100.0 | 100.0 | 100.0 |
| exp prop continuous | unequal, low | 100.0 | 100.0 | 100.0 |
| exp prop discrete | unequal, low | 100.0 | 100.0 | 100.0 |
| logn continuous | equal | 100.0 | 100.0 | 100.0 |
| logn discrete | equal | 100.0 | 100.0 | 100.0 |
| logn continuous | unequal, high | 100.0 | 100.0 | 100.0 |
| logn discrete | unequal, high | 100.0 | 100.0 | 100.0 |
| logn continuous | unequal, low | 100.0 | 100.0 | 100.0 |
| logn discrete | unequal, low | 100.0 | 100.0 | 100.0 |
| pwExp continuous | equal | 100.0 | 100.0 | 100.0 |
| pwExp discrete | equal | 100.0 | 100.0 | 100.0 |
| pwExp continuous | unequal, high | 99.3 | 99.3 | 99.4 |
| pwExp discrete | unequal, high | 100.0 | 100.0 | 100.0 |
| pwExp continuous | unequal, low | 100.0 | 100.0 | 100.0 |
| pwExp discrete | unequal, low | 100.0 | 100.0 | 100.0 |
| Weib late continuous | equal | 100.0 | 100.0 | 100.0 |
| Weib late discrete | equal | 100.0 | 100.0 | 100.0 |
| Weib late continuous | unequal, high | 100.0 | 100.0 | 100.0 |
| Weib late discrete | unequal, high | 100.0 | 100.0 | 100.0 |
| Weib late continuous | unequal, low | 100.0 | 100.0 | 100.0 |
| Weib late discrete | unequal, low | 100.0 | 100.0 | 100.0 |
| Weib prop continuous | equal | 100.0 | 100.0 | 100.0 |
| Weib prop discrete | equal | 100.0 | 100.0 | 100.0 |
| Weib prop continuous | unequal, high | 100.0 | 100.0 | 100.0 |
| Weib prop discrete | unequal, high | 100.0 | 100.0 | 100.0 |
| Weib prop continuous | unequal, low | 100.0 | 100.0 | 100.0 |
| Weib prop discrete | unequal, low | 100.0 | 100.0 | 100.0 |
| Weib scale continuous | equal | 100.0 | 100.0 | 100.0 |
| Weib scale discrete | equal | 100.0 | 100.0 | 100.0 |
| Weib scale continuous | unequal, high | 100.0 | 100.0 | 100.0 |
| Weib scale discrete | unequal, high | 100.0 | 100.0 | 100.0 |
| Weib scale continuous | unequal, low | 100.0 | 100.0 | 100.0 |
| Weib scale discrete | unequal, low | 100.0 | 100.0 | 100.0 |
| Weib shape continuous | equal | 100.0 | 100.0 | 100.0 |
| Weib shape discrete | equal | 100.0 | 100.0 | 100.0 |
| Weib shape continuous | unequal, high | 100.0 | 100.0 | 100.0 |
| Weib shape discrete | unequal, high | 100.0 | 100.0 | 100.0 |
| Weib shape continuous | unequal, low | 100.0 | 100.0 | 100.0 |
| Weib shape discrete | unequal, low | 100.0 | 100.0 | 100.0 |

Table S20: Rejection rates in percent for the 2-by-2 design with $\delta = 1.5$ and balanced medium sample sizes.

| Distribution | Censoring distribution | asymptotic | asymptotic_bonf | permutation_bonf |
|---|---|---|---|---|
| exp early continuous | equal | 49.4 | 48.5 | 48.0 |
| exp early discrete | equal | 55.1 | 53.7 | 56.5 |
| exp early continuous | unequal, high | 40.6 | 39.1 | 39.2 |
| exp early discrete | unequal, high | 42.6 | 41.7 | 42.6 |
| exp early continuous | unequal, low | 46.4 | 45.6 | 46.6 |
| exp early discrete | unequal, low | 52.9 | 52.0 | 51.3 |
| exp late continuous | equal | 52.2 | 50.6 | 51.0 |
| exp late discrete | equal | 59.0 | 56.2 | 58.0 |
| exp late continuous | unequal, high | 41.9 | 40.0 | 39.2 |
| exp late discrete | unequal, high | 45.1 | 43.1 | 43.3 |
| exp late continuous | unequal, low | 50.8 | 49.7 | 48.0 |
| exp late discrete | unequal, low | 55.1 | 54.2 | 53.9 |
| exp prop continuous | equal | 50.2 | 49.3 | 48.7 |
| exp prop discrete | equal | 54.8 | 53.5 | 53.8 |
| exp prop continuous | unequal, high | 40.9 | 39.5 | 39.3 |
| exp prop discrete | unequal, high | 44.2 | 42.8 | 44.3 |
| exp prop continuous | unequal, low | 49.3 | 48.3 | 48.6 |
| exp prop discrete | unequal, low | 52.6 | 51.4 | 51.4 |
| logn continuous | equal | 94.2 | 93.8 | 93.7 |
| logn discrete | equal | 97.9 | 97.4 | 96.7 |
| logn continuous | unequal, high | 73.5 | 72.2 | 71.6 |
| logn discrete | unequal, high | 80.8 | 80.0 | 79.1 |
| logn continuous | unequal, low | 92.2 | 91.9 | 91.7 |
| logn discrete | unequal, low | 97.1 | 97.0 | 96.5 |
| pwExp continuous | equal | 48.3 | 47.2 | 45.6 |
| pwExp discrete | equal | 56.0 | 54.6 | 54.4 |
| pwExp continuous | unequal, high | 40.8 | 39.3 | 39.1 |
| pwExp discrete | unequal, high | 43.9 | 42.9 | 41.7 |
| pwExp continuous | unequal, low | 47.2 | 45.9 | 46.2 |
| pwExp discrete | unequal, low | 52.4 | 51.1 | 50.9 |
| Weib late continuous | equal | 94.8 | 94.4 | 94.1 |
| Weib late discrete | equal | 97.0 | 96.7 | 96.3 |
| Weib late continuous | unequal, high | 77.4 | 77.0 | 75.2 |
| Weib late discrete | unequal, high | 81.7 | 80.9 | 81.8 |
| Weib late continuous | unequal, low | 94.5 | 94.3 | 94.3 |
| Weib late discrete | unequal, low | 97.4 | 97.1 | 96.4 |
| Weib prop continuous | equal | 93.1 | 92.8 | 92.4 |
| Weib prop discrete | equal | 96.7 | 96.5 | 95.9 |
| Weib prop continuous | unequal, high | 76.0 | 74.9 | 74.4 |
| Weib prop discrete | unequal, high | 82.3 | 81.4 | 81.1 |
| Weib prop continuous | unequal, low | 92.8 | 92.2 | 92.9 |
| Weib prop discrete | unequal, low | 96.6 | 96.5 | 96.9 |
| Weib scale continuous | equal | 87.2 | 86.6 | 86.8 |
| Weib scale discrete | equal | 92.3 | 91.4 | 92.0 |
| Weib scale continuous | unequal, high | 70.1 | 69.1 | 68.5 |
| Weib scale discrete | unequal, high | 76.3 | 75.5 | 76.3 |
| Weib scale continuous | unequal, low | 86.9 | 86.3 | 87.1 |
| Weib scale discrete | unequal, low | 92.0 | 91.8 | 91.7 |
| Weib shape continuous | equal | 78.8 | 77.5 | 79.1 |
| Weib shape discrete | equal | 85.5 | 85.0 | 85.1 |
| Weib shape continuous | unequal, high | 65.0 | 63.4 | 62.7 |
| Weib shape discrete | unequal, high | 71.1 | 70.0 | 70.1 |
| Weib shape continuous | unequal, low | 77.7 | 76.9 | 77.1 |
| Weib shape discrete | unequal, low | 85.4 | 84.7 | 85.6 |

Table S21: Rejection rates in percent for the 2-by-2 design with $\delta = 1.5$ and balanced small sample sizes.

| Distribution | Censoring distribution | asymptotic | asymptotic_bonf | permutation_bonf |
|---|---|---|---|---|
| exp early continuous | equal | 13.0 | 12.1 | 11.8 |
| exp early discrete | equal | 12.3 | 12.1 | 10.7 |
| exp early continuous | unequal, high | 12.4 | 11.8 | 10.0 |
| exp early discrete | unequal, high | 11.3 | 10.5 | 9.3 |
| exp early continuous | unequal, low | 12.9 | 12.2 | 11.3 |
| exp early discrete | unequal, low | 13.9 | 13.1 | 11.3 |
| exp late continuous | equal | 14.1 | 13.3 | 12.5 |
| exp late discrete | equal | 14.5 | 14.0 | 14.2 |
| exp late continuous | unequal, high | 13.2 | 12.8 | 10.8 |
| exp late discrete | unequal, high | 13.2 | 12.2 | 10.3 |
| exp late continuous | unequal, low | 13.8 | 12.9 | 12.3 |
| exp late discrete | unequal, low | 15.0 | 13.9 | 13.7 |
| exp prop continuous | equal | 12.7 | 12.2 | 10.6 |
| exp prop discrete | equal | 13.2 | 12.8 | 12.0 |
| exp prop continuous | unequal, high | 11.8 | 11.4 | 9.2 |
| exp prop discrete | unequal, high | 12.3 | 12.3 | 10.2 |
| exp prop continuous | unequal, low | 13.2 | 12.5 | 11.9 |
| exp prop discrete | unequal, low | 13.2 | 12.7 | 12.3 |
| logn continuous | equal | 29.3 | 28.2 | 27.2 |
| logn discrete | equal | 35.5 | 34.7 | 31.7 |
| logn continuous | unequal, high | 22.4 | 21.8 | 19.1 |
| logn discrete | unequal, high | 26.6 | 25.8 | 23.1 |
| logn continuous | unequal, low | 29.1 | 28.4 | 26.2 |
| logn discrete | unequal, low | 33.1 | 32.5 | 31.7 |
| pwExp continuous | equal | 13.6 | 13.1 | 11.3 |
| pwExp discrete | equal | 14.1 | 13.8 | 12.5 |
| pwExp continuous | unequal, high | 12.8 | 12.2 | 11.4 |
| pwExp discrete | unequal, high | 13.8 | 12.9 | 10.5 |
| pwExp continuous | unequal, low | 12.8 | 12.4 | 10.7 |
| pwExp discrete | unequal, low | 14.3 | 13.5 | 13.1 |
| Weib late continuous | equal | 30.0 | 29.6 | 27.5 |
| Weib late discrete | equal | 32.6 | 32.0 | 30.3 |
| Weib late continuous | unequal, high | 22.1 | 21.9 | 19.0 |
| Weib late discrete | unequal, high | 24.0 | 23.3 | 21.4 |
| Weib late continuous | unequal, low | 29.2 | 28.7 | 27.0 |
| Weib late discrete | unequal, low | 32.7 | 31.7 | 29.8 |
| Weib prop continuous | equal | 28.4 | 27.6 | 25.1 |
| Weib prop discrete | equal | 30.6 | 29.7 | 28.2 |
| Weib prop continuous | unequal, high | 21.6 | 21.1 | 19.6 |
| Weib prop discrete | unequal, high | 23.7 | 23.1 | 20.7 |
| Weib prop continuous | unequal, low | 28.0 | 27.3 | 25.9 |
| Weib prop discrete | unequal, low | 31.7 | 30.9 | 28.9 |
| Weib scale continuous | equal | 23.9 | 23.1 | 21.7 |
| Weib scale discrete | equal | 26.5 | 25.5 | 23.2 |
| Weib scale continuous | unequal, high | 19.2 | 18.2 | 15.8 |
| Weib scale discrete | unequal, high | 20.5 | 19.8 | 17.5 |
| Weib scale continuous | unequal, low | 24.3 | 24.0 | 21.3 |
| Weib scale discrete | unequal, low | 26.2 | 24.9 | 25.1 |
| Weib shape continuous | equal | 19.0 | 18.2 | 17.0 |
| Weib shape discrete | equal | 20.7 | 19.0 | 18.8 |
| Weib shape continuous | unequal, high | 16.3 | 16.0 | 13.9 |
| Weib shape discrete | unequal, high | 17.8 | 17.4 | 15.4 |
| Weib shape continuous | unequal, low | 18.1 | 17.3 | 17.0 |
| Weib shape discrete | unequal, low | 21.0 | 19.5 | 19.3 |

Table S22: Rejection rates in percent for the 2-by-2 design with $\delta = 1.5$ and unbalanced large sample sizes.

| Distribution | Censoring distribution | asymptotic | asymptotic_bonf | permutation_bonf |
|---|---|---|---|---|
| exp early continuous | equal | 93.4 | 92.3 | 91.6 |
| exp early discrete | equal | 96.8 | 96.1 | 95.4 |
| exp early continuous | unequal, high | 85.2 | 83.4 | 83.7 |
| exp early discrete | unequal, high | 87.7 | 86.2 | 85.1 |
| exp early continuous | unequal, low | 93.3 | 91.9 | 91.5 |
| exp early discrete | unequal, low | 97.0 | 96.0 | 95.5 |
| exp late continuous | equal | 97.3 | 96.9 | 96.9 |
| exp late discrete | equal | 98.8 | 98.2 | 98.4 |
| exp late continuous | unequal, high | 89.6 | 88.3 | 88.2 |
| exp late discrete | unequal, high | 93.5 | 91.6 | 90.8 |
| exp late continuous | unequal, low | 97.9 | 97.4 | 96.9 |
| exp late discrete | unequal, low | 99.1 | 98.6 | 98.5 |
| exp prop continuous | equal | 94.8 | 94.0 | 94.3 |
| exp prop discrete | equal | 98.0 | 97.0 | 96.4 |
| exp prop continuous | unequal, high | 85.8 | 83.9 | 83.3 |
| exp prop discrete | unequal, high | 89.9 | 88.1 | 86.8 |
| exp prop continuous | unequal, low | 95.5 | 94.0 | 93.4 |
| exp prop discrete | unequal, low | 98.0 | 97.4 | 96.4 |
| logn continuous | equal | 100.0 | 100.0 | 100.0 |
| logn discrete | equal | 100.0 | 100.0 | 100.0 |
| logn continuous | unequal, high | 99.8 | 99.8 | 99.8 |
| logn discrete | unequal, high | 99.9 | 99.9 | 99.9 |
| logn continuous | unequal, low | 100.0 | 100.0 | 100.0 |
| logn discrete | unequal, low | 100.0 | 100.0 | 100.0 |
| pwExp continuous | equal | 92.6 | 90.6 | 91.1 |
| pwExp discrete | equal | 95.6 | 94.5 | 95.2 |
| pwExp continuous | unequal, high | 85.0 | 83.0 | 83.4 |
| pwExp discrete | unequal, high | 87.6 | 85.7 | 84.9 |
| pwExp continuous | unequal, low | 92.7 | 90.4 | 90.7 |
| pwExp discrete | unequal, low | 95.9 | 95.1 | 95.0 |
| Weib late continuous | equal | 100.0 | 100.0 | 100.0 |
| Weib late discrete | equal | 99.8 | 99.6 | 99.7 |
| Weib late continuous | unequal, high | 100.0 | 100.0 | 99.9 |
| Weib late discrete | unequal, high | 99.2 | 99.1 | 99.0 |
| Weib late continuous | unequal, low | 100.0 | 100.0 | 100.0 |
| Weib late discrete | unequal, low | 99.9 | 99.9 | 99.9 |
| Weib prop continuous | equal | 100.0 | 100.0 | 100.0 |
| Weib prop discrete | equal | 100.0 | 100.0 | 100.0 |
| Weib prop continuous | unequal, high | 99.8 | 99.8 | 99.8 |
| Weib prop discrete | unequal, high | 100.0 | 100.0 | 99.9 |
| Weib prop continuous | unequal, low | 100.0 | 100.0 | 100.0 |
| Weib prop discrete | unequal, low | 100.0 | 100.0 | 100.0 |
| Weib scale continuous | equal | 100.0 | 100.0 | 100.0 |
| Weib scale discrete | equal | 100.0 | 100.0 | 100.0 |
| Weib scale continuous | unequal, high | 99.2 | 98.6 | 98.2 |
| Weib scale discrete | unequal, high | 99.9 | 99.8 | 99.4 |
| Weib scale continuous | unequal, low | 100.0 | 100.0 | 100.0 |
| Weib scale discrete | unequal, low | 100.0 | 100.0 | 100.0 |
| Weib shape continuous | equal | 99.3 | 98.9 | 98.6 |
| Weib shape discrete | equal | 99.6 | 99.5 | 99.6 |
| Weib shape continuous | unequal, high | 97.2 | 96.2 | 95.3 |
| Weib shape discrete | unequal, high | 98.0 | 97.4 | 97.6 |
| Weib shape continuous | unequal, low | 99.1 | 98.8 | 98.3 |
| Weib shape discrete | unequal, low | 99.8 | 99.6 | 99.6 |

Table S23: Rejection rates in percent for the 2-by-2 design with $\delta = 1.5$ and unbalanced medium sample sizes.

| Distribution | Censoring distribution | asymptotic | asymptotic_bonf | permutation_bonf |
|---|---|---|---|---|
| exp early continuous | equal | 22.1 | 19.6 | 16.2 |
| exp early discrete | equal | 24.7 | 22.3 | 19.3 |
| exp early continuous | unequal, high | 20.3 | 18.3 | 14.8 |
| exp early discrete | unequal, high | 20.9 | 19.4 | 15.0 |
| exp early continuous | unequal, low | 20.9 | 19.1 | 15.6 |
| exp early discrete | unequal, low | 24.0 | 21.4 | 18.4 |
| exp late continuous | equal | 26.7 | 23.5 | 19.5 |
| exp late discrete | equal | 27.6 | 25.4 | 22.1 |
| exp late continuous | unequal, high | 21.2 | 19.6 | 15.5 |
| exp late discrete | unequal, high | 22.3 | 20.0 | 16.8 |
| exp late continuous | unequal, low | 24.5 | 22.2 | 18.9 |
| exp late discrete | unequal, low | 27.6 | 25.5 | 21.0 |
| exp prop continuous | equal | 22.9 | 20.6 | 17.8 |
| exp prop discrete | equal | 25.3 | 23.2 | 20.9 |
| exp prop continuous | unequal, high | 19.2 | 17.5 | 14.0 |
| exp prop discrete | unequal, high | 21.3 | 19.7 | 15.5 |
| exp prop continuous | unequal, low | 23.6 | 21.2 | 17.8 |
| exp prop discrete | unequal, low | 26.5 | 23.9 | 21.2 |
| logn continuous | equal | 50.4 | 47.3 | 40.8 |
| logn discrete | equal | 58.6 | 54.3 | 48.0 |
| logn continuous | unequal, high | 36.1 | 33.5 | 25.3 |
| logn discrete | unequal, high | 40.3 | 36.1 | 26.5 |
| logn continuous | unequal, low | 53.9 | 50.4 | 43.0 |
| logn discrete | unequal, low | 61.7 | 58.1 | 49.2 |
| pwExp continuous | equal | 21.3 | 17.9 | 16.8 |
| pwExp discrete | equal | 24.3 | 21.0 | 19.0 |
| pwExp continuous | unequal, high | 18.9 | 17.0 | 14.3 |
| pwExp discrete | unequal, high | 20.1 | 18.5 | 15.0 |
| pwExp continuous | unequal, low | 21.0 | 18.6 | 14.9 |
| pwExp discrete | unequal, low | 23.5 | 21.2 | 17.8 |
| Weib late continuous | equal | 56.0 | 52.0 | 45.0 |
| Weib late discrete | equal | 62.9 | 59.6 | 51.4 |
| Weib late continuous | unequal, high | 38.6 | 35.2 | 26.4 |
| Weib late discrete | unequal, high | 45.3 | 40.8 | 29.4 |
| Weib late continuous | unequal, low | 60.8 | 56.3 | 48.7 |
| Weib late discrete | unequal, low | 69.3 | 65.1 | 55.2 |
| Weib prop continuous | equal | 50.9 | 47.1 | 39.6 |
| Weib prop discrete | equal | 57.6 | 52.8 | 44.6 |
| Weib prop continuous | unequal, high | 35.4 | 32.5 | 24.6 |
| Weib prop discrete | unequal, high | 41.4 | 38.7 | 26.9 |
| Weib prop continuous | unequal, low | 54.3 | 51.0 | 43.4 |
| Weib prop discrete | unequal, low | 62.7 | 58.8 | 49.7 |
| Weib scale continuous | equal | 39.4 | 35.0 | 28.9 |
| Weib scale discrete | equal | 43.4 | 40.4 | 33.5 |
| Weib scale continuous | unequal, high | 28.7 | 26.1 | 18.9 |
| Weib scale discrete | unequal, high | 32.2 | 29.1 | 19.3 |
| Weib scale continuous | unequal, low | 40.2 | 36.7 | 30.7 |
| Weib scale discrete | unequal, low | 46.5 | 42.8 | 34.3 |
| Weib shape continuous | equal | 25.6 | 20.9 | 16.8 |
| Weib shape discrete | equal | 31.4 | 26.8 | 21.9 |
| Weib shape continuous | unequal, high | 22.5 | 20.1 | 14.9 |
| Weib shape discrete | unequal, high | 24.7 | 22.3 | 15.2 |
| Weib shape continuous | unequal, low | 26.1 | 21.0 | 16.2 |
| Weib shape discrete | unequal, low | 33.3 | 28.9 | 20.9 |

Table S24: Rejection rates in percent for the 2-by-2 design with $\delta = 1.5$ and unbalanced small sample sizes.

| Distribution | Censoring distribution | asymptotic | asymptotic_bonf | permutation_bonf |
|---|---|---|---|---|
| exp early continuous | equal | 9.1 | 8.2 | 3.5 |
| exp early discrete | equal | 8.9 | 8.2 | 3.7 |
| exp early continuous | unequal, high | 10.4 | 9.7 | 2.3 |
| exp early discrete | unequal, high | 10.7 | 9.8 | 2.7 |
| exp early continuous | unequal, low | 9.1 | 8.0 | 3.4 |
| exp early discrete | unequal, low | 8.6 | 7.9 | 3.0 |
| exp late continuous | equal | 9.4 | 8.5 | 3.5 |
| exp late discrete | equal | 10.6 | 9.4 | 4.1 |
| exp late continuous | unequal, high | 10.6 | 9.6 | 2.2 |
| exp late discrete | unequal, high | 11.3 | 10.4 | 2.4 |
| exp late continuous | unequal, low | 8.9 | 7.3 | 3.1 |
| exp late discrete | unequal, low | 9.6 | 8.1 | 2.8 |
| exp prop continuous | equal | 10.4 | 9.6 | 3.9 |
| exp prop discrete | equal | 10.8 | 9.9 | 4.8 |
| exp prop continuous | unequal, high | 10.7 | 9.9 | 3.0 |
| exp prop discrete | unequal, high | 9.9 | 9.1 | 3.6 |
| exp prop continuous | unequal, low | 9.9 | 9.2 | 4.0 |
| exp prop discrete | unequal, low | 10.7 | 9.8 | 3.7 |
| logn continuous | equal | 12.0 | 11.4 | 3.4 |
| logn discrete | equal | 13.1 | 11.6 | 4.2 |
| logn continuous | unequal, high | 13.1 | 12.0 | 4.5 |
| logn discrete | unequal, high | 15.0 | 13.7 | 4.7 |
| logn continuous | unequal, low | 11.7 | 10.4 | 3.2 |
| logn discrete | unequal, low | 12.8 | 11.7 | 3.2 |
| pwExp continuous | equal | 8.6 | 8.0 | 3.3 |
| pwExp discrete | equal | 9.5 | 8.2 | 3.5 |
| pwExp continuous | unequal, high | 9.4 | 8.8 | 2.1 |
| pwExp discrete | unequal, high | 10.2 | 9.3 | 2.6 |
| pwExp continuous | unequal, low | 8.9 | 7.2 | 2.9 |
| pwExp discrete | unequal, low | 8.7 | 7.6 | 2.9 |
| Weib late continuous | equal | 13.1 | 11.5 | 3.9 |
| Weib late discrete | equal | 14.3 | 12.4 | 4.5 |
| Weib late continuous | unequal, high | 14.6 | 13.1 | 3.8 |
| Weib late discrete | unequal, high | 15.6 | 14.1 | 4.1 |
| Weib late continuous | unequal, low | 14.0 | 12.9 | 3.8 |
| Weib late discrete | unequal, low | 16.1 | 13.9 | 4.4 |
| Weib prop continuous | equal | 12.1 | 10.8 | 3.5 |
| Weib prop discrete | equal | 13.4 | 11.9 | 3.8 |
| Weib prop continuous | unequal, high | 13.7 | 12.4 | 3.5 |
| Weib prop discrete | unequal, high | 13.9 | 12.3 | 4.0 |
| Weib prop continuous | unequal, low | 12.7 | 10.8 | 3.6 |
| Weib prop discrete | unequal, low | 13.3 | 12.0 | 3.8 |
| Weib scale continuous | equal | 11.4 | 10.2 | 3.6 |
| Weib scale discrete | equal | 11.3 | 10.2 | 3.9 |
| Weib scale continuous | unequal, high | 13.2 | 12.1 | 3.9 |
| Weib scale discrete | unequal, high | 13.5 | 11.4 | 3.6 |
| Weib scale continuous | unequal, low | 9.8 | 8.4 | 3.3 |
| Weib scale discrete | unequal, low | 10.0 | 8.5 | 3.2 |
| Weib shape continuous | equal | 15.0 | 13.2 | 6.6 |
| Weib shape discrete | equal | 14.8 | 13.3 | 5.7 |
| Weib shape continuous | unequal, high | 12.9 | 11.3 | 3.6 |
| Weib shape discrete | unequal, high | 13.7 | 12.1 | 3.6 |
| Weib shape continuous | unequal, low | 12.9 | 12.1 | 5.6 |
| Weib shape discrete | unequal, low | 12.8 | 12.0 | 5.0 |

Table S25: Rejection rates in percent for the Dunnett-type contrast matrix with $\delta = 1.5$ and balanced large sample sizes.

| Distribution | Censoring distribution | asymptotic | asymptotic_bonf | permutation_bonf |
|---|---|---|---|---|
| exp early continuous | equal | 100.0 | 100.0 | 100.0 |
| exp early discrete | equal | 100.0 | 100.0 | 100.0 |
| exp early continuous | unequal, high | 99.9 | 99.9 | 99.9 |
| exp early discrete | unequal, high | 100.0 | 100.0 | 100.0 |
| exp early continuous | unequal, low | 100.0 | 100.0 | 100.0 |
| exp early discrete | unequal, low | 100.0 | 100.0 | 100.0 |
| exp late continuous | equal | 100.0 | 100.0 | 100.0 |
| exp late discrete | equal | 100.0 | 100.0 | 100.0 |
| exp late continuous | unequal, high | 100.0 | 100.0 | 100.0 |
| exp late discrete | unequal, high | 99.9 | 99.9 | 99.9 |
| exp late continuous | unequal, low | 100.0 | 100.0 | 100.0 |
| exp late discrete | unequal, low | 100.0 | 100.0 | 100.0 |
| exp prop continuous | equal | 100.0 | 100.0 | 100.0 |
| exp prop discrete | equal | 100.0 | 100.0 | 100.0 |
| exp prop continuous | unequal, high | 100.0 | 100.0 | 100.0 |
| exp prop discrete | unequal, high | 100.0 | 100.0 | 100.0 |
| exp prop continuous | unequal, low | 100.0 | 100.0 | 100.0 |
| exp prop discrete | unequal, low | 100.0 | 100.0 | 100.0 |
| logn continuous | equal | 100.0 | 100.0 | 100.0 |
| logn discrete | equal | 100.0 | 100.0 | 100.0 |
| logn continuous | unequal, high | 100.0 | 100.0 | 100.0 |
| logn discrete | unequal, high | 100.0 | 100.0 | 100.0 |
| logn continuous | unequal, low | 100.0 | 100.0 | 100.0 |
| logn discrete | unequal, low | 100.0 | 100.0 | 100.0 |
| pwExp continuous | equal | 100.0 | 100.0 | 100.0 |
| pwExp discrete | equal | 100.0 | 100.0 | 100.0 |
| pwExp continuous | unequal, high | 100.0 | 100.0 | 100.0 |
| pwExp discrete | unequal, high | 100.0 | 100.0 | 100.0 |
| pwExp continuous | unequal, low | 100.0 | 100.0 | 100.0 |
| pwExp discrete | unequal, low | 100.0 | 100.0 | 100.0 |
| Weib late continuous | equal | 100.0 | 100.0 | 100.0 |
| Weib late discrete | equal | 100.0 | 100.0 | 100.0 |
| Weib late continuous | unequal, high | 100.0 | 100.0 | 100.0 |
| Weib late discrete | unequal, high | 100.0 | 100.0 | 100.0 |
| Weib late continuous | unequal, low | 100.0 | 100.0 | 100.0 |
| Weib late discrete | unequal, low | 100.0 | 100.0 | 100.0 |
| Weib prop continuous | equal | 100.0 | 100.0 | 100.0 |
| Weib prop discrete | equal | 100.0 | 100.0 | 100.0 |
| Weib prop continuous | unequal, high | 100.0 | 100.0 | 100.0 |
| Weib prop discrete | unequal, high | 100.0 | 100.0 | 100.0 |
| Weib prop continuous | unequal, low | 100.0 | 100.0 | 100.0 |
| Weib prop discrete | unequal, low | 100.0 | 100.0 | 100.0 |
| Weib scale continuous | equal | 100.0 | 100.0 | 100.0 |
| Weib scale discrete | equal | 100.0 | 100.0 | 100.0 |
| Weib scale continuous | unequal, high | 100.0 | 100.0 | 100.0 |
| Weib scale discrete | unequal, high | 100.0 | 100.0 | 100.0 |
| Weib scale continuous | unequal, low | 100.0 | 100.0 | 100.0 |
| Weib scale discrete | unequal, low | 100.0 | 100.0 | 100.0 |
| Weib shape continuous | equal | 100.0 | 100.0 | 100.0 |
| Weib shape discrete | equal | 100.0 | 100.0 | 100.0 |
| Weib shape continuous | unequal, high | 100.0 | 100.0 | 100.0 |
| Weib shape discrete | unequal, high | 100.0 | 100.0 | 100.0 |
| Weib shape continuous | unequal, low | 100.0 | 100.0 | 100.0 |
| Weib shape discrete | unequal, low | 100.0 | 100.0 | 100.0 |

Table S26: Rejection rates in percent for the Dunnett-type contrast matrix with $\delta = 1.5$ and balanced medium sample sizes.

| Distribution | Censoring distribution | asymptotic | asymptotic_bonf | permutation_bonf |
|---|---|---|---|---|
| exp early continuous | equal | 46.5 | 43.8 | 42.7 |
| exp early discrete | equal | 53.9 | 50.7 | 50.7 |
| exp early continuous | unequal, high | 37.0 | 34.2 | 33.2 |
| exp early discrete | unequal, high | 39.3 | 37.5 | 37.0 |
| exp early continuous | unequal, low | 45.7 | 42.2 | 42.9 |
| exp early discrete | unequal, low | 52.9 | 49.7 | 50.1 |
| exp late continuous | equal | 50.7 | 46.7 | 47.4 |
| exp late discrete | equal | 58.9 | 55.6 | 55.7 |
| exp late continuous | unequal, high | 38.3 | 36.5 | 35.4 |
| exp late discrete | unequal, high | 42.3 | 39.4 | 39.2 |
| exp late continuous | unequal, low | 50.4 | 47.2 | 47.4 |
| exp late discrete | unequal, low | 57.6 | 54.8 | 55.2 |
| exp prop continuous | equal | 49.8 | 47.4 | 45.9 |
| exp prop discrete | equal | 55.6 | 52.3 | 50.9 |
| exp prop continuous | unequal, high | 40.7 | 37.6 | 37.3 |
| exp prop discrete | unequal, high | 43.2 | 40.8 | 40.6 |
| exp prop continuous | unequal, low | 47.4 | 44.7 | 45.4 |
| exp prop discrete | unequal, low | 53.1 | 50.8 | 49.6 |
| logn continuous | equal | 93.9 | 93.1 | 93.2 |
| logn discrete | equal | 96.8 | 96.3 | 96.3 |
| logn continuous | unequal, high | 74.7 | 72.6 | 71.5 |
| logn discrete | unequal, high | 82.8 | 81.7 | 80.3 |
| logn continuous | unequal, low | 94.8 | 93.5 | 93.3 |
| logn discrete | unequal, low | 96.9 | 96.3 | 96.4 |
| pwExp continuous | equal | 45.5 | 43.4 | 42.9 |
| pwExp discrete | equal | 52.4 | 49.8 | 48.2 |
| pwExp continuous | unequal, high | 36.2 | 34.5 | 33.0 |
| pwExp discrete | unequal, high | 40.5 | 38.4 | 38.5 |
| pwExp continuous | unequal, low | 46.5 | 43.8 | 43.9 |
| pwExp discrete | unequal, low | 52.7 | 49.9 | 49.5 |
| Weib late continuous | equal | 95.2 | 94.2 | 93.4 |
| Weib late discrete | equal | 96.6 | 95.7 | 95.6 |
| Weib late continuous | unequal, high | 78.8 | 77.8 | 76.0 |
| Weib late discrete | unequal, high | 82.6 | 81.0 | 79.2 |
| Weib late continuous | unequal, low | 96.9 | 95.6 | 95.2 |
| Weib late discrete | unequal, low | 97.8 | 97.7 | 97.2 |
| Weib prop continuous | equal | 92.7 | 92.1 | 91.5 |
| Weib prop discrete | equal | 97.0 | 96.6 | 96.7 |
| Weib prop continuous | unequal, high | 76.2 | 74.8 | 73.1 |
| Weib prop discrete | unequal, high | 84.8 | 83.2 | 82.3 |
| Weib prop continuous | unequal, low | 95.2 | 94.2 | 93.6 |
| Weib prop discrete | unequal, low | 98.0 | 97.4 | 96.7 |
| Weib scale continuous | equal | 86.0 | 84.5 | 84.2 |
| Weib scale discrete | equal | 92.2 | 91.4 | 91.4 |
| Weib scale continuous | unequal, high | 69.2 | 67.2 | 66.2 |
| Weib scale discrete | unequal, high | 78.7 | 77.1 | 75.7 |
| Weib scale continuous | unequal, low | 89.0 | 87.2 | 85.9 |
| Weib scale discrete | unequal, low | 93.7 | 92.8 | 91.6 |
| Weib shape continuous | equal | 72.0 | 69.9 | 68.2 |
| Weib shape discrete | equal | 81.8 | 80.4 | 79.6 |
| Weib shape continuous | unequal, high | 60.8 | 58.6 | 56.4 |
| Weib shape discrete | unequal, high | 67.2 | 66.3 | 64.9 |
| Weib shape continuous | unequal, low | 71.5 | 69.7 | 69.8 |
| Weib shape discrete | unequal, low | 83.0 | 81.1 | 80.8 |

Table S27: Rejection rates in percent for the Dunnett-type contrast matrix with $\delta = 1.5$ and balanced small sample sizes.

| Distribution | Censoring distribution | asymptotic | asymptotic_bonf | permutation_bonf |
|---|---|---|---|---|
| exp early continuous | equal | 11.6 | 10.1 | 8.0 |
| exp early discrete | equal | 10.1 | 9.5 | 8.7 |
| exp early continuous | unequal, high | 11.7 | 10.6 | 8.7 |
| exp early discrete | unequal, high | 9.7 | 8.6 | 7.3 |
| exp early continuous | unequal, low | 11.5 | 10.2 | 9.0 |
| exp early discrete | unequal, low | 10.4 | 9.2 | 7.9 |
| exp late continuous | equal | 12.2 | 10.5 | 8.3 |
| exp late discrete | equal | 12.0 | 10.5 | 9.5 |
| exp late continuous | unequal, high | 12.6 | 11.3 | 8.8 |
| exp late discrete | unequal, high | 12.3 | 10.9 | 8.8 |
| exp late continuous | unequal, low | 12.0 | 10.7 | 9.5 |
| exp late discrete | unequal, low | 13.2 | 11.6 | 10.4 |
| exp prop continuous | equal | 10.0 | 9.7 | 7.3 |
| exp prop discrete | equal | 12.0 | 11.2 | 10.0 |
| exp prop continuous | unequal, high | 11.1 | 10.6 | 8.1 |
| exp prop discrete | unequal, high | 10.5 | 9.4 | 7.8 |
| exp prop continuous | unequal, low | 10.4 | 9.2 | 7.9 |
| exp prop discrete | unequal, low | 11.5 | 10.8 | 8.9 |
| logn continuous | equal | 24.1 | 22.2 | 20.1 |
| logn discrete | equal | 27.4 | 25.8 | 23.3 |
| logn continuous | unequal, high | 19.2 | 18.0 | 16.0 |
| logn discrete | unequal, high | 20.9 | 19.9 | 17.5 |
| logn continuous | unequal, low | 25.4 | 23.7 | 20.9 |
| logn discrete | unequal, low | 28.8 | 26.8 | 25.7 |
| pwExp continuous | equal | 10.9 | 9.5 | 7.3 |
| pwExp discrete | equal | 10.9 | 9.7 | 7.6 |
| pwExp continuous | unequal, high | 11.3 | 10.4 | 7.5 |
| pwExp discrete | unequal, high | 12.4 | 11.3 | 9.1 |
| pwExp continuous | unequal, low | 10.5 | 8.9 | 8.1 |
| pwExp discrete | unequal, low | 11.4 | 10.1 | 8.8 |
| Weib late continuous | equal | 23.4 | 21.9 | 19.9 |
| Weib late discrete | equal | 27.1 | 24.9 | 22.1 |
| Weib late continuous | unequal, high | 18.4 | 17.5 | 13.3 |
| Weib late discrete | unequal, high | 21.3 | 20.2 | 16.1 |
| Weib late continuous | unequal, low | 27.5 | 26.3 | 22.6 |
| Weib late discrete | unequal, low | 30.6 | 29.1 | 25.9 |
| Weib prop continuous | equal | 21.8 | 20.7 | 18.4 |
| Weib prop discrete | equal | 24.4 | 23.1 | 20.0 |
| Weib prop continuous | unequal, high | 17.8 | 16.7 | 12.8 |
| Weib prop discrete | unequal, high | 20.0 | 19.0 | 14.5 |
| Weib prop continuous | unequal, low | 25.6 | 23.5 | 21.7 |
| Weib prop discrete | unequal, low | 29.3 | 26.9 | 24.6 |
| Weib scale continuous | equal | 16.8 | 15.6 | 14.1 |
| Weib scale discrete | equal | 17.5 | 15.9 | 15.3 |
| Weib scale continuous | unequal, high | 13.3 | 12.6 | 10.7 |
| Weib scale discrete | unequal, high | 14.6 | 13.5 | 11.3 |
| Weib scale continuous | unequal, low | 18.3 | 16.7 | 15.8 |
| Weib scale discrete | unequal, low | 20.4 | 18.7 | 17.2 |
| Weib shape continuous | equal | 12.2 | 11.2 | 10.9 |
| Weib shape discrete | equal | 13.5 | 12.3 | 11.6 |
| Weib shape continuous | unequal, high | 10.8 | 10.1 | 8.7 |
| Weib shape discrete | unequal, high | 11.6 | 10.7 | 9.7 |
| Weib shape continuous | unequal, low | 12.1 | 11.2 | 11.7 |
| Weib shape discrete | unequal, low | 13.2 | 12.7 | 13.1 |

Table S28: Rejection rates in percent for the Dunnett-type contrast matrix with $\delta = 1.5$ and unbalanced large sample sizes.

| Distribution | Censoring distribution | asymptotic | asymptotic_bonf | permutation_bonf |
|---|---|---|---|---|
| exp early continuous | equal | 89.8 | 89.1 | 85.7 |
| exp early discrete | equal | 94.7 | 93.8 | 92.2 |
| exp early continuous | unequal, high | 83.5 | 82.2 | 78.1 |
| exp early discrete | unequal, high | 86.9 | 86.0 | 83.3 |
| exp early continuous | unequal, low | 90.1 | 88.8 | 88.0 |
| exp early discrete | unequal, low | 94.7 | 93.8 | 92.4 |
| exp late continuous | equal | 97.1 | 96.9 | 94.9 |
| exp late discrete | equal | 98.6 | 98.5 | 97.1 |
| exp late continuous | unequal, high | 89.2 | 88.1 | 84.5 |
| exp late discrete | unequal, high | 92.6 | 91.9 | 88.5 |
| exp late continuous | unequal, low | 97.7 | 97.5 | 96.0 |
| exp late discrete | unequal, low | 99.3 | 99.2 | 98.8 |
| exp prop continuous | equal | 94.7 | 94.3 | 92.3 |
| exp prop discrete | equal | 98.0 | 97.7 | 96.3 |
| exp prop continuous | unequal, high | 84.3 | 82.9 | 80.2 |
| exp prop discrete | unequal, high | 89.4 | 88.6 | 83.8 |
| exp prop continuous | unequal, low | 94.6 | 94.1 | 92.8 |
| exp prop discrete | unequal, low | 97.9 | 97.8 | 96.2 |
| logn continuous | equal | 100.0 | 100.0 | 100.0 |
| logn discrete | equal | 100.0 | 100.0 | 100.0 |
| logn continuous | unequal, high | 99.8 | 99.8 | 99.7 |
| logn discrete | unequal, high | 99.9 | 99.9 | 99.6 |
| logn continuous | unequal, low | 100.0 | 100.0 | 100.0 |
| logn discrete | unequal, low | 100.0 | 100.0 | 100.0 |
| pwExp continuous | equal | 88.8 | 88.1 | 85.3 |
| pwExp discrete | equal | 93.5 | 92.9 | 91.3 |
| pwExp continuous | unequal, high | 81.5 | 80.8 | 78.0 |
| pwExp discrete | unequal, high | 85.5 | 84.4 | 82.3 |
| pwExp continuous | unequal, low | 88.4 | 87.4 | 85.0 |
| pwExp discrete | unequal, low | 94.5 | 94.4 | 92.2 |
| Weib late continuous | equal | 100.0 | 100.0 | 100.0 |
| Weib late discrete | equal | 99.7 | 99.7 | 99.5 |
| Weib late continuous | unequal, high | 100.0 | 100.0 | 99.9 |
| Weib late discrete | unequal, high | 98.8 | 98.8 | 98.7 |
| Weib late continuous | unequal, low | 100.0 | 100.0 | 100.0 |
| Weib late discrete | unequal, low | 99.9 | 99.9 | 99.9 |
| Weib prop continuous | equal | 100.0 | 100.0 | 100.0 |
| Weib prop discrete | equal | 100.0 | 100.0 | 100.0 |
| Weib prop continuous | unequal, high | 100.0 | 100.0 | 99.5 |
| Weib prop discrete | unequal, high | 100.0 | 100.0 | 99.8 |
| Weib prop continuous | unequal, low | 100.0 | 100.0 | 100.0 |
| Weib prop discrete | unequal, low | 100.0 | 100.0 | 100.0 |
| Weib scale continuous | equal | 99.9 | 99.9 | 99.7 |
| Weib scale discrete | equal | 100.0 | 100.0 | 100.0 |
| Weib scale continuous | unequal, high | 98.5 | 98.2 | 97.2 |
| Weib scale discrete | unequal, high | 99.7 | 99.7 | 98.8 |
| Weib scale continuous | unequal, low | 100.0 | 100.0 | 99.8 |
| Weib scale discrete | unequal, low | 100.0 | 100.0 | 100.0 |
| Weib shape continuous | equal | 97.2 | 96.9 | 95.3 |
| Weib shape discrete | equal | 99.2 | 99.1 | 98.3 |
| Weib shape continuous | unequal, high | 94.1 | 93.8 | 90.6 |
| Weib shape discrete | unequal, high | 97.1 | 96.9 | 94.0 |
| Weib shape continuous | unequal, low | 96.9 | 96.6 | 95.5 |
| Weib shape discrete | unequal, low | 98.9 | 98.9 | 98.1 |

Table S29: Rejection rates in percent for the Dunnett-type contrast matrix with $\delta = 1.5$ and unbalanced medium sample sizes.

| Distribution | Censoring distribution | asymptotic | asymptotic_bonf | permutation_bonf |
|---|---|---|---|---|
| exp early continuous | equal | 14.1 | 13.8 | 8.5 |
| exp early discrete | equal | 16.0 | 15.0 | 9.9 |
| exp early continuous | unequal, high | 12.9 | 12.2 | 7.1 |
| exp early discrete | unequal, high | 13.5 | 12.9 | 6.5 |
| exp early continuous | unequal, low | 14.5 | 13.5 | 9.5 |
| exp early discrete | unequal, low | 17.2 | 15.9 | 10.3 |
| exp late continuous | equal | 17.3 | 16.4 | 11.0 |
| exp late discrete | equal | 19.1 | 17.8 | 12.3 |
| exp late continuous | unequal, high | 13.5 | 12.9 | 7.2 |
| exp late discrete | unequal, high | 14.3 | 13.4 | 7.9 |
| exp late continuous | unequal, low | 16.8 | 15.9 | 10.9 |
| exp late discrete | unequal, low | 19.6 | 18.2 | 11.7 |
| exp prop continuous | equal | 15.2 | 14.2 | 9.9 |
| exp prop discrete | equal | 17.0 | 16.2 | 9.6 |
| exp prop continuous | unequal, high | 12.5 | 11.9 | 7.7 |
| exp prop discrete | unequal, high | 15.2 | 14.3 | 8.6 |
| exp prop continuous | unequal, low | 14.7 | 13.8 | 10.1 |
| exp prop discrete | unequal, low | 17.9 | 16.4 | 10.4 |
| logn continuous | equal | 36.7 | 35.6 | 18.5 |
| logn discrete | equal | 40.4 | 39.4 | 19.0 |
| logn continuous | unequal, high | 23.9 | 23.1 | 8.0 |
| logn discrete | unequal, high | 26.1 | 25.5 | 7.6 |
| logn continuous | unequal, low | 40.6 | 39.5 | 18.1 |
| logn discrete | unequal, low | 45.8 | 45.2 | 20.6 |
| pwExp continuous | equal | 13.1 | 12.8 | 8.9 |
| pwExp discrete | equal | 15.4 | 14.6 | 10.2 |
| pwExp continuous | unequal, high | 12.3 | 11.4 | 6.6 |
| pwExp discrete | unequal, high | 12.8 | 12.0 | 7.5 |
| pwExp continuous | unequal, low | 14.1 | 13.0 | 8.9 |
| pwExp discrete | unequal, low | 15.6 | 14.7 | 9.4 |
| Weib late continuous | equal | 41.8 | 40.2 | 22.0 |
| Weib late discrete | equal | 48.9 | 46.9 | 26.1 |
| Weib late continuous | unequal, high | 28.5 | 26.7 | 10.8 |
| Weib late discrete | unequal, high | 31.7 | 30.3 | 10.5 |
| Weib late continuous | unequal, low | 47.4 | 45.8 | 24.6 |
| Weib late discrete | unequal, low | 55.5 | 54.2 | 27.2 |
| Weib prop continuous | equal | 36.9 | 35.7 | 18.9 |
| Weib prop discrete | equal | 42.0 | 41.1 | 19.6 |
| Weib prop continuous | unequal, high | 25.7 | 25.0 | 10.4 |
| Weib prop discrete | unequal, high | 29.9 | 28.4 | 9.8 |
| Weib prop continuous | unequal, low | 41.8 | 40.5 | 19.5 |
| Weib prop discrete | unequal, low | 48.7 | 47.6 | 21.6 |
| Weib scale continuous | equal | 25.3 | 24.2 | 12.4 |
| Weib scale discrete | equal | 28.6 | 27.7 | 12.8 |
| Weib scale continuous | unequal, high | 20.0 | 19.2 | 8.0 |
| Weib scale discrete | unequal, high | 21.6 | 20.8 | 7.0 |
| Weib scale continuous | unequal, low | 25.6 | 24.8 | 11.1 |
| Weib scale discrete | unequal, low | 30.1 | 29.2 | 12.3 |
| Weib shape continuous | equal | 14.6 | 13.9 | 7.3 |
| Weib shape discrete | equal | 17.7 | 16.8 | 7.2 |
| Weib shape continuous | unequal, high | 15.9 | 15.0 | 6.9 |
| Weib shape discrete | unequal, high | 17.2 | 17.0 | 6.6 |
| Weib shape continuous | unequal, low | 14.2 | 13.8 | 7.0 |
| Weib shape discrete | unequal, low | 17.8 | 17.2 | 7.5 |

Table S30: Rejection rates in percent for the Dunnett-type contrast matrix with $\delta = 1.5$ and unbalanced small sample sizes.

| Distribution | Censoring distribution | asymptotic | asymptotic_bonf | permutation_bonf |
|---|---|---|---|---|
| exp early continuous | equal | 11.8 | 11.4 | 4.2 |
| exp early discrete | equal | 11.3 | 10.8 | 3.1 |
| exp early continuous | unequal, high | 13.2 | 12.8 | 3.6 |
| exp early discrete | unequal, high | 16.0 | 15.6 | 4.0 |
| exp early continuous | unequal, low | 12.0 | 11.7 | 3.8 |
| exp early discrete | unequal, low | 12.3 | 11.6 | 3.0 |
| exp late continuous | equal | 11.9 | 10.9 | 3.8 |
| exp late discrete | equal | 11.7 | 11.1 | 3.1 |
| exp late continuous | unequal, high | 12.9 | 12.5 | 3.5 |
| exp late discrete | unequal, high | 15.3 | 14.6 | 4.0 |
| exp late continuous | unequal, low | 11.8 | 11.3 | 3.9 |
| exp late discrete | unequal, low | 12.2 | 11.3 | 3.1 |
| exp prop continuous | equal | 12.2 | 11.7 | 5.1 |
| exp prop discrete | equal | 12.7 | 12.4 | 4.9 |
| exp prop continuous | unequal, high | 15.7 | 15.2 | 5.0 |
| exp prop discrete | unequal, high | 16.3 | 15.8 | 5.0 |
| exp prop continuous | unequal, low | 11.8 | 11.0 | 4.5 |
| exp prop discrete | unequal, low | 12.3 | 11.8 | 4.1 |
| logn continuous | equal | 17.2 | 16.9 | 5.0 |
| logn discrete | equal | 17.8 | 17.3 | 4.6 |
| logn continuous | unequal, high | 29.0 | 28.7 | 5.8 |
| logn discrete | unequal, high | 31.1 | 30.6 | 5.5 |
| logn continuous | unequal, low | 16.2 | 15.8 | 3.8 |
| logn discrete | unequal, low | 16.4 | 16.2 | 4.0 |
| pwExp continuous | equal | 11.0 | 10.2 | 3.9 |
| pwExp discrete | equal | 10.9 | 10.5 | 3.7 |
| pwExp continuous | unequal, high | 12.9 | 12.4 | 3.6 |
| pwExp discrete | unequal, high | 14.9 | 14.4 | 4.0 |
| pwExp continuous | unequal, low | 11.1 | 10.7 | 3.7 |
| pwExp discrete | unequal, low | 11.3 | 10.7 | 3.7 |
| Weib late continuous | equal | 13.5 | 13.1 | 4.1 |
| Weib late discrete | equal | 15.5 | 15.0 | 3.6 |
| Weib late continuous | unequal, high | 21.5 | 20.8 | 4.5 |
| Weib late discrete | unequal, high | 25.8 | 25.1 | 4.6 |
| Weib late continuous | unequal, low | 14.3 | 13.9 | 3.7 |
| Weib late discrete | unequal, low | 16.1 | 16.0 | 4.1 |
| Weib prop continuous | equal | 14.0 | 13.6 | 4.1 |
| Weib prop discrete | equal | 15.7 | 15.3 | 4.6 |
| Weib prop continuous | unequal, high | 22.6 | 21.8 | 4.8 |
| Weib prop discrete | unequal, high | 27.4 | 26.8 | 5.1 |
| Weib prop continuous | unequal, low | 13.7 | 13.3 | 3.5 |
| Weib prop discrete | unequal, low | 15.6 | 15.3 | 3.1 |
| Weib scale continuous | equal | 15.9 | 15.6 | 4.7 |
| Weib scale discrete | equal | 18.3 | 18.2 | 5.0 |
| Weib scale continuous | unequal, high | 26.9 | 26.6 | 4.4 |
| Weib scale discrete | unequal, high | 30.7 | 30.2 | 4.8 |
| Weib scale continuous | unequal, low | 15.9 | 15.6 | 3.7 |
| Weib scale discrete | unequal, low | 17.1 | 16.8 | 3.3 |
| Weib shape continuous | equal | 23.1 | 22.8 | 6.8 |
| Weib shape discrete | equal | 24.0 | 23.9 | 6.2 |
| Weib shape continuous | unequal, high | 29.7 | 29.5 | 4.8 |
| Weib shape discrete | unequal, high | 34.5 | 34.3 | 6.0 |
| Weib shape continuous | unequal, low | 24.1 | 23.9 | 5.7 |
| Weib shape discrete | unequal, low | 24.7 | 24.6 | 5.4 |

Table S31: Rejection rates in percent for the Tukey-type contrast matrix with $\delta = 1.5$ and balanced large sample sizes.

| Distribution | Censoring distribution | asymptotic | asymptotic_bonf | permutation_bonf |
|---|---|---|---|---|
| exp early continuous | equal | 100.0 | 100.0 | 100.0 |
| exp early discrete | equal | 100.0 | 100.0 | 100.0 |
| exp early continuous | unequal, high | 100.0 | 100.0 | 100.0 |
| exp early discrete | unequal, high | 100.0 | 100.0 | 100.0 |
| exp early continuous | unequal, low | 100.0 | 100.0 | 100.0 |
| exp early discrete | unequal, low | 100.0 | 100.0 | 100.0 |
| exp late continuous | equal | 100.0 | 100.0 | 100.0 |
| exp late discrete | equal | 100.0 | 100.0 | 100.0 |
| exp late continuous | unequal, high | 100.0 | 100.0 | 100.0 |
| exp late discrete | unequal, high | 99.9 | 99.9 | 99.9 |
| exp late continuous | unequal, low | 100.0 | 100.0 | 100.0 |
| exp late discrete | unequal, low | 100.0 | 100.0 | 100.0 |
| exp prop continuous | equal | 100.0 | 100.0 | 100.0 |
| exp prop discrete | equal | 100.0 | 100.0 | 100.0 |
| exp prop continuous | unequal, high | 100.0 | 100.0 | 100.0 |
| exp prop discrete | unequal, high | 100.0 | 100.0 | 100.0 |
| exp prop continuous | unequal, low | 100.0 | 100.0 | 100.0 |
| exp prop discrete | unequal, low | 100.0 | 100.0 | 100.0 |
| logn continuous | equal | 100.0 | 100.0 | 100.0 |
| logn discrete | equal | 100.0 | 100.0 | 100.0 |
| logn continuous | unequal, high | 100.0 | 100.0 | 100.0 |
| logn discrete | unequal, high | 100.0 | 100.0 | 100.0 |
| logn continuous | unequal, low | 100.0 | 100.0 | 100.0 |
| logn discrete | unequal, low | 100.0 | 100.0 | 100.0 |
| pwExp continuous | equal | 100.0 | 100.0 | 100.0 |
| pwExp discrete | equal | 100.0 | 100.0 | 100.0 |
| pwExp continuous | unequal, high | 100.0 | 100.0 | 100.0 |
| pwExp discrete | unequal, high | 100.0 | 100.0 | 100.0 |
| pwExp continuous | unequal, low | 100.0 | 100.0 | 100.0 |
| pwExp discrete | unequal, low | 100.0 | 100.0 | 100.0 |
| Weib late continuous | equal | 100.0 | 100.0 | 100.0 |
| Weib late discrete | equal | 100.0 | 100.0 | 100.0 |
| Weib late continuous | unequal, high | 100.0 | 100.0 | 100.0 |
| Weib late discrete | unequal, high | 100.0 | 100.0 | 100.0 |
| Weib late continuous | unequal, low | 100.0 | 100.0 | 100.0 |
| Weib late discrete | unequal, low | 100.0 | 100.0 | 100.0 |
| Weib prop continuous | equal | 100.0 | 100.0 | 100.0 |
| Weib prop discrete | equal | 100.0 | 100.0 | 100.0 |
| Weib prop continuous | unequal, high | 100.0 | 100.0 | 100.0 |
| Weib prop discrete | unequal, high | 100.0 | 100.0 | 100.0 |
| Weib prop continuous | unequal, low | 100.0 | 100.0 | 100.0 |
| Weib prop discrete | unequal, low | 100.0 | 100.0 | 100.0 |
| Weib scale continuous | equal | 100.0 | 100.0 | 100.0 |
| Weib scale discrete | equal | 100.0 | 100.0 | 100.0 |
| Weib scale continuous | unequal, high | 100.0 | 100.0 | 100.0 |
| Weib scale discrete | unequal, high | 100.0 | 100.0 | 100.0 |
| Weib scale continuous | unequal, low | 100.0 | 100.0 | 100.0 |
| Weib scale discrete | unequal, low | 100.0 | 100.0 | 100.0 |
| Weib shape continuous | equal | 100.0 | 100.0 | 100.0 |
| Weib shape discrete | equal | 100.0 | 100.0 | 100.0 |
| Weib shape continuous | unequal, high | 100.0 | 100.0 | 100.0 |
| Weib shape discrete | unequal, high | 100.0 | 100.0 | 100.0 |
| Weib shape continuous | unequal, low | 100.0 | 100.0 | 100.0 |
| Weib shape discrete | unequal, low | 100.0 | 100.0 | 100.0 |

Table S32: Rejection rates in percent for the Tukey-type contrast matrix with $\delta = 1.5$ and balanced medium sample sizes.

| Distribution | Censoring distribution | asymptotic | asymptotic_bonf | permutation_bonf |
|---|---|---|---|---|
| exp early continuous | equal | 65.9 | 61.6 | 60.0 |
| exp early discrete | equal | 74.4 | 70.4 | 67.7 |
| exp early continuous | unequal, high | 54.1 | 50.2 | 49.3 |
| exp early discrete | unequal, high | 59.9 | 55.5 | 53.8 |
| exp early continuous | unequal, low | 66.6 | 63.1 | 60.4 |
| exp early discrete | unequal, low | 74.0 | 69.9 | 67.0 |
| exp late continuous | equal | 72.1 | 67.9 | 66.4 |
| exp late discrete | equal | 79.5 | 75.9 | 74.9 |
| exp late continuous | unequal, high | 59.0 | 54.5 | 52.8 |
| exp late discrete | unequal, high | 63.6 | 58.9 | 57.9 |
| exp late continuous | unequal, low | 73.4 | 70.1 | 67.0 |
| exp late discrete | unequal, low | 79.6 | 76.0 | 73.8 |
| exp prop continuous | equal | 66.1 | 62.0 | 61.2 |
| exp prop discrete | equal | 74.4 | 71.0 | 68.7 |
| exp prop continuous | unequal, high | 56.0 | 51.7 | 51.1 |
| exp prop discrete | unequal, high | 61.7 | 57.3 | 54.5 |
| exp prop continuous | unequal, low | 66.5 | 63.0 | 60.0 |
| exp prop discrete | unequal, low | 73.9 | 70.0 | 67.0 |
| logn continuous | equal | 99.3 | 99.2 | 98.8 |
| logn discrete | equal | 99.8 | 99.6 | 99.5 |
| logn continuous | unequal, high | 89.8 | 87.8 | 85.6 |
| logn discrete | unequal, high | 94.4 | 92.9 | 92.2 |
| logn continuous | unequal, low | 99.4 | 99.2 | 98.9 |
| logn discrete | unequal, low | 99.9 | 99.8 | 99.7 |
| pwExp continuous | equal | 63.8 | 59.1 | 59.3 |
| pwExp discrete | equal | 70.1 | 66.5 | 66.0 |
| pwExp continuous | unequal, high | 51.8 | 48.9 | 48.4 |
| pwExp discrete | unequal, high | 59.0 | 55.2 | 52.4 |
| pwExp continuous | unequal, low | 63.4 | 59.2 | 57.2 |
| pwExp discrete | unequal, low | 69.9 | 66.5 | 64.7 |
| Weib late continuous | equal | 99.5 | 99.3 | 99.2 |
| Weib late discrete | equal | 99.3 | 98.9 | 98.6 |
| Weib late continuous | unequal, high | 93.6 | 92.2 | 90.4 |
| Weib late discrete | unequal, high | 95.8 | 94.4 | 92.7 |
| Weib late continuous | unequal, low | 99.6 | 99.5 | 99.4 |
| Weib late discrete | unequal, low | 99.3 | 99.3 | 99.0 |
| Weib prop continuous | equal | 99.1 | 98.7 | 98.5 |
| Weib prop discrete | equal | 99.9 | 99.8 | 99.7 |
| Weib prop continuous | unequal, high | 92.8 | 90.8 | 88.6 |
| Weib prop discrete | unequal, high | 95.8 | 94.9 | 93.9 |
| Weib prop continuous | unequal, low | 99.5 | 99.0 | 99.0 |
| Weib prop discrete | unequal, low | 100.0 | 99.9 | 99.8 |
| Weib scale continuous | equal | 95.1 | 94.2 | 93.6 |
| Weib scale discrete | equal | 97.8 | 97.2 | 96.5 |
| Weib scale continuous | unequal, high | 84.5 | 81.6 | 80.8 |
| Weib scale discrete | unequal, high | 91.2 | 88.7 | 87.2 |
| Weib scale continuous | unequal, low | 95.7 | 95.1 | 94.2 |
| Weib scale discrete | unequal, low | 98.7 | 98.4 | 98.1 |
| Weib shape continuous | equal | 84.5 | 81.0 | 80.2 |
| Weib shape discrete | equal | 91.5 | 89.4 | 88.0 |
| Weib shape continuous | unequal, high | 75.0 | 72.2 | 71.2 |
| Weib shape discrete | unequal, high | 81.7 | 79.8 | 77.6 |
| Weib shape continuous | unequal, low | 83.4 | 81.3 | 80.4 |
| Weib shape discrete | unequal, low | 91.4 | 88.6 | 88.8 |

Table S33: Rejection rates in percent for the Tukey-type contrast matrix with $\delta = 1.5$ and balanced small sample sizes.

| Distribution | Censoring distribution | asymptotic | asymptotic_bonf | permutation_bonf |
|---|---|---|---|---|
| exp early continuous | equal | 14.9 | 12.5 | 9.9 |
| exp early discrete | equal | 14.3 | 12.4 | 10.3 |
| exp early continuous | unequal, high | 16.4 | 13.5 | 11.4 |
| exp early discrete | unequal, high | 13.5 | 11.8 | 9.8 |
| exp early continuous | unequal, low | 16.3 | 14.5 | 10.3 |
| exp early discrete | unequal, low | 14.8 | 12.2 | 10.9 |
| exp late continuous | equal | 17.4 | 15.7 | 12.7 |
| exp late discrete | equal | 18.0 | 15.9 | 14.5 |
| exp late continuous | unequal, high | 17.8 | 15.3 | 10.6 |
| exp late discrete | unequal, high | 17.7 | 15.4 | 10.9 |
| exp late continuous | unequal, low | 18.1 | 16.7 | 12.9 |
| exp late discrete | unequal, low | 20.0 | 18.2 | 14.0 |
| exp prop continuous | equal | 14.2 | 12.5 | 10.9 |
| exp prop discrete | equal | 15.7 | 14.1 | 12.1 |
| exp prop continuous | unequal, high | 14.0 | 12.5 | 10.6 |
| exp prop discrete | unequal, high | 15.6 | 13.4 | 10.0 |
| exp prop continuous | unequal, low | 14.6 | 12.7 | 10.0 |
| exp prop discrete | unequal, low | 16.9 | 14.9 | 12.3 |
| logn continuous | equal | 32.8 | 29.5 | 26.8 |
| logn discrete | equal | 38.7 | 34.6 | 31.6 |
| logn continuous | unequal, high | 27.0 | 24.0 | 19.5 |
| logn discrete | unequal, high | 30.2 | 27.2 | 22.2 |
| logn continuous | unequal, low | 36.2 | 32.6 | 28.1 |
| logn discrete | unequal, low | 43.1 | 38.7 | 34.9 |
| pwExp continuous | equal | 13.5 | 11.4 | 9.7 |
| pwExp discrete | equal | 14.9 | 12.5 | 10.4 |
| pwExp continuous | unequal, high | 15.1 | 13.0 | 9.4 |
| pwExp discrete | unequal, high | 16.6 | 14.0 | 11.3 |
| pwExp continuous | unequal, low | 14.6 | 13.3 | 9.9 |
| pwExp discrete | unequal, low | 16.9 | 14.3 | 11.0 |
| Weib late continuous | equal | 34.9 | 31.9 | 26.9 |
| Weib late discrete | equal | 40.0 | 36.1 | 30.7 |
| Weib late continuous | unequal, high | 29.0 | 25.6 | 19.4 |
| Weib late discrete | unequal, high | 30.7 | 28.1 | 22.1 |
| Weib late continuous | unequal, low | 40.4 | 37.0 | 30.4 |
| Weib late discrete | unequal, low | 46.3 | 42.9 | 37.5 |
| Weib prop continuous | equal | 30.9 | 28.7 | 25.0 |
| Weib prop discrete | equal | 36.4 | 32.0 | 29.6 |
| Weib prop continuous | unequal, high | 26.8 | 23.7 | 17.9 |
| Weib prop discrete | unequal, high | 27.8 | 25.5 | 19.9 |
| Weib prop continuous | unequal, low | 36.6 | 32.4 | 27.9 |
| Weib prop discrete | unequal, low | 42.6 | 39.1 | 33.6 |
| Weib scale continuous | equal | 22.2 | 18.9 | 16.9 |
| Weib scale discrete | equal | 25.3 | 22.0 | 19.0 |
| Weib scale continuous | unequal, high | 20.2 | 16.9 | 14.1 |
| Weib scale discrete | unequal, high | 20.8 | 17.8 | 14.8 |
| Weib scale continuous | unequal, low | 23.6 | 21.0 | 18.1 |
| Weib scale discrete | unequal, low | 29.0 | 25.6 | 22.2 |
| Weib shape continuous | equal | 13.4 | 11.2 | 11.6 |
| Weib shape discrete | equal | 15.1 | 13.0 | 12.2 |
| Weib shape continuous | unequal, high | 14.9 | 12.8 | 10.9 |
| Weib shape discrete | unequal, high | 16.0 | 13.8 | 12.3 |
| Weib shape continuous | unequal, low | 15.4 | 13.2 | 12.4 |
| Weib shape discrete | unequal, low | 18.0 | 15.6 | 14.2 |

Table S34: Rejection rates in percent for the Tukey-type contrast matrix with $\delta = 1.5$ and unbalanced large sample sizes.

| Distribution | Censoring distribution | asymptotic | asymptotic_bonf | permutation_bonf |
|---|---|---|---|---|
| exp early continuous | equal | 92.2 | 89.3 | 85.1 |
| exp early discrete | equal | 95.6 | 93.7 | 91.7 |
| exp early continuous | unequal, high | 85.1 | 80.7 | 76.9 |
| exp early discrete | unequal, high | 88.3 | 85.5 | 81.1 |
| exp early continuous | unequal, low | 91.7 | 89.5 | 86.6 |
| exp early discrete | unequal, low | 96.1 | 93.7 | 92.5 |
| exp late continuous | equal | 98.4 | 95.9 | 94.3 |
| exp late discrete | equal | 99.0 | 98.2 | 97.2 |
| exp late continuous | unequal, high | 91.7 | 87.5 | 83.9 |
| exp late discrete | unequal, high | 95.0 | 92.1 | 88.3 |
| exp late continuous | unequal, low | 98.3 | 96.9 | 96.4 |
| exp late discrete | unequal, low | 99.6 | 99.0 | 98.3 |
| exp prop continuous | equal | 95.9 | 93.1 | 91.7 |
| exp prop discrete | equal | 98.3 | 96.9 | 95.0 |
| exp prop continuous | unequal, high | 88.6 | 84.6 | 80.3 |
| exp prop discrete | unequal, high | 91.3 | 88.0 | 84.9 |
| exp prop continuous | unequal, low | 96.2 | 93.5 | 92.7 |
| exp prop discrete | unequal, low | 98.9 | 97.6 | 96.8 |
| logn continuous | equal | 100.0 | 100.0 | 100.0 |
| logn discrete | equal | 100.0 | 100.0 | 100.0 |
| logn continuous | unequal, high | 99.9 | 99.8 | 99.2 |
| logn discrete | unequal, high | 99.9 | 99.9 | 99.8 |
| logn continuous | unequal, low | 100.0 | 100.0 | 100.0 |
| logn discrete | unequal, low | 100.0 | 100.0 | 100.0 |
| pwExp continuous | equal | 90.4 | 87.2 | 85.8 |
| pwExp discrete | equal | 95.5 | 92.9 | 91.2 |
| pwExp continuous | unequal, high | 84.6 | 81.5 | 77.6 |
| pwExp discrete | unequal, high | 88.8 | 84.6 | 81.7 |
| pwExp continuous | unequal, low | 91.0 | 87.5 | 86.3 |
| pwExp discrete | unequal, low | 95.3 | 93.7 | 91.6 |
| Weib late continuous | equal | 100.0 | 100.0 | 100.0 |
| Weib late discrete | equal | 99.7 | 99.7 | 99.6 |
| Weib late continuous | unequal, high | 100.0 | 100.0 | 100.0 |
| Weib late discrete | unequal, high | 99.1 | 98.7 | 98.5 |
| Weib late continuous | unequal, low | 100.0 | 100.0 | 100.0 |
| Weib late discrete | unequal, low | 99.9 | 99.9 | 99.9 |
| Weib prop continuous | equal | 100.0 | 100.0 | 100.0 |
| Weib prop discrete | equal | 100.0 | 100.0 | 100.0 |
| Weib prop continuous | unequal, high | 99.9 | 99.9 | 99.9 |
| Weib prop discrete | unequal, high | 100.0 | 100.0 | 99.8 |
| Weib prop continuous | unequal, low | 100.0 | 100.0 | 100.0 |
| Weib prop discrete | unequal, low | 100.0 | 100.0 | 100.0 |
| Weib scale continuous | equal | 100.0 | 100.0 | 99.9 |
| Weib scale discrete | equal | 100.0 | 100.0 | 99.9 |
| Weib scale continuous | unequal, high | 99.0 | 98.6 | 97.0 |
| Weib scale discrete | unequal, high | 99.7 | 99.3 | 98.5 |
| Weib scale continuous | unequal, low | 100.0 | 99.9 | 99.8 |
| Weib scale discrete | unequal, low | 100.0 | 100.0 | 100.0 |
| Weib shape continuous | equal | 97.7 | 96.5 | 95.0 |
| Weib shape discrete | equal | 99.4 | 99.0 | 98.3 |
| Weib shape continuous | unequal, high | 95.9 | 94.3 | 89.3 |
| Weib shape discrete | unequal, high | 97.6 | 96.6 | 93.1 |
| Weib shape continuous | unequal, low | 97.9 | 97.2 | 95.1 |
| Weib shape discrete | unequal, low | 99.3 | 99.0 | 98.1 |

Table S35: Rejection rates in percent for the Tukey-type contrast matrix with $\delta = 1.5$ and unbalanced medium sample sizes.

| Distribution | Censoring distribution | asymptotic | asymptotic_bonf | permutation_bonf |
|---|---|---|---|---|
| exp early continuous | equal | 16.1 | 12.9 | 9.2 |
| exp early discrete | equal | 19.3 | 15.6 | 10.7 |
| exp early continuous | unequal, high | 13.8 | 11.4 | 6.7 |
| exp early discrete | unequal, high | 15.1 | 12.4 | 6.6 |
| exp early continuous | unequal, low | 15.7 | 12.9 | 8.9 |
| exp early discrete | unequal, low | 19.1 | 15.4 | 11.5 |
| exp late continuous | equal | 19.4 | 15.3 | 11.0 |
| exp late discrete | equal | 21.6 | 18.1 | 11.8 |
| exp late continuous | unequal, high | 16.0 | 12.3 | 7.3 |
| exp late discrete | unequal, high | 16.8 | 13.8 | 7.5 |
| exp late continuous | unequal, low | 18.5 | 15.8 | 10.1 |
| exp late discrete | unequal, low | 22.5 | 17.8 | 12.8 |
| exp prop continuous | equal | 15.8 | 12.8 | 8.9 |
| exp prop discrete | equal | 18.6 | 15.6 | 10.1 |
| exp prop continuous | unequal, high | 15.0 | 12.7 | 8.0 |
| exp prop discrete | unequal, high | 16.3 | 14.2 | 8.6 |
| exp prop continuous | unequal, low | 16.8 | 13.2 | 9.5 |
| exp prop discrete | unequal, low | 19.7 | 16.5 | 9.6 |
| logn continuous | equal | 42.0 | 36.2 | 19.7 |
| logn discrete | equal | 46.3 | 39.8 | 23.1 |
| logn continuous | unequal, high | 28.3 | 24.0 | 8.5 |
| logn discrete | unequal, high | 30.4 | 27.3 | 9.7 |
| logn continuous | unequal, low | 45.5 | 40.7 | 20.6 |
| logn discrete | unequal, low | 52.2 | 46.9 | 24.5 |
| pwExp continuous | equal | 15.3 | 11.9 | 8.3 |
| pwExp discrete | equal | 17.9 | 14.4 | 9.7 |
| pwExp continuous | unequal, high | 13.6 | 11.9 | 6.2 |
| pwExp discrete | unequal, high | 14.7 | 12.0 | 6.4 |
| pwExp continuous | unequal, low | 14.5 | 12.0 | 8.7 |
| pwExp discrete | unequal, low | 17.3 | 13.7 | 9.3 |
| Weib late continuous | equal | 48.5 | 40.7 | 23.9 |
| Weib late discrete | equal | 54.5 | 48.0 | 27.5 |
| Weib late continuous | unequal, high | 31.1 | 26.5 | 11.2 |
| Weib late discrete | unequal, high | 36.3 | 31.2 | 11.7 |
| Weib late continuous | unequal, low | 54.5 | 49.3 | 25.0 |
| Weib late discrete | unequal, low | 61.4 | 55.9 | 30.7 |
| Weib prop continuous | equal | 42.3 | 36.1 | 20.2 |
| Weib prop discrete | equal | 47.1 | 39.8 | 21.4 |
| Weib prop continuous | unequal, high | 29.3 | 24.6 | 10.9 |
| Weib prop discrete | unequal, high | 33.0 | 27.4 | 10.4 |
| Weib prop continuous | unequal, low | 47.0 | 40.4 | 20.2 |
| Weib prop discrete | unequal, low | 55.0 | 48.4 | 24.6 |
| Weib scale continuous | equal | 27.9 | 23.4 | 12.4 |
| Weib scale discrete | equal | 31.6 | 27.0 | 14.4 |
| Weib scale continuous | unequal, high | 21.7 | 17.7 | 7.9 |
| Weib scale discrete | unequal, high | 23.7 | 19.4 | 8.1 |
| Weib scale continuous | unequal, low | 27.0 | 21.6 | 11.7 |
| Weib scale discrete | unequal, low | 34.7 | 27.7 | 12.1 |
| Weib shape continuous | equal | 16.1 | 12.9 | 6.9 |
| Weib shape discrete | equal | 19.2 | 15.8 | 8.3 |
| Weib shape continuous | unequal, high | 16.6 | 13.3 | 7.0 |
| Weib shape discrete | unequal, high | 17.4 | 14.9 | 6.0 |
| Weib shape continuous | unequal, low | 16.5 | 12.4 | 6.4 |
| Weib shape discrete | unequal, low | 19.3 | 15.6 | 7.6 |

Table S36: Rejection rates in percent for the Tukey-type contrast matrix with $\delta = 1.5$ and unbalanced small sample sizes.

| Distribution | Censoring distribution | asymptotic | asymptotic_bonf | permutation_bonf |
|---|---|---|---|---|
| exp early continuous | equal | 11.5 | 10.1 | 3.4 |
| exp early discrete | equal | 11.0 | 9.6 | 3.2 |
| exp early continuous | unequal, high | 13.2 | 11.8 | 3.6 |
| exp early discrete | unequal, high | 15.7 | 14.5 | 3.9 |
| exp early continuous | unequal, low | 11.1 | 10.0 | 2.2 |
| exp early discrete | unequal, low | 11.3 | 10.3 | 2.8 |
| exp late continuous | equal | 11.9 | 9.4 | 2.5 |
| exp late discrete | equal | 11.8 | 10.4 | 2.6 |
| exp late continuous | unequal, high | 13.0 | 11.9 | 3.3 |
| exp late discrete | unequal, high | 14.7 | 13.4 | 3.4 |
| exp late continuous | unequal, low | 11.1 | 9.6 | 2.0 |
| exp late discrete | unequal, low | 12.1 | 10.3 | 2.2 |
| exp prop continuous | equal | 12.4 | 11.1 | 3.4 |
| exp prop discrete | equal | 13.3 | 11.0 | 3.8 |
| exp prop continuous | unequal, high | 15.4 | 13.6 | 4.3 |
| exp prop discrete | unequal, high | 16.1 | 14.7 | 3.6 |
| exp prop continuous | unequal, low | 11.4 | 9.6 | 3.5 |
| exp prop discrete | unequal, low | 12.1 | 10.1 | 2.9 |
| logn continuous | equal | 16.3 | 14.8 | 3.9 |
| logn discrete | equal | 18.1 | 16.4 | 3.6 |
| logn continuous | unequal, high | 28.5 | 26.6 | 4.7 |
| logn discrete | unequal, high | 30.5 | 27.7 | 5.0 |
| logn continuous | unequal, low | 15.1 | 13.0 | 3.3 |
| logn discrete | unequal, low | 15.8 | 13.5 | 2.9 |
| pwExp continuous | equal | 11.7 | 9.6 | 2.9 |
| pwExp discrete | equal | 11.2 | 9.7 | 2.6 |
| pwExp continuous | unequal, high | 13.4 | 12.0 | 3.1 |
| pwExp discrete | unequal, high | 14.5 | 13.4 | 3.5 |
| pwExp continuous | unequal, low | 10.6 | 9.3 | 2.0 |
| pwExp discrete | unequal, low | 11.3 | 9.9 | 2.5 |
| Weib late continuous | equal | 14.2 | 12.9 | 3.5 |
| Weib late discrete | equal | 16.6 | 14.4 | 3.3 |
| Weib late continuous | unequal, high | 20.2 | 18.2 | 4.8 |
| Weib late discrete | unequal, high | 24.9 | 22.3 | 4.4 |
| Weib late continuous | unequal, low | 14.5 | 12.7 | 2.5 |
| Weib late discrete | unequal, low | 16.2 | 14.1 | 3.3 |
| Weib prop continuous | equal | 14.5 | 13.3 | 3.5 |
| Weib prop discrete | equal | 17.0 | 15.0 | 3.4 |
| Weib prop continuous | unequal, high | 21.3 | 18.5 | 4.4 |
| Weib prop discrete | unequal, high | 25.7 | 22.8 | 4.1 |
| Weib prop continuous | unequal, low | 14.2 | 12.6 | 2.4 |
| Weib prop discrete | unequal, low | 15.6 | 13.8 | 3.2 |
| Weib scale continuous | equal | 15.9 | 15.3 | 3.5 |
| Weib scale discrete | equal | 18.6 | 17.3 | 3.8 |
| Weib scale continuous | unequal, high | 26.4 | 24.4 | 4.4 |
| Weib scale discrete | unequal, high | 29.3 | 27.1 | 4.1 |
| Weib scale continuous | unequal, low | 15.6 | 13.8 | 2.8 |
| Weib scale discrete | unequal, low | 17.3 | 15.3 | 3.4 |
| Weib shape continuous | equal | 23.2 | 22.2 | 6.6 |
| Weib shape discrete | equal | 24.3 | 22.9 | 5.6 |
| Weib shape continuous | unequal, high | 28.8 | 26.7 | 4.6 |
| Weib shape discrete | unequal, high | 33.5 | 31.3 | 4.7 |
| Weib shape continuous | unequal, low | 23.6 | 22.1 | 5.3 |
| Weib shape discrete | unequal, low | 24.5 | 22.8 | 5.5 |

Table S37: Rejection rates in percent for the 2-by-2 design with $\delta = 1.5$ and balanced large sample sizes under equal censoring.

| Distribution | Method | $\mathcal{H}_{0,4}$ | $\mathcal{H}_{0,5}$ | $\mathcal{H}_{0,6}$ | $\mathcal{H}_{0,1}$ | $\mathcal{H}_{0,2}$ | $\mathcal{H}_{0,3}$ | $\mathcal{H}_{0,7}$ | $\mathcal{H}_{0,8}$ | $\mathcal{H}_{0,9}$ |
|---|---|---|---|---|---|---|---|---|---|---|
| exp early continuous | asymptotic | 50.5 | 32.8 | 69.2 | 49.6 | 33.6 | 72.1 | 49.1 | 32.4 | 69.5 |
|  | asymptotic_bonf | 49.9 | 32.0 | 68.6 | 48.9 | 33.0 | 71.4 | 48.3 | 31.8 | 69.2 |
|  | permutation_bonf | 50.0 | 31.9 | 68.7 | 48.0 | 32.3 | 70.2 | 48.5 | 31.9 | 69.6 |
| exp early discrete | asymptotic | 59.1 | 37.9 | 74.8 | 56.4 | 38.0 | 77.0 | 55.4 | 37.8 | 77.4 |
|  | asymptotic_bonf | 58.0 | 37.7 | 74.1 | 55.8 | 37.3 | 76.7 | 55.1 | 37.0 | 77.1 |
|  | permutation_bonf | 58.3 | 37.0 | 73.9 | 55.2 | 37.0 | 76.0 | 54.7 | 37.7 | 76.1 |
| exp late continuous | asymptotic | 52.8 | 34.9 | 72.2 | 52.5 | 34.6 | 73.9 | 50.9 | 33.5 | 74.0 |
|  | asymptotic_bonf | 52.3 | 33.9 | 71.9 | 51.4 | 33.4 | 73.5 | 49.7 | 32.7 | 72.8 |
|  | permutation_bonf | 52.3 | 33.0 | 70.9 | 50.8 | 33.1 | 72.7 | 49.7 | 32.7 | 72.6 |
| exp late discrete | asymptotic | 61.5 | 41.2 | 78.7 | 59.6 | 40.5 | 79.8 | 59.0 | 39.7 | 79.7 |
|  | asymptotic_bonf | 60.6 | 40.4 | 78.2 | 59.1 | 39.7 | 79.4 | 58.3 | 39.3 | 79.0 |
|  | permutation_bonf | 60.3 | 39.7 | 77.7 | 57.8 | 38.2 | 79.1 | 57.0 | 38.4 | 78.1 |
| exp prop continuous | asymptotic | 51.1 | 35.6 | 70.1 | 51.3 | 34.6 | 71.0 | 51.1 | 36.1 | 73.0 |
|  | asymptotic_bonf | 50.0 | 35.2 | 69.5 | 50.4 | 33.7 | 70.5 | 50.1 | 35.2 | 72.3 |
|  | permutation_bonf | 51.2 | 34.5 | 69.3 | 50.1 | 34.7 | 69.6 | 49.8 | 34.7 | 72.1 |
| exp prop discrete | asymptotic | 57.8 | 41.2 | 77.2 | 57.2 | 39.4 | 77.9 | 58.2 | 40.7 | 78.7 |
|  | asymptotic_bonf | 57.2 | 40.4 | 76.7 | 56.4 | 38.7 | 77.2 | 57.2 | 40.4 | 78.3 |
|  | permutation_bonf | 56.7 | 40.1 | 76.3 | 57.5 | 38.1 | 76.7 | 58.0 | 39.3 | 77.6 |
| logn continuous | asymptotic | 95.6 | 83.0 | 99.3 | 95.3 | 84.0 | 99.3 | 95.7 | 84.2 | 99.0 |
|  | asymptotic_bonf | 95.4 | 82.9 | 99.2 | 95.2 | 83.7 | 99.3 | 95.6 | 83.7 | 99.0 |
|  | permutation_bonf | 95.1 | 82.4 | 99.4 | 95.2 | 83.5 | 99.6 | 95.4 | 83.1 | 99.0 |
| logn discrete | asymptotic | 97.4 | 89.1 | 99.8 | 98.1 | 90.2 | 99.8 | 97.3 | 90.1 | 99.9 |
|  | asymptotic_bonf | 97.4 | 89.1 | 99.8 | 98.0 | 89.8 | 99.8 | 97.3 | 89.8 | 99.8 |
|  | permutation_bonf | 97.7 | 88.8 | 99.7 | 97.8 | 90.3 | 99.8 | 97.1 | 89.7 | 99.7 |
| pwExp continuous | asymptotic | 51.2 | 33.2 | 69.4 | 49.4 | 32.4 | 71.3 | 48.4 | 31.8 | 70.5 |
|  | asymptotic_bonf | 50.2 | 32.6 | 68.6 | 49.1 | 32.1 | 70.9 | 48.0 | 31.2 | 70.2 |
|  | permutation_bonf | 48.7 | 32.3 | 68.7 | 47.2 | 32.3 | 70.3 | 48.4 | 31.2 | 68.7 |
| pwExp discrete | asymptotic | 57.8 | 39.4 | 76.6 | 57.6 | 38.4 | 77.5 | 54.6 | 37.9 | 76.0 |
|  | asymptotic_bonf | 57.1 | 39.0 | 76.2 | 57.0 | 37.7 | 77.1 | 53.9 | 37.5 | 75.7 |
|  | permutation_bonf | 56.7 | 38.2 | 75.6 | 56.1 | 38.1 | 75.7 | 54.1 | 38.0 | 76.0 |
| Weib late continuous | asymptotic | 95.4 | 80.7 | 99.3 | 95.0 | 81.4 | 99.4 | 94.4 | 83.5 | 99.0 |
|  | asymptotic_bonf | 95.1 | 80.0 | 99.3 | 94.8 | 81.4 | 99.4 | 94.0 | 82.7 | 99.0 |
|  | permutation_bonf | 95.0 | 80.2 | 99.2 | 94.7 | 80.2 | 99.1 | 93.8 | 82.5 | 98.9 |
| Weib late discrete | asymptotic | 97.6 | 88.1 | 99.7 | 97.9 | 89.7 | 100.0 | 98.0 | 88.9 | 99.6 |
|  | asymptotic_bonf | 97.5 | 87.8 | 99.7 | 97.9 | 89.4 | 100.0 | 98.0 | 88.6 | 99.6 |
|  | permutation_bonf | 97.4 | 87.6 | 99.9 | 97.7 | 88.7 | 100.0 | 97.7 | 88.2 | 99.6 |
| Weib prop continuous | asymptotic | 94.3 | 80.1 | 99.1 | 94.8 | 80.2 | 99.0 | 93.9 | 82.2 | 99.0 |
|  | asymptotic_bonf | 94.3 | 79.6 | 99.0 | 94.8 | 79.8 | 99.0 | 93.7 | 81.9 | 99.0 |
|  | permutation_bonf | 93.8 | 79.5 | 98.8 | 94.5 | 79.7 | 99.0 | 93.2 | 82.0 | 99.0 |
| Weib prop discrete | asymptotic | 97.1 | 86.8 | 99.8 | 97.7 | 88.4 | 99.9 | 97.3 | 87.7 | 99.6 |
|  | asymptotic_bonf | 97.0 | 86.7 | 99.8 | 97.7 | 88.2 | 99.8 | 97.2 | 87.5 | 99.6 |
|  | permutation_bonf | 97.0 | 86.3 | 99.9 | 97.4 | 87.8 | 99.9 | 97.1 | 86.9 | 99.6 |
| Weib scale continuous | asymptotic | 92.3 | 76.6 | 98.2 | 92.5 | 76.7 | 98.1 | 91.3 | 78.2 | 98.4 |
|  | asymptotic_bonf | 92.3 | 76.4 | 98.1 | 92.3 | 76.6 | 98.0 | 91.0 | 77.9 | 98.4 |
|  | permutation_bonf | 91.8 | 75.6 | 97.9 | 92.0 | 76.4 | 98.1 | 90.8 | 77.8 | 98.6 |
| Weib scale discrete | asymptotic | 95.9 | 84.6 | 99.4 | 96.9 | 85.6 | 99.5 | 95.9 | 85.1 | 99.1 |
|  | asymptotic_bonf | 95.8 | 84.3 | 99.3 | 96.8 | 85.4 | 99.5 | 95.8 | 84.9 | 99.1 |
|  | permutation_bonf | 95.6 | 83.9 | 99.3 | 96.3 | 85.9 | 99.5 | 94.8 | 84.3 | 99.3 |
| Weib shape continuous | asymptotic | 87.9 | 70.7 | 96.1 | 88.3 | 71.4 | 95.9 | 87.4 | 73.7 | 96.4 |
|  | asymptotic_bonf | 87.3 | 70.4 | 95.9 | 87.9 | 71.0 | 95.7 | 87.2 | 73.0 | 96.4 |
|  | permutation_bonf | 87.5 | 69.5 | 96.1 | 88.0 | 71.0 | 96.0 | 87.3 | 73.2 | 96.4 |
| Weib shape discrete | asymptotic | 94.1 | 79.9 | 98.6 | 94.1 | 80.4 | 98.7 | 93.6 | 81.1 | 98.3 |
|  | asymptotic_bonf | 93.9 | 79.5 | 98.5 | 93.9 | 80.0 | 98.7 | 93.6 | 80.8 | 98.3 |
|  | permutation_bonf | 93.4 | 79.2 | 98.3 | 94.1 | 80.3 | 98.8 | 93.7 | 79.7 | 97.9 |

Table S38: Rejection rates in percent for the 2-by-2 design with $\delta = 1.5$ and balanced medium sample sizes under equal censoring.

| Distribution | Method | $\mathcal{H}_{0,4}$ | $\mathcal{H}_{0,5}$ | $\mathcal{H}_{0,6}$ | $\mathcal{H}_{0,1}$ | $\mathcal{H}_{0,2}$ | $\mathcal{H}_{0,3}$ | $\mathcal{H}_{0,7}$ | $\mathcal{H}_{0,8}$ | $\mathcal{H}_{0,9}$ |
|---|---|---|---|---|---|---|---|---|---|---|
| exp early continuous | asymptotic | 7.4 | 5.7 | 9.4 | 7.2 | 3.3 | 10.4 | 6.1 | 3.0 | 11.0 |
| | asymptotic_bonf | 7.2 | 5.6 | 9.2 | 7.0 | 3.3 | 9.8 | 6.0 | 3.0 | 10.8 |
| | permutation_bonf | 7.1 | 5.3 | 9.2 | 6.8 | 3.3 | 9.3 | 5.9 | 3.0 | 10.1 |
| exp early discrete | asymptotic | 8.2 | 6.3 | 12.0 | 8.2 | 3.9 | 12.1 | 7.4 | 3.4 | 12.6 |
| | asymptotic_bonf | 8.0 | 6.0 | 11.2 | 8.1 | 3.9 | 11.6 | 7.3 | 3.3 | 12.4 |
| | permutation_bonf | 8.5 | 5.6 | 11.8 | 8.1 | 4.4 | 12.0 | 7.1 | 3.0 | 12.8 |
| exp late continuous | asymptotic | 6.8 | 6.2 | 9.5 | 7.0 | 3.4 | 10.8 | 6.5 | 3.2 | 11.7 |
| | asymptotic_bonf | 6.7 | 5.6 | 9.1 | 6.6 | 3.2 | 10.4 | 6.3 | 3.2 | 11.4 |
| | permutation_bonf | 6.9 | 6.0 | 10.0 | 6.8 | 3.5 | 10.2 | 6.2 | 3.3 | 10.4 |
| exp late discrete | asymptotic | 7.7 | 6.6 | 11.9 | 8.2 | 4.6 | 13.0 | 8.3 | 3.5 | 13.1 |
| | asymptotic_bonf | 7.4 | 6.3 | 11.0 | 7.9 | 4.1 | 12.5 | 7.8 | 3.5 | 12.7 |
| | permutation_bonf | 7.9 | 6.0 | 11.7 | 8.3 | 4.5 | 12.3 | 7.8 | 3.3 | 13.2 |
| exp prop continuous | asymptotic | 7.4 | 5.1 | 9.4 | 6.4 | 4.7 | 8.9 | 7.0 | 4.9 | 9.1 |
| | asymptotic_bonf | 7.4 | 4.9 | 9.1 | 6.3 | 4.5 | 8.8 | 6.4 | 4.7 | 9.1 |
| | permutation_bonf | 6.8 | 4.6 | 9.0 | 6.8 | 4.4 | 8.7 | 6.9 | 4.8 | 9.0 |
| exp prop discrete | asymptotic | 8.3 | 5.6 | 10.9 | 7.1 | 5.3 | 10.2 | 7.9 | 5.4 | 11.5 |
| | asymptotic_bonf | 8.3 | 5.3 | 10.9 | 7.0 | 5.3 | 10.0 | 7.4 | 5.2 | 10.7 |
| | permutation_bonf | 8.3 | 5.4 | 11.0 | 6.8 | 5.2 | 10.0 | 7.8 | 5.5 | 10.5 |
| logn continuous | asymptotic | 19.9 | 13.2 | 33.1 | 22.0 | 14.2 | 33.4 | 23.1 | 13.5 | 33.3 |
| | asymptotic_bonf | 19.2 | 12.7 | 32.6 | 21.6 | 13.9 | 32.8 | 22.6 | 13.2 | 32.6 |
| | permutation_bonf | 18.5 | 12.9 | 32.1 | 23.1 | 14.2 | 33.1 | 21.9 | 12.9 | 31.7 |
| logn discrete | asymptotic | 25.4 | 16.4 | 41.3 | 26.4 | 17.7 | 40.5 | 26.7 | 17.3 | 39.3 |
| | asymptotic_bonf | 24.7 | 16.2 | 40.4 | 26.2 | 17.2 | 39.6 | 26.0 | 16.9 | 39.2 |
| | permutation_bonf | 24.4 | 15.5 | 39.4 | 25.3 | 17.4 | 39.4 | 27.3 | 16.6 | 38.7 |
| pwExp continuous | asymptotic | 7.2 | 5.3 | 9.4 | 6.7 | 3.9 | 9.4 | 5.6 | 2.7 | 9.9 |
| | asymptotic_bonf | 6.9 | 5.3 | 9.0 | 6.6 | 3.6 | 9.1 | 5.4 | 2.6 | 9.6 |
| | permutation_bonf | 7.0 | 5.1 | 8.6 | 5.6 | 3.7 | 8.8 | 5.9 | 2.4 | 9.1 |
| pwExp discrete | asymptotic | 8.7 | 6.2 | 11.7 | 7.2 | 4.5 | 11.5 | 7.4 | 3.4 | 12.1 |
| | asymptotic_bonf | 8.7 | 5.8 | 11.3 | 7.1 | 4.4 | 11.0 | 7.2 | 3.0 | 11.8 |
| | permutation_bonf | 7.9 | 6.0 | 11.0 | 6.8 | 4.6 | 11.1 | 7.0 | 3.4 | 12.0 |
| Weib late continuous | asymptotic | 19.3 | 12.9 | 35.3 | 20.7 | 13.6 | 31.3 | 22.0 | 11.8 | 34.1 |
| | asymptotic_bonf | 18.8 | 12.6 | 34.8 | 20.3 | 13.3 | 31.0 | 21.7 | 11.3 | 33.4 |
| | permutation_bonf | 18.2 | 12.4 | 34.6 | 21.0 | 13.7 | 30.9 | 21.8 | 12.0 | 34.1 |
| Weib late discrete | asymptotic | 23.1 | 15.3 | 40.0 | 25.3 | 16.4 | 36.8 | 27.4 | 14.3 | 38.7 |
| | asymptotic_bonf | 23.0 | 14.9 | 39.6 | 24.4 | 16.2 | 36.1 | 26.4 | 14.0 | 38.4 |
| | permutation_bonf | 22.3 | 14.7 | 39.2 | 24.9 | 15.2 | 35.6 | 25.9 | 14.3 | 38.3 |
| Weib prop continuous | asymptotic | 19.0 | 12.5 | 33.9 | 19.7 | 13.1 | 29.9 | 21.9 | 11.7 | 33.2 |
| | asymptotic_bonf | 18.5 | 12.1 | 33.8 | 19.4 | 12.9 | 29.5 | 21.5 | 11.1 | 32.8 |
| | permutation_bonf | 18.1 | 12.4 | 33.5 | 20.4 | 13.0 | 29.9 | 20.8 | 11.6 | 33.6 |
| Weib prop discrete | asymptotic | 22.9 | 16.0 | 40.9 | 25.7 | 15.2 | 36.9 | 26.5 | 14.9 | 39.0 |
| | asymptotic_bonf | 22.7 | 15.7 | 40.3 | 25.1 | 14.9 | 36.4 | 25.9 | 14.5 | 38.1 |
| | permutation_bonf | 21.4 | 15.5 | 40.9 | 25.3 | 15.8 | 36.3 | 26.1 | 14.4 | 38.0 |
| Weib scale continuous | asymptotic | 16.6 | 11.2 | 30.3 | 18.3 | 11.3 | 26.3 | 18.5 | 10.5 | 28.8 |
| | asymptotic_bonf | 16.6 | 10.7 | 29.6 | 17.4 | 11.0 | 25.9 | 18.0 | 10.2 | 28.5 |
| | permutation_bonf | 16.4 | 11.7 | 30.1 | 17.5 | 11.6 | 26.3 | 18.2 | 10.7 | 28.2 |
| Weib scale discrete | asymptotic | 20.0 | 14.2 | 37.0 | 22.8 | 14.1 | 31.5 | 23.3 | 13.8 | 34.3 |
| | asymptotic_bonf | 19.6 | 14.1 | 35.8 | 22.6 | 13.9 | 31.1 | 22.6 | 13.4 | 33.7 |
| | permutation_bonf | 19.4 | 14.5 | 37.0 | 22.8 | 13.6 | 31.2 | 22.8 | 13.2 | 32.7 |
| Weib shape continuous | asymptotic | 13.5 | 9.4 | 24.9 | 15.3 | 9.4 | 21.8 | 15.9 | 8.3 | 24.2 |
| | asymptotic_bonf | 13.0 | 8.8 | 24.3 | 14.7 | 9.3 | 21.1 | 15.5 | 8.1 | 23.6 |
| | permutation_bonf | 13.6 | 9.9 | 24.7 | 14.4 | 9.3 | 20.9 | 15.0 | 8.7 | 24.1 |
| Weib shape discrete | asymptotic | 16.8 | 12.3 | 31.9 | 18.5 | 10.9 | 27.2 | 20.1 | 11.1 | 29.5 |
| | asymptotic_bonf | 16.5 | 11.9 | 31.2 | 18.1 | 10.5 | 26.3 | 19.6 | 10.7 | 28.8 |
| | permutation_bonf | 16.2 | 11.9 | 31.7 | 18.2 | 11.2 | 26.3 | 18.9 | 11.3 | 28.5 |

Table S39: Rejection rates in percent for the 2-by-2 design with $\delta = 1.5$ and balanced small sample sizes under equal censoring.

| Distribution | Method | $\mathcal{H}_{0,4}$ | $\mathcal{H}_{0,5}$ | $\mathcal{H}_{0,6}$ | $\mathcal{H}_{0,1}$ | $\mathcal{H}_{0,2}$ | $\mathcal{H}_{0,3}$ | $\mathcal{H}_{0,7}$ | $\mathcal{H}_{0,8}$ | $\mathcal{H}_{0,9}$ |
|---|---|---|---|---|---|---|---|---|---|---|
| exp early continuous | asymptotic | 2.1 | 1.3 | 1.7 | 2.0 | 0.8 | 1.7 | 1.3 | 1.6 | 1.2 |
|  | asymptotic_bonf | 2.0 | 1.3 | 1.4 | 1.9 | 0.8 | 1.7 | 1.2 | 1.4 | 1.1 |
|  | permutation_bonf | 1.7 | 1.2 | 1.5 | 1.7 | 0.8 | 1.5 | 1.4 | 1.7 | 0.9 |
| exp early discrete | asymptotic | 2.0 | 1.4 | 1.2 | 1.0 | 1.6 | 1.5 | 2.0 | 0.8 | 1.7 |
|  | asymptotic_bonf | 1.9 | 1.4 | 1.2 | 1.0 | 1.6 | 1.5 | 1.9 | 0.8 | 1.7 |
|  | permutation_bonf | 1.6 | 1.5 | 1.5 | 0.8 | 1.0 | 1.2 | 1.7 | 0.5 | 1.5 |
| exp late continuous | asymptotic | 2.2 | 1.2 | 1.6 | 2.2 | 1.0 | 1.8 | 2.1 | 1.3 | 1.4 |
|  | asymptotic_bonf | 1.9 | 1.2 | 1.4 | 2.1 | 0.9 | 1.8 | 2.0 | 1.2 | 1.3 |
|  | permutation_bonf | 1.5 | 1.2 | 1.6 | 2.0 | 1.1 | 1.6 | 1.5 | 1.5 | 1.1 |
| exp late discrete | asymptotic | 2.2 | 1.2 | 1.4 | 2.2 | 1.0 | 2.3 | 2.1 | 1.6 | 1.3 |
|  | asymptotic_bonf | 2.0 | 1.2 | 1.4 | 2.0 | 1.0 | 2.2 | 2.0 | 1.5 | 1.3 |
|  | permutation_bonf | 1.9 | 1.0 | 1.7 | 1.9 | 1.1 | 2.1 | 2.3 | 1.6 | 1.4 |
| exp prop continuous | asymptotic | 1.2 | 1.3 | 1.9 | 1.8 | 1.4 | 2.1 | 0.8 | 1.0 | 2.0 |
|  | asymptotic_bonf | 1.2 | 1.3 | 1.8 | 1.8 | 1.4 | 2.0 | 0.8 | 0.9 | 1.8 |
|  | permutation_bonf | 0.9 | 0.8 | 2.0 | 1.3 | 1.3 | 1.9 | 0.7 | 0.9 | 1.5 |
| exp prop discrete | asymptotic | 2.0 | 0.9 | 2.1 | 1.0 | 0.8 | 2.5 | 1.3 | 1.2 | 2.5 |
|  | asymptotic_bonf | 2.0 | 0.9 | 2.0 | 1.0 | 0.8 | 2.5 | 1.1 | 1.1 | 2.5 |
|  | permutation_bonf | 2.0 | 1.0 | 1.8 | 0.9 | 0.8 | 1.9 | 1.2 | 1.0 | 2.2 |
| logn continuous | asymptotic | 4.1 | 1.9 | 3.9 | 4.0 | 3.4 | 7.0 | 3.0 | 1.4 | 3.8 |
|  | asymptotic_bonf | 3.7 | 1.7 | 3.9 | 3.8 | 3.4 | 6.8 | 2.9 | 1.4 | 3.7 |
|  | permutation_bonf | 4.0 | 1.7 | 4.1 | 3.5 | 3.3 | 6.1 | 2.7 | 1.5 | 3.8 |
| logn discrete | asymptotic | 4.2 | 2.2 | 5.3 | 4.5 | 4.0 | 7.6 | 4.0 | 2.0 | 6.0 |
|  | asymptotic_bonf | 4.1 | 2.2 | 5.1 | 4.5 | 3.7 | 7.5 | 4.0 | 1.8 | 5.8 |
|  | permutation_bonf | 3.8 | 2.1 | 4.7 | 4.1 | 3.7 | 7.3 | 3.3 | 1.6 | 4.7 |
| pwExp continuous | asymptotic | 1.9 | 1.1 | 1.6 | 2.1 | 0.9 | 1.9 | 1.9 | 1.5 | 1.4 |
|  | asymptotic_bonf | 1.8 | 1.0 | 1.5 | 2.1 | 0.8 | 1.7 | 1.9 | 1.5 | 1.4 |
|  | permutation_bonf | 1.6 | 1.1 | 1.3 | 1.9 | 0.7 | 1.6 | 1.2 | 1.4 | 1.0 |
| pwExp discrete | asymptotic | 2.2 | 1.0 | 1.6 | 2.4 | 0.9 | 1.9 | 1.9 | 1.5 | 1.6 |
|  | asymptotic_bonf | 2.2 | 1.0 | 1.6 | 2.3 | 0.9 | 1.9 | 1.9 | 1.4 | 1.5 |
|  | permutation_bonf | 1.7 | 1.1 | 1.5 | 1.9 | 0.7 | 1.8 | 1.6 | 1.5 | 1.4 |
| Weib late continuous | asymptotic | 4.1 | 2.4 | 4.8 | 3.0 | 3.2 | 5.0 | 4.2 | 2.8 | 4.8 |
|  | asymptotic_bonf | 4.0 | 2.4 | 4.6 | 3.0 | 3.1 | 5.0 | 4.1 | 2.6 | 4.7 |
|  | permutation_bonf | 3.6 | 2.0 | 4.1 | 2.7 | 2.7 | 4.8 | 3.7 | 2.7 | 4.4 |
| Weib late discrete | asymptotic | 4.9 | 3.0 | 5.9 | 3.7 | 3.4 | 5.2 | 3.9 | 2.5 | 5.8 |
|  | asymptotic_bonf | 4.8 | 2.9 | 5.8 | 3.5 | 3.4 | 4.9 | 3.8 | 2.5 | 5.7 |
|  | permutation_bonf | 4.0 | 2.4 | 5.6 | 3.1 | 3.3 | 4.6 | 3.2 | 2.5 | 5.6 |
| Weib prop continuous | asymptotic | 4.0 | 2.0 | 4.6 | 2.9 | 3.1 | 4.3 | 3.7 | 2.7 | 5.1 |
|  | asymptotic_bonf | 3.9 | 2.0 | 4.3 | 2.8 | 3.1 | 4.1 | 3.5 | 2.5 | 5.0 |
|  | permutation_bonf | 3.3 | 1.6 | 4.0 | 2.4 | 2.6 | 4.1 | 3.5 | 2.6 | 4.0 |
| Weib prop discrete | asymptotic | 4.5 | 2.7 | 5.4 | 3.4 | 3.3 | 4.6 | 3.5 | 2.4 | 5.7 |
|  | asymptotic_bonf | 4.5 | 2.3 | 5.2 | 3.3 | 3.3 | 4.5 | 3.4 | 2.4 | 5.5 |
|  | permutation_bonf | 3.9 | 2.3 | 4.7 | 3.3 | 2.8 | 4.2 | 3.6 | 2.6 | 4.6 |
| Weib scale continuous | asymptotic | 2.9 | 2.0 | 4.4 | 2.7 | 2.4 | 3.5 | 2.8 | 1.9 | 4.3 |
|  | asymptotic_bonf | 2.6 | 1.8 | 4.1 | 2.5 | 2.4 | 3.3 | 2.8 | 1.8 | 4.3 |
|  | permutation_bonf | 2.5 | 1.8 | 4.4 | 1.8 | 2.2 | 3.6 | 2.6 | 2.3 | 3.3 |
| Weib scale discrete | asymptotic | 2.9 | 1.9 | 4.6 | 2.8 | 3.2 | 3.8 | 2.9 | 2.7 | 5.4 |
|  | asymptotic_bonf | 2.9 | 1.8 | 4.4 | 2.7 | 3.0 | 3.6 | 2.7 | 2.5 | 5.1 |
|  | permutation_bonf | 2.5 | 1.7 | 4.1 | 2.5 | 2.9 | 3.4 | 2.4 | 2.3 | 4.4 |
| Weib shape continuous | asymptotic | 1.9 | 1.8 | 2.8 | 1.3 | 2.3 | 3.1 | 1.9 | 1.7 | 4.1 |
|  | asymptotic_bonf | 1.8 | 1.6 | 2.8 | 1.2 | 2.2 | 2.9 | 1.8 | 1.7 | 3.6 |
|  | permutation_bonf | 1.6 | 1.5 | 3.0 | 1.4 | 1.9 | 2.5 | 1.9 | 1.7 | 3.1 |
| Weib shape discrete | asymptotic | 2.5 | 1.6 | 3.6 | 1.8 | 2.4 | 2.9 | 2.0 | 1.9 | 4.1 |
|  | asymptotic_bonf | 2.1 | 1.3 | 3.4 | 1.5 | 2.3 | 2.7 | 1.9 | 1.8 | 3.8 |
|  | permutation_bonf | 2.2 | 1.6 | 3.6 | 1.3 | 2.1 | 2.0 | 2.2 | 2.0 | 3.6 |

Table S40: Rejection rates in percent for the 2-by-2 design with $\delta = 1.5$ and unbalanced large sample sizes under equal censoring.

| Distribution | Method | $\mathcal{H}_{0,4}$ | $\mathcal{H}_{0,5}$ | $\mathcal{H}_{0,6}$ | $\mathcal{H}_{0,1}$ | $\mathcal{H}_{0,2}$ | $\mathcal{H}_{0,3}$ | $\mathcal{H}_{0,7}$ | $\mathcal{H}_{0,8}$ | $\mathcal{H}_{0,9}$ |
|---|---|---|---|---|---|---|---|---|---|---|
| exp early continuous | asymptotic | 25.0 | 13.5 | 39.6 | 25.4 | 13.2 | 38.0 | 24.2 | 13.6 | 37.5 |
| | asymptotic_bonf | 23.2 | 12.2 | 37.6 | 23.8 | 12.2 | 36.2 | 22.4 | 12.3 | 35.9 |
| | permutation_bonf | 23.1 | 12.1 | 36.7 | 22.7 | 11.8 | 35.4 | 21.8 | 12.1 | 35.1 |
| exp early discrete | asymptotic | 28.8 | 15.4 | 45.8 | 28.9 | 15.7 | 45.3 | 28.6 | 16.5 | 43.1 |
| | asymptotic_bonf | 26.8 | 14.4 | 43.7 | 26.9 | 14.8 | 42.6 | 27.3 | 14.9 | 41.3 |
| | permutation_bonf | 26.8 | 14.0 | 42.2 | 25.9 | 14.6 | 42.1 | 26.5 | 15.0 | 40.6 |
| exp late continuous | asymptotic | 27.0 | 14.4 | 45.2 | 26.3 | 15.2 | 43.3 | 26.4 | 15.3 | 42.7 |
| | asymptotic_bonf | 25.3 | 13.2 | 43.6 | 24.7 | 13.7 | 40.9 | 25.1 | 13.5 | 40.7 |
| | permutation_bonf | 24.9 | 13.1 | 42.8 | 25.0 | 13.1 | 40.4 | 24.2 | 14.3 | 40.5 |
| exp late discrete | asymptotic | 31.3 | 18.1 | 52.7 | 30.2 | 17.3 | 51.0 | 31.2 | 18.1 | 47.4 |
| | asymptotic_bonf | 29.4 | 16.8 | 49.4 | 29.2 | 15.6 | 49.0 | 29.6 | 16.9 | 46.1 |
| | permutation_bonf | 28.4 | 15.7 | 49.1 | 29.2 | 15.3 | 47.3 | 29.2 | 16.5 | 45.7 |
| exp prop continuous | asymptotic | 28.5 | 17.0 | 38.1 | 26.0 | 16.0 | 40.4 | 26.2 | 15.4 | 36.2 |
| | asymptotic_bonf | 26.6 | 16.2 | 36.2 | 24.3 | 15.2 | 38.0 | 24.2 | 14.2 | 34.8 |
| | permutation_bonf | 27.4 | 14.9 | 35.1 | 24.3 | 13.0 | 38.0 | 24.3 | 14.1 | 34.8 |
| exp prop discrete | asymptotic | 32.3 | 20.3 | 44.6 | 28.8 | 18.6 | 45.8 | 31.1 | 17.8 | 43.6 |
| | asymptotic_bonf | 30.9 | 18.8 | 42.0 | 27.2 | 17.6 | 44.6 | 29.0 | 16.5 | 41.5 |
| | permutation_bonf | 30.6 | 18.0 | 40.2 | 27.6 | 16.3 | 43.0 | 28.2 | 16.4 | 41.1 |
| logn continuous | asymptotic | 69.1 | 48.7 | 84.3 | 69.4 | 52.8 | 84.3 | 68.9 | 49.1 | 82.9 |
| | asymptotic_bonf | 67.1 | 46.0 | 82.6 | 67.2 | 49.7 | 83.0 | 66.7 | 46.5 | 81.8 |
| | permutation_bonf | 66.5 | 45.4 | 81.8 | 65.6 | 49.2 | 82.2 | 66.5 | 45.1 | 81.2 |
| logn discrete | asymptotic | 77.3 | 58.0 | 89.8 | 76.8 | 58.8 | 88.9 | 76.5 | 57.9 | 88.9 |
| | asymptotic_bonf | 75.8 | 55.1 | 88.0 | 75.4 | 57.0 | 87.6 | 75.0 | 54.5 | 88.1 |
| | permutation_bonf | 74.9 | 53.9 | 87.2 | 74.3 | 55.8 | 87.3 | 74.3 | 53.9 | 87.9 |
| pwExp continuous | asymptotic | 23.9 | 13.3 | 39.6 | 22.9 | 12.8 | 38.9 | 22.9 | 13.1 | 37.7 |
| | asymptotic_bonf | 22.6 | 11.9 | 36.8 | 21.6 | 11.7 | 35.8 | 21.4 | 12.4 | 35.1 |
| | permutation_bonf | 22.1 | 11.5 | 37.2 | 20.9 | 11.2 | 35.1 | 22.0 | 12.0 | 34.6 |
| pwExp discrete | asymptotic | 26.9 | 16.1 | 46.1 | 26.6 | 14.9 | 44.8 | 26.4 | 16.0 | 43.0 |
| | asymptotic_bonf | 25.8 | 14.5 | 44.3 | 24.6 | 13.8 | 43.0 | 25.3 | 14.9 | 41.3 |
| | permutation_bonf | 26.8 | 14.5 | 44.9 | 24.5 | 13.7 | 42.2 | 25.4 | 14.6 | 40.2 |
| Weib late continuous | asymptotic | 70.4 | 46.9 | 86.6 | 70.9 | 45.8 | 88.4 | 70.5 | 48.8 | 87.1 |
| | asymptotic_bonf | 68.2 | 44.7 | 85.6 | 69.2 | 43.9 | 87.5 | 68.5 | 46.8 | 85.6 |
| | permutation_bonf | 66.7 | 44.5 | 85.7 | 68.4 | 42.6 | 86.7 | 68.1 | 45.7 | 85.5 |
| Weib late discrete | asymptotic | 76.1 | 54.5 | 90.5 | 74.1 | 54.8 | 92.2 | 76.4 | 54.1 | 90.2 |
| | asymptotic_bonf | 75.0 | 51.8 | 89.0 | 72.7 | 51.8 | 90.8 | 73.9 | 52.8 | 89.6 |
| | permutation_bonf | 74.1 | 50.9 | 88.5 | 72.3 | 50.4 | 90.2 | 73.3 | 50.7 | 89.4 |
| Weib prop continuous | asymptotic | 67.5 | 45.8 | 84.3 | 67.8 | 44.7 | 86.1 | 67.7 | 47.0 | 84.3 |
| | asymptotic_bonf | 65.7 | 43.0 | 83.3 | 66.1 | 42.5 | 84.7 | 65.3 | 44.4 | 82.8 |
| | permutation_bonf | 63.6 | 42.3 | 82.8 | 65.4 | 41.1 | 84.9 | 66.0 | 43.9 | 83.2 |
| Weib prop discrete | asymptotic | 76.9 | 54.1 | 90.4 | 75.7 | 54.1 | 91.6 | 77.2 | 53.9 | 90.3 |
| | asymptotic_bonf | 74.7 | 51.8 | 89.4 | 73.1 | 51.9 | 91.2 | 76.0 | 51.7 | 89.6 |
| | permutation_bonf | 74.2 | 50.6 | 88.8 | 72.6 | 50.1 | 90.2 | 75.3 | 51.1 | 89.6 |
| Weib scale continuous | asymptotic | 58.5 | 38.3 | 76.6 | 58.9 | 39.5 | 77.4 | 59.0 | 39.0 | 76.2 |
| | asymptotic_bonf | 57.0 | 36.4 | 74.8 | 56.9 | 37.5 | 75.6 | 56.6 | 36.4 | 74.3 |
| | permutation_bonf | 54.3 | 34.8 | 73.6 | 55.8 | 35.0 | 74.3 | 55.0 | 36.2 | 73.1 |
| Weib scale discrete | asymptotic | 66.4 | 46.2 | 84.2 | 67.3 | 46.4 | 85.3 | 67.7 | 46.9 | 83.8 |
| | asymptotic_bonf | 64.4 | 44.8 | 82.5 | 64.6 | 44.2 | 84.2 | 65.4 | 43.3 | 82.8 |
| | permutation_bonf | 63.7 | 43.5 | 81.3 | 64.4 | 43.2 | 82.5 | 64.2 | 42.3 | 82.6 |
| Weib shape continuous | asymptotic | 45.5 | 29.0 | 63.0 | 49.0 | 30.5 | 64.9 | 45.7 | 29.8 | 65.1 |
| | asymptotic_bonf | 41.7 | 25.5 | 59.8 | 46.0 | 27.8 | 62.0 | 42.2 | 27.0 | 61.0 |
| | permutation_bonf | 41.5 | 24.9 | 58.9 | 44.5 | 25.7 | 61.2 | 42.3 | 26.9 | 60.3 |
| Weib shape discrete | asymptotic | 55.5 | 35.9 | 75.0 | 56.7 | 37.4 | 75.5 | 56.0 | 36.7 | 74.9 |
| | asymptotic_bonf | 52.6 | 33.9 | 71.5 | 53.3 | 34.6 | 73.0 | 51.8 | 34.5 | 72.2 |
| | permutation_bonf | 51.8 | 32.5 | 70.6 | 53.4 | 33.4 | 71.8 | 51.8 | 33.5 | 71.9 |

Table S41: Rejection rates in percent for the 2-by-2 design with $\delta = 1.5$ and unbalanced medium sample sizes under equal censoring.

| Distribution | Method | $\mathcal{H}_{0,4}$ | $\mathcal{H}_{0,5}$ | $\mathcal{H}_{0,6}$ | $\mathcal{H}_{0,1}$ | $\mathcal{H}_{0,2}$ | $\mathcal{H}_{0,3}$ | $\mathcal{H}_{0,7}$ | $\mathcal{H}_{0,8}$ | $\mathcal{H}_{0,9}$ |
|---|---|---|---|---|---|---|---|---|---|---|
| exp early continuous | asymptotic | 3.0 | 1.9 | 4.5 | 3.3 | 1.5 | 4.0 | 3.3 | 1.9 | 4.2 |
|  | asymptotic_bonf | 2.4 | 1.5 | 4.1 | 2.8 | 1.3 | 3.6 | 2.8 | 1.8 | 3.7 |
|  | permutation_bonf | 1.6 | 1.6 | 3.1 | 2.4 | 0.9 | 3.2 | 2.4 | 1.3 | 3.3 |
| exp early discrete | asymptotic | 3.3 | 2.0 | 5.0 | 3.6 | 1.7 | 4.2 | 3.6 | 2.3 | 5.4 |
|  | asymptotic_bonf | 2.9 | 1.8 | 4.5 | 3.5 | 1.4 | 4.1 | 2.9 | 1.9 | 5.0 |
|  | permutation_bonf | 1.8 | 1.7 | 4.1 | 2.8 | 0.9 | 3.2 | 2.7 | 1.8 | 4.3 |
| exp late continuous | asymptotic | 2.8 | 2.1 | 4.7 | 4.1 | 2.3 | 4.6 | 4.1 | 2.1 | 5.3 |
|  | asymptotic_bonf | 2.2 | 2.0 | 4.3 | 3.8 | 1.6 | 4.1 | 3.8 | 1.9 | 4.3 |
|  | permutation_bonf | 2.3 | 1.9 | 3.7 | 3.3 | 1.1 | 3.4 | 2.8 | 1.4 | 3.5 |
| exp late discrete | asymptotic | 2.9 | 2.2 | 5.2 | 4.5 | 2.1 | 5.2 | 4.1 | 2.2 | 5.9 |
|  | asymptotic_bonf | 2.4 | 2.0 | 4.8 | 4.3 | 1.9 | 4.4 | 3.9 | 2.0 | 5.2 |
|  | permutation_bonf | 2.2 | 2.1 | 4.0 | 3.1 | 1.3 | 3.8 | 3.4 | 1.8 | 4.8 |
| exp prop continuous | asymptotic | 3.4 | 2.5 | 4.2 | 2.3 | 1.9 | 4.6 | 3.2 | 2.1 | 4.8 |
|  | asymptotic_bonf | 3.0 | 2.2 | 3.7 | 1.9 | 1.7 | 4.3 | 3.0 | 1.8 | 4.3 |
|  | permutation_bonf | 2.5 | 1.7 | 2.9 | 1.9 | 1.2 | 3.7 | 2.2 | 1.8 | 4.4 |
| exp prop discrete | asymptotic | 3.4 | 3.0 | 4.6 | 2.8 | 2.4 | 5.3 | 3.3 | 1.8 | 5.4 |
|  | asymptotic_bonf | 2.8 | 2.9 | 4.4 | 2.4 | 2.1 | 4.7 | 3.1 | 1.7 | 4.7 |
|  | permutation_bonf | 2.8 | 2.1 | 3.6 | 2.0 | 1.3 | 3.8 | 2.8 | 1.7 | 4.9 |
| logn continuous | asymptotic | 8.3 | 5.6 | 12.6 | 6.8 | 4.4 | 12.7 | 6.6 | 6.0 | 13.6 |
|  | asymptotic_bonf | 7.3 | 5.1 | 11.5 | 5.7 | 3.7 | 12.2 | 5.9 | 5.4 | 12.5 |
|  | permutation_bonf | 6.0 | 3.5 | 9.6 | 5.2 | 2.4 | 9.8 | 5.0 | 3.9 | 11.8 |
| logn discrete | asymptotic | 10.3 | 6.3 | 16.4 | 8.0 | 5.1 | 14.7 | 8.5 | 6.7 | 16.3 |
|  | asymptotic_bonf | 9.1 | 5.3 | 14.5 | 7.4 | 4.4 | 13.9 | 7.9 | 6.1 | 14.6 |
|  | permutation_bonf | 7.1 | 4.2 | 11.4 | 6.4 | 2.6 | 12.0 | 6.3 | 5.2 | 14.1 |
| pwExp continuous | asymptotic | 1.9 | 2.2 | 4.0 | 2.9 | 1.8 | 3.9 | 3.1 | 1.7 | 4.5 |
|  | asymptotic_bonf | 1.6 | 1.7 | 3.5 | 2.2 | 1.2 | 3.3 | 2.6 | 1.6 | 4.2 |
|  | permutation_bonf | 1.5 | 1.8 | 3.5 | 2.2 | 1.4 | 2.8 | 2.4 | 1.1 | 3.8 |
| pwExp discrete | asymptotic | 2.1 | 2.3 | 4.9 | 3.4 | 2.0 | 4.5 | 3.3 | 1.8 | 5.7 |
|  | asymptotic_bonf | 1.7 | 2.2 | 4.3 | 2.8 | 1.6 | 3.9 | 2.8 | 1.6 | 4.7 |
|  | permutation_bonf | 1.4 | 1.8 | 3.7 | 2.4 | 1.7 | 3.3 | 2.7 | 1.3 | 4.6 |
| Weib late continuous | asymptotic | 8.8 | 4.7 | 14.0 | 7.2 | 5.1 | 14.4 | 9.9 | 4.7 | 14.7 |
|  | asymptotic_bonf | 7.8 | 4.1 | 12.4 | 6.3 | 4.4 | 13.2 | 8.8 | 3.8 | 13.7 |
|  | permutation_bonf | 6.6 | 2.9 | 10.4 | 5.1 | 2.6 | 11.0 | 6.7 | 3.2 | 13.0 |
| Weib late discrete | asymptotic | 9.5 | 6.4 | 16.2 | 9.6 | 6.3 | 16.8 | 12.7 | 5.5 | 17.6 |
|  | asymptotic_bonf | 8.1 | 5.6 | 14.9 | 8.3 | 6.0 | 15.3 | 11.6 | 5.4 | 15.3 |
|  | permutation_bonf | 6.8 | 3.6 | 13.0 | 6.3 | 3.4 | 12.5 | 9.7 | 4.5 | 13.7 |
| Weib prop continuous | asymptotic | 7.6 | 4.3 | 12.4 | 6.3 | 4.8 | 12.8 | 8.7 | 4.3 | 13.9 |
|  | asymptotic_bonf | 7.0 | 3.7 | 11.4 | 5.7 | 4.2 | 11.6 | 7.7 | 3.6 | 12.6 |
|  | permutation_bonf | 6.2 | 2.5 | 9.7 | 4.5 | 1.8 | 9.5 | 6.3 | 2.9 | 11.8 |
| Weib prop discrete | asymptotic | 9.1 | 5.8 | 14.5 | 8.6 | 5.7 | 15.6 | 10.4 | 5.0 | 15.5 |
|  | asymptotic_bonf | 7.9 | 4.7 | 13.8 | 7.9 | 5.0 | 13.7 | 8.8 | 4.6 | 14.4 |
|  | permutation_bonf | 6.4 | 2.9 | 11.4 | 5.6 | 3.3 | 11.6 | 7.6 | 3.5 | 12.7 |
| Weib scale continuous | asymptotic | 6.1 | 3.1 | 8.9 | 5.7 | 3.6 | 9.9 | 6.4 | 3.1 | 10.3 |
|  | asymptotic_bonf | 5.7 | 2.4 | 7.5 | 4.6 | 2.9 | 8.4 | 5.7 | 2.6 | 8.8 |
|  | permutation_bonf | 4.5 | 1.8 | 6.2 | 3.1 | 1.3 | 6.9 | 5.3 | 2.1 | 7.7 |
| Weib scale discrete | asymptotic | 7.2 | 3.2 | 11.0 | 6.7 | 3.9 | 11.8 | 7.8 | 3.7 | 11.6 |
|  | asymptotic_bonf | 6.3 | 3.0 | 9.9 | 5.7 | 3.5 | 10.5 | 6.9 | 3.4 | 10.3 |
|  | permutation_bonf | 5.3 | 1.8 | 7.6 | 4.1 | 1.9 | 8.5 | 6.2 | 2.6 | 8.7 |
| Weib shape continuous | asymptotic | 4.0 | 1.8 | 5.9 | 3.5 | 2.3 | 6.0 | 4.5 | 2.5 | 5.7 |
|  | asymptotic_bonf | 2.8 | 1.2 | 5.2 | 2.7 | 1.7 | 4.8 | 3.3 | 1.9 | 4.6 |
|  | permutation_bonf | 2.0 | 0.9 | 4.3 | 1.5 | 1.2 | 4.0 | 2.8 | 1.3 | 3.9 |
| Weib shape discrete | asymptotic | 4.9 | 1.8 | 7.6 | 5.6 | 2.5 | 7.1 | 5.7 | 3.0 | 8.1 |
|  | asymptotic_bonf | 4.1 | 1.3 | 6.3 | 4.0 | 2.1 | 6.0 | 4.4 | 2.6 | 7.0 |
|  | permutation_bonf | 2.7 | 0.7 | 5.0 | 2.3 | 1.3 | 4.9 | 4.4 | 2.0 | 6.2 |

Table S42: Rejection rates in percent for the 2-by-2 design with $\delta = 1.5$ and unbalanced small sample sizes under equal censoring.

| Distribution | Method | $\mathcal{H}_{0,4}$ | $\mathcal{H}_{0,5}$ | $\mathcal{H}_{0,6}$ | $\mathcal{H}_{0,1}$ | $\mathcal{H}_{0,2}$ | $\mathcal{H}_{0,3}$ | $\mathcal{H}_{0,7}$ | $\mathcal{H}_{0,8}$ | $\mathcal{H}_{0,9}$ |
|---|---|---|---|---|---|---|---|---|---|---|
| exp early continuous | asymptotic | 1.3 | 1.4 | 1.9 | 1.2 | 1.2 | 1.0 | 0.6 | 1.0 | 2.2 |
| | asymptotic_bonf | 1.2 | 1.3 | 1.4 | 1.0 | 1.2 | 0.9 | 0.5 | 1.0 | 1.9 |
| | permutation_bonf | 0.3 | 0.5 | 0.5 | 0.1 | 0.4 | 0.4 | 0.3 | 0.4 | 1.2 |
| exp early discrete | asymptotic | 1.4 | 1.1 | 1.6 | 1.4 | 1.2 | 1.0 | 0.8 | 1.3 | 2.0 |
| | asymptotic_bonf | 1.2 | 1.1 | 1.5 | 1.2 | 1.1 | 1.0 | 0.7 | 0.9 | 2.0 |
| | permutation_bonf | 0.2 | 0.3 | 0.6 | 0.2 | 0.4 | 0.5 | 0.3 | 0.6 | 1.0 |
| exp late continuous | asymptotic | 1.6 | 0.9 | 1.5 | 1.2 | 1.0 | 1.3 | 0.7 | 1.0 | 2.4 |
| | asymptotic_bonf | 1.4 | 0.9 | 1.4 | 0.8 | 0.8 | 1.2 | 0.5 | 1.0 | 2.3 |
| | permutation_bonf | 0.2 | 0.3 | 0.3 | 0.4 | 0.3 | 0.4 | 0.2 | 0.4 | 1.3 |
| exp late discrete | asymptotic | 1.9 | 1.0 | 1.7 | 0.9 | 1.2 | 1.7 | 0.9 | 1.1 | 2.6 |
| | asymptotic_bonf | 1.5 | 0.9 | 1.2 | 0.9 | 1.0 | 1.7 | 0.6 | 1.0 | 2.4 |
| | permutation_bonf | 0.2 | 0.0 | 0.8 | 0.4 | 0.2 | 0.6 | 0.1 | 0.5 | 1.7 |
| exp prop continuous | asymptotic | 1.4 | 1.6 | 2.2 | 1.1 | 1.1 | 2.0 | 1.2 | 0.8 | 1.8 |
| | asymptotic_bonf | 1.4 | 1.3 | 2.0 | 1.1 | 1.1 | 1.9 | 1.2 | 0.8 | 1.4 |
| | permutation_bonf | 0.8 | 0.4 | 0.5 | 0.4 | 0.0 | 0.9 | 0.6 | 0.3 | 0.7 |
| exp prop discrete | asymptotic | 1.4 | 1.4 | 2.1 | 1.1 | 1.0 | 2.5 | 1.1 | 1.0 | 1.8 |
| | asymptotic_bonf | 1.3 | 1.4 | 2.0 | 0.9 | 0.9 | 2.3 | 1.0 | 0.7 | 1.8 |
| | permutation_bonf | 0.4 | 0.6 | 1.1 | 0.5 | 0.1 | 1.0 | 0.5 | 0.6 | 0.7 |
| logn continuous | asymptotic | 1.5 | 1.0 | 2.7 | 1.9 | 1.9 | 2.0 | 1.9 | 1.4 | 2.6 |
| | asymptotic_bonf | 1.3 | 0.7 | 2.6 | 1.8 | 1.6 | 2.0 | 1.9 | 0.9 | 2.4 |
| | permutation_bonf | 0.3 | 0.3 | 1.0 | 0.2 | 0.3 | 0.6 | 0.7 | 0.2 | 0.8 |
| logn discrete | asymptotic | 2.1 | 1.4 | 2.8 | 2.2 | 1.6 | 2.3 | 1.9 | 1.3 | 2.8 |
| | asymptotic_bonf | 1.9 | 1.1 | 2.6 | 2.0 | 1.2 | 1.9 | 1.7 | 0.9 | 2.5 |
| | permutation_bonf | 0.4 | 0.2 | 1.2 | 0.3 | 0.3 | 0.6 | 0.8 | 0.2 | 1.1 |
| pwExp continuous | asymptotic | 0.8 | 1.4 | 1.6 | 0.5 | 1.1 | 1.4 | 0.6 | 1.0 | 1.9 |
| | asymptotic_bonf | 0.7 | 1.3 | 1.4 | 0.5 | 0.9 | 1.4 | 0.6 | 1.0 | 1.8 |
| | permutation_bonf | 0.2 | 0.3 | 0.5 | 0.1 | 0.3 | 0.6 | 0.1 | 0.5 | 1.0 |
| pwExp discrete | asymptotic | 1.2 | 1.4 | 1.8 | 0.7 | 0.9 | 1.7 | 0.6 | 1.2 | 2.2 |
| | asymptotic_bonf | 0.8 | 1.3 | 1.5 | 0.5 | 0.8 | 1.7 | 0.6 | 1.0 | 1.8 |
| | permutation_bonf | 0.2 | 0.5 | 0.6 | 0.2 | 0.2 | 0.5 | 0.2 | 0.5 | 1.1 |
| Weib late continuous | asymptotic | 1.9 | 1.4 | 3.4 | 1.9 | 1.6 | 2.3 | 1.7 | 1.2 | 2.3 |
| | asymptotic_bonf | 1.7 | 1.3 | 2.9 | 1.6 | 1.4 | 1.9 | 1.3 | 1.1 | 2.2 |
| | permutation_bonf | 0.4 | 0.4 | 0.7 | 0.1 | 0.1 | 0.7 | 0.6 | 0.5 | 1.1 |
| Weib late discrete | asymptotic | 2.3 | 1.5 | 3.2 | 2.0 | 1.6 | 2.5 | 1.7 | 1.3 | 3.0 |
| | asymptotic_bonf | 1.9 | 1.5 | 2.8 | 1.7 | 1.1 | 2.2 | 1.6 | 1.2 | 2.7 |
| | permutation_bonf | 0.8 | 0.6 | 0.9 | 0.4 | 0.2 | 0.4 | 0.7 | 0.5 | 0.9 |
| Weib prop continuous | asymptotic | 1.8 | 1.5 | 2.8 | 1.4 | 1.7 | 2.2 | 1.3 | 1.5 | 1.9 |
| | asymptotic_bonf | 1.6 | 1.4 | 2.3 | 1.1 | 1.6 | 1.9 | 1.2 | 1.1 | 1.8 |
| | permutation_bonf | 0.3 | 0.4 | 0.9 | 0.2 | 0.1 | 0.7 | 0.5 | 0.3 | 0.9 |
| Weib prop discrete | asymptotic | 2.0 | 1.4 | 2.8 | 1.8 | 1.7 | 2.8 | 1.3 | 1.2 | 2.2 |
| | asymptotic_bonf | 1.6 | 1.4 | 2.5 | 1.5 | 1.4 | 2.5 | 1.0 | 0.9 | 1.9 |
| | permutation_bonf | 0.4 | 0.5 | 1.2 | 0.3 | 0.1 | 0.5 | 0.3 | 0.3 | 1.0 |
| Weib scale continuous | asymptotic | 1.6 | 1.3 | 2.6 | 1.5 | 1.8 | 2.0 | 1.1 | 1.3 | 1.8 |
| | asymptotic_bonf | 1.5 | 1.3 | 2.1 | 1.4 | 1.6 | 1.8 | 1.1 | 0.8 | 1.7 |
| | permutation_bonf | 0.3 | 0.3 | 0.8 | 0.2 | 0.3 | 0.8 | 0.4 | 0.1 | 0.7 |
| Weib scale discrete | asymptotic | 1.5 | 1.5 | 2.3 | 1.4 | 2.4 | 2.3 | 0.8 | 0.9 | 2.0 |
| | asymptotic_bonf | 1.2 | 1.4 | 2.0 | 1.3 | 2.1 | 2.2 | 0.8 | 0.7 | 1.8 |
| | permutation_bonf | 0.5 | 0.6 | 0.8 | 0.4 | 0.3 | 0.4 | 0.7 | 0.1 | 0.7 |
| Weib shape continuous | asymptotic | 2.6 | 1.6 | 2.4 | 2.0 | 3.4 | 1.7 | 2.2 | 2.4 | 1.7 |
| | asymptotic_bonf | 2.2 | 1.4 | 2.1 | 1.9 | 3.0 | 1.6 | 1.8 | 1.8 | 1.5 |
| | permutation_bonf | 0.9 | 0.5 | 1.0 | 0.4 | 0.8 | 0.9 | 1.1 | 0.5 | 0.9 |
| Weib shape discrete | asymptotic | 2.4 | 1.7 | 2.5 | 1.8 | 3.1 | 2.6 | 2.2 | 1.7 | 1.8 |
| | asymptotic_bonf | 2.1 | 1.6 | 2.3 | 1.6 | 2.5 | 2.0 | 2.0 | 1.4 | 1.7 |
| | permutation_bonf | 0.7 | 0.7 | 0.9 | 0.3 | 0.5 | 0.5 | 1.1 | 0.5 | 0.8 |

Table S43: Rejection rates in percent for the 2-by-2 design with $\delta = 1.5$ and balanced large sample sizes under unequal, high censoring.

| Distribution | Method | $\mathcal{H}_{0,4}$ | $\mathcal{H}_{0,5}$ | $\mathcal{H}_{0,6}$ | $\mathcal{H}_{0,1}$ | $\mathcal{H}_{0,2}$ | $\mathcal{H}_{0,3}$ | $\mathcal{H}_{0,7}$ | $\mathcal{H}_{0,8}$ | $\mathcal{H}_{0,9}$ |
|---|---|---|---|---|---|---|---|---|---|---|
| exp early continuous | asymptotic | 38.0 | 25.0 | 55.3 | 35.6 | 24.5 | 55.1 | 38.9 | 22.5 | 55.6 |
|  | asymptotic_bonf | 37.4 | 24.5 | 54.8 | 35.0 | 24.0 | 54.5 | 38.3 | 22.3 | 55.2 |
|  | permutation_bonf | 38.1 | 23.3 | 53.7 | 35.7 | 23.9 | 54.5 | 37.2 | 22.8 | 54.2 |
| exp early discrete | asymptotic | 41.8 | 27.3 | 60.0 | 39.5 | 27.6 | 60.1 | 41.6 | 26.1 | 61.0 |
|  | asymptotic_bonf | 41.7 | 26.9 | 59.5 | 38.9 | 27.0 | 59.6 | 41.4 | 25.2 | 60.7 |
|  | permutation_bonf | 42.3 | 26.5 | 59.2 | 37.2 | 27.3 | 60.0 | 39.6 | 25.4 | 59.4 |
| exp late continuous | asymptotic | 38.6 | 24.5 | 56.8 | 36.9 | 24.1 | 55.6 | 39.4 | 24.5 | 57.3 |
|  | asymptotic_bonf | 37.9 | 24.0 | 55.9 | 36.5 | 23.3 | 55.0 | 38.8 | 23.8 | 56.8 |
|  | permutation_bonf | 37.7 | 24.5 | 55.1 | 36.2 | 24.5 | 54.2 | 38.9 | 24.2 | 55.9 |
| exp late discrete | asymptotic | 43.7 | 28.3 | 61.6 | 40.8 | 27.6 | 61.4 | 42.7 | 26.1 | 62.1 |
|  | asymptotic_bonf | 43.2 | 27.5 | 61.1 | 40.1 | 27.2 | 60.4 | 41.7 | 25.5 | 61.2 |
|  | permutation_bonf | 43.1 | 26.6 | 59.8 | 39.4 | 26.9 | 60.9 | 41.1 | 26.2 | 60.8 |
| exp prop continuous | asymptotic | 39.2 | 25.7 | 54.0 | 38.9 | 23.2 | 55.4 | 38.2 | 25.3 | 54.5 |
|  | asymptotic_bonf | 38.6 | 25.2 | 53.6 | 38.2 | 22.7 | 54.3 | 37.5 | 24.9 | 53.7 |
|  | permutation_bonf | 38.8 | 24.9 | 54.2 | 37.7 | 23.3 | 53.7 | 36.3 | 24.2 | 53.5 |
| exp prop discrete | asymptotic | 42.5 | 27.2 | 58.8 | 41.7 | 25.9 | 59.0 | 42.1 | 27.9 | 58.0 |
|  | asymptotic_bonf | 42.2 | 26.5 | 57.7 | 41.3 | 25.5 | 58.1 | 41.3 | 27.3 | 57.7 |
|  | permutation_bonf | 42.1 | 27.4 | 59.0 | 41.2 | 25.4 | 58.1 | 41.1 | 27.4 | 58.4 |
| logn continuous | asymptotic | 79.1 | 58.9 | 90.4 | 80.5 | 60.4 | 92.0 | 80.1 | 59.4 | 91.9 |
|  | asymptotic_bonf | 78.9 | 58.3 | 90.4 | 79.9 | 60.1 | 91.7 | 79.8 | 58.9 | 91.7 |
|  | permutation_bonf | 78.3 | 58.4 | 90.9 | 78.7 | 59.6 | 90.8 | 79.2 | 59.2 | 91.6 |
| logn discrete | asymptotic | 86.8 | 67.4 | 95.6 | 87.4 | 71.7 | 96.5 | 85.8 | 69.6 | 95.6 |
|  | asymptotic_bonf | 86.4 | 67.1 | 95.5 | 86.7 | 71.3 | 96.3 | 85.5 | 69.5 | 95.5 |
|  | permutation_bonf | 86.7 | 67.4 | 95.4 | 86.6 | 70.0 | 95.7 | 85.8 | 69.1 | 94.9 |
| pwExp continuous | asymptotic | 36.8 | 24.1 | 54.6 | 35.0 | 23.8 | 55.4 | 37.3 | 23.1 | 55.9 |
|  | asymptotic_bonf | 36.2 | 23.7 | 54.3 | 34.2 | 23.3 | 54.7 | 36.5 | 22.4 | 55.2 |
|  | permutation_bonf | 35.9 | 24.1 | 53.2 | 35.1 | 22.8 | 54.4 | 36.6 | 23.0 | 54.5 |
| pwExp discrete | asymptotic | 41.4 | 25.8 | 58.2 | 38.7 | 25.9 | 59.2 | 40.0 | 26.4 | 59.8 |
|  | asymptotic_bonf | 40.6 | 25.2 | 57.5 | 38.1 | 25.5 | 58.8 | 39.6 | 26.1 | 59.1 |
|  | permutation_bonf | 40.7 | 25.8 | 58.3 | 38.5 | 26.0 | 59.2 | 39.1 | 25.7 | 58.6 |
| Weib late continuous | asymptotic | 78.5 | 60.1 | 93.5 | 78.6 | 59.4 | 91.1 | 78.3 | 60.4 | 91.8 |
|  | asymptotic_bonf | 77.6 | 59.3 | 93.3 | 78.1 | 58.8 | 90.9 | 77.7 | 59.8 | 91.6 |
|  | permutation_bonf | 76.4 | 59.3 | 92.7 | 78.6 | 59.3 | 90.9 | 77.9 | 59.3 | 90.8 |
| Weib late discrete | asymptotic | 84.6 | 67.8 | 96.4 | 85.0 | 67.5 | 95.8 | 85.2 | 70.9 | 95.9 |
|  | asymptotic_bonf | 84.1 | 67.2 | 96.4 | 84.8 | 67.1 | 95.7 | 84.8 | 70.4 | 95.8 |
|  | permutation_bonf | 83.0 | 66.7 | 96.2 | 85.5 | 67.2 | 95.1 | 84.3 | 69.9 | 95.5 |
| Weib prop continuous | asymptotic | 76.8 | 59.3 | 92.4 | 77.2 | 58.7 | 90.6 | 77.9 | 59.5 | 91.5 |
|  | asymptotic_bonf | 76.4 | 58.7 | 92.2 | 77.0 | 57.7 | 90.3 | 77.6 | 58.5 | 91.2 |
|  | permutation_bonf | 76.5 | 58.7 | 92.0 | 77.6 | 58.1 | 90.7 | 77.6 | 58.5 | 91.2 |
| Weib prop discrete | asymptotic | 84.0 | 66.8 | 96.0 | 84.5 | 67.0 | 95.6 | 84.4 | 69.2 | 95.2 |
|  | asymptotic_bonf | 83.8 | 66.5 | 95.9 | 84.5 | 66.7 | 95.5 | 84.0 | 68.8 | 95.2 |
|  | permutation_bonf | 83.3 | 66.0 | 95.9 | 85.3 | 66.5 | 94.8 | 84.0 | 68.7 | 94.8 |
| Weib scale continuous | asymptotic | 74.1 | 56.7 | 90.6 | 75.2 | 56.2 | 89.4 | 73.8 | 56.9 | 88.9 |
|  | asymptotic_bonf | 73.2 | 56.0 | 90.4 | 75.0 | 55.4 | 89.1 | 73.2 | 56.4 | 88.6 |
|  | permutation_bonf | 72.3 | 55.7 | 89.8 | 74.1 | 54.7 | 88.5 | 73.1 | 56.6 | 89.0 |
| Weib scale discrete | asymptotic | 81.2 | 64.8 | 95.0 | 82.6 | 64.2 | 93.6 | 82.6 | 66.4 | 93.6 |
|  | asymptotic_bonf | 80.8 | 64.3 | 94.7 | 82.3 | 63.7 | 93.3 | 82.2 | 65.8 | 93.5 |
|  | permutation_bonf | 80.6 | 63.8 | 94.0 | 81.7 | 63.9 | 92.9 | 81.9 | 66.8 | 93.4 |
| Weib shape continuous | asymptotic | 70.9 | 53.9 | 88.3 | 71.6 | 52.8 | 87.0 | 72.3 | 53.2 | 86.9 |
|  | asymptotic_bonf | 70.4 | 53.3 | 87.8 | 71.0 | 52.3 | 86.8 | 71.8 | 52.7 | 86.9 |
|  | permutation_bonf | 70.4 | 53.2 | 88.3 | 71.6 | 52.4 | 86.2 | 71.4 | 53.2 | 86.4 |
| Weib shape discrete | asymptotic | 78.9 | 63.0 | 92.0 | 79.3 | 62.1 | 92.2 | 80.4 | 62.7 | 91.8 |
|  | asymptotic_bonf | 78.5 | 62.5 | 91.9 | 78.9 | 61.9 | 92.0 | 80.0 | 62.2 | 91.5 |
|  | permutation_bonf | 77.9 | 61.7 | 92.0 | 80.3 | 61.1 | 92.3 | 79.8 | 62.7 | 91.5 |

Table S44: Rejection rates in percent for the 2-by-2 design with $\delta = 1.5$ and balanced medium sample sizes under unequal, high censoring.

| Distribution | Method | $\mathcal{H}_{0,4}$ | $\mathcal{H}_{0,5}$ | $\mathcal{H}_{0,6}$ | $\mathcal{H}_{0,1}$ | $\mathcal{H}_{0,2}$ | $\mathcal{H}_{0,3}$ | $\mathcal{H}_{0,7}$ | $\mathcal{H}_{0,8}$ | $\mathcal{H}_{0,9}$ |
|---|---|---|---|---|---|---|---|---|---|---|
| exp early continuous | asymptotic | 4.7 | 3.9 | 6.9 | 5.2 | 2.9 | 8.4 | 5.6 | 2.5 | 7.0 |
| | asymptotic_bonf | 4.5 | 3.6 | 6.6 | 5.0 | 2.8 | 8.1 | 5.3 | 2.4 | 6.9 |
| | permutation_bonf | 4.9 | 3.8 | 7.3 | 4.6 | 2.8 | 8.3 | 5.4 | 2.6 | 6.4 |
| exp early discrete | asymptotic | 6.7 | 3.9 | 7.7 | 5.4 | 3.3 | 8.8 | 5.8 | 2.7 | 8.0 |
| | asymptotic_bonf | 6.3 | 3.8 | 7.7 | 5.3 | 3.2 | 8.5 | 5.6 | 2.5 | 7.7 |
| | permutation_bonf | 5.8 | 4.1 | 7.7 | 5.4 | 4.2 | 8.7 | 5.6 | 2.2 | 7.3 |
| exp late continuous | asymptotic | 5.1 | 3.7 | 7.2 | 4.7 | 3.4 | 8.7 | 5.4 | 3.0 | 7.2 |
| | asymptotic_bonf | 4.7 | 3.6 | 7.0 | 4.5 | 3.1 | 8.5 | 5.3 | 3.0 | 6.6 |
| | permutation_bonf | 5.0 | 3.8 | 7.6 | 4.4 | 2.9 | 8.6 | 5.0 | 2.8 | 6.0 |
| exp late discrete | asymptotic | 6.4 | 4.3 | 7.8 | 5.7 | 3.7 | 8.9 | 6.3 | 2.6 | 8.4 |
| | asymptotic_bonf | 6.1 | 3.8 | 7.6 | 5.4 | 3.6 | 8.8 | 5.9 | 2.5 | 8.0 |
| | permutation_bonf | 5.9 | 3.8 | 7.9 | 5.2 | 3.6 | 8.3 | 5.9 | 2.4 | 7.1 |
| exp prop continuous | asymptotic | 6.0 | 4.7 | 7.4 | 4.9 | 3.4 | 6.2 | 4.8 | 4.0 | 7.2 |
| | asymptotic_bonf | 5.9 | 4.4 | 7.1 | 4.7 | 3.2 | 6.2 | 4.5 | 3.9 | 6.8 |
| | permutation_bonf | 5.2 | 4.1 | 6.8 | 5.3 | 3.1 | 6.2 | 4.7 | 3.7 | 7.1 |
| exp prop discrete | asymptotic | 5.9 | 5.3 | 8.4 | 5.0 | 4.2 | 7.3 | 5.3 | 4.1 | 7.7 |
| | asymptotic_bonf | 5.7 | 5.1 | 8.1 | 4.8 | 4.0 | 7.0 | 5.0 | 4.0 | 7.7 |
| | permutation_bonf | 6.2 | 5.4 | 8.2 | 5.7 | 4.4 | 6.5 | 4.8 | 3.9 | 8.4 |
| logn continuous | asymptotic | 12.0 | 7.8 | 20.3 | 12.6 | 7.4 | 18.5 | 14.0 | 8.0 | 19.2 |
| | asymptotic_bonf | 11.2 | 7.5 | 19.8 | 12.5 | 7.3 | 17.8 | 13.7 | 7.5 | 18.6 |
| | permutation_bonf | 11.2 | 7.5 | 18.9 | 12.3 | 7.0 | 17.6 | 12.6 | 8.1 | 17.7 |
| logn discrete | asymptotic | 13.2 | 9.7 | 23.5 | 14.3 | 8.9 | 22.8 | 15.9 | 10.0 | 23.3 |
| | asymptotic_bonf | 12.9 | 9.7 | 22.8 | 14.0 | 8.8 | 21.7 | 15.4 | 9.4 | 22.7 |
| | permutation_bonf | 12.7 | 9.8 | 22.8 | 15.0 | 8.6 | 22.4 | 15.9 | 9.9 | 22.4 |
| pwExp continuous | asymptotic | 4.4 | 3.8 | 7.1 | 4.5 | 3.5 | 7.8 | 5.5 | 2.7 | 6.8 |
| | asymptotic_bonf | 4.4 | 3.7 | 6.7 | 4.1 | 3.4 | 7.3 | 5.4 | 2.6 | 6.8 |
| | permutation_bonf | 4.6 | 4.3 | 6.8 | 3.8 | 3.3 | 7.1 | 5.6 | 2.5 | 6.9 |
| pwExp discrete | asymptotic | 5.8 | 3.8 | 8.2 | 4.8 | 3.9 | 8.7 | 6.3 | 2.3 | 8.0 |
| | asymptotic_bonf | 5.7 | 3.8 | 8.1 | 4.8 | 3.8 | 8.5 | 5.9 | 2.2 | 8.0 |
| | permutation_bonf | 5.6 | 3.9 | 7.6 | 4.8 | 3.4 | 8.0 | 5.6 | 2.3 | 7.8 |
| Weib late continuous | asymptotic | 11.2 | 7.5 | 20.7 | 12.1 | 8.6 | 19.4 | 12.1 | 7.8 | 19.7 |
| | asymptotic_bonf | 10.9 | 7.4 | 20.4 | 11.7 | 8.5 | 18.8 | 12.0 | 7.7 | 19.0 |
| | permutation_bonf | 10.8 | 6.8 | 20.0 | 11.7 | 8.4 | 18.0 | 11.5 | 7.6 | 18.5 |
| Weib late discrete | asymptotic | 13.2 | 8.5 | 24.3 | 14.6 | 8.9 | 23.2 | 13.7 | 8.9 | 22.3 |
| | asymptotic_bonf | 13.0 | 7.8 | 23.6 | 14.2 | 8.8 | 22.7 | 13.4 | 8.5 | 21.9 |
| | permutation_bonf | 12.8 | 8.4 | 22.4 | 13.8 | 8.8 | 22.1 | 13.3 | 9.1 | 22.7 |
| Weib prop continuous | asymptotic | 11.1 | 6.7 | 20.1 | 11.6 | 8.5 | 18.9 | 12.2 | 7.8 | 19.1 |
| | asymptotic_bonf | 10.6 | 6.6 | 19.8 | 10.9 | 8.5 | 18.4 | 11.7 | 7.8 | 18.8 |
| | permutation_bonf | 10.4 | 6.9 | 19.0 | 11.3 | 7.9 | 18.0 | 10.8 | 7.9 | 18.0 |
| Weib prop discrete | asymptotic | 13.5 | 8.0 | 24.2 | 15.4 | 9.0 | 23.8 | 13.7 | 9.6 | 22.9 |
| | asymptotic_bonf | 13.1 | 7.6 | 23.5 | 14.8 | 8.4 | 23.0 | 13.4 | 9.3 | 22.4 |
| | permutation_bonf | 12.4 | 7.8 | 23.4 | 14.7 | 9.0 | 22.0 | 12.6 | 8.9 | 22.9 |
| Weib scale continuous | asymptotic | 9.9 | 6.8 | 17.9 | 11.4 | 7.1 | 16.1 | 10.5 | 6.7 | 17.8 |
| | asymptotic_bonf | 9.8 | 6.5 | 17.6 | 11.1 | 6.8 | 15.8 | 10.0 | 6.5 | 17.2 |
| | permutation_bonf | 9.7 | 6.4 | 16.8 | 11.0 | 7.1 | 15.9 | 10.2 | 6.7 | 16.5 |
| Weib scale discrete | asymptotic | 12.6 | 7.5 | 22.0 | 13.0 | 8.1 | 21.7 | 12.7 | 8.0 | 20.7 |
| | asymptotic_bonf | 12.4 | 7.2 | 21.9 | 12.7 | 8.0 | 21.1 | 12.2 | 8.0 | 20.4 |
| | permutation_bonf | 12.1 | 7.4 | 21.4 | 12.7 | 8.5 | 20.1 | 12.3 | 8.8 | 19.3 |
| Weib shape continuous | asymptotic | 9.7 | 6.0 | 16.6 | 9.0 | 6.9 | 14.7 | 9.9 | 6.0 | 17.0 |
| | asymptotic_bonf | 9.3 | 5.7 | 16.2 | 8.7 | 6.6 | 14.5 | 9.8 | 5.9 | 16.2 |
| | permutation_bonf | 9.4 | 5.7 | 16.0 | 9.1 | 6.4 | 14.2 | 9.7 | 6.7 | 15.3 |
| Weib shape discrete | asymptotic | 10.4 | 7.1 | 20.2 | 11.2 | 7.0 | 18.4 | 11.5 | 7.3 | 19.7 |
| | asymptotic_bonf | 10.2 | 6.5 | 19.8 | 11.0 | 6.8 | 18.1 | 11.1 | 7.2 | 19.3 |
| | permutation_bonf | 10.6 | 6.8 | 20.0 | 10.5 | 7.2 | 17.3 | 10.7 | 7.3 | 19.1 |

Table S45: Rejection rates in percent for the 2-by-2 design with $\delta = 1.5$ and balanced small sample sizes under unequal, high censoring.

| Distribution | Method | $\mathcal{H}_{0,4}$ | $\mathcal{H}_{0,5}$ | $\mathcal{H}_{0,6}$ | $\mathcal{H}_{0,1}$ | $\mathcal{H}_{0,2}$ | $\mathcal{H}_{0,3}$ | $\mathcal{H}_{0,7}$ | $\mathcal{H}_{0,8}$ | $\mathcal{H}_{0,9}$ |
|---|---|---|---|---|---|---|---|---|---|---|
| exp early continuous | asymptotic | 1.3 | 1.1 | 1.9 | 2.0 | 1.0 | 1.7 | 1.5 | 1.5 | 1.5 |
| | asymptotic_bonf | 1.3 | 1.1 | 1.7 | 2.0 | 1.0 | 1.7 | 1.4 | 1.3 | 1.4 |
| | permutation_bonf | 1.2 | 1.1 | 1.7 | 1.6 | 0.5 | 1.0 | 1.1 | 1.3 | 1.2 |
| exp early discrete | asymptotic | 1.5 | 1.0 | 1.7 | 0.7 | 1.2 | 1.5 | 1.1 | 1.1 | 2.1 |
| | asymptotic_bonf | 1.4 | 0.9 | 1.3 | 0.7 | 1.2 | 1.3 | 1.1 | 1.1 | 2.1 |
| | permutation_bonf | 1.3 | 1.0 | 1.4 | 0.9 | 1.1 | 1.0 | 1.0 | 0.9 | 1.3 |
| exp late continuous | asymptotic | 1.1 | 1.1 | 2.2 | 2.1 | 0.9 | 1.3 | 2.2 | 1.4 | 1.8 |
| | asymptotic_bonf | 1.1 | 1.1 | 2.1 | 2.0 | 0.9 | 1.3 | 2.0 | 1.3 | 1.7 |
| | permutation_bonf | 1.1 | 1.0 | 1.9 | 1.7 | 0.5 | 1.1 | 1.6 | 1.0 | 1.4 |
| exp late discrete | asymptotic | 1.5 | 1.4 | 2.3 | 2.2 | 1.0 | 1.7 | 1.6 | 0.8 | 1.6 |
| | asymptotic_bonf | 1.2 | 1.0 | 2.3 | 2.2 | 1.0 | 1.6 | 1.6 | 0.7 | 1.4 |
| | permutation_bonf | 0.9 | 1.2 | 1.7 | 1.8 | 1.0 | 1.3 | 1.3 | 0.6 | 1.3 |
| exp prop continuous | asymptotic | 2.0 | 1.3 | 2.0 | 1.0 | 1.0 | 1.6 | 1.1 | 0.7 | 1.9 |
| | asymptotic_bonf | 1.9 | 1.3 | 2.0 | 0.9 | 1.0 | 1.5 | 1.1 | 0.6 | 1.7 |
| | permutation_bonf | 1.8 | 1.0 | 1.6 | 0.9 | 0.6 | 1.2 | 1.0 | 0.6 | 1.1 |
| exp prop discrete | asymptotic | 1.3 | 0.7 | 2.6 | 1.0 | 0.7 | 1.8 | 1.4 | 1.4 | 2.1 |
| | asymptotic_bonf | 1.3 | 0.7 | 2.6 | 1.0 | 0.7 | 1.7 | 1.4 | 1.4 | 2.1 |
| | permutation_bonf | 1.4 | 0.5 | 2.3 | 0.9 | 0.5 | 1.6 | 0.7 | 1.3 | 1.6 |
| logn continuous | asymptotic | 2.4 | 1.5 | 3.3 | 3.0 | 2.4 | 4.1 | 2.6 | 1.6 | 3.7 |
| | asymptotic_bonf | 2.4 | 1.4 | 3.2 | 2.8 | 2.4 | 4.0 | 2.5 | 1.6 | 3.6 |
| | permutation_bonf | 2.5 | 1.4 | 3.2 | 2.6 | 1.9 | 2.7 | 1.9 | 1.5 | 3.2 |
| logn discrete | asymptotic | 2.7 | 1.7 | 4.1 | 3.6 | 3.5 | 5.3 | 2.9 | 2.1 | 3.7 |
| | asymptotic_bonf | 2.7 | 1.7 | 3.8 | 3.5 | 3.3 | 5.3 | 2.7 | 2.0 | 3.7 |
| | permutation_bonf | 2.9 | 1.6 | 3.6 | 3.3 | 2.9 | 4.1 | 2.3 | 1.8 | 3.2 |
| pwExp continuous | asymptotic | 1.3 | 1.3 | 1.7 | 1.7 | 1.0 | 1.4 | 2.1 | 1.4 | 1.6 |
| | asymptotic_bonf | 1.2 | 1.1 | 1.7 | 1.6 | 1.0 | 1.3 | 2.1 | 1.3 | 1.6 |
| | permutation_bonf | 1.1 | 1.2 | 1.8 | 1.5 | 0.7 | 1.0 | 1.7 | 1.1 | 1.6 |
| pwExp discrete | asymptotic | 1.5 | 1.3 | 2.2 | 2.0 | 1.1 | 1.4 | 2.0 | 1.6 | 1.5 |
| | asymptotic_bonf | 1.5 | 1.3 | 2.0 | 1.9 | 1.1 | 1.1 | 1.8 | 1.3 | 1.5 |
| | permutation_bonf | 0.9 | 1.2 | 2.1 | 1.6 | 0.9 | 1.0 | 1.2 | 0.8 | 1.2 |
| Weib late continuous | asymptotic | 2.2 | 2.2 | 3.3 | 2.1 | 2.1 | 2.8 | 3.1 | 2.2 | 4.1 |
| | asymptotic_bonf | 2.2 | 2.2 | 3.2 | 2.1 | 2.1 | 2.8 | 3.1 | 2.2 | 4.0 |
| | permutation_bonf | 1.8 | 2.3 | 2.9 | 1.6 | 1.9 | 2.0 | 2.5 | 2.2 | 3.2 |
| Weib late discrete | asymptotic | 2.1 | 2.6 | 3.6 | 1.8 | 2.8 | 3.2 | 3.5 | 2.5 | 4.1 |
| | asymptotic_bonf | 2.1 | 2.5 | 3.4 | 1.7 | 2.8 | 3.0 | 3.4 | 2.3 | 3.8 |
| | permutation_bonf | 2.0 | 2.5 | 3.2 | 1.9 | 2.1 | 2.6 | 3.1 | 2.1 | 3.1 |
| Weib prop continuous | asymptotic | 2.5 | 1.8 | 3.1 | 2.1 | 2.3 | 2.7 | 2.9 | 2.2 | 3.8 |
| | asymptotic_bonf | 2.5 | 1.8 | 3.1 | 2.1 | 2.2 | 2.3 | 2.9 | 2.2 | 3.8 |
| | permutation_bonf | 1.9 | 2.2 | 3.0 | 1.9 | 1.9 | 2.0 | 2.6 | 2.2 | 3.5 |
| Weib prop discrete | asymptotic | 2.3 | 2.4 | 4.1 | 1.9 | 2.6 | 3.1 | 3.3 | 2.3 | 3.7 |
| | asymptotic_bonf | 2.1 | 2.3 | 4.0 | 1.8 | 2.5 | 3.1 | 3.2 | 2.2 | 3.6 |
| | permutation_bonf | 1.9 | 2.3 | 3.5 | 1.7 | 2.1 | 2.5 | 3.2 | 2.0 | 3.3 |
| Weib scale continuous | asymptotic | 1.6 | 1.8 | 3.1 | 2.0 | 2.3 | 2.3 | 2.9 | 2.4 | 2.6 |
| | asymptotic_bonf | 1.5 | 1.8 | 3.0 | 1.9 | 2.2 | 2.1 | 2.8 | 2.0 | 2.4 |
| | permutation_bonf | 1.4 | 1.7 | 2.4 | 1.5 | 2.0 | 1.9 | 2.3 | 1.8 | 2.1 |
| Weib scale discrete | asymptotic | 1.6 | 2.1 | 3.9 | 1.7 | 2.6 | 2.7 | 2.9 | 2.0 | 3.1 |
| | asymptotic_bonf | 1.5 | 1.8 | 3.6 | 1.7 | 2.6 | 2.7 | 2.9 | 1.9 | 3.1 |
| | permutation_bonf | 1.6 | 1.7 | 3.1 | 1.6 | 2.0 | 2.3 | 2.4 | 1.9 | 2.4 |
| Weib shape continuous | asymptotic | 1.2 | 1.6 | 2.9 | 1.1 | 1.8 | 1.9 | 2.6 | 1.8 | 2.9 |
| | asymptotic_bonf | 1.2 | 1.6 | 2.9 | 1.1 | 1.8 | 1.9 | 2.4 | 1.5 | 2.8 |
| | permutation_bonf | 1.2 | 1.2 | 2.4 | 1.2 | 1.6 | 1.6 | 2.1 | 1.6 | 2.3 |
| Weib shape discrete | asymptotic | 1.5 | 1.9 | 3.5 | 1.3 | 2.2 | 2.1 | 2.8 | 1.9 | 2.9 |
| | asymptotic_bonf | 1.4 | 1.9 | 3.3 | 1.3 | 2.2 | 2.0 | 2.8 | 1.8 | 2.9 |
| | permutation_bonf | 1.1 | 1.9 | 2.8 | 1.3 | 2.1 | 1.8 | 2.3 | 1.8 | 2.2 |

Table S46: Rejection rates in percent for the 2-by-2 design with $\delta = 1.5$ and unbalanced large sample sizes under unequal, high censoring.

| Distribution | Method | $\mathcal{H}_{0,4}$ | $\mathcal{H}_{0,5}$ | $\mathcal{H}_{0,6}$ | $\mathcal{H}_{0,1}$ | $\mathcal{H}_{0,2}$ | $\mathcal{H}_{0,3}$ | $\mathcal{H}_{0,7}$ | $\mathcal{H}_{0,8}$ | $\mathcal{H}_{0,9}$ |
|---|---|---|---|---|---|---|---|---|---|---|
| exp early continuous | asymptotic | 18.8 | 10.8 | 29.9 | 17.8 | 9.9 | 31.3 | 18.3 | 10.7 | 29.0 |
|  | asymptotic_bonf | 18.0 | 10.2 | 28.7 | 17.1 | 9.1 | 29.2 | 17.3 | 9.5 | 27.8 |
|  | permutation_bonf | 17.7 | 9.6 | 29.8 | 16.7 | 9.1 | 27.9 | 17.1 | 9.1 | 27.0 |
| exp early discrete | asymptotic | 20.6 | 12.1 | 34.5 | 18.7 | 11.7 | 32.2 | 20.7 | 12.0 | 30.9 |
|  | asymptotic_bonf | 19.3 | 11.6 | 32.9 | 17.5 | 11.1 | 30.9 | 19.5 | 11.1 | 29.6 |
|  | permutation_bonf | 19.4 | 10.9 | 32.8 | 17.0 | 10.2 | 30.3 | 18.8 | 10.4 | 29.3 |
| exp late continuous | asymptotic | 18.9 | 11.7 | 32.4 | 18.8 | 11.2 | 31.5 | 18.9 | 10.8 | 30.8 |
|  | asymptotic_bonf | 18.3 | 10.9 | 31.2 | 17.5 | 10.6 | 30.1 | 17.9 | 9.6 | 29.5 |
|  | permutation_bonf | 18.5 | 10.3 | 30.7 | 16.6 | 10.2 | 29.2 | 18.2 | 9.0 | 30.0 |
| exp late discrete | asymptotic | 21.4 | 12.4 | 37.4 | 20.6 | 12.8 | 34.4 | 21.1 | 12.8 | 32.6 |
|  | asymptotic_bonf | 19.7 | 11.6 | 35.8 | 19.2 | 12.1 | 32.9 | 19.5 | 11.6 | 30.5 |
|  | permutation_bonf | 19.7 | 10.5 | 35.1 | 18.5 | 11.3 | 32.9 | 20.2 | 11.4 | 30.5 |
| exp prop continuous | asymptotic | 19.5 | 12.4 | 26.1 | 16.8 | 10.2 | 29.8 | 19.1 | 12.8 | 28.3 |
|  | asymptotic_bonf | 18.4 | 11.2 | 24.7 | 15.3 | 9.1 | 28.5 | 17.9 | 12.0 | 26.8 |
|  | permutation_bonf | 19.0 | 10.1 | 23.8 | 15.2 | 8.5 | 28.0 | 17.1 | 11.0 | 26.0 |
| exp prop discrete | asymptotic | 21.4 | 15.1 | 30.4 | 18.0 | 11.0 | 34.1 | 20.3 | 14.1 | 30.1 |
|  | asymptotic_bonf | 20.6 | 13.8 | 28.6 | 16.8 | 10.3 | 32.1 | 19.5 | 13.9 | 28.9 |
|  | permutation_bonf | 19.9 | 13.2 | 27.6 | 17.5 | 10.0 | 31.2 | 19.4 | 12.6 | 28.5 |
| logn continuous | asymptotic | 49.8 | 29.3 | 61.5 | 48.5 | 32.8 | 63.7 | 47.1 | 32.3 | 62.1 |
|  | asymptotic_bonf | 48.2 | 27.0 | 59.7 | 46.5 | 30.8 | 61.5 | 45.0 | 30.2 | 60.3 |
|  | permutation_bonf | 46.3 | 26.6 | 58.1 | 45.4 | 29.4 | 60.3 | 44.5 | 29.3 | 60.0 |
| logn discrete | asymptotic | 58.5 | 35.9 | 69.4 | 55.1 | 39.5 | 70.5 | 53.0 | 36.1 | 70.0 |
|  | asymptotic_bonf | 56.1 | 34.2 | 67.2 | 54.0 | 37.3 | 69.4 | 51.6 | 34.7 | 68.7 |
|  | permutation_bonf | 54.7 | 31.3 | 66.5 | 51.5 | 35.4 | 68.4 | 50.4 | 32.9 | 67.0 |
| pwExp continuous | asymptotic | 16.4 | 9.8 | 29.9 | 17.4 | 9.2 | 31.5 | 17.3 | 10.7 | 30.1 |
|  | asymptotic_bonf | 15.9 | 9.1 | 28.1 | 16.2 | 7.9 | 30.4 | 16.0 | 9.9 | 28.1 |
|  | permutation_bonf | 16.0 | 9.2 | 29.1 | 17.0 | 9.3 | 29.6 | 16.0 | 9.2 | 27.5 |
| pwExp discrete | asymptotic | 20.2 | 10.8 | 33.7 | 18.8 | 11.8 | 32.7 | 18.5 | 11.0 | 32.1 |
|  | asymptotic_bonf | 19.1 | 10.2 | 32.5 | 17.3 | 10.0 | 31.5 | 17.6 | 10.1 | 30.4 |
|  | permutation_bonf | 17.8 | 9.6 | 31.4 | 16.9 | 9.9 | 31.2 | 17.5 | 9.5 | 30.5 |
| Weib late continuous | asymptotic | 47.8 | 31.5 | 65.3 | 48.6 | 29.4 | 67.2 | 46.5 | 28.5 | 65.8 |
|  | asymptotic_bonf | 46.1 | 29.3 | 63.5 | 47.5 | 27.8 | 65.6 | 44.7 | 27.5 | 64.0 |
|  | permutation_bonf | 44.9 | 26.3 | 62.1 | 45.3 | 26.0 | 64.6 | 42.6 | 27.0 | 63.0 |
| Weib late discrete | asymptotic | 53.8 | 34.5 | 72.2 | 52.6 | 33.5 | 71.9 | 53.8 | 33.2 | 70.4 |
|  | asymptotic_bonf | 51.3 | 32.7 | 71.3 | 50.9 | 32.0 | 70.5 | 52.0 | 31.2 | 68.4 |
|  | permutation_bonf | 49.8 | 31.2 | 69.2 | 48.6 | 30.3 | 69.8 | 50.8 | 30.7 | 68.2 |
| Weib prop continuous | asymptotic | 46.2 | 28.2 | 63.9 | 46.0 | 28.2 | 65.2 | 43.6 | 28.6 | 63.3 |
|  | asymptotic_bonf | 43.8 | 26.9 | 61.1 | 44.5 | 25.8 | 63.1 | 41.5 | 27.1 | 61.2 |
|  | permutation_bonf | 42.7 | 25.0 | 59.4 | 43.7 | 25.3 | 60.9 | 41.8 | 26.7 | 61.0 |
| Weib prop discrete | asymptotic | 53.8 | 33.6 | 72.6 | 53.1 | 34.7 | 72.2 | 53.9 | 33.9 | 71.1 |
|  | asymptotic_bonf | 51.6 | 31.6 | 71.1 | 51.8 | 33.1 | 71.2 | 50.9 | 32.2 | 69.6 |
|  | permutation_bonf | 50.1 | 30.6 | 69.2 | 49.4 | 30.7 | 70.8 | 48.8 | 30.6 | 68.9 |
| Weib scale continuous | asymptotic | 40.2 | 25.9 | 57.2 | 41.9 | 25.0 | 60.0 | 39.8 | 25.8 | 56.9 |
|  | asymptotic_bonf | 37.2 | 23.6 | 55.2 | 40.3 | 23.3 | 58.2 | 37.4 | 24.0 | 54.8 |
|  | permutation_bonf | 36.7 | 22.6 | 54.1 | 37.8 | 21.2 | 56.3 | 36.7 | 22.4 | 54.7 |
| Weib scale discrete | asymptotic | 46.7 | 29.9 | 65.3 | 48.2 | 30.4 | 66.5 | 46.5 | 29.9 | 64.1 |
|  | asymptotic_bonf | 44.6 | 27.5 | 63.4 | 45.9 | 28.6 | 64.6 | 43.6 | 28.6 | 61.7 |
|  | permutation_bonf | 43.4 | 25.3 | 61.5 | 43.7 | 27.0 | 62.4 | 43.4 | 27.3 | 61.3 |
| Weib shape continuous | asymptotic | 35.1 | 22.3 | 50.5 | 36.0 | 21.5 | 53.2 | 33.6 | 22.9 | 50.7 |
|  | asymptotic_bonf | 33.2 | 20.5 | 48.0 | 33.7 | 19.8 | 50.8 | 31.3 | 20.5 | 49.0 |
|  | permutation_bonf | 32.0 | 20.0 | 47.2 | 31.8 | 18.9 | 50.4 | 30.4 | 19.3 | 47.9 |
| Weib shape discrete | asymptotic | 41.0 | 26.0 | 58.3 | 42.4 | 26.4 | 59.3 | 41.4 | 25.6 | 58.0 |
|  | asymptotic_bonf | 38.6 | 23.7 | 55.9 | 40.0 | 25.1 | 57.2 | 38.7 | 24.4 | 55.3 |
|  | permutation_bonf | 37.3 | 22.7 | 54.7 | 37.3 | 23.5 | 56.6 | 37.3 | 23.9 | 55.4 |

Table S47: Rejection rates in percent for the 2-by-2 design with $\delta = 1.5$ and unbalanced medium sample sizes under unequal, high censoring.

| Distribution | Method | $\mathcal{H}_{0,4}$ | $\mathcal{H}_{0,5}$ | $\mathcal{H}_{0,6}$ | $\mathcal{H}_{0,1}$ | $\mathcal{H}_{0,2}$ | $\mathcal{H}_{0,3}$ | $\mathcal{H}_{0,7}$ | $\mathcal{H}_{0,8}$ | $\mathcal{H}_{0,9}$ |
|---|---|---|---|---|---|---|---|---|---|---|
| exp early continuous | asymptotic | 2.7 | 1.7 | 3.8 | 2.6 | 1.7 | 3.6 | 3.0 | 1.3 | 3.9 |
| | asymptotic_bonf | 2.2 | 1.4 | 3.5 | 2.4 | 1.2 | 3.2 | 2.7 | 1.3 | 3.7 |
| | permutation_bonf | 1.6 | 1.2 | 3.1 | 1.8 | 0.9 | 2.3 | 2.0 | 1.3 | 3.1 |
| exp early discrete | asymptotic | 3.1 | 1.7 | 4.5 | 2.3 | 1.4 | 4.2 | 2.7 | 1.4 | 4.1 |
| | asymptotic_bonf | 2.9 | 1.4 | 3.8 | 2.2 | 1.4 | 4.0 | 2.6 | 1.3 | 3.7 |
| | permutation_bonf | 2.0 | 1.0 | 3.2 | 1.4 | 0.6 | 3.1 | 1.9 | 1.0 | 3.5 |
| exp late continuous | asymptotic | 2.2 | 2.3 | 4.0 | 2.6 | 1.5 | 3.5 | 3.2 | 1.8 | 3.7 |
| | asymptotic_bonf | 1.8 | 2.0 | 3.7 | 2.5 | 1.3 | 3.4 | 3.0 | 1.6 | 3.5 |
| | permutation_bonf | 1.0 | 1.6 | 3.1 | 1.7 | 1.1 | 2.5 | 2.8 | 0.8 | 2.9 |
| exp late discrete | asymptotic | 3.0 | 2.0 | 3.8 | 2.2 | 1.7 | 4.7 | 3.3 | 1.5 | 4.3 |
| | asymptotic_bonf | 2.9 | 1.8 | 3.7 | 2.1 | 1.4 | 3.6 | 2.8 | 1.2 | 3.9 |
| | permutation_bonf | 2.0 | 1.1 | 3.3 | 2.0 | 0.8 | 3.4 | 2.1 | 1.1 | 4.1 |
| exp prop continuous | asymptotic | 2.7 | 2.3 | 3.0 | 2.2 | 1.5 | 3.4 | 2.5 | 2.0 | 3.3 |
| | asymptotic_bonf | 2.5 | 2.0 | 2.7 | 1.7 | 1.5 | 3.3 | 2.2 | 1.8 | 3.2 |
| | permutation_bonf | 2.2 | 1.3 | 2.3 | 1.4 | 0.6 | 2.8 | 1.7 | 1.0 | 2.8 |
| exp prop discrete | asymptotic | 3.2 | 2.3 | 3.9 | 2.3 | 1.4 | 3.8 | 2.9 | 2.8 | 4.1 |
| | asymptotic_bonf | 2.8 | 2.2 | 3.7 | 2.1 | 1.3 | 3.4 | 2.5 | 2.4 | 3.7 |
| | permutation_bonf | 2.4 | 1.1 | 2.8 | 1.9 | 0.7 | 2.9 | 2.1 | 1.8 | 3.0 |
| logn continuous | asymptotic | 5.2 | 3.4 | 8.2 | 3.7 | 3.4 | 7.2 | 5.2 | 3.8 | 8.9 |
| | asymptotic_bonf | 4.8 | 3.2 | 7.3 | 3.5 | 3.1 | 6.7 | 4.1 | 3.6 | 8.4 |
| | permutation_bonf | 3.8 | 1.8 | 5.0 | 2.5 | 1.2 | 4.8 | 3.6 | 2.8 | 6.7 |
| logn discrete | asymptotic | 6.4 | 3.8 | 9.0 | 5.5 | 3.4 | 8.5 | 5.6 | 4.4 | 10.4 |
| | asymptotic_bonf | 6.1 | 3.6 | 7.9 | 4.4 | 3.0 | 7.6 | 4.4 | 3.7 | 9.2 |
| | permutation_bonf | 4.1 | 1.8 | 6.0 | 2.7 | 1.3 | 5.9 | 3.6 | 2.4 | 7.3 |
| pwExp continuous | asymptotic | 1.4 | 1.9 | 4.0 | 2.6 | 1.2 | 3.3 | 2.2 | 2.0 | 3.5 |
| | asymptotic_bonf | 1.3 | 1.8 | 3.9 | 2.3 | 1.0 | 3.0 | 1.9 | 1.7 | 3.1 |
| | permutation_bonf | 1.0 | 1.5 | 3.3 | 1.8 | 0.7 | 2.1 | 1.9 | 1.2 | 2.8 |
| pwExp discrete | asymptotic | 2.1 | 1.9 | 4.1 | 2.4 | 1.4 | 3.7 | 2.9 | 2.2 | 3.8 |
| | asymptotic_bonf | 2.0 | 1.9 | 3.9 | 2.2 | 1.3 | 3.2 | 2.6 | 2.0 | 3.3 |
| | permutation_bonf | 1.3 | 1.3 | 3.6 | 1.7 | 0.8 | 2.8 | 2.0 | 1.0 | 3.1 |
| Weib late continuous | asymptotic | 5.6 | 3.1 | 7.8 | 4.3 | 2.5 | 8.5 | 6.3 | 3.6 | 8.0 |
| | asymptotic_bonf | 4.8 | 2.6 | 6.9 | 3.6 | 2.3 | 7.9 | 5.5 | 3.2 | 7.7 |
| | permutation_bonf | 3.7 | 0.9 | 5.1 | 2.2 | 0.7 | 6.4 | 4.3 | 1.9 | 6.4 |
| Weib late discrete | asymptotic | 6.6 | 3.8 | 9.7 | 4.6 | 3.3 | 10.7 | 6.5 | 3.5 | 9.9 |
| | asymptotic_bonf | 5.9 | 2.9 | 8.5 | 3.9 | 3.2 | 9.2 | 6.0 | 2.9 | 8.8 |
| | permutation_bonf | 3.8 | 1.7 | 6.0 | 2.6 | 1.3 | 7.1 | 4.2 | 1.9 | 7.8 |
| Weib prop continuous | asymptotic | 4.7 | 2.7 | 6.9 | 4.1 | 2.5 | 8.0 | 5.4 | 3.0 | 8.2 |
| | asymptotic_bonf | 4.3 | 2.0 | 6.4 | 3.5 | 2.2 | 7.1 | 5.0 | 2.9 | 7.1 |
| | permutation_bonf | 3.0 | 0.8 | 5.0 | 2.2 | 0.7 | 5.6 | 4.1 | 1.8 | 6.0 |
| Weib prop discrete | asymptotic | 5.9 | 2.9 | 9.3 | 4.5 | 3.3 | 8.9 | 6.5 | 3.4 | 8.6 |
| | asymptotic_bonf | 5.3 | 2.5 | 8.9 | 4.0 | 2.9 | 8.6 | 5.6 | 2.8 | 8.3 |
| | permutation_bonf | 4.0 | 0.8 | 6.2 | 2.8 | 0.7 | 6.0 | 4.0 | 1.6 | 7.0 |
| Weib scale continuous | asymptotic | 3.7 | 2.2 | 6.0 | 3.4 | 2.5 | 5.7 | 4.1 | 2.5 | 5.7 |
| | asymptotic_bonf | 3.2 | 1.9 | 5.4 | 2.9 | 2.1 | 4.7 | 3.5 | 2.2 | 5.1 |
| | permutation_bonf | 2.7 | 1.1 | 3.8 | 1.8 | 0.6 | 3.1 | 2.8 | 1.9 | 4.3 |
| Weib scale discrete | asymptotic | 5.7 | 2.6 | 6.4 | 4.1 | 2.3 | 6.8 | 5.0 | 2.6 | 6.8 |
| | asymptotic_bonf | 4.9 | 2.3 | 6.0 | 3.6 | 2.0 | 5.9 | 4.7 | 2.4 | 5.9 |
| | permutation_bonf | 3.3 | 1.0 | 3.8 | 1.9 | 0.6 | 3.7 | 2.9 | 1.8 | 5.2 |
| Weib shape continuous | asymptotic | 3.8 | 1.9 | 4.7 | 2.8 | 1.8 | 4.2 | 3.3 | 1.4 | 4.5 |
| | asymptotic_bonf | 3.3 | 1.4 | 4.5 | 2.0 | 1.5 | 3.8 | 2.7 | 1.4 | 3.7 |
| | permutation_bonf | 2.1 | 0.7 | 3.3 | 1.6 | 0.6 | 2.7 | 2.2 | 1.0 | 3.0 |
| Weib shape discrete | asymptotic | 3.7 | 2.1 | 4.9 | 3.1 | 2.4 | 5.0 | 4.4 | 1.9 | 4.9 |
| | asymptotic_bonf | 3.4 | 1.8 | 4.1 | 2.6 | 1.8 | 4.8 | 4.0 | 1.5 | 4.2 |
| | permutation_bonf | 2.3 | 1.0 | 2.3 | 1.8 | 0.5 | 2.6 | 2.7 | 1.2 | 3.7 |

Table S48: Rejection rates in percent for the 2-by-2 design with $\delta = 1.5$ and unbalanced small sample sizes under unequal, high censoring.

| Distribution | Method | $\mathcal{H}_{0,4}$ | $\mathcal{H}_{0,5}$ | $\mathcal{H}_{0,6}$ | $\mathcal{H}_{0,1}$ | $\mathcal{H}_{0,2}$ | $\mathcal{H}_{0,3}$ | $\mathcal{H}_{0,7}$ | $\mathcal{H}_{0,8}$ | $\mathcal{H}_{0,9}$ |
|---|---|---|---|---|---|---|---|---|---|---|
| exp early continuous | asymptotic | 1.4 | 1.1 | 1.9 | 1.0 | 1.7 | 1.0 | 1.1 | 1.2 | 2.1 |
| | asymptotic_bonf | 1.3 | 1.1 | 1.7 | 1.0 | 1.4 | 1.0 | 1.0 | 1.1 | 2.0 |
| | permutation_bonf | 0.1 | 0.2 | 0.5 | 0.1 | 0.3 | 0.0 | 0.2 | 0.4 | 0.6 |
| exp early discrete | asymptotic | 1.7 | 1.3 | 1.6 | 1.0 | 1.7 | 1.2 | 1.2 | 1.4 | 1.8 |
| | asymptotic_bonf | 1.3 | 1.1 | 1.2 | 0.9 | 1.7 | 1.0 | 1.2 | 1.2 | 1.8 |
| | permutation_bonf | 0.3 | 0.2 | 0.2 | 0.2 | 0.4 | 0.2 | 0.3 | 0.5 | 0.6 |
| exp late continuous | asymptotic | 1.9 | 1.0 | 1.7 | 1.2 | 1.6 | 1.1 | 0.8 | 1.2 | 1.8 |
| | asymptotic_bonf | 1.6 | 0.9 | 1.4 | 1.2 | 1.3 | 0.8 | 0.8 | 1.2 | 1.6 |
| | permutation_bonf | 0.1 | 0.2 | 0.3 | 0.3 | 0.1 | 0.1 | 0.2 | 0.5 | 0.4 |
| exp late discrete | asymptotic | 1.8 | 1.0 | 1.7 | 1.4 | 1.3 | 1.2 | 1.6 | 1.4 | 2.3 |
| | asymptotic_bonf | 1.6 | 0.8 | 1.5 | 1.3 | 1.2 | 1.1 | 1.5 | 1.3 | 1.9 |
| | permutation_bonf | 0.2 | 0.0 | 0.3 | 0.3 | 0.3 | 0.2 | 0.3 | 0.5 | 0.3 |
| exp prop continuous | asymptotic | 1.2 | 1.5 | 2.3 | 1.4 | 1.4 | 1.7 | 1.4 | 1.0 | 1.6 |
| | asymptotic_bonf | 1.2 | 1.4 | 2.1 | 1.2 | 1.2 | 1.6 | 1.2 | 1.0 | 1.5 |
| | permutation_bonf | 0.5 | 0.4 | 0.4 | 0.3 | 0.0 | 0.4 | 0.3 | 0.5 | 0.8 |
| exp prop discrete | asymptotic | 1.7 | 1.6 | 1.6 | 1.4 | 1.4 | 1.4 | 1.6 | 1.1 | 1.6 |
| | asymptotic_bonf | 1.6 | 1.4 | 1.5 | 1.3 | 1.3 | 1.4 | 1.5 | 1.1 | 1.4 |
| | permutation_bonf | 0.5 | 0.3 | 0.7 | 0.4 | 0.0 | 0.6 | 0.5 | 0.5 | 0.4 |
| logn continuous | asymptotic | 1.6 | 1.7 | 1.7 | 2.3 | 2.6 | 2.0 | 1.8 | 1.3 | 2.8 |
| | asymptotic_bonf | 1.3 | 1.4 | 1.4 | 2.1 | 2.3 | 2.0 | 1.7 | 1.0 | 2.5 |
| | permutation_bonf | 0.5 | 0.5 | 0.4 | 0.3 | 0.8 | 0.8 | 0.6 | 0.5 | 0.6 |
| logn discrete | asymptotic | 1.9 | 2.1 | 2.0 | 2.6 | 2.3 | 2.4 | 1.9 | 1.4 | 3.4 |
| | asymptotic_bonf | 1.8 | 1.9 | 1.7 | 2.0 | 2.0 | 2.1 | 1.7 | 1.4 | 3.0 |
| | permutation_bonf | 0.5 | 0.5 | 0.5 | 0.3 | 0.6 | 0.8 | 0.7 | 0.7 | 0.9 |
| pwExp continuous | asymptotic | 1.0 | 1.1 | 1.8 | 1.2 | 0.9 | 1.3 | 0.7 | 0.6 | 2.1 |
| | asymptotic_bonf | 0.9 | 1.0 | 1.7 | 1.2 | 0.7 | 1.1 | 0.7 | 0.6 | 1.8 |
| | permutation_bonf | 0.2 | 0.3 | 0.3 | 0.4 | 0.1 | 0.2 | 0.2 | 0.3 | 0.3 |
| pwExp discrete | asymptotic | 1.2 | 1.3 | 2.0 | 1.0 | 1.1 | 1.4 | 1.2 | 0.9 | 2.1 |
| | asymptotic_bonf | 0.8 | 1.3 | 1.6 | 1.0 | 1.0 | 1.3 | 1.2 | 0.8 | 1.9 |
| | permutation_bonf | 0.3 | 0.3 | 0.4 | 0.2 | 0.2 | 0.2 | 0.5 | 0.5 | 0.3 |
| Weib late continuous | asymptotic | 2.3 | 1.6 | 2.6 | 2.2 | 2.4 | 2.8 | 2.2 | 1.0 | 2.5 |
| | asymptotic_bonf | 2.0 | 1.4 | 2.2 | 1.9 | 2.3 | 2.6 | 1.9 | 1.0 | 2.2 |
| | permutation_bonf | 0.2 | 0.2 | 0.7 | 0.4 | 0.4 | 0.8 | 0.6 | 0.4 | 0.7 |
| Weib late discrete | asymptotic | 2.2 | 1.7 | 3.3 | 2.3 | 2.3 | 3.0 | 2.7 | 1.4 | 2.3 |
| | asymptotic_bonf | 2.1 | 1.5 | 2.8 | 2.0 | 1.9 | 3.0 | 2.4 | 1.4 | 2.1 |
| | permutation_bonf | 0.1 | 0.3 | 0.6 | 0.4 | 0.4 | 0.8 | 0.6 | 0.9 | 0.7 |
| Weib prop continuous | asymptotic | 2.0 | 1.6 | 2.4 | 2.0 | 2.2 | 2.5 | 2.0 | 1.2 | 1.9 |
| | asymptotic_bonf | 1.5 | 1.5 | 2.1 | 1.7 | 2.2 | 2.1 | 1.7 | 1.1 | 1.8 |
| | permutation_bonf | 0.2 | 0.3 | 0.5 | 0.3 | 0.5 | 0.7 | 0.4 | 0.3 | 0.8 |
| Weib prop discrete | asymptotic | 1.8 | 1.7 | 3.1 | 2.2 | 2.2 | 2.5 | 2.4 | 0.8 | 2.1 |
| | asymptotic_bonf | 1.6 | 1.6 | 2.5 | 1.9 | 1.7 | 2.3 | 2.2 | 0.7 | 1.8 |
| | permutation_bonf | 0.2 | 0.4 | 0.7 | 0.4 | 0.5 | 0.8 | 0.4 | 0.3 | 0.9 |
| Weib scale continuous | asymptotic | 1.7 | 2.3 | 2.4 | 2.0 | 2.5 | 2.5 | 1.4 | 1.6 | 1.6 |
| | asymptotic_bonf | 1.2 | 2.1 | 2.2 | 1.8 | 2.3 | 2.2 | 1.1 | 1.4 | 1.4 |
| | permutation_bonf | 0.2 | 0.4 | 0.3 | 0.3 | 0.5 | 0.7 | 0.5 | 0.5 | 0.7 |
| Weib scale discrete | asymptotic | 1.9 | 2.0 | 2.2 | 2.4 | 1.8 | 2.3 | 2.0 | 1.3 | 1.1 |
| | asymptotic_bonf | 1.8 | 1.8 | 1.9 | 1.7 | 1.4 | 2.0 | 1.7 | 1.1 | 1.1 |
| | permutation_bonf | 0.3 | 0.6 | 0.4 | 0.5 | 0.3 | 0.5 | 0.5 | 0.5 | 0.4 |
| Weib shape continuous | asymptotic | 2.0 | 1.6 | 2.0 | 1.6 | 3.0 | 2.5 | 1.1 | 0.8 | 1.1 |
| | asymptotic_bonf | 1.7 | 1.5 | 1.8 | 1.4 | 2.5 | 1.9 | 0.9 | 0.6 | 0.9 |
| | permutation_bonf | 0.4 | 0.3 | 0.5 | 0.0 | 0.5 | 0.8 | 0.4 | 0.3 | 0.5 |
| Weib shape discrete | asymptotic | 2.1 | 2.0 | 2.6 | 2.0 | 1.8 | 2.4 | 1.5 | 1.0 | 1.0 |
| | asymptotic_bonf | 2.1 | 1.5 | 2.1 | 1.7 | 1.5 | 2.2 | 1.2 | 0.7 | 0.9 |
| | permutation_bonf | 0.3 | 0.4 | 0.6 | 0.1 | 0.5 | 0.5 | 0.4 | 0.4 | 0.4 |

Table S49: Rejection rates in percent for the 2-by-2 design with $\delta = 1.5$ and balanced large sample sizes under unequal, low censoring.

| Distribution | Method | $\mathcal{H}_{0,4}$ | $\mathcal{H}_{0,5}$ | $\mathcal{H}_{0,6}$ | $\mathcal{H}_{0,1}$ | $\mathcal{H}_{0,2}$ | $\mathcal{H}_{0,3}$ | $\mathcal{H}_{0,7}$ | $\mathcal{H}_{0,8}$ | $\mathcal{H}_{0,9}$ |
|---|---|---|---|---|---|---|---|---|---|---|
| exp early continuous | asymptotic | 49.5 | 31.2 | 67.4 | 47.7 | 32.9 | 68.2 | 47.5 | 30.7 | 68.3 |
| | asymptotic_bonf | 48.5 | 30.3 | 66.7 | 46.9 | 32.1 | 67.6 | 46.3 | 29.8 | 67.3 |
| | permutation_bonf | 48.3 | 29.2 | 66.4 | 46.4 | 31.0 | 67.9 | 46.5 | 29.8 | 66.4 |
| exp early discrete | asymptotic | 56.5 | 36.4 | 73.5 | 54.7 | 36.0 | 75.1 | 54.1 | 35.7 | 75.7 |
| | asymptotic_bonf | 56.0 | 35.5 | 73.4 | 53.5 | 35.5 | 74.7 | 53.6 | 35.0 | 74.9 |
| | permutation_bonf | 54.7 | 34.6 | 71.9 | 53.1 | 34.8 | 74.3 | 53.1 | 34.6 | 74.5 |
| exp late continuous | asymptotic | 51.9 | 33.1 | 70.2 | 50.2 | 34.1 | 71.1 | 48.8 | 31.8 | 70.9 |
| | asymptotic_bonf | 51.3 | 32.4 | 69.7 | 49.2 | 33.4 | 70.7 | 48.6 | 31.5 | 69.9 |
| | permutation_bonf | 51.1 | 31.1 | 68.6 | 47.9 | 32.2 | 69.1 | 48.2 | 30.8 | 70.8 |
| exp late discrete | asymptotic | 59.0 | 39.6 | 76.2 | 56.2 | 38.2 | 77.9 | 55.6 | 37.9 | 78.1 |
| | asymptotic_bonf | 58.5 | 38.9 | 75.7 | 55.5 | 37.2 | 77.2 | 54.9 | 36.4 | 77.6 |
| | permutation_bonf | 57.2 | 37.5 | 74.9 | 54.6 | 36.0 | 76.1 | 55.0 | 35.3 | 76.8 |
| exp prop continuous | asymptotic | 50.1 | 34.1 | 68.6 | 49.9 | 33.2 | 68.6 | 48.5 | 34.3 | 69.2 |
| | asymptotic_bonf | 49.1 | 32.9 | 67.6 | 49.3 | 32.9 | 68.0 | 47.6 | 33.4 | 68.1 |
| | permutation_bonf | 50.1 | 32.7 | 67.9 | 48.7 | 32.7 | 68.0 | 48.4 | 32.9 | 68.2 |
| exp prop discrete | asymptotic | 55.4 | 40.3 | 74.8 | 56.0 | 37.0 | 75.1 | 56.4 | 38.7 | 76.7 |
| | asymptotic_bonf | 54.8 | 39.4 | 74.2 | 55.4 | 36.5 | 74.8 | 55.2 | 37.8 | 76.1 |
| | permutation_bonf | 55.0 | 38.1 | 74.2 | 55.4 | 36.5 | 74.1 | 55.6 | 38.3 | 75.2 |
| logn continuous | asymptotic | 94.3 | 81.0 | 98.9 | 93.3 | 81.6 | 99.7 | 93.9 | 83.9 | 98.7 |
| | asymptotic_bonf | 94.1 | 80.7 | 98.8 | 93.3 | 81.4 | 99.7 | 93.9 | 83.6 | 98.5 |
| | permutation_bonf | 94.5 | 81.3 | 98.5 | 93.6 | 80.8 | 99.5 | 93.3 | 82.7 | 98.3 |
| logn discrete | asymptotic | 97.2 | 88.0 | 99.9 | 97.0 | 89.1 | 99.8 | 97.0 | 90.7 | 99.6 |
| | asymptotic_bonf | 97.1 | 87.5 | 99.9 | 97.0 | 89.0 | 99.8 | 97.0 | 90.5 | 99.6 |
| | permutation_bonf | 96.9 | 87.9 | 99.9 | 96.6 | 88.4 | 99.8 | 96.8 | 89.6 | 99.5 |
| pwExp continuous | asymptotic | 48.4 | 30.9 | 66.6 | 47.0 | 32.1 | 68.3 | 45.9 | 29.9 | 67.7 |
| | asymptotic_bonf | 47.6 | 30.3 | 66.2 | 46.3 | 31.7 | 67.7 | 45.6 | 29.4 | 67.1 |
| | permutation_bonf | 47.7 | 30.5 | 66.5 | 45.7 | 32.5 | 67.7 | 45.6 | 29.7 | 66.4 |
| pwExp discrete | asymptotic | 55.0 | 37.0 | 73.5 | 53.9 | 35.8 | 74.7 | 52.7 | 34.6 | 74.9 |
| | asymptotic_bonf | 54.1 | 36.4 | 72.9 | 52.9 | 35.4 | 74.2 | 52.4 | 33.9 | 74.4 |
| | permutation_bonf | 54.6 | 37.6 | 72.0 | 53.7 | 35.8 | 72.9 | 52.5 | 33.6 | 73.3 |
| Weib late continuous | asymptotic | 94.4 | 78.5 | 98.8 | 94.1 | 80.4 | 98.8 | 94.3 | 82.2 | 99.1 |
| | asymptotic_bonf | 94.4 | 77.7 | 98.8 | 94.0 | 80.0 | 98.7 | 94.3 | 82.0 | 99.0 |
| | permutation_bonf | 94.8 | 78.6 | 98.5 | 94.6 | 79.9 | 98.5 | 94.2 | 81.3 | 98.7 |
| Weib late discrete | asymptotic | 97.4 | 86.8 | 99.7 | 97.8 | 88.6 | 99.9 | 97.3 | 88.8 | 99.6 |
| | asymptotic_bonf | 97.3 | 86.6 | 99.7 | 97.8 | 88.5 | 99.9 | 97.1 | 88.5 | 99.5 |
| | permutation_bonf | 97.4 | 87.2 | 99.7 | 97.8 | 88.7 | 99.8 | 96.4 | 88.7 | 99.4 |
| Weib prop continuous | asymptotic | 94.1 | 77.8 | 98.6 | 93.6 | 79.4 | 98.7 | 93.5 | 81.0 | 98.8 |
| | asymptotic_bonf | 94.0 | 77.2 | 98.6 | 93.6 | 78.9 | 98.7 | 93.4 | 80.3 | 98.8 |
| | permutation_bonf | 94.0 | 77.4 | 98.4 | 94.0 | 78.5 | 98.6 | 93.3 | 80.7 | 98.7 |
| Weib prop discrete | asymptotic | 97.3 | 85.8 | 99.6 | 97.2 | 87.7 | 99.7 | 96.7 | 88.1 | 99.5 |
| | asymptotic_bonf | 97.3 | 85.2 | 99.5 | 97.1 | 87.4 | 99.7 | 96.6 | 87.9 | 99.4 |
| | permutation_bonf | 97.1 | 85.8 | 99.8 | 96.9 | 87.0 | 99.7 | 96.2 | 87.6 | 99.4 |
| Weib scale continuous | asymptotic | 91.1 | 74.6 | 97.8 | 91.6 | 75.8 | 97.8 | 90.8 | 77.1 | 98.4 |
| | asymptotic_bonf | 90.8 | 74.1 | 97.8 | 91.4 | 75.1 | 97.7 | 90.6 | 76.3 | 98.4 |
| | permutation_bonf | 90.8 | 73.5 | 97.6 | 91.7 | 74.9 | 97.3 | 90.3 | 76.4 | 98.3 |
| Weib scale discrete | asymptotic | 95.3 | 82.7 | 99.0 | 95.6 | 83.7 | 99.2 | 95.0 | 83.9 | 99.2 |
| | asymptotic_bonf | 95.2 | 82.3 | 99.0 | 95.5 | 83.5 | 99.1 | 94.8 | 83.8 | 99.1 |
| | permutation_bonf | 94.9 | 82.0 | 99.1 | 95.4 | 83.8 | 98.8 | 94.8 | 82.7 | 99.3 |
| Weib shape continuous | asymptotic | 85.7 | 67.5 | 96.0 | 87.1 | 69.3 | 95.1 | 85.9 | 70.6 | 95.8 |
| | asymptotic_bonf | 85.3 | 67.3 | 95.9 | 87.0 | 68.6 | 94.7 | 85.7 | 70.5 | 95.8 |
| | permutation_bonf | 85.5 | 67.6 | 96.0 | 86.4 | 68.9 | 94.9 | 86.1 | 70.1 | 95.8 |
| Weib shape discrete | asymptotic | 92.6 | 78.2 | 97.8 | 92.9 | 78.3 | 98.0 | 92.0 | 80.1 | 98.7 |
| | asymptotic_bonf | 92.4 | 77.6 | 97.8 | 92.8 | 78.1 | 97.8 | 91.7 | 79.9 | 98.7 |
| | permutation_bonf | 92.4 | 77.5 | 97.6 | 92.9 | 78.0 | 97.8 | 91.7 | 79.4 | 98.5 |

Table S50: Rejection rates in percent for the 2-by-2 design with $\delta = 1.5$ and balanced medium sample sizes under unequal, low censoring.

| Distribution | Method | $\mathcal{H}_{0,4}$ | $\mathcal{H}_{0,5}$ | $\mathcal{H}_{0,6}$ | $\mathcal{H}_{0,1}$ | $\mathcal{H}_{0,2}$ | $\mathcal{H}_{0,3}$ | $\mathcal{H}_{0,7}$ | $\mathcal{H}_{0,8}$ | $\mathcal{H}_{0,9}$ |
|---|---|---|---|---|---|---|---|---|---|---|
| exp early continuous | asymptotic | 5.6 | 5.4 | 9.2 | 6.9 | 3.1 | 8.9 | 6.5 | 2.7 | 9.8 |
|  | asymptotic_bonf | 5.5 | 5.4 | 8.9 | 6.5 | 3.0 | 8.8 | 6.2 | 2.7 | 9.6 |
|  | permutation_bonf | 6.0 | 5.7 | 8.7 | 6.7 | 2.9 | 9.1 | 6.3 | 2.8 | 9.3 |
| exp early discrete | asymptotic | 7.1 | 6.3 | 10.6 | 7.4 | 3.6 | 11.7 | 6.7 | 3.1 | 11.3 |
|  | asymptotic_bonf | 7.0 | 6.2 | 10.2 | 7.1 | 3.5 | 11.5 | 6.6 | 3.1 | 11.1 |
|  | permutation_bonf | 7.9 | 6.0 | 10.1 | 7.1 | 3.9 | 11.2 | 6.7 | 2.7 | 10.8 |
| exp late continuous | asymptotic | 5.8 | 5.9 | 9.9 | 7.5 | 3.3 | 9.9 | 7.6 | 3.4 | 10.4 |
|  | asymptotic_bonf | 5.7 | 5.8 | 9.9 | 7.0 | 3.1 | 9.8 | 7.3 | 3.2 | 10.0 |
|  | permutation_bonf | 6.0 | 6.0 | 9.3 | 7.1 | 3.7 | 9.7 | 6.6 | 2.7 | 9.1 |
| exp late discrete | asymptotic | 7.6 | 6.0 | 11.0 | 8.3 | 4.0 | 12.1 | 7.7 | 3.5 | 11.9 |
|  | asymptotic_bonf | 7.2 | 6.0 | 10.7 | 8.0 | 3.6 | 11.9 | 7.6 | 3.5 | 11.5 |
|  | permutation_bonf | 7.6 | 5.6 | 11.1 | 7.6 | 4.1 | 11.9 | 7.1 | 3.1 | 12.1 |
| exp prop continuous | asymptotic | 7.3 | 4.7 | 9.0 | 6.0 | 4.8 | 8.7 | 6.6 | 4.8 | 9.7 |
|  | asymptotic_bonf | 7.2 | 4.7 | 8.8 | 5.9 | 4.6 | 8.5 | 6.4 | 4.5 | 9.2 |
|  | permutation_bonf | 6.9 | 4.3 | 9.5 | 6.1 | 4.3 | 8.3 | 6.4 | 4.9 | 9.5 |
| exp prop discrete | asymptotic | 7.6 | 5.1 | 10.2 | 6.9 | 5.1 | 9.2 | 7.0 | 5.3 | 11.1 |
|  | asymptotic_bonf | 7.4 | 5.1 | 9.8 | 6.8 | 5.0 | 9.2 | 6.9 | 5.1 | 10.5 |
|  | permutation_bonf | 7.8 | 5.6 | 10.0 | 6.4 | 4.6 | 8.9 | 7.6 | 5.2 | 9.5 |
| logn continuous | asymptotic | 18.9 | 12.3 | 30.3 | 21.0 | 13.9 | 32.8 | 21.6 | 12.7 | 32.0 |
|  | asymptotic_bonf | 18.8 | 11.9 | 30.0 | 20.5 | 13.8 | 32.8 | 20.9 | 12.3 | 31.6 |
|  | permutation_bonf | 18.7 | 12.2 | 30.2 | 19.9 | 13.8 | 31.3 | 20.9 | 12.1 | 31.4 |
| logn discrete | asymptotic | 23.9 | 15.1 | 40.8 | 27.0 | 18.5 | 38.0 | 25.9 | 16.6 | 38.2 |
|  | asymptotic_bonf | 23.3 | 14.7 | 39.7 | 26.8 | 18.3 | 37.7 | 25.3 | 16.3 | 37.9 |
|  | permutation_bonf | 23.3 | 15.1 | 39.8 | 26.4 | 17.7 | 37.3 | 25.6 | 16.3 | 37.8 |
| pwExp continuous | asymptotic | 5.8 | 5.5 | 9.2 | 6.9 | 4.0 | 8.9 | 6.6 | 2.6 | 8.9 |
|  | asymptotic_bonf | 5.6 | 5.4 | 8.8 | 6.3 | 3.8 | 8.8 | 6.4 | 2.5 | 8.8 |
|  | permutation_bonf | 6.2 | 5.3 | 8.9 | 6.2 | 3.9 | 8.8 | 6.6 | 2.7 | 9.2 |
| pwExp discrete | asymptotic | 6.9 | 5.8 | 11.1 | 6.8 | 4.6 | 11.0 | 7.2 | 3.5 | 10.9 |
|  | asymptotic_bonf | 6.7 | 5.4 | 10.8 | 6.8 | 4.3 | 10.8 | 7.0 | 3.3 | 10.7 |
|  | permutation_bonf | 6.9 | 5.6 | 10.7 | 6.6 | 4.2 | 11.2 | 6.5 | 3.1 | 11.0 |
| Weib late continuous | asymptotic | 18.8 | 12.7 | 33.4 | 20.4 | 13.5 | 30.0 | 20.9 | 11.9 | 32.6 |
|  | asymptotic_bonf | 18.5 | 12.3 | 32.8 | 20.0 | 13.3 | 29.5 | 20.6 | 11.4 | 32.4 |
|  | permutation_bonf | 18.1 | 12.6 | 33.6 | 20.9 | 13.6 | 29.3 | 19.8 | 11.2 | 33.0 |
| Weib late discrete | asymptotic | 22.6 | 16.1 | 40.4 | 25.8 | 16.2 | 36.0 | 26.4 | 15.3 | 39.5 |
|  | asymptotic_bonf | 22.1 | 15.6 | 39.4 | 24.9 | 15.6 | 35.6 | 25.7 | 14.6 | 38.9 |
|  | permutation_bonf | 22.3 | 16.5 | 38.8 | 25.0 | 15.3 | 35.2 | 26.0 | 15.2 | 38.3 |
| Weib prop continuous | asymptotic | 17.8 | 12.4 | 32.5 | 19.4 | 12.8 | 29.6 | 20.1 | 11.3 | 31.6 |
|  | asymptotic_bonf | 17.2 | 12.2 | 32.2 | 19.2 | 12.4 | 29.2 | 19.9 | 10.9 | 31.2 |
|  | permutation_bonf | 17.0 | 12.7 | 32.5 | 19.4 | 12.6 | 28.4 | 19.4 | 10.8 | 31.0 |
| Weib prop discrete | asymptotic | 22.4 | 15.5 | 39.0 | 25.4 | 15.9 | 35.9 | 26.4 | 14.6 | 38.4 |
|  | asymptotic_bonf | 22.0 | 15.0 | 38.4 | 24.8 | 15.6 | 35.5 | 26.1 | 14.2 | 37.7 |
|  | permutation_bonf | 22.4 | 15.0 | 38.6 | 24.9 | 16.3 | 35.6 | 25.4 | 14.2 | 38.5 |
| Weib scale continuous | asymptotic | 15.9 | 11.2 | 28.7 | 16.9 | 11.0 | 25.9 | 18.1 | 9.7 | 27.9 |
|  | asymptotic_bonf | 15.7 | 11.0 | 28.2 | 16.7 | 10.8 | 25.1 | 17.7 | 9.5 | 27.8 |
|  | permutation_bonf | 16.1 | 11.2 | 27.9 | 16.8 | 11.3 | 25.6 | 17.3 | 9.2 | 26.4 |
| Weib scale discrete | asymptotic | 20.2 | 13.7 | 34.7 | 21.8 | 13.6 | 31.6 | 23.2 | 12.9 | 33.3 |
|  | asymptotic_bonf | 19.6 | 13.7 | 34.1 | 21.5 | 13.1 | 31.1 | 22.9 | 12.3 | 32.8 |
|  | permutation_bonf | 19.4 | 13.4 | 34.6 | 22.3 | 13.4 | 31.9 | 22.8 | 12.6 | 32.5 |
| Weib shape continuous | asymptotic | 12.8 | 9.0 | 24.6 | 14.5 | 9.5 | 20.5 | 14.2 | 8.1 | 23.8 |
|  | asymptotic_bonf | 12.6 | 8.7 | 24.0 | 13.7 | 9.3 | 20.1 | 13.8 | 8.0 | 23.4 |
|  | permutation_bonf | 13.3 | 8.3 | 23.4 | 14.1 | 9.5 | 20.0 | 13.4 | 7.7 | 23.5 |
| Weib shape discrete | asymptotic | 16.7 | 11.8 | 29.0 | 18.1 | 11.3 | 27.5 | 19.0 | 11.4 | 27.2 |
|  | asymptotic_bonf | 16.4 | 11.6 | 28.4 | 17.3 | 10.9 | 27.1 | 18.3 | 11.0 | 26.7 |
|  | permutation_bonf | 16.5 | 11.2 | 28.8 | 17.8 | 11.4 | 27.6 | 18.7 | 11.1 | 27.7 |

Table S51: Rejection rates in percent for the 2-by-2 design with $\delta = 1.5$ and balanced small sample sizes under unequal, low censoring.

| Distribution | Method | $\mathcal{H}_{0,4}$ | $\mathcal{H}_{0,5}$ | $\mathcal{H}_{0,6}$ | $\mathcal{H}_{0,1}$ | $\mathcal{H}_{0,2}$ | $\mathcal{H}_{0,3}$ | $\mathcal{H}_{0,7}$ | $\mathcal{H}_{0,8}$ | $\mathcal{H}_{0,9}$ |
|---|---|---|---|---|---|---|---|---|---|---|
| exp early continuous | asymptotic | 2.1 | 1.2 | 1.4 | 1.7 | 0.8 | 1.8 | 1.7 | 1.6 | 1.4 |
| | asymptotic_bonf | 1.8 | 1.2 | 1.4 | 1.6 | 0.8 | 1.7 | 1.6 | 1.4 | 1.3 |
| | permutation_bonf | 1.7 | 1.0 | 1.2 | 1.4 | 0.9 | 1.6 | 1.1 | 1.6 | 1.3 |
| exp early discrete | asymptotic | 2.1 | 1.5 | 1.4 | 1.2 | 1.6 | 1.6 | 2.3 | 1.0 | 1.9 |
| | asymptotic_bonf | 1.9 | 1.5 | 1.2 | 1.0 | 1.5 | 1.4 | 2.3 | 1.0 | 1.8 |
| | permutation_bonf | 1.6 | 1.2 | 1.3 | 1.0 | 1.3 | 1.4 | 1.8 | 0.6 | 1.5 |
| exp late continuous | asymptotic | 1.5 | 1.4 | 1.6 | 1.9 | 0.9 | 1.7 | 1.9 | 1.8 | 1.5 |
| | asymptotic_bonf | 1.3 | 1.3 | 1.4 | 1.9 | 0.9 | 1.6 | 1.8 | 1.6 | 1.5 |
| | permutation_bonf | 1.6 | 1.1 | 1.2 | 1.6 | 1.2 | 1.4 | 1.6 | 1.6 | 1.2 |
| exp late discrete | asymptotic | 1.6 | 1.4 | 1.6 | 1.8 | 1.3 | 2.1 | 2.4 | 1.7 | 1.9 |
| | asymptotic_bonf | 1.5 | 1.2 | 1.4 | 1.7 | 1.3 | 2.0 | 2.0 | 1.6 | 1.9 |
| | permutation_bonf | 1.6 | 0.9 | 1.6 | 1.6 | 1.4 | 1.9 | 1.9 | 1.7 | 1.4 |
| exp prop continuous | asymptotic | 1.3 | 1.3 | 2.0 | 1.6 | 1.5 | 2.0 | 1.1 | 1.3 | 1.8 |
| | asymptotic_bonf | 1.2 | 1.1 | 2.0 | 1.5 | 1.4 | 2.0 | 1.1 | 1.2 | 1.7 |
| | permutation_bonf | 1.1 | 0.9 | 1.8 | 1.5 | 1.4 | 1.8 | 0.9 | 1.6 | 1.6 |
| exp prop discrete | asymptotic | 2.2 | 0.7 | 2.3 | 1.4 | 1.1 | 2.6 | 1.1 | 1.1 | 2.4 |
| | asymptotic_bonf | 2.1 | 0.7 | 2.2 | 1.3 | 0.9 | 2.5 | 1.1 | 1.1 | 2.3 |
| | permutation_bonf | 1.9 | 1.0 | 1.9 | 1.4 | 1.1 | 1.9 | 1.1 | 0.6 | 2.2 |
| logn continuous | asymptotic | 3.9 | 2.0 | 4.3 | 3.5 | 4.1 | 6.2 | 3.1 | 1.0 | 4.4 |
| | asymptotic_bonf | 3.9 | 2.0 | 4.2 | 3.4 | 3.9 | 6.0 | 3.0 | 1.0 | 4.3 |
| | permutation_bonf | 3.9 | 1.7 | 4.0 | 2.8 | 3.6 | 5.4 | 2.9 | 1.0 | 4.0 |
| logn discrete | asymptotic | 4.8 | 2.4 | 4.9 | 4.4 | 4.3 | 7.3 | 3.8 | 1.4 | 4.8 |
| | asymptotic_bonf | 4.5 | 2.3 | 4.8 | 4.4 | 4.1 | 7.3 | 3.8 | 1.2 | 4.8 |
| | permutation_bonf | 4.7 | 2.4 | 4.4 | 4.0 | 4.1 | 7.1 | 3.8 | 1.2 | 4.7 |
| pwExp continuous | asymptotic | 1.6 | 1.2 | 1.3 | 1.7 | 0.9 | 1.9 | 1.9 | 1.6 | 1.3 |
| | asymptotic_bonf | 1.5 | 1.2 | 1.2 | 1.6 | 0.9 | 1.7 | 1.9 | 1.6 | 1.3 |
| | permutation_bonf | 1.5 | 1.1 | 1.2 | 1.0 | 0.9 | 1.5 | 1.0 | 1.7 | 1.2 |
| pwExp discrete | asymptotic | 2.0 | 1.3 | 1.4 | 1.8 | 1.1 | 2.1 | 1.9 | 1.6 | 1.9 |
| | asymptotic_bonf | 1.7 | 1.3 | 1.3 | 1.8 | 1.0 | 2.0 | 1.7 | 1.5 | 1.9 |
| | permutation_bonf | 1.9 | 1.3 | 1.6 | 1.3 | 1.2 | 1.9 | 1.0 | 1.9 | 1.6 |
| Weib late continuous | asymptotic | 3.5 | 2.4 | 4.6 | 2.9 | 3.6 | 4.1 | 3.4 | 3.1 | 5.0 |
| | asymptotic_bonf | 3.4 | 2.4 | 4.5 | 2.9 | 3.4 | 4.1 | 3.3 | 3.1 | 4.9 |
| | permutation_bonf | 3.4 | 2.3 | 4.4 | 2.4 | 3.1 | 3.7 | 3.2 | 2.9 | 5.2 |
| Weib late discrete | asymptotic | 3.8 | 3.1 | 5.5 | 2.9 | 3.6 | 5.0 | 4.3 | 3.5 | 5.8 |
| | asymptotic_bonf | 3.6 | 3.0 | 5.4 | 2.9 | 3.5 | 4.8 | 4.2 | 3.3 | 5.6 |
| | permutation_bonf | 3.3 | 2.3 | 4.8 | 2.9 | 3.4 | 4.7 | 3.7 | 3.1 | 5.3 |
| Weib prop continuous | asymptotic | 3.6 | 2.3 | 4.5 | 2.7 | 3.2 | 3.9 | 3.2 | 3.3 | 4.8 |
| | asymptotic_bonf | 3.6 | 2.3 | 4.3 | 2.6 | 3.0 | 3.9 | 3.1 | 3.3 | 4.6 |
| | permutation_bonf | 3.0 | 2.1 | 4.2 | 2.2 | 2.9 | 3.7 | 3.0 | 3.0 | 5.1 |
| Weib prop discrete | asymptotic | 4.0 | 2.6 | 5.6 | 3.2 | 3.3 | 4.8 | 4.2 | 3.2 | 5.7 |
| | asymptotic_bonf | 3.8 | 2.6 | 5.4 | 3.2 | 3.2 | 4.5 | 4.0 | 3.2 | 5.6 |
| | permutation_bonf | 3.5 | 2.4 | 5.4 | 2.9 | 3.5 | 4.3 | 3.5 | 2.9 | 5.3 |
| Weib scale continuous | asymptotic | 2.6 | 2.0 | 4.3 | 2.3 | 2.7 | 3.5 | 2.6 | 2.4 | 4.3 |
| | asymptotic_bonf | 2.6 | 1.9 | 4.1 | 2.2 | 2.7 | 3.5 | 2.6 | 2.3 | 4.2 |
| | permutation_bonf | 2.3 | 2.0 | 3.6 | 1.8 | 2.2 | 3.2 | 2.3 | 2.5 | 4.0 |
| Weib scale discrete | asymptotic | 2.7 | 2.2 | 4.8 | 2.3 | 3.2 | 4.1 | 3.0 | 2.5 | 4.8 |
| | asymptotic_bonf | 2.6 | 2.2 | 4.6 | 2.2 | 2.9 | 3.8 | 2.8 | 2.1 | 4.7 |
| | permutation_bonf | 2.7 | 2.3 | 4.6 | 2.2 | 2.7 | 3.6 | 2.7 | 2.4 | 4.5 |
| Weib shape continuous | asymptotic | 2.0 | 1.4 | 3.1 | 1.1 | 2.0 | 2.6 | 1.9 | 1.9 | 3.6 |
| | asymptotic_bonf | 1.9 | 1.4 | 2.9 | 1.1 | 1.9 | 2.5 | 1.8 | 1.8 | 3.4 |
| | permutation_bonf | 1.8 | 1.7 | 2.8 | 0.9 | 1.8 | 2.1 | 2.2 | 1.7 | 3.4 |
| Weib shape discrete | asymptotic | 2.3 | 1.8 | 3.5 | 1.4 | 2.3 | 3.3 | 2.1 | 2.1 | 4.2 |
| | asymptotic_bonf | 1.9 | 1.7 | 3.4 | 1.4 | 2.1 | 3.2 | 1.9 | 2.0 | 3.9 |
| | permutation_bonf | 2.1 | 1.6 | 3.4 | 1.3 | 2.1 | 3.0 | 2.0 | 1.9 | 3.6 |

Table S52: Rejection rates in percent for the 2-by-2 design with $\delta = 1.5$ and unbalanced large sample sizes under unequal, low censoring.

| Distribution | Method | $\mathcal{H}_{0,4}$ | $\mathcal{H}_{0,5}$ | $\mathcal{H}_{0,6}$ | $\mathcal{H}_{0,1}$ | $\mathcal{H}_{0,2}$ | $\mathcal{H}_{0,3}$ | $\mathcal{H}_{0,7}$ | $\mathcal{H}_{0,8}$ | $\mathcal{H}_{0,9}$ |
|---|---|---|---|---|---|---|---|---|---|---|
| exp early continuous | asymptotic | 24.2 | 13.3 | 39.7 | 23.5 | 13.1 | 37.4 | 24.2 | 13.0 | 37.8 |
| | asymptotic_bonf | 22.7 | 12.5 | 38.4 | 22.7 | 12.0 | 35.9 | 22.7 | 12.2 | 35.5 |
| | permutation_bonf | 22.2 | 11.7 | 37.7 | 21.4 | 11.7 | 34.8 | 22.4 | 12.7 | 35.8 |
| exp early discrete | asymptotic | 27.5 | 15.7 | 44.8 | 27.5 | 15.1 | 44.9 | 28.6 | 15.6 | 43.8 |
| | asymptotic_bonf | 25.6 | 14.6 | 42.7 | 25.7 | 14.2 | 42.8 | 27.1 | 14.1 | 41.8 |
| | permutation_bonf | 26.2 | 13.8 | 42.3 | 25.9 | 13.1 | 42.9 | 26.8 | 14.8 | 40.4 |
| exp late continuous | asymptotic | 26.1 | 14.5 | 45.9 | 26.5 | 14.5 | 44.1 | 26.8 | 14.6 | 41.8 |
| | asymptotic_bonf | 25.0 | 13.3 | 44.2 | 24.4 | 13.3 | 42.3 | 25.0 | 13.2 | 40.1 |
| | permutation_bonf | 25.0 | 13.3 | 42.6 | 24.2 | 12.5 | 40.5 | 25.6 | 13.8 | 39.4 |
| exp late discrete | asymptotic | 30.3 | 17.5 | 51.8 | 31.2 | 17.8 | 50.4 | 31.5 | 17.6 | 50.0 |
| | asymptotic_bonf | 28.8 | 16.5 | 49.2 | 29.6 | 16.4 | 48.9 | 30.1 | 16.3 | 47.4 |
| | permutation_bonf | 29.4 | 15.7 | 48.2 | 28.2 | 16.0 | 49.2 | 28.9 | 16.7 | 46.2 |
| exp prop continuous | asymptotic | 28.2 | 16.5 | 37.5 | 25.5 | 14.9 | 39.6 | 26.5 | 15.0 | 36.4 |
| | asymptotic_bonf | 26.3 | 14.8 | 35.7 | 23.6 | 14.0 | 37.5 | 24.2 | 14.3 | 34.8 |
| | permutation_bonf | 26.7 | 14.4 | 34.3 | 23.0 | 13.0 | 36.6 | 23.6 | 14.6 | 33.9 |
| exp prop discrete | asymptotic | 31.2 | 20.0 | 42.6 | 28.4 | 18.0 | 44.5 | 30.1 | 17.9 | 43.5 |
| | asymptotic_bonf | 29.4 | 18.1 | 41.0 | 26.7 | 16.5 | 42.8 | 28.7 | 17.0 | 40.8 |
| | permutation_bonf | 29.0 | 18.0 | 39.4 | 26.5 | 15.1 | 43.3 | 28.1 | 16.2 | 39.4 |
| logn continuous | asymptotic | 72.5 | 49.2 | 84.8 | 70.1 | 52.3 | 84.0 | 69.4 | 50.4 | 84.3 |
| | asymptotic_bonf | 70.3 | 47.1 | 84.0 | 68.2 | 50.3 | 82.6 | 68.0 | 48.3 | 83.0 |
| | permutation_bonf | 69.4 | 46.0 | 82.6 | 67.0 | 48.2 | 82.7 | 67.4 | 47.3 | 82.5 |
| logn discrete | asymptotic | 80.5 | 58.2 | 91.2 | 78.3 | 61.6 | 89.7 | 78.1 | 59.7 | 89.9 |
| | asymptotic_bonf | 79.1 | 56.2 | 90.2 | 76.9 | 59.5 | 88.4 | 76.8 | 57.3 | 89.0 |
| | permutation_bonf | 78.1 | 54.3 | 89.1 | 75.2 | 57.4 | 88.0 | 76.5 | 55.9 | 88.9 |
| pwExp continuous | asymptotic | 22.4 | 13.8 | 39.4 | 22.9 | 12.7 | 36.6 | 22.5 | 12.4 | 36.5 |
| | asymptotic_bonf | 20.6 | 12.4 | 37.4 | 21.0 | 11.8 | 34.7 | 21.6 | 11.3 | 35.0 |
| | permutation_bonf | 21.2 | 11.2 | 37.0 | 20.5 | 11.4 | 34.7 | 21.3 | 11.0 | 35.5 |
| pwExp discrete | asymptotic | 26.6 | 15.3 | 45.4 | 26.1 | 15.2 | 43.9 | 26.1 | 14.9 | 43.3 |
| | asymptotic_bonf | 25.3 | 14.6 | 43.4 | 25.0 | 13.5 | 42.1 | 25.1 | 14.0 | 41.1 |
| | permutation_bonf | 25.0 | 14.6 | 43.3 | 24.0 | 13.8 | 41.6 | 25.4 | 13.4 | 41.4 |
| Weib late continuous | asymptotic | 72.0 | 50.3 | 88.6 | 70.9 | 47.6 | 88.9 | 72.5 | 50.1 | 88.2 |
| | asymptotic_bonf | 70.3 | 48.3 | 87.3 | 69.4 | 46.0 | 88.0 | 71.1 | 48.6 | 87.2 |
| | permutation_bonf | 69.4 | 47.8 | 85.8 | 68.5 | 44.0 | 87.8 | 70.6 | 47.8 | 87.3 |
| Weib late discrete | asymptotic | 81.1 | 58.1 | 92.3 | 79.1 | 58.1 | 93.6 | 80.0 | 58.1 | 93.3 |
| | asymptotic_bonf | 80.0 | 56.3 | 91.8 | 77.8 | 55.8 | 93.0 | 78.7 | 56.3 | 93.2 |
| | permutation_bonf | 78.5 | 55.1 | 90.7 | 76.7 | 54.8 | 92.7 | 79.1 | 55.3 | 92.8 |
| Weib prop continuous | asymptotic | 68.9 | 47.6 | 86.6 | 68.3 | 45.4 | 86.6 | 70.9 | 48.5 | 85.8 |
| | asymptotic_bonf | 67.1 | 45.4 | 85.0 | 67.2 | 44.1 | 86.0 | 69.2 | 46.6 | 84.9 |
| | permutation_bonf | 66.2 | 44.0 | 83.4 | 65.9 | 42.0 | 85.4 | 67.9 | 45.2 | 84.8 |
| Weib prop discrete | asymptotic | 79.3 | 55.6 | 91.5 | 76.9 | 55.6 | 93.1 | 78.2 | 55.2 | 92.5 |
| | asymptotic_bonf | 77.2 | 53.9 | 90.9 | 75.8 | 53.3 | 92.3 | 77.1 | 53.4 | 91.5 |
| | permutation_bonf | 77.0 | 53.5 | 89.8 | 74.5 | 52.1 | 92.1 | 76.6 | 52.8 | 90.7 |
| Weib scale continuous | asymptotic | 58.6 | 38.8 | 77.4 | 60.4 | 39.0 | 78.2 | 59.4 | 38.6 | 76.8 |
| | asymptotic_bonf | 56.8 | 36.9 | 74.5 | 58.2 | 36.7 | 76.6 | 57.1 | 36.3 | 75.0 |
| | permutation_bonf | 56.1 | 35.3 | 74.5 | 56.8 | 35.1 | 76.0 | 57.1 | 35.7 | 74.1 |
| Weib scale discrete | asymptotic | 68.1 | 46.2 | 86.1 | 67.5 | 45.9 | 86.0 | 69.1 | 46.7 | 85.3 |
| | asymptotic_bonf | 66.7 | 44.7 | 84.9 | 66.0 | 44.1 | 84.7 | 67.3 | 44.5 | 84.3 |
| | permutation_bonf | 65.3 | 43.2 | 83.2 | 65.2 | 43.6 | 84.2 | 65.3 | 43.4 | 83.8 |
| Weib shape continuous | asymptotic | 45.3 | 28.3 | 61.8 | 46.3 | 29.1 | 63.6 | 45.1 | 28.9 | 64.3 |
| | asymptotic_bonf | 41.7 | 25.2 | 59.3 | 43.7 | 26.6 | 61.1 | 41.9 | 27.2 | 61.3 |
| | permutation_bonf | 41.5 | 24.6 | 58.0 | 43.8 | 25.0 | 60.2 | 41.0 | 25.8 | 61.1 |
| Weib shape discrete | asymptotic | 54.7 | 36.2 | 72.7 | 57.0 | 37.1 | 74.5 | 55.4 | 36.7 | 74.0 |
| | asymptotic_bonf | 51.9 | 34.4 | 70.2 | 53.7 | 34.5 | 72.0 | 51.7 | 34.7 | 71.4 |
| | permutation_bonf | 51.8 | 32.2 | 69.2 | 53.6 | 32.8 | 71.0 | 52.3 | 33.3 | 71.1 |

Table S53: Rejection rates in percent for the 2-by-2 design with $\delta = 1.5$ and unbalanced medium sample sizes under unequal, low censoring.

| Distribution | Method | $\mathcal{H}_{0,4}$ | $\mathcal{H}_{0,5}$ | $\mathcal{H}_{0,6}$ | $\mathcal{H}_{0,1}$ | $\mathcal{H}_{0,2}$ | $\mathcal{H}_{0,3}$ | $\mathcal{H}_{0,7}$ | $\mathcal{H}_{0,8}$ | $\mathcal{H}_{0,9}$ |
|---|---|---|---|---|---|---|---|---|---|---|
| exp early continuous | asymptotic | 2.6 | 1.7 | 4.6 | 2.9 | 1.5 | 3.7 | 2.7 | 2.0 | 4.4 |
| | asymptotic_bonf | 2.2 | 1.5 | 4.3 | 2.9 | 1.3 | 3.3 | 2.3 | 1.8 | 3.8 |
| | permutation_bonf | 1.6 | 1.1 | 3.3 | 2.2 | 1.0 | 2.9 | 2.2 | 1.7 | 3.0 |
| exp early discrete | asymptotic | 3.0 | 1.9 | 4.8 | 3.5 | 1.8 | 4.7 | 3.5 | 2.2 | 5.0 |
| | asymptotic_bonf | 2.9 | 1.5 | 4.4 | 3.4 | 1.4 | 4.0 | 2.8 | 2.1 | 4.5 |
| | permutation_bonf | 1.7 | 1.6 | 4.0 | 2.4 | 1.2 | 3.7 | 2.3 | 1.6 | 4.5 |
| exp late continuous | asymptotic | 2.6 | 2.2 | 4.7 | 3.7 | 1.7 | 4.4 | 3.2 | 2.2 | 5.0 |
| | asymptotic_bonf | 2.2 | 1.9 | 4.1 | 3.4 | 1.5 | 3.9 | 2.9 | 2.1 | 4.4 |
| | permutation_bonf | 1.9 | 1.5 | 3.8 | 2.7 | 1.0 | 3.1 | 3.0 | 1.8 | 3.5 |
| exp late discrete | asymptotic | 3.4 | 2.0 | 5.1 | 4.0 | 2.3 | 5.0 | 3.3 | 2.5 | 6.1 |
| | asymptotic_bonf | 3.0 | 2.0 | 4.4 | 3.7 | 2.1 | 4.7 | 3.2 | 2.3 | 5.4 |
| | permutation_bonf | 1.8 | 1.7 | 4.3 | 2.9 | 1.5 | 3.8 | 2.7 | 1.9 | 4.9 |
| exp prop continuous | asymptotic | 3.6 | 2.5 | 3.8 | 2.5 | 1.9 | 4.6 | 2.8 | 2.3 | 5.3 |
| | asymptotic_bonf | 3.1 | 2.3 | 3.4 | 2.2 | 1.7 | 4.1 | 2.3 | 2.2 | 4.9 |
| | permutation_bonf | 2.4 | 1.6 | 3.3 | 1.3 | 1.2 | 3.1 | 2.4 | 1.7 | 4.2 |
| exp prop discrete | asymptotic | 3.7 | 2.9 | 4.6 | 3.3 | 2.1 | 5.2 | 3.2 | 2.3 | 5.5 |
| | asymptotic_bonf | 3.5 | 2.3 | 4.1 | 2.7 | 1.9 | 4.6 | 2.8 | 2.2 | 5.0 |
| | permutation_bonf | 3.5 | 1.8 | 3.2 | 2.6 | 1.3 | 3.6 | 2.6 | 2.0 | 4.8 |
| logn continuous | asymptotic | 8.1 | 5.9 | 13.7 | 7.2 | 5.1 | 12.6 | 7.4 | 6.2 | 13.4 |
| | asymptotic_bonf | 7.3 | 5.3 | 12.4 | 6.5 | 4.6 | 11.8 | 6.6 | 5.8 | 12.9 |
| | permutation_bonf | 5.7 | 3.6 | 10.4 | 5.2 | 3.2 | 10.5 | 5.5 | 4.5 | 11.1 |
| logn discrete | asymptotic | 10.3 | 6.8 | 17.5 | 8.7 | 5.7 | 16.0 | 9.3 | 6.9 | 16.9 |
| | asymptotic_bonf | 9.3 | 6.0 | 16.2 | 7.8 | 5.1 | 14.7 | 8.6 | 6.6 | 16.3 |
| | permutation_bonf | 7.0 | 4.3 | 12.7 | 6.2 | 3.5 | 12.4 | 7.2 | 4.7 | 13.4 |
| pwExp continuous | asymptotic | 1.9 | 2.2 | 4.5 | 2.8 | 1.6 | 3.5 | 2.6 | 1.7 | 4.7 |
| | asymptotic_bonf | 1.8 | 1.6 | 4.0 | 2.3 | 1.5 | 3.0 | 2.4 | 1.6 | 4.1 |
| | permutation_bonf | 1.4 | 1.8 | 3.5 | 1.8 | 0.9 | 2.5 | 1.8 | 1.0 | 3.4 |
| pwExp discrete | asymptotic | 2.2 | 2.3 | 4.4 | 3.3 | 2.3 | 4.4 | 2.9 | 1.9 | 5.3 |
| | asymptotic_bonf | 1.8 | 1.9 | 4.1 | 2.8 | 2.0 | 3.8 | 2.7 | 1.9 | 4.8 |
| | permutation_bonf | 1.5 | 1.7 | 3.9 | 2.2 | 1.6 | 3.3 | 1.9 | 1.3 | 3.9 |
| Weib late continuous | asymptotic | 9.9 | 5.1 | 13.8 | 8.0 | 5.6 | 14.0 | 10.4 | 4.9 | 16.1 |
| | asymptotic_bonf | 9.2 | 4.2 | 12.5 | 7.7 | 5.0 | 13.1 | 9.2 | 4.8 | 14.7 |
| | permutation_bonf | 7.2 | 2.9 | 11.2 | 5.6 | 2.3 | 11.1 | 8.0 | 4.4 | 13.0 |
| Weib late discrete | asymptotic | 12.1 | 6.2 | 17.0 | 10.3 | 6.3 | 17.9 | 14.1 | 5.9 | 19.1 |
| | asymptotic_bonf | 11.0 | 5.6 | 15.9 | 9.4 | 5.5 | 16.8 | 12.8 | 5.1 | 17.7 |
| | permutation_bonf | 7.5 | 3.3 | 13.1 | 7.3 | 3.7 | 14.3 | 9.9 | 4.1 | 14.8 |
| Weib prop continuous | asymptotic | 8.9 | 4.1 | 11.9 | 7.5 | 5.2 | 12.5 | 9.2 | 4.0 | 14.3 |
| | asymptotic_bonf | 8.2 | 3.8 | 11.1 | 6.8 | 4.5 | 11.4 | 8.3 | 3.8 | 13.2 |
| | permutation_bonf | 6.7 | 2.7 | 9.8 | 5.1 | 1.9 | 9.1 | 7.1 | 3.7 | 11.5 |
| Weib prop discrete | asymptotic | 11.0 | 5.1 | 14.9 | 10.1 | 5.7 | 16.2 | 11.4 | 5.1 | 17.3 |
| | asymptotic_bonf | 10.0 | 4.6 | 13.9 | 9.0 | 5.0 | 14.7 | 10.5 | 4.5 | 15.8 |
| | permutation_bonf | 7.9 | 2.9 | 12.3 | 6.1 | 2.4 | 11.8 | 8.9 | 3.8 | 13.5 |
| Weib scale continuous | asymptotic | 6.5 | 2.8 | 8.7 | 4.9 | 3.7 | 8.4 | 7.2 | 3.4 | 9.7 |
| | asymptotic_bonf | 5.9 | 2.3 | 8.1 | 4.2 | 3.2 | 7.5 | 6.1 | 2.8 | 8.9 |
| | permutation_bonf | 5.0 | 1.5 | 7.2 | 3.3 | 1.8 | 6.1 | 5.0 | 2.4 | 7.9 |
| Weib scale discrete | asymptotic | 8.3 | 3.5 | 10.5 | 6.9 | 4.4 | 10.7 | 8.6 | 3.9 | 11.9 |
| | asymptotic_bonf | 7.5 | 3.0 | 8.9 | 5.9 | 3.8 | 9.7 | 7.8 | 3.3 | 10.6 |
| | permutation_bonf | 5.9 | 1.9 | 7.8 | 4.1 | 1.7 | 7.5 | 6.3 | 2.4 | 9.4 |
| Weib shape continuous | asymptotic | 4.2 | 1.3 | 5.9 | 3.4 | 2.2 | 5.3 | 4.3 | 2.6 | 6.0 |
| | asymptotic_bonf | 3.3 | 0.9 | 5.1 | 2.6 | 1.7 | 4.6 | 3.4 | 1.7 | 4.8 |
| | permutation_bonf | 2.2 | 0.5 | 4.0 | 1.9 | 0.9 | 4.0 | 2.6 | 1.7 | 3.8 |
| Weib shape discrete | asymptotic | 5.8 | 2.1 | 7.3 | 4.5 | 3.3 | 6.9 | 6.0 | 3.3 | 8.2 |
| | asymptotic_bonf | 4.6 | 1.7 | 6.3 | 3.9 | 2.5 | 6.2 | 4.7 | 2.5 | 6.8 |
| | permutation_bonf | 3.2 | 0.5 | 5.1 | 2.0 | 0.8 | 4.3 | 3.7 | 2.1 | 5.1 |

Table S54: Rejection rates in percent for the 2-by-2 design with $\delta = 1.5$ and unbalanced small sample sizes under unequal, low censoring.

| Distribution | Method | $\mathcal{H}_{0,4}$ | $\mathcal{H}_{0,5}$ | $\mathcal{H}_{0,6}$ | $\mathcal{H}_{0,1}$ | $\mathcal{H}_{0,2}$ | $\mathcal{H}_{0,3}$ | $\mathcal{H}_{0,7}$ | $\mathcal{H}_{0,8}$ | $\mathcal{H}_{0,9}$ |
|---|---|---|---|---|---|---|---|---|---|---|
| exp early continuous | asymptotic | 1.4 | 0.8 | 2.1 | 1.5 | 1.3 | 0.8 | 0.7 | 1.2 | 1.6 |
| | asymptotic_bonf | 1.2 | 0.8 | 1.6 | 1.2 | 1.2 | 0.7 | 0.6 | 1.0 | 1.6 |
| | permutation_bonf | 0.1 | 0.3 | 0.6 | 0.1 | 0.6 | 0.3 | 0.2 | 0.5 | 1.1 |
| exp early discrete | asymptotic | 1.4 | 0.9 | 1.9 | 1.0 | 1.2 | 1.0 | 0.6 | 1.4 | 1.8 |
| | asymptotic_bonf | 1.1 | 0.8 | 1.7 | 1.0 | 1.2 | 1.0 | 0.6 | 1.1 | 1.6 |
| | permutation_bonf | 0.2 | 0.2 | 0.7 | 0.1 | 0.3 | 0.5 | 0.2 | 0.5 | 0.7 |
| exp late continuous | asymptotic | 1.4 | 0.7 | 1.9 | 1.0 | 1.1 | 1.1 | 0.9 | 1.1 | 2.1 |
| | asymptotic_bonf | 1.2 | 0.7 | 1.2 | 0.9 | 1.0 | 1.0 | 0.7 | 0.9 | 1.7 |
| | permutation_bonf | 0.0 | 0.2 | 0.3 | 0.3 | 0.3 | 0.3 | 0.2 | 0.4 | 1.4 |
| exp late discrete | asymptotic | 1.5 | 1.0 | 1.6 | 1.3 | 1.0 | 1.6 | 0.9 | 1.2 | 2.1 |
| | asymptotic_bonf | 1.1 | 0.6 | 1.2 | 1.1 | 0.9 | 1.4 | 0.9 | 1.0 | 2.1 |
| | permutation_bonf | 0.2 | 0.1 | 0.5 | 0.2 | 0.1 | 0.5 | 0.2 | 0.4 | 1.0 |
| exp prop continuous | asymptotic | 1.5 | 1.2 | 1.8 | 0.9 | 0.9 | 2.1 | 1.2 | 1.0 | 1.6 |
| | asymptotic_bonf | 1.3 | 1.1 | 1.7 | 0.9 | 0.7 | 2.1 | 1.0 | 1.0 | 1.4 |
| | permutation_bonf | 0.3 | 0.4 | 1.3 | 0.2 | 0.0 | 0.9 | 0.6 | 0.2 | 0.6 |
| exp prop discrete | asymptotic | 1.5 | 1.3 | 2.0 | 0.8 | 0.8 | 2.3 | 1.1 | 1.2 | 1.6 |
| | asymptotic_bonf | 1.4 | 1.3 | 1.8 | 0.7 | 0.7 | 1.9 | 0.9 | 1.1 | 1.5 |
| | permutation_bonf | 0.4 | 0.5 | 1.0 | 0.2 | 0.0 | 0.7 | 0.5 | 0.2 | 0.6 |
| logn continuous | asymptotic | 1.3 | 1.1 | 2.4 | 1.4 | 1.4 | 2.1 | 2.0 | 0.9 | 2.0 |
| | asymptotic_bonf | 1.2 | 1.0 | 2.3 | 1.2 | 1.3 | 1.7 | 1.7 | 0.8 | 1.8 |
| | permutation_bonf | 0.1 | 0.2 | 0.6 | 0.2 | 0.4 | 0.5 | 0.5 | 0.4 | 0.8 |
| logn discrete | asymptotic | 1.2 | 1.4 | 2.4 | 1.6 | 1.2 | 2.2 | 2.1 | 1.2 | 2.4 |
| | asymptotic_bonf | 1.0 | 1.3 | 2.2 | 1.5 | 1.2 | 2.1 | 1.9 | 0.9 | 2.3 |
| | permutation_bonf | 0.1 | 0.3 | 0.4 | 0.1 | 0.3 | 0.3 | 0.7 | 0.4 | 0.8 |
| pwExp continuous | asymptotic | 1.0 | 1.1 | 1.9 | 0.8 | 0.8 | 1.5 | 0.8 | 1.0 | 2.0 |
| | asymptotic_bonf | 0.7 | 1.0 | 1.7 | 0.5 | 0.7 | 1.2 | 0.4 | 1.0 | 1.6 |
| | permutation_bonf | 0.1 | 0.3 | 0.6 | 0.2 | 0.3 | 0.3 | 0.3 | 0.2 | 0.9 |
| pwExp discrete | asymptotic | 0.9 | 1.4 | 1.9 | 0.7 | 0.6 | 1.6 | 0.7 | 1.1 | 2.2 |
| | asymptotic_bonf | 0.7 | 1.2 | 1.7 | 0.7 | 0.5 | 1.4 | 0.6 | 0.9 | 1.5 |
| | permutation_bonf | 0.0 | 0.3 | 0.6 | 0.1 | 0.2 | 0.5 | 0.2 | 0.4 | 0.9 |
| Weib late continuous | asymptotic | 1.9 | 1.1 | 3.3 | 1.9 | 1.6 | 2.7 | 1.1 | 1.4 | 3.0 |
| | asymptotic_bonf | 1.6 | 1.1 | 3.0 | 1.7 | 1.5 | 2.6 | 1.0 | 1.1 | 2.9 |
| | permutation_bonf | 0.4 | 0.2 | 0.9 | 0.2 | 0.0 | 0.7 | 0.2 | 0.5 | 1.2 |
| Weib late discrete | asymptotic | 2.3 | 1.0 | 3.7 | 2.1 | 1.7 | 3.2 | 2.0 | 1.7 | 2.8 |
| | asymptotic_bonf | 2.1 | 0.8 | 3.3 | 2.0 | 1.4 | 2.7 | 1.5 | 1.3 | 2.5 |
| | permutation_bonf | 0.4 | 0.3 | 1.2 | 0.3 | 0.2 | 0.7 | 0.4 | 0.5 | 1.3 |
| Weib prop continuous | asymptotic | 1.5 | 1.2 | 2.8 | 1.7 | 1.7 | 2.4 | 0.9 | 1.3 | 2.8 |
| | asymptotic_bonf | 1.3 | 1.1 | 2.6 | 1.5 | 1.4 | 2.0 | 0.8 | 1.1 | 2.2 |
| | permutation_bonf | 0.2 | 0.3 | 0.9 | 0.2 | 0.1 | 0.7 | 0.3 | 0.3 | 1.1 |
| Weib prop discrete | asymptotic | 1.7 | 1.1 | 3.2 | 1.8 | 1.7 | 2.7 | 1.4 | 1.2 | 2.5 |
| | asymptotic_bonf | 1.6 | 0.9 | 2.7 | 1.7 | 1.4 | 2.5 | 1.1 | 1.0 | 2.4 |
| | permutation_bonf | 0.4 | 0.1 | 1.1 | 0.2 | 0.1 | 0.6 | 0.3 | 0.3 | 1.3 |
| Weib scale continuous | asymptotic | 1.1 | 1.1 | 1.7 | 1.5 | 1.4 | 2.1 | 0.9 | 0.6 | 1.7 |
| | asymptotic_bonf | 0.8 | 0.9 | 1.5 | 1.2 | 1.3 | 2.1 | 0.7 | 0.6 | 1.4 |
| | permutation_bonf | 0.3 | 0.3 | 0.6 | 0.2 | 0.2 | 0.5 | 0.5 | 0.3 | 0.6 |
| Weib scale discrete | asymptotic | 1.2 | 1.1 | 2.0 | 1.6 | 1.3 | 2.2 | 0.9 | 0.8 | 1.8 |
| | asymptotic_bonf | 1.1 | 0.9 | 1.8 | 1.1 | 1.3 | 1.8 | 0.8 | 0.6 | 1.5 |
| | permutation_bonf | 0.4 | 0.3 | 0.7 | 0.2 | 0.1 | 0.4 | 0.4 | 0.3 | 0.5 |
| Weib shape continuous | asymptotic | 2.5 | 1.6 | 2.1 | 2.0 | 2.2 | 1.7 | 1.9 | 1.1 | 1.7 |
| | asymptotic_bonf | 2.2 | 1.4 | 1.7 | 1.7 | 2.0 | 1.7 | 1.9 | 1.0 | 1.7 |
| | permutation_bonf | 0.5 | 0.5 | 0.7 | 0.2 | 0.5 | 0.7 | 0.8 | 0.7 | 1.0 |
| Weib shape discrete | asymptotic | 2.9 | 1.6 | 2.2 | 1.5 | 2.1 | 2.0 | 1.9 | 1.3 | 1.9 |
| | asymptotic_bonf | 2.4 | 1.4 | 2.1 | 1.3 | 1.9 | 1.9 | 1.5 | 1.2 | 1.6 |
| | permutation_bonf | 0.7 | 0.5 | 0.7 | 0.3 | 0.4 | 0.4 | 0.7 | 0.6 | 0.8 |

Table S55: Rejection rates in percent for the Dunnett-type contrast matrix with $\delta = 1.5$ and balanced large sample sizes under equal censoring.

| Distribution | Method | $\mathcal{H}_{0,7}$ | $\mathcal{H}_{0,8}$ | $\mathcal{H}_{0,9}$ |
|---|---|---|---|---|
| exp early continuous | asymptotic | 85.6 | 65.4 | 94.9 |
|  | asymptotic_bonf | 84.4 | 63.7 | 94.6 |
|  | permutation_bonf | 83.1 | 63.6 | 94.2 |
| exp early discrete | asymptotic | 90.7 | 71.9 | 97.3 |
|  | asymptotic_bonf | 89.9 | 69.9 | 97.0 |
|  | permutation_bonf | 89.7 | 69.1 | 96.5 |
| exp late continuous | asymptotic | 88.2 | 68.9 | 96.3 |
|  | asymptotic_bonf | 87.3 | 66.9 | 96.0 |
|  | permutation_bonf | 86.7 | 66.9 | 96.0 |
| exp late discrete | asymptotic | 92.7 | 74.8 | 98.1 |
|  | asymptotic_bonf | 91.7 | 73.6 | 98.0 |
|  | permutation_bonf | 91.9 | 73.4 | 98.0 |
| exp prop continuous | asymptotic | 86.8 | 69.9 | 96.4 |
|  | asymptotic_bonf | 85.8 | 68.7 | 96.0 |
|  | permutation_bonf | 84.9 | 67.5 | 96.1 |
| exp prop discrete | asymptotic | 90.3 | 76.5 | 97.7 |
|  | asymptotic_bonf | 90.0 | 75.5 | 97.6 |
|  | permutation_bonf | 89.6 | 75.9 | 97.6 |
| logn continuous | asymptotic | 100.0 | 98.9 | 100.0 |
|  | asymptotic_bonf | 100.0 | 98.9 | 100.0 |
|  | permutation_bonf | 100.0 | 98.8 | 100.0 |
| logn discrete | asymptotic | 100.0 | 99.6 | 100.0 |
|  | asymptotic_bonf | 100.0 | 99.5 | 100.0 |
|  | permutation_bonf | 100.0 | 99.5 | 100.0 |
| pwExp continuous | asymptotic | 84.0 | 61.9 | 95.2 |
|  | asymptotic_bonf | 82.7 | 61.2 | 94.5 |
|  | permutation_bonf | 83.0 | 61.9 | 94.4 |
| pwExp discrete | asymptotic | 89.0 | 69.7 | 96.9 |
|  | asymptotic_bonf | 88.4 | 68.2 | 96.8 |
|  | permutation_bonf | 87.9 | 67.6 | 96.4 |
| Weib late continuous | asymptotic | 99.9 | 99.2 | 100.0 |
|  | asymptotic_bonf | 99.9 | 99.1 | 100.0 |
|  | permutation_bonf | 99.9 | 98.9 | 100.0 |
| Weib late discrete | asymptotic | 100.0 | 99.5 | 100.0 |
|  | asymptotic_bonf | 100.0 | 99.5 | 100.0 |
|  | permutation_bonf | 100.0 | 99.5 | 100.0 |
| Weib prop continuous | asymptotic | 99.9 | 98.7 | 100.0 |
|  | asymptotic_bonf | 99.9 | 98.7 | 100.0 |
|  | permutation_bonf | 99.9 | 98.5 | 100.0 |
| Weib prop discrete | asymptotic | 100.0 | 99.6 | 100.0 |
|  | asymptotic_bonf | 100.0 | 99.4 | 100.0 |
|  | permutation_bonf | 100.0 | 99.4 | 100.0 |
| Weib scale continuous | asymptotic | 99.5 | 97.7 | 100.0 |
|  | asymptotic_bonf | 99.4 | 97.3 | 100.0 |
|  | permutation_bonf | 99.4 | 97.3 | 100.0 |
| Weib scale discrete | asymptotic | 99.9 | 98.9 | 100.0 |
|  | asymptotic_bonf | 99.9 | 98.9 | 100.0 |
|  | permutation_bonf | 99.9 | 98.8 | 100.0 |
| Weib shape continuous | asymptotic | 98.7 | 94.1 | 99.7 |
|  | asymptotic_bonf | 98.7 | 93.4 | 99.7 |
|  | permutation_bonf | 98.7 | 92.8 | 99.9 |
| Weib shape discrete | asymptotic | 99.8 | 97.2 | 100.0 |
|  | asymptotic_bonf | 99.7 | 97.1 | 100.0 |
|  | permutation_bonf | 99.6 | 97.0 | 100.0 |

Table S56: Rejection rates in percent for the Dunnett-type contrast matrix with $\delta = 1.5$ and balanced medium sample sizes under equal censoring.

| Distribution | Method | $\mathcal{H}_{0,7}$ | $\mathcal{H}_{0,8}$ | $\mathcal{H}_{0,9}$ |
|---|---|---|---|---|
| exp early continuous | asymptotic | 15.5 | 10.0 | 22.0 |
| | asymptotic_bonf | 14.4 | 9.5 | 20.6 |
| | permutation_bonf | 14.2 | 9.2 | 20.1 |
| exp early discrete | asymptotic | 18.7 | 11.3 | 26.5 |
| | asymptotic_bonf | 17.5 | 10.7 | 24.6 |
| | permutation_bonf | 17.9 | 11.0 | 23.9 |
| exp late continuous | asymptotic | 17.2 | 11.4 | 23.7 |
| | asymptotic_bonf | 15.6 | 10.3 | 22.2 |
| | permutation_bonf | 15.5 | 10.3 | 22.6 |
| exp late discrete | asymptotic | 20.7 | 13.3 | 29.1 |
| | asymptotic_bonf | 18.8 | 12.4 | 27.8 |
| | permutation_bonf | 19.8 | 12.4 | 27.6 |
| exp prop continuous | asymptotic | 17.4 | 11.6 | 22.1 |
| | asymptotic_bonf | 16.5 | 10.7 | 20.8 |
| | permutation_bonf | 15.7 | 11.3 | 20.0 |
| exp prop discrete | asymptotic | 19.5 | 13.0 | 25.5 |
| | asymptotic_bonf | 18.2 | 11.8 | 24.2 |
| | permutation_bonf | 17.3 | 12.0 | 23.2 |
| logn continuous | asymptotic | 44.5 | 30.5 | 65.4 |
| | asymptotic_bonf | 43.1 | 29.6 | 64.3 |
| | permutation_bonf | 43.8 | 29.0 | 64.3 |
| logn discrete | asymptotic | 51.8 | 38.0 | 71.7 |
| | asymptotic_bonf | 50.4 | 36.5 | 70.9 |
| | permutation_bonf | 51.4 | 36.2 | 70.5 |
| pwExp continuous | asymptotic | 15.1 | 10.2 | 21.5 |
| | asymptotic_bonf | 14.3 | 9.4 | 20.7 |
| | permutation_bonf | 14.0 | 9.5 | 20.5 |
| pwExp discrete | asymptotic | 17.8 | 11.9 | 25.4 |
| | asymptotic_bonf | 16.6 | 11.3 | 23.7 |
| | permutation_bonf | 15.9 | 10.8 | 23.5 |
| Weib late continuous | asymptotic | 45.3 | 31.4 | 65.7 |
| | asymptotic_bonf | 43.6 | 29.8 | 64.3 |
| | permutation_bonf | 42.5 | 29.3 | 63.1 |
| Weib late discrete | asymptotic | 51.9 | 36.2 | 71.1 |
| | asymptotic_bonf | 50.3 | 35.0 | 69.7 |
| | permutation_bonf | 50.2 | 34.5 | 69.1 |
| Weib prop continuous | asymptotic | 41.9 | 28.1 | 62.6 |
| | asymptotic_bonf | 41.0 | 27.2 | 61.9 |
| | permutation_bonf | 39.7 | 27.6 | 61.1 |
| Weib prop discrete | asymptotic | 51.8 | 35.9 | 71.0 |
| | asymptotic_bonf | 50.5 | 34.3 | 70.1 |
| | permutation_bonf | 50.5 | 34.6 | 68.5 |
| Weib scale continuous | asymptotic | 35.8 | 23.8 | 53.7 |
| | asymptotic_bonf | 33.9 | 22.8 | 52.4 |
| | permutation_bonf | 33.7 | 24.4 | 52.6 |
| Weib scale discrete | asymptotic | 44.6 | 29.4 | 61.4 |
| | asymptotic_bonf | 43.5 | 28.5 | 59.5 |
| | permutation_bonf | 42.4 | 28.5 | 59.3 |
| Weib shape continuous | asymptotic | 27.1 | 17.1 | 41.2 |
| | asymptotic_bonf | 25.7 | 16.3 | 40.1 |
| | permutation_bonf | 24.2 | 17.5 | 38.6 |
| Weib shape discrete | asymptotic | 33.3 | 22.9 | 50.3 |
| | asymptotic_bonf | 32.5 | 21.8 | 48.9 |
| | permutation_bonf | 32.6 | 21.9 | 47.5 |

Table S57: Rejection rates in percent for the Dunnett-type contrast matrix with $\delta = 1.5$ and balanced small sample sizes under equal censoring.

| Distribution | Method | $\mathcal{H}_{0,7}$ | $\mathcal{H}_{0,8}$ | $\mathcal{H}_{0,9}$ |
|---|---|---|---|---|
| exp early continuous | asymptotic | 3.1 | 1.3 | 3.2 |
| | asymptotic_bonf | 2.7 | 1.2 | 3.0 |
| | permutation_bonf | 2.4 | 1.0 | 2.2 |
| exp early discrete | asymptotic | 3.0 | 2.3 | 2.6 |
| | asymptotic_bonf | 2.8 | 2.2 | 2.5 |
| | permutation_bonf | 2.6 | 1.5 | 2.6 |
| exp late continuous | asymptotic | 3.4 | 1.6 | 3.6 |
| | asymptotic_bonf | 3.0 | 1.6 | 3.1 |
| | permutation_bonf | 2.6 | 1.0 | 2.6 |
| exp late discrete | asymptotic | 3.6 | 1.6 | 4.1 |
| | asymptotic_bonf | 3.5 | 1.1 | 3.8 |
| | permutation_bonf | 2.9 | 1.4 | 3.3 |
| exp prop continuous | asymptotic | 2.0 | 1.8 | 3.2 |
| | asymptotic_bonf | 1.9 | 1.7 | 3.1 |
| | permutation_bonf | 1.4 | 1.3 | 2.5 |
| exp prop discrete | asymptotic | 2.8 | 1.6 | 4.9 |
| | asymptotic_bonf | 2.5 | 1.5 | 4.5 |
| | permutation_bonf | 2.2 | 1.4 | 4.0 |
| logn continuous | asymptotic | 6.4 | 4.6 | 11.0 |
| | asymptotic_bonf | 6.0 | 4.1 | 10.4 |
| | permutation_bonf | 6.2 | 3.5 | 9.0 |
| logn discrete | asymptotic | 8.4 | 5.3 | 12.7 |
| | asymptotic_bonf | 7.9 | 5.1 | 11.6 |
| | permutation_bonf | 7.2 | 4.4 | 10.7 |
| pwExp continuous | asymptotic | 2.5 | 1.3 | 3.1 |
| | asymptotic_bonf | 2.5 | 1.3 | 2.5 |
| | permutation_bonf | 1.9 | 0.9 | 1.7 |
| pwExp discrete | asymptotic | 2.6 | 1.3 | 3.6 |
| | asymptotic_bonf | 2.4 | 1.2 | 3.4 |
| | permutation_bonf | 2.2 | 1.0 | 2.3 |
| Weib late continuous | asymptotic | 5.8 | 5.7 | 8.9 |
| | asymptotic_bonf | 5.5 | 5.4 | 8.3 |
| | permutation_bonf | 5.1 | 5.1 | 7.0 |
| Weib late discrete | asymptotic | 7.1 | 6.7 | 10.7 |
| | asymptotic_bonf | 6.8 | 6.1 | 9.7 |
| | permutation_bonf | 6.4 | 5.4 | 8.5 |
| Weib prop continuous | asymptotic | 5.5 | 5.5 | 8.0 |
| | asymptotic_bonf | 5.2 | 5.4 | 7.6 |
| | permutation_bonf | 5.0 | 4.4 | 6.0 |
| Weib prop discrete | asymptotic | 6.3 | 6.3 | 9.5 |
| | asymptotic_bonf | 6.1 | 5.8 | 9.0 |
| | permutation_bonf | 5.6 | 5.2 | 7.3 |
| Weib scale continuous | asymptotic | 4.3 | 3.7 | 5.8 |
| | asymptotic_bonf | 4.2 | 3.5 | 5.2 |
| | permutation_bonf | 3.3 | 3.4 | 4.0 |
| Weib scale discrete | asymptotic | 4.1 | 4.5 | 5.9 |
| | asymptotic_bonf | 3.8 | 4.1 | 5.3 |
| | permutation_bonf | 3.9 | 4.2 | 5.2 |
| Weib shape continuous | asymptotic | 1.7 | 2.8 | 3.6 |
| | asymptotic_bonf | 1.7 | 2.4 | 3.3 |
| | permutation_bonf | 1.6 | 2.4 | 3.2 |
| Weib shape discrete | asymptotic | 2.7 | 3.1 | 4.3 |
| | asymptotic_bonf | 2.3 | 3.0 | 3.9 |
| | permutation_bonf | 2.0 | 3.1 | 3.1 |

Table S58: Rejection rates in percent for the Dunnett-type contrast matrix with $\delta = 1.5$ and unbalanced large sample sizes under equal censoring.

| Distribution | Method | $\mathcal{H}_{0,7}$ | $\mathcal{H}_{0,8}$ | $\mathcal{H}_{0,9}$ |
|---|---|---|---|---|
| exp early continuous | asymptotic | 36.8 | 20.8 | 59.5 |
| | asymptotic_bonf | 36.1 | 20.1 | 58.4 |
| | permutation_bonf | 32.4 | 18.1 | 57.2 |
| exp early discrete | asymptotic | 42.7 | 25.5 | 65.7 |
| | asymptotic_bonf | 41.8 | 24.5 | 64.5 |
| | permutation_bonf | 38.4 | 22.0 | 64.2 |
| exp late continuous | asymptotic | 44.5 | 25.3 | 67.9 |
| | asymptotic_bonf | 42.6 | 24.2 | 67.3 |
| | permutation_bonf | 39.0 | 21.4 | 66.5 |
| exp late discrete | asymptotic | 51.1 | 30.4 | 74.1 |
| | asymptotic_bonf | 50.0 | 29.1 | 73.5 |
| | permutation_bonf | 46.6 | 25.0 | 71.0 |
| exp prop continuous | asymptotic | 42.0 | 25.2 | 60.8 |
| | asymptotic_bonf | 41.4 | 24.7 | 60.0 |
| | permutation_bonf | 39.0 | 22.2 | 58.1 |
| exp prop discrete | asymptotic | 48.8 | 30.6 | 68.2 |
| | asymptotic_bonf | 47.8 | 29.8 | 67.5 |
| | permutation_bonf | 45.6 | 26.7 | 65.1 |
| logn continuous | asymptotic | 88.4 | 68.3 | 95.5 |
| | asymptotic_bonf | 88.3 | 67.3 | 95.4 |
| | permutation_bonf | 85.9 | 64.2 | 94.9 |
| logn discrete | asymptotic | 92.7 | 76.6 | 98.0 |
| | asymptotic_bonf | 92.5 | 75.9 | 97.8 |
| | permutation_bonf | 90.2 | 71.5 | 97.1 |
| pwExp continuous | asymptotic | 34.7 | 19.8 | 59.6 |
| | asymptotic_bonf | 33.8 | 19.1 | 58.6 |
| | permutation_bonf | 32.5 | 17.0 | 57.0 |
| pwExp discrete | asymptotic | 42.0 | 23.7 | 66.1 |
| | asymptotic_bonf | 40.9 | 23.3 | 65.3 |
| | permutation_bonf | 38.6 | 20.8 | 64.1 |
| Weib late continuous | asymptotic | 88.1 | 67.8 | 97.4 |
| | asymptotic_bonf | 87.8 | 67.1 | 97.3 |
| | permutation_bonf | 85.6 | 62.5 | 96.8 |
| Weib late discrete | asymptotic | 89.9 | 73.2 | 97.1 |
| | asymptotic_bonf | 89.7 | 72.9 | 97.1 |
| | permutation_bonf | 88.2 | 69.4 | 96.4 |
| Weib prop continuous | asymptotic | 85.6 | 64.5 | 96.2 |
| | asymptotic_bonf | 85.2 | 63.4 | 95.9 |
| | permutation_bonf | 83.3 | 58.8 | 95.6 |
| Weib prop discrete | asymptotic | 90.3 | 72.6 | 97.9 |
| | asymptotic_bonf | 90.0 | 72.3 | 97.8 |
| | permutation_bonf | 87.8 | 67.9 | 97.8 |
| Weib scale continuous | asymptotic | 74.8 | 53.2 | 89.5 |
| | asymptotic_bonf | 74.4 | 52.3 | 89.2 |
| | permutation_bonf | 72.4 | 47.1 | 87.4 |
| Weib scale discrete | asymptotic | 81.7 | 60.4 | 95.2 |
| | asymptotic_bonf | 81.7 | 60.0 | 95.1 |
| | permutation_bonf | 78.4 | 55.7 | 93.2 |
| Weib shape continuous | asymptotic | 60.0 | 34.8 | 76.9 |
| | asymptotic_bonf | 59.1 | 34.5 | 76.4 |
| | permutation_bonf | 53.5 | 27.9 | 73.6 |
| Weib shape discrete | asymptotic | 70.5 | 47.1 | 85.1 |
| | asymptotic_bonf | 70.0 | 46.1 | 85.1 |
| | permutation_bonf | 65.2 | 39.3 | 83.4 |

Table S59: Rejection rates in percent for the Dunnett-type contrast matrix with $\delta = 1.5$ and unbalanced medium sample sizes under equal censoring.

| Distribution | Method | $\mathcal{H}_{0,7}$ | $\mathcal{H}_{0,8}$ | $\mathcal{H}_{0,9}$ |
|---|---|---|---|---|
| exp early continuous | asymptotic | 3.1 | 2.3 | 5.2 |
| | asymptotic_bonf | 3.0 | 2.3 | 5.0 |
| | permutation_bonf | 1.3 | 0.9 | 3.1 |
| exp early discrete | asymptotic | 3.6 | 2.6 | 6.3 |
| | asymptotic_bonf | 3.3 | 2.3 | 6.1 |
| | permutation_bonf | 1.7 | 0.8 | 4.1 |
| exp late continuous | asymptotic | 4.2 | 2.2 | 7.1 |
| | asymptotic_bonf | 3.6 | 2.2 | 6.9 |
| | permutation_bonf | 2.0 | 0.9 | 4.5 |
| exp late discrete | asymptotic | 5.2 | 2.6 | 7.7 |
| | asymptotic_bonf | 4.6 | 2.6 | 7.1 |
| | permutation_bonf | 2.5 | 1.1 | 5.3 |
| exp prop continuous | asymptotic | 4.3 | 2.4 | 5.2 |
| | asymptotic_bonf | 4.1 | 2.0 | 4.8 |
| | permutation_bonf | 2.4 | 1.0 | 3.2 |
| exp prop discrete | asymptotic | 4.9 | 2.6 | 6.2 |
| | asymptotic_bonf | 4.6 | 2.4 | 6.1 |
| | permutation_bonf | 2.4 | 0.7 | 3.7 |
| logn continuous | asymptotic | 9.9 | 6.3 | 18.2 |
| | asymptotic_bonf | 9.5 | 6.3 | 17.9 |
| | permutation_bonf | 3.8 | 0.7 | 10.2 |
| logn discrete | asymptotic | 12.1 | 7.1 | 21.5 |
| | asymptotic_bonf | 11.7 | 7.0 | 20.9 |
| | permutation_bonf | 3.9 | 0.7 | 11.3 |
| pwExp continuous | asymptotic | 3.1 | 2.1 | 4.3 |
| | asymptotic_bonf | 2.8 | 2.1 | 4.3 |
| | permutation_bonf | 1.3 | 0.8 | 3.3 |
| pwExp discrete | asymptotic | 3.9 | 2.2 | 5.9 |
| | asymptotic_bonf | 3.6 | 2.1 | 5.6 |
| | permutation_bonf | 1.6 | 0.8 | 3.9 |
| Weib late continuous | asymptotic | 12.1 | 5.8 | 22.1 |
| | asymptotic_bonf | 11.6 | 5.5 | 21.3 |
| | permutation_bonf | 5.0 | 0.7 | 13.2 |
| Weib late discrete | asymptotic | 14.0 | 7.2 | 26.6 |
| | asymptotic_bonf | 13.6 | 6.4 | 25.8 |
| | permutation_bonf | 6.4 | 0.9 | 15.9 |
| Weib prop continuous | asymptotic | 10.3 | 4.8 | 19.5 |
| | asymptotic_bonf | 9.8 | 4.6 | 18.9 |
| | permutation_bonf | 4.3 | 0.5 | 10.8 |
| Weib prop discrete | asymptotic | 11.6 | 5.9 | 22.9 |
| | asymptotic_bonf | 11.2 | 5.9 | 22.5 |
| | permutation_bonf | 4.5 | 0.5 | 11.8 |
| Weib scale continuous | asymptotic | 7.7 | 2.7 | 11.6 |
| | asymptotic_bonf | 7.6 | 2.4 | 11.1 |
| | permutation_bonf | 2.0 | 0.2 | 6.5 |
| Weib scale discrete | asymptotic | 8.6 | 3.9 | 13.3 |
| | asymptotic_bonf | 8.4 | 3.7 | 12.9 |
| | permutation_bonf | 2.0 | 0.2 | 7.3 |
| Weib shape continuous | asymptotic | 3.7 | 1.2 | 5.5 |
| | asymptotic_bonf | 3.5 | 1.2 | 5.3 |
| | permutation_bonf | 0.9 | 0.1 | 2.6 |
| Weib shape discrete | asymptotic | 4.3 | 1.9 | 7.5 |
| | asymptotic_bonf | 4.2 | 1.8 | 7.1 |
| | permutation_bonf | 0.8 | 0.0 | 3.0 |

Table S60: Rejection rates in percent for the Dunnett-type contrast matrix with $\delta = 1.5$ and unbalanced small sample sizes under equal censoring.

| Distribution | Method | $\mathcal{H}_{0,7}$ | $\mathcal{H}_{0,8}$ | $\mathcal{H}_{0,9}$ |
|---|---|---|---|---|
| exp early continuous | asymptotic | 1.1 | 3.3 | 2.3 |
| | asymptotic_bonf | 1.0 | 3.2 | 2.3 |
| | permutation_bonf | 0.2 | 0.2 | 0.2 |
| exp early discrete | asymptotic | 1.2 | 3.1 | 2.4 |
| | asymptotic_bonf | 1.1 | 3.1 | 2.3 |
| | permutation_bonf | 0.1 | 0.3 | 0.2 |
| exp late continuous | asymptotic | 1.7 | 2.2 | 2.9 |
| | asymptotic_bonf | 1.4 | 2.2 | 2.7 |
| | permutation_bonf | 0.1 | 0.1 | 0.1 |
| exp late discrete | asymptotic | 1.8 | 2.2 | 3.2 |
| | asymptotic_bonf | 1.7 | 2.1 | 3.0 |
| | permutation_bonf | 0.0 | 0.2 | 0.2 |
| exp prop continuous | asymptotic | 1.8 | 2.1 | 2.5 |
| | asymptotic_bonf | 1.6 | 2.1 | 2.3 |
| | permutation_bonf | 0.2 | 0.3 | 0.4 |
| exp prop discrete | asymptotic | 1.8 | 2.6 | 2.6 |
| | asymptotic_bonf | 1.7 | 2.6 | 2.6 |
| | permutation_bonf | 0.1 | 0.3 | 0.4 |
| logn continuous | asymptotic | 3.4 | 4.8 | 3.1 |
| | asymptotic_bonf | 3.3 | 4.7 | 3.1 |
| | permutation_bonf | 0.3 | 0.3 | 0.2 |
| logn discrete | asymptotic | 3.5 | 5.3 | 3.0 |
| | asymptotic_bonf | 3.4 | 5.3 | 3.0 |
| | permutation_bonf | 0.2 | 0.2 | 0.1 |
| pwExp continuous | asymptotic | 1.4 | 2.4 | 2.3 |
| | asymptotic_bonf | 1.2 | 2.4 | 2.0 |
| | permutation_bonf | 0.1 | 0.1 | 0.3 |
| pwExp discrete | asymptotic | 1.3 | 2.6 | 2.5 |
| | asymptotic_bonf | 1.3 | 2.6 | 2.4 |
| | permutation_bonf | 0.1 | 0.1 | 0.4 |
| Weib late continuous | asymptotic | 2.8 | 2.2 | 3.2 |
| | asymptotic_bonf | 2.7 | 2.2 | 3.2 |
| | permutation_bonf | 0.1 | 0.1 | 0.3 |
| Weib late discrete | asymptotic | 2.9 | 3.2 | 4.0 |
| | asymptotic_bonf | 2.8 | 3.2 | 3.8 |
| | permutation_bonf | 0.0 | 0.1 | 0.2 |
| Weib prop continuous | asymptotic | 2.5 | 3.3 | 3.2 |
| | asymptotic_bonf | 2.5 | 3.3 | 3.0 |
| | permutation_bonf | 0.1 | 0.1 | 0.3 |
| Weib prop discrete | asymptotic | 2.8 | 3.7 | 3.4 |
| | asymptotic_bonf | 2.6 | 3.6 | 3.4 |
| | permutation_bonf | 0.1 | 0.1 | 0.2 |
| Weib scale continuous | asymptotic | 3.3 | 4.6 | 2.5 |
| | asymptotic_bonf | 3.3 | 4.5 | 2.5 |
| | permutation_bonf | 0.2 | 0.4 | 0.5 |
| Weib scale discrete | asymptotic | 3.5 | 5.7 | 2.9 |
| | asymptotic_bonf | 3.5 | 5.7 | 2.9 |
| | permutation_bonf | 0.2 | 0.5 | 0.3 |
| Weib shape continuous | asymptotic | 6.2 | 9.3 | 3.5 |
| | asymptotic_bonf | 6.1 | 9.3 | 3.3 |
| | permutation_bonf | 1.3 | 1.0 | 1.2 |
| Weib shape discrete | asymptotic | 6.3 | 9.5 | 3.7 |
| | asymptotic_bonf | 6.3 | 9.5 | 3.5 |
| | permutation_bonf | 0.7 | 1.0 | 1.0 |

Table S61: Rejection rates in percent for the Dunnett-type contrast matrix with $\delta = 1.5$ and balanced large sample sizes under unequal, high censoring.

| Distribution | Method | $\mathcal{H}_{0,7}$ | $\mathcal{H}_{0,8}$ | $\mathcal{H}_{0,9}$ |
|---|---|---|---|---|
| exp early continuous | asymptotic | 74.9 | 56.0 | 89.5 |
| | asymptotic_bonf | 73.2 | 54.1 | 88.8 |
| | permutation_bonf | 72.8 | 52.6 | 88.5 |
| exp early discrete | asymptotic | 78.5 | 59.2 | 91.4 |
| | asymptotic_bonf | 77.1 | 57.0 | 91.0 |
| | permutation_bonf | 77.1 | 56.9 | 90.5 |
| exp late continuous | asymptotic | 77.3 | 57.1 | 90.3 |
| | asymptotic_bonf | 75.3 | 54.6 | 89.5 |
| | permutation_bonf | 73.8 | 53.5 | 89.5 |
| exp late discrete | asymptotic | 79.7 | 60.8 | 92.1 |
| | asymptotic_bonf | 78.5 | 59.5 | 91.6 |
| | permutation_bonf | 78.4 | 58.5 | 91.5 |
| exp prop continuous | asymptotic | 75.1 | 57.2 | 89.7 |
| | asymptotic_bonf | 74.0 | 55.5 | 89.3 |
| | permutation_bonf | 73.5 | 54.6 | 89.0 |
| exp prop discrete | asymptotic | 79.3 | 61.5 | 92.6 |
| | asymptotic_bonf | 78.2 | 59.2 | 91.6 |
| | permutation_bonf | 78.5 | 58.7 | 92.0 |
| logn continuous | asymptotic | 99.2 | 93.3 | 99.8 |
| | asymptotic_bonf | 98.7 | 92.9 | 99.8 |
| | permutation_bonf | 98.7 | 92.8 | 99.7 |
| logn discrete | asymptotic | 99.6 | 96.5 | 100.0 |
| | asymptotic_bonf | 99.5 | 96.1 | 100.0 |
| | permutation_bonf | 99.8 | 96.4 | 100.0 |
| pwExp continuous | asymptotic | 73.2 | 53.7 | 89.2 |
| | asymptotic_bonf | 71.1 | 51.3 | 88.5 |
| | permutation_bonf | 70.4 | 51.7 | 88.2 |
| pwExp discrete | asymptotic | 77.8 | 57.8 | 91.6 |
| | asymptotic_bonf | 75.7 | 56.6 | 90.6 |
| | permutation_bonf | 75.1 | 55.4 | 90.7 |
| Weib late continuous | asymptotic | 98.9 | 93.1 | 100.0 |
| | asymptotic_bonf | 98.8 | 92.7 | 100.0 |
| | permutation_bonf | 98.6 | 92.3 | 100.0 |
| Weib late discrete | asymptotic | 99.7 | 96.4 | 100.0 |
| | asymptotic_bonf | 99.6 | 96.4 | 100.0 |
| | permutation_bonf | 99.5 | 96.1 | 99.9 |
| Weib prop continuous | asymptotic | 98.5 | 92.3 | 100.0 |
| | asymptotic_bonf | 98.3 | 91.8 | 100.0 |
| | permutation_bonf | 98.4 | 91.9 | 100.0 |
| Weib prop discrete | asymptotic | 99.5 | 95.9 | 100.0 |
| | asymptotic_bonf | 99.5 | 95.9 | 100.0 |
| | permutation_bonf | 99.5 | 95.6 | 100.0 |
| Weib scale continuous | asymptotic | 98.2 | 89.4 | 99.8 |
| | asymptotic_bonf | 97.9 | 88.9 | 99.8 |
| | permutation_bonf | 97.5 | 89.1 | 99.8 |
| Weib scale discrete | asymptotic | 99.2 | 94.5 | 100.0 |
| | asymptotic_bonf | 99.0 | 93.9 | 100.0 |
| | permutation_bonf | 98.9 | 94.1 | 100.0 |
| Weib shape continuous | asymptotic | 97.0 | 85.8 | 99.4 |
| | asymptotic_bonf | 96.4 | 85.0 | 99.3 |
| | permutation_bonf | 96.3 | 85.0 | 99.3 |
| Weib shape discrete | asymptotic | 98.5 | 92.0 | 99.8 |
| | asymptotic_bonf | 98.4 | 91.2 | 99.8 |
| | permutation_bonf | 97.9 | 90.8 | 99.7 |

Table S62: Rejection rates in percent for the Dunnett-type contrast matrix with $\delta = 1.5$ and balanced medium sample sizes under unequal, high censoring.

| Distribution | Method | $\mathcal{H}_{0,7}$ | $\mathcal{H}_{0,8}$ | $\mathcal{H}_{0,9}$ |
|---|---|---|---|---|
| exp early continuous | asymptotic | 13.0 | 7.7 | 16.4 |
| | asymptotic_bonf | 12.3 | 6.9 | 15.2 |
| | permutation_bonf | 11.9 | 6.8 | 14.6 |
| exp early discrete | asymptotic | 13.9 | 7.4 | 18.6 |
| | asymptotic_bonf | 13.3 | 6.7 | 17.5 |
| | permutation_bonf | 13.3 | 7.1 | 17.3 |
| exp late continuous | asymptotic | 12.0 | 8.2 | 17.7 |
| | asymptotic_bonf | 11.5 | 7.8 | 16.9 |
| | permutation_bonf | 11.2 | 7.5 | 16.6 |
| exp late discrete | asymptotic | 15.1 | 7.6 | 19.7 |
| | asymptotic_bonf | 13.8 | 7.0 | 18.4 |
| | permutation_bonf | 13.6 | 7.0 | 18.5 |
| exp prop continuous | asymptotic | 13.2 | 9.2 | 17.7 |
| | asymptotic_bonf | 12.4 | 8.6 | 16.1 |
| | permutation_bonf | 11.9 | 8.7 | 16.2 |
| exp prop discrete | asymptotic | 14.9 | 9.7 | 19.0 |
| | asymptotic_bonf | 14.1 | 9.1 | 17.9 |
| | permutation_bonf | 14.2 | 8.7 | 17.6 |
| logn continuous | asymptotic | 27.6 | 18.6 | 46.1 |
| | asymptotic_bonf | 26.2 | 17.2 | 45.5 |
| | permutation_bonf | 26.0 | 16.6 | 44.0 |
| logn discrete | asymptotic | 33.6 | 21.4 | 54.5 |
| | asymptotic_bonf | 32.3 | 20.8 | 53.2 |
| | permutation_bonf | 32.0 | 20.9 | 51.3 |
| pwExp continuous | asymptotic | 11.6 | 7.4 | 17.1 |
| | asymptotic_bonf | 10.5 | 7.1 | 16.4 |
| | permutation_bonf | 10.2 | 7.0 | 15.3 |
| pwExp discrete | asymptotic | 14.0 | 7.7 | 18.9 |
| | asymptotic_bonf | 13.1 | 7.4 | 17.8 |
| | permutation_bonf | 13.7 | 7.5 | 17.7 |
| Weib late continuous | asymptotic | 30.2 | 18.5 | 46.3 |
| | asymptotic_bonf | 29.1 | 17.9 | 44.7 |
| | permutation_bonf | 27.9 | 17.3 | 45.0 |
| Weib late discrete | asymptotic | 34.5 | 22.2 | 50.8 |
| | asymptotic_bonf | 33.1 | 20.7 | 49.5 |
| | permutation_bonf | 32.5 | 19.8 | 48.2 |
| Weib prop continuous | asymptotic | 29.6 | 17.2 | 43.9 |
| | asymptotic_bonf | 28.4 | 16.1 | 43.0 |
| | permutation_bonf | 26.1 | 17.1 | 41.8 |
| Weib prop discrete | asymptotic | 34.7 | 21.8 | 52.6 |
| | asymptotic_bonf | 33.1 | 20.4 | 51.8 |
| | permutation_bonf | 32.4 | 20.9 | 50.7 |
| Weib scale continuous | asymptotic | 24.4 | 15.3 | 39.6 |
| | asymptotic_bonf | 23.2 | 14.7 | 38.7 |
| | permutation_bonf | 22.7 | 14.2 | 38.4 |
| Weib scale discrete | asymptotic | 29.6 | 18.6 | 47.8 |
| | asymptotic_bonf | 28.4 | 18.1 | 46.2 |
| | permutation_bonf | 27.8 | 18.0 | 46.0 |
| Weib shape continuous | asymptotic | 20.5 | 13.0 | 34.0 |
| | asymptotic_bonf | 19.8 | 11.6 | 33.0 |
| | permutation_bonf | 18.5 | 12.3 | 31.4 |
| Weib shape discrete | asymptotic | 24.8 | 15.1 | 39.1 |
| | asymptotic_bonf | 24.2 | 14.5 | 38.3 |
| | permutation_bonf | 23.2 | 14.4 | 38.2 |

Table S63: Rejection rates in percent for the Dunnett-type contrast matrix with $\delta = 1.5$ and balanced small sample sizes under unequal, high censoring.

| Distribution | Method | $\mathcal{H}_{0,7}$ | $\mathcal{H}_{0,8}$ | $\mathcal{H}_{0,9}$ |
|---|---|---|---|---|
| exp early continuous | asymptotic | 3.0 | 1.7 | 3.6 |
| | asymptotic_bonf | 2.6 | 1.7 | 3.4 |
| | permutation_bonf | 2.3 | 1.4 | 3.0 |
| exp early discrete | asymptotic | 2.3 | 1.9 | 2.2 |
| | asymptotic_bonf | 2.1 | 1.7 | 2.0 |
| | permutation_bonf | 1.9 | 1.6 | 1.7 |
| exp late continuous | asymptotic | 2.8 | 2.1 | 4.4 |
| | asymptotic_bonf | 2.7 | 1.7 | 3.9 |
| | permutation_bonf | 2.0 | 1.6 | 2.7 |
| exp late discrete | asymptotic | 3.3 | 1.7 | 3.8 |
| | asymptotic_bonf | 3.1 | 1.6 | 3.5 |
| | permutation_bonf | 2.5 | 1.3 | 2.9 |
| exp prop continuous | asymptotic | 1.8 | 2.0 | 3.7 |
| | asymptotic_bonf | 1.6 | 2.0 | 3.5 |
| | permutation_bonf | 1.0 | 1.9 | 2.3 |
| exp prop discrete | asymptotic | 2.6 | 1.7 | 3.8 |
| | asymptotic_bonf | 2.4 | 1.3 | 3.5 |
| | permutation_bonf | 1.6 | 1.2 | 2.9 |
| logn continuous | asymptotic | 4.7 | 3.1 | 8.0 |
| | asymptotic_bonf | 4.5 | 2.7 | 7.6 |
| | permutation_bonf | 4.4 | 2.3 | 6.5 |
| logn discrete | asymptotic | 5.6 | 3.9 | 8.2 |
| | asymptotic_bonf | 5.4 | 3.5 | 8.0 |
| | permutation_bonf | 4.5 | 3.5 | 6.5 |
| pwExp continuous | asymptotic | 2.3 | 1.8 | 3.7 |
| | asymptotic_bonf | 2.2 | 1.8 | 3.4 |
| | permutation_bonf | 1.7 | 1.5 | 2.2 |
| pwExp discrete | asymptotic | 2.8 | 1.9 | 4.2 |
| | asymptotic_bonf | 2.6 | 1.9 | 4.0 |
| | permutation_bonf | 2.1 | 1.6 | 3.0 |
| Weib late continuous | asymptotic | 3.7 | 4.4 | 7.6 |
| | asymptotic_bonf | 3.5 | 4.3 | 7.2 |
| | permutation_bonf | 2.1 | 3.5 | 5.7 |
| Weib late discrete | asymptotic | 4.4 | 5.4 | 9.0 |
| | asymptotic_bonf | 4.0 | 5.2 | 8.6 |
| | permutation_bonf | 2.6 | 4.6 | 7.3 |
| Weib prop continuous | asymptotic | 3.2 | 4.5 | 7.5 |
| | asymptotic_bonf | 2.9 | 4.4 | 6.9 |
| | permutation_bonf | 2.5 | 3.2 | 5.1 |
| Weib prop discrete | asymptotic | 4.1 | 5.0 | 8.6 |
| | asymptotic_bonf | 3.7 | 4.8 | 8.4 |
| | permutation_bonf | 2.6 | 3.7 | 6.4 |
| Weib scale continuous | asymptotic | 2.1 | 3.4 | 4.9 |
| | asymptotic_bonf | 2.0 | 3.3 | 4.8 |
| | permutation_bonf | 1.5 | 3.2 | 3.7 |
| Weib scale discrete | asymptotic | 2.3 | 4.1 | 5.5 |
| | asymptotic_bonf | 2.1 | 3.7 | 5.1 |
| | permutation_bonf | 1.6 | 3.5 | 4.3 |
| Weib shape continuous | asymptotic | 1.6 | 2.3 | 3.8 |
| | asymptotic_bonf | 1.5 | 2.1 | 3.6 |
| | permutation_bonf | 1.4 | 1.8 | 2.8 |
| Weib shape discrete | asymptotic | 1.3 | 2.7 | 4.4 |
| | asymptotic_bonf | 1.3 | 2.4 | 4.0 |
| | permutation_bonf | 1.0 | 2.2 | 3.7 |

Table S64: Rejection rates in percent for the Dunnett-type contrast matrix with $\delta = 1.5$ and unbalanced large sample sizes under unequal, high censoring.

| Distribution | Method | $\mathcal{H}_{0,7}$ | $\mathcal{H}_{0,8}$ | $\mathcal{H}_{0,9}$ |
|---|---|---|---|---|
| exp early continuous | asymptotic | 30.9 | 17.7 | 53.0 |
| | asymptotic_bonf | 30.1 | 16.9 | 52.0 |
| | permutation_bonf | 28.0 | 14.3 | 50.3 |
| exp early discrete | asymptotic | 34.0 | 19.8 | 56.3 |
| | asymptotic_bonf | 33.2 | 19.0 | 55.8 |
| | permutation_bonf | 31.5 | 17.4 | 54.0 |
| exp late continuous | asymptotic | 32.8 | 18.7 | 57.5 |
| | asymptotic_bonf | 31.7 | 17.6 | 56.0 |
| | permutation_bonf | 30.2 | 15.5 | 54.3 |
| exp late discrete | asymptotic | 37.0 | 20.5 | 61.1 |
| | asymptotic_bonf | 36.3 | 19.9 | 60.2 |
| | permutation_bonf | 33.6 | 16.9 | 57.9 |
| exp prop continuous | asymptotic | 32.2 | 20.4 | 49.3 |
| | asymptotic_bonf | 31.1 | 19.5 | 48.4 |
| | permutation_bonf | 28.8 | 18.3 | 46.8 |
| exp prop discrete | asymptotic | 35.5 | 22.7 | 54.1 |
| | asymptotic_bonf | 34.5 | 22.5 | 53.1 |
| | permutation_bonf | 32.1 | 18.7 | 50.7 |
| logn continuous | asymptotic | 71.7 | 47.6 | 84.4 |
| | asymptotic_bonf | 70.9 | 47.1 | 84.0 |
| | permutation_bonf | 66.3 | 39.0 | 81.9 |
| logn discrete | asymptotic | 77.1 | 54.7 | 90.7 |
| | asymptotic_bonf | 76.7 | 53.9 | 90.6 |
| | permutation_bonf | 71.6 | 44.2 | 86.6 |
| pwExp continuous | asymptotic | 28.7 | 16.6 | 51.4 |
| | asymptotic_bonf | 28.0 | 15.5 | 50.7 |
| | permutation_bonf | 26.5 | 13.4 | 49.0 |
| pwExp discrete | asymptotic | 33.3 | 18.2 | 56.5 |
| | asymptotic_bonf | 32.5 | 17.6 | 55.6 |
| | permutation_bonf | 30.0 | 14.4 | 54.6 |
| Weib late continuous | asymptotic | 71.7 | 48.6 | 87.7 |
| | asymptotic_bonf | 70.6 | 47.5 | 87.1 |
| | permutation_bonf | 65.2 | 38.8 | 85.9 |
| Weib late discrete | asymptotic | 74.6 | 52.4 | 89.3 |
| | asymptotic_bonf | 74.2 | 51.7 | 88.9 |
| | permutation_bonf | 68.4 | 42.3 | 87.3 |
| Weib prop continuous | asymptotic | 68.0 | 46.1 | 85.1 |
| | asymptotic_bonf | 67.3 | 44.9 | 85.0 |
| | permutation_bonf | 62.5 | 36.0 | 83.2 |
| Weib prop discrete | asymptotic | 75.6 | 51.9 | 90.5 |
| | asymptotic_bonf | 74.8 | 50.9 | 90.2 |
| | permutation_bonf | 69.9 | 42.7 | 88.3 |
| Weib scale continuous | asymptotic | 60.8 | 39.0 | 78.0 |
| | asymptotic_bonf | 59.9 | 38.4 | 77.2 |
| | permutation_bonf | 55.8 | 30.8 | 74.1 |
| Weib scale discrete | asymptotic | 67.2 | 44.2 | 83.5 |
| | asymptotic_bonf | 66.7 | 43.4 | 83.2 |
| | permutation_bonf | 60.4 | 35.2 | 80.3 |
| Weib shape continuous | asymptotic | 50.9 | 30.2 | 69.3 |
| | asymptotic_bonf | 50.6 | 29.6 | 68.8 |
| | permutation_bonf | 45.1 | 22.7 | 64.2 |
| Weib shape discrete | asymptotic | 58.4 | 36.5 | 75.6 |
| | asymptotic_bonf | 57.7 | 36.0 | 74.6 |
| | permutation_bonf | 50.1 | 26.3 | 70.7 |

Table S65: Rejection rates in percent for the Dunnett-type contrast matrix with $\delta = 1.5$ and unbalanced medium sample sizes under unequal, high censoring.

| Distribution | Method | $\mathcal{H}_{0,7}$ | $\mathcal{H}_{0,8}$ | $\mathcal{H}_{0,9}$ |
|---|---|---|---|---|
| exp early continuous | asymptotic | 2.8 | 2.3 | 4.4 |
| | asymptotic_bonf | 2.6 | 2.2 | 4.3 |
| | permutation_bonf | 0.6 | 0.4 | 2.7 |
| exp early discrete | asymptotic | 3.1 | 1.7 | 5.3 |
| | asymptotic_bonf | 3.1 | 1.6 | 4.9 |
| | permutation_bonf | 0.7 | 0.3 | 2.9 |
| exp late continuous | asymptotic | 3.7 | 2.1 | 4.4 |
| | asymptotic_bonf | 3.6 | 2.0 | 4.3 |
| | permutation_bonf | 0.7 | 0.4 | 3.0 |
| exp late discrete | asymptotic | 3.3 | 1.8 | 5.6 |
| | asymptotic_bonf | 3.1 | 1.7 | 5.1 |
| | permutation_bonf | 1.2 | 0.3 | 3.3 |
| exp prop continuous | asymptotic | 3.1 | 2.1 | 3.9 |
| | asymptotic_bonf | 2.8 | 1.9 | 3.9 |
| | permutation_bonf | 1.4 | 0.5 | 2.5 |
| exp prop discrete | asymptotic | 3.6 | 2.4 | 4.8 |
| | asymptotic_bonf | 3.4 | 2.2 | 4.5 |
| | permutation_bonf | 1.5 | 0.3 | 2.2 |
| logn continuous | asymptotic | 5.4 | 4.0 | 10.9 |
| | asymptotic_bonf | 5.2 | 3.9 | 10.6 |
| | permutation_bonf | 1.1 | 0.0 | 3.4 |
| logn discrete | asymptotic | 6.4 | 4.2 | 12.6 |
| | asymptotic_bonf | 6.2 | 4.2 | 12.1 |
| | permutation_bonf | 0.8 | 0.0 | 3.4 |
| pwExp continuous | asymptotic | 2.4 | 1.8 | 4.8 |
| | asymptotic_bonf | 2.3 | 1.8 | 4.3 |
| | permutation_bonf | 0.5 | 0.6 | 2.5 |
| pwExp discrete | asymptotic | 2.5 | 1.5 | 5.2 |
| | asymptotic_bonf | 2.4 | 1.5 | 4.6 |
| | permutation_bonf | 0.8 | 0.4 | 3.1 |
| Weib late continuous | asymptotic | 7.3 | 3.3 | 12.8 |
| | asymptotic_bonf | 6.6 | 3.0 | 12.4 |
| | permutation_bonf | 1.4 | 0.0 | 4.2 |
| Weib late discrete | asymptotic | 7.7 | 3.2 | 16.6 |
| | asymptotic_bonf | 7.6 | 3.1 | 15.4 |
| | permutation_bonf | 0.7 | 0.1 | 5.3 |
| Weib prop continuous | asymptotic | 6.8 | 2.2 | 11.4 |
| | asymptotic_bonf | 6.7 | 2.0 | 11.3 |
| | permutation_bonf | 1.2 | 0.0 | 4.0 |
| Weib prop discrete | asymptotic | 7.8 | 2.9 | 14.6 |
| | asymptotic_bonf | 7.3 | 2.5 | 14.1 |
| | permutation_bonf | 1.0 | 0.1 | 3.7 |
| Weib scale continuous | asymptotic | 5.0 | 1.9 | 7.6 |
| | asymptotic_bonf | 4.7 | 1.7 | 7.6 |
| | permutation_bonf | 0.8 | 0.1 | 1.9 |
| Weib scale discrete | asymptotic | 5.3 | 1.7 | 9.1 |
| | asymptotic_bonf | 5.2 | 1.6 | 8.5 |
| | permutation_bonf | 0.5 | 0.1 | 1.5 |
| Weib shape continuous | asymptotic | 3.1 | 1.8 | 5.3 |
| | asymptotic_bonf | 3.0 | 1.8 | 4.9 |
| | permutation_bonf | 0.4 | 0.0 | 1.6 |
| Weib shape discrete | asymptotic | 3.6 | 1.9 | 5.9 |
| | asymptotic_bonf | 3.6 | 1.9 | 5.6 |
| | permutation_bonf | 0.3 | 0.0 | 1.2 |

Table S66: Rejection rates in percent for the Dunnett-type contrast matrix with $\delta = 1.5$ and unbalanced small sample sizes under unequal, high censoring.

| Distribution | Method | $\mathcal{H}_{0,7}$ | $\mathcal{H}_{0,8}$ | $\mathcal{H}_{0,9}$ |
|---|---|---|---|---|
| exp early continuous | asymptotic | 1.3 | 4.5 | 2.0 |
| | asymptotic_bonf | 1.3 | 4.3 | 1.9 |
| | permutation_bonf | 0.2 | 0.2 | 0.1 |
| exp early discrete | asymptotic | 2.2 | 5.5 | 2.1 |
| | asymptotic_bonf | 2.2 | 5.5 | 2.0 |
| | permutation_bonf | 0.3 | 0.1 | 0.1 |
| exp late continuous | asymptotic | 1.8 | 4.1 | 2.0 |
| | asymptotic_bonf | 1.7 | 4.0 | 1.9 |
| | permutation_bonf | 0.2 | 0.1 | 0.1 |
| exp late discrete | asymptotic | 2.1 | 4.5 | 2.7 |
| | asymptotic_bonf | 2.0 | 4.3 | 2.5 |
| | permutation_bonf | 0.3 | 0.0 | 0.1 |
| exp prop continuous | asymptotic | 2.0 | 4.8 | 2.6 |
| | asymptotic_bonf | 2.0 | 4.6 | 2.6 |
| | permutation_bonf | 0.3 | 0.4 | 0.2 |
| exp prop discrete | asymptotic | 2.8 | 5.2 | 2.7 |
| | asymptotic_bonf | 2.8 | 5.2 | 2.7 |
| | permutation_bonf | 0.4 | 0.2 | 0.2 |
| logn continuous | asymptotic | 5.5 | 11.5 | 3.9 |
| | asymptotic_bonf | 5.5 | 11.2 | 3.9 |
| | permutation_bonf | 0.5 | 0.6 | 0.2 |
| logn discrete | asymptotic | 6.1 | 12.1 | 4.6 |
| | asymptotic_bonf | 6.0 | 12.1 | 4.4 |
| | permutation_bonf | 0.4 | 0.8 | 0.2 |
| pwExp continuous | asymptotic | 1.5 | 4.0 | 2.0 |
| | asymptotic_bonf | 1.4 | 3.9 | 1.8 |
| | permutation_bonf | 0.0 | 0.2 | 0.2 |
| pwExp discrete | asymptotic | 1.8 | 4.8 | 2.0 |
| | asymptotic_bonf | 1.7 | 4.8 | 1.9 |
| | permutation_bonf | 0.0 | 0.1 | 0.1 |
| Weib late continuous | asymptotic | 5.2 | 6.9 | 4.1 |
| | asymptotic_bonf | 5.2 | 6.8 | 3.6 |
| | permutation_bonf | 0.2 | 0.2 | 0.5 |
| Weib late discrete | asymptotic | 7.0 | 7.9 | 4.9 |
| | asymptotic_bonf | 6.9 | 7.5 | 4.8 |
| | permutation_bonf | 0.2 | 0.4 | 0.4 |
| Weib prop continuous | asymptotic | 5.9 | 7.3 | 3.5 |
| | asymptotic_bonf | 5.5 | 7.3 | 3.3 |
| | permutation_bonf | 0.3 | 0.1 | 0.5 |
| Weib prop discrete | asymptotic | 7.2 | 9.0 | 5.0 |
| | asymptotic_bonf | 7.1 | 8.6 | 4.8 |
| | permutation_bonf | 0.3 | 0.2 | 0.6 |
| Weib scale continuous | asymptotic | 6.8 | 10.5 | 4.4 |
| | asymptotic_bonf | 6.8 | 10.5 | 4.4 |
| | permutation_bonf | 0.4 | 0.4 | 0.4 |
| Weib scale discrete | asymptotic | 8.4 | 11.5 | 4.5 |
| | asymptotic_bonf | 8.2 | 11.4 | 4.4 |
| | permutation_bonf | 0.2 | 0.2 | 0.5 |
| Weib shape continuous | asymptotic | 7.7 | 12.4 | 4.0 |
| | asymptotic_bonf | 7.7 | 12.4 | 4.0 |
| | permutation_bonf | 0.9 | 0.4 | 0.4 |
| Weib shape discrete | asymptotic | 9.3 | 14.1 | 5.3 |
| | asymptotic_bonf | 9.3 | 14.0 | 5.2 |
| | permutation_bonf | 0.7 | 0.5 | 1.1 |

Table S67: Rejection rates in percent for the Dunnett-type contrast matrix with $\delta = 1.5$ and balanced large sample sizes under unequal, low censoring.

| Distribution | Method | $\mathcal{H}_{0,7}$ | $\mathcal{H}_{0,8}$ | $\mathcal{H}_{0,9}$ |
|---|---|---|---|---|
| exp early continuous | asymptotic | 84.3 | 64.0 | 94.3 |
| | asymptotic_bonf | 83.7 | 61.8 | 94.0 |
| | permutation_bonf | 83.0 | 62.4 | 93.6 |
| exp early discrete | asymptotic | 89.0 | 71.3 | 96.4 |
| | asymptotic_bonf | 88.2 | 69.5 | 96.1 |
| | permutation_bonf | 88.1 | 69.6 | 95.6 |
| exp late continuous | asymptotic | 86.6 | 68.1 | 95.9 |
| | asymptotic_bonf | 85.9 | 66.0 | 95.6 |
| | permutation_bonf | 85.5 | 65.4 | 95.8 |
| exp late discrete | asymptotic | 91.5 | 74.7 | 97.7 |
| | asymptotic_bonf | 90.7 | 72.6 | 97.5 |
| | permutation_bonf | 89.9 | 73.0 | 97.4 |
| exp prop continuous | asymptotic | 85.3 | 69.1 | 96.1 |
| | asymptotic_bonf | 84.2 | 67.4 | 95.6 |
| | permutation_bonf | 82.8 | 67.1 | 95.7 |
| exp prop discrete | asymptotic | 90.0 | 75.5 | 97.9 |
| | asymptotic_bonf | 89.4 | 74.5 | 97.6 |
| | permutation_bonf | 88.7 | 74.5 | 97.6 |
| logn continuous | asymptotic | 100.0 | 98.7 | 100.0 |
| | asymptotic_bonf | 100.0 | 98.4 | 100.0 |
| | permutation_bonf | 100.0 | 98.3 | 100.0 |
| logn discrete | asymptotic | 100.0 | 99.8 | 100.0 |
| | asymptotic_bonf | 100.0 | 99.7 | 100.0 |
| | permutation_bonf | 100.0 | 99.8 | 100.0 |
| pwExp continuous | asymptotic | 83.7 | 62.2 | 94.7 |
| | asymptotic_bonf | 82.4 | 60.2 | 94.1 |
| | permutation_bonf | 81.6 | 59.9 | 93.9 |
| pwExp discrete | asymptotic | 87.6 | 69.6 | 96.2 |
| | asymptotic_bonf | 86.9 | 68.3 | 96.0 |
| | permutation_bonf | 86.6 | 68.7 | 96.2 |
| Weib late continuous | asymptotic | 100.0 | 99.4 | 100.0 |
| | asymptotic_bonf | 100.0 | 99.2 | 100.0 |
| | permutation_bonf | 100.0 | 99.2 | 100.0 |
| Weib late discrete | asymptotic | 100.0 | 99.8 | 100.0 |
| | asymptotic_bonf | 100.0 | 99.8 | 100.0 |
| | permutation_bonf | 100.0 | 99.8 | 100.0 |
| Weib prop continuous | asymptotic | 100.0 | 99.0 | 100.0 |
| | asymptotic_bonf | 99.9 | 98.8 | 100.0 |
| | permutation_bonf | 99.9 | 98.6 | 100.0 |
| Weib prop discrete | asymptotic | 100.0 | 99.8 | 100.0 |
| | asymptotic_bonf | 100.0 | 99.8 | 100.0 |
| | permutation_bonf | 100.0 | 99.8 | 100.0 |
| Weib scale continuous | asymptotic | 99.6 | 97.8 | 100.0 |
| | asymptotic_bonf | 99.6 | 97.7 | 100.0 |
| | permutation_bonf | 99.6 | 97.6 | 100.0 |
| Weib scale discrete | asymptotic | 100.0 | 99.3 | 100.0 |
| | asymptotic_bonf | 99.9 | 99.3 | 100.0 |
| | permutation_bonf | 99.8 | 99.1 | 100.0 |
| Weib shape continuous | asymptotic | 98.9 | 92.7 | 99.8 |
| | asymptotic_bonf | 98.6 | 92.4 | 99.8 |
| | permutation_bonf | 98.5 | 91.9 | 99.7 |
| Weib shape discrete | asymptotic | 99.6 | 97.4 | 100.0 |
| | asymptotic_bonf | 99.4 | 97.2 | 100.0 |
| | permutation_bonf | 99.6 | 97.4 | 100.0 |

Table S68: Rejection rates in percent for the Dunnett-type contrast matrix with $\delta = 1.5$ and balanced medium sample sizes under unequal, low censoring.

| Distribution | Method | $\mathcal{H}_{0,7}$ | $\mathcal{H}_{0,8}$ | $\mathcal{H}_{0,9}$ |
|---|---|---|---|---|
| exp early continuous | asymptotic | 14.6 | 10.3 | 21.0 |
| | asymptotic_bonf | 13.5 | 9.5 | 19.0 |
| | permutation_bonf | 13.8 | 9.6 | 19.9 |
| exp early discrete | asymptotic | 17.6 | 12.0 | 25.6 |
| | asymptotic_bonf | 16.3 | 10.9 | 24.1 |
| | permutation_bonf | 17.6 | 11.0 | 23.1 |
| exp late continuous | asymptotic | 15.9 | 11.1 | 23.8 |
| | asymptotic_bonf | 15.0 | 10.7 | 21.9 |
| | permutation_bonf | 15.5 | 11.0 | 22.2 |
| exp late discrete | asymptotic | 19.9 | 12.6 | 28.3 |
| | asymptotic_bonf | 18.5 | 12.0 | 26.6 |
| | permutation_bonf | 19.2 | 12.3 | 26.7 |
| exp prop continuous | asymptotic | 16.2 | 10.9 | 21.1 |
| | asymptotic_bonf | 15.3 | 10.2 | 19.9 |
| | permutation_bonf | 15.3 | 10.8 | 19.9 |
| exp prop discrete | asymptotic | 18.1 | 12.5 | 24.8 |
| | asymptotic_bonf | 17.7 | 11.7 | 23.7 |
| | permutation_bonf | 16.6 | 11.3 | 23.2 |
| logn continuous | asymptotic | 45.0 | 32.4 | 66.6 |
| | asymptotic_bonf | 42.6 | 31.2 | 64.5 |
| | permutation_bonf | 43.7 | 30.5 | 64.4 |
| logn discrete | asymptotic | 53.5 | 39.1 | 72.9 |
| | asymptotic_bonf | 52.2 | 37.5 | 71.9 |
| | permutation_bonf | 52.5 | 37.1 | 71.5 |
| pwExp continuous | asymptotic | 14.8 | 10.5 | 21.9 |
| | asymptotic_bonf | 13.9 | 9.7 | 20.9 |
| | permutation_bonf | 13.8 | 10.2 | 20.4 |
| pwExp discrete | asymptotic | 17.6 | 12.5 | 24.9 |
| | asymptotic_bonf | 16.7 | 11.5 | 23.4 |
| | permutation_bonf | 16.5 | 11.2 | 23.4 |
| Weib late continuous | asymptotic | 47.2 | 33.4 | 66.4 |
| | asymptotic_bonf | 44.5 | 31.6 | 65.4 |
| | permutation_bonf | 44.6 | 32.1 | 63.5 |
| Weib late discrete | asymptotic | 54.9 | 40.2 | 73.7 |
| | asymptotic_bonf | 53.3 | 39.1 | 72.7 |
| | permutation_bonf | 52.5 | 38.5 | 71.7 |
| Weib prop continuous | asymptotic | 44.1 | 31.3 | 64.0 |
| | asymptotic_bonf | 42.8 | 30.5 | 62.4 |
| | permutation_bonf | 43.0 | 30.1 | 61.0 |
| Weib prop discrete | asymptotic | 52.9 | 38.3 | 71.7 |
| | asymptotic_bonf | 51.8 | 36.8 | 70.5 |
| | permutation_bonf | 51.8 | 36.9 | 69.2 |
| Weib scale continuous | asymptotic | 37.5 | 25.4 | 54.6 |
| | asymptotic_bonf | 36.0 | 24.6 | 53.5 |
| | permutation_bonf | 34.7 | 23.9 | 52.9 |
| Weib scale discrete | asymptotic | 46.2 | 30.8 | 62.9 |
| | asymptotic_bonf | 44.5 | 30.1 | 61.6 |
| | permutation_bonf | 44.5 | 29.0 | 60.9 |
| Weib shape continuous | asymptotic | 25.6 | 18.0 | 39.9 |
| | asymptotic_bonf | 25.0 | 16.7 | 38.9 |
| | permutation_bonf | 24.4 | 17.7 | 39.3 |
| Weib shape discrete | asymptotic | 34.7 | 23.2 | 51.2 |
| | asymptotic_bonf | 33.0 | 22.7 | 49.6 |
| | permutation_bonf | 32.7 | 22.5 | 48.9 |

Table S69: Rejection rates in percent for the Dunnett-type contrast matrix with $\delta = 1.5$ and balanced small sample sizes under unequal, low censoring.

| Distribution | Method | $\mathcal{H}_{0,7}$ | $\mathcal{H}_{0,8}$ | $\mathcal{H}_{0,9}$ |
|---|---|---|---|---|
| exp early continuous | asymptotic | 3.3 | 1.3 | 3.2 |
| | asymptotic_bonf | 3.1 | 1.3 | 2.8 |
| | permutation_bonf | 2.7 | 1.3 | 2.5 |
| exp early discrete | asymptotic | 3.0 | 2.4 | 2.8 |
| | asymptotic_bonf | 2.8 | 2.0 | 2.3 |
| | permutation_bonf | 2.3 | 1.9 | 2.1 |
| exp late continuous | asymptotic | 3.3 | 1.7 | 3.5 |
| | asymptotic_bonf | 3.1 | 1.4 | 3.3 |
| | permutation_bonf | 2.5 | 1.7 | 2.5 |
| exp late discrete | asymptotic | 3.7 | 2.0 | 4.2 |
| | asymptotic_bonf | 3.4 | 1.8 | 3.8 |
| | permutation_bonf | 3.2 | 1.8 | 3.0 |
| exp prop continuous | asymptotic | 1.9 | 1.8 | 3.4 |
| | asymptotic_bonf | 1.6 | 1.5 | 3.3 |
| | permutation_bonf | 1.4 | 1.3 | 2.9 |
| exp prop discrete | asymptotic | 3.1 | 1.7 | 4.4 |
| | asymptotic_bonf | 3.1 | 1.6 | 4.0 |
| | permutation_bonf | 2.6 | 1.4 | 3.2 |
| logn continuous | asymptotic | 7.5 | 4.9 | 11.5 |
| | asymptotic_bonf | 7.0 | 4.6 | 10.8 |
| | permutation_bonf | 6.2 | 4.4 | 9.1 |
| logn discrete | asymptotic | 9.0 | 6.1 | 13.2 |
| | asymptotic_bonf | 8.3 | 5.6 | 12.2 |
| | permutation_bonf | 7.9 | 5.7 | 11.4 |
| pwExp continuous | asymptotic | 2.2 | 1.6 | 3.0 |
| | asymptotic_bonf | 2.0 | 1.3 | 2.5 |
| | permutation_bonf | 1.8 | 1.5 | 1.9 |
| pwExp discrete | asymptotic | 2.8 | 1.5 | 3.8 |
| | asymptotic_bonf | 2.6 | 1.5 | 3.3 |
| | permutation_bonf | 2.4 | 1.6 | 2.4 |
| Weib late continuous | asymptotic | 7.2 | 6.6 | 11.1 |
| | asymptotic_bonf | 7.0 | 6.2 | 10.5 |
| | permutation_bonf | 5.7 | 5.5 | 9.0 |
| Weib late discrete | asymptotic | 8.2 | 7.4 | 13.2 |
| | asymptotic_bonf | 8.1 | 7.0 | 12.4 |
| | permutation_bonf | 7.0 | 6.1 | 11.2 |
| Weib prop continuous | asymptotic | 6.6 | 6.4 | 9.8 |
| | asymptotic_bonf | 6.3 | 5.4 | 9.1 |
| | permutation_bonf | 5.5 | 5.3 | 8.1 |
| Weib prop discrete | asymptotic | 8.2 | 6.9 | 12.4 |
| | asymptotic_bonf | 7.5 | 6.3 | 11.4 |
| | permutation_bonf | 7.3 | 6.0 | 9.7 |
| Weib scale continuous | asymptotic | 4.2 | 4.1 | 7.1 |
| | asymptotic_bonf | 3.9 | 3.7 | 6.4 |
| | permutation_bonf | 3.4 | 4.2 | 5.5 |
| Weib scale discrete | asymptotic | 5.1 | 4.8 | 8.1 |
| | asymptotic_bonf | 4.9 | 4.5 | 7.2 |
| | permutation_bonf | 4.6 | 4.4 | 6.4 |
| Weib shape continuous | asymptotic | 1.7 | 2.5 | 3.9 |
| | asymptotic_bonf | 1.6 | 2.1 | 3.5 |
| | permutation_bonf | 1.7 | 2.4 | 3.4 |
| Weib shape discrete | asymptotic | 2.3 | 2.8 | 4.7 |
| | asymptotic_bonf | 2.2 | 2.8 | 4.5 |
| | permutation_bonf | 2.3 | 3.1 | 4.5 |

Table S70: Rejection rates in percent for the Dunnett-type contrast matrix with $\delta = 1.5$ and unbalanced large sample sizes under unequal, low censoring.

| Distribution | Method | $\mathcal{H}_{0,7}$ | $\mathcal{H}_{0,8}$ | $\mathcal{H}_{0,9}$ |
|---|---|---|---|---|
| exp early continuous | asymptotic | 37.4 | 21.4 | 59.5 |
|  | asymptotic_bonf | 36.4 | 20.5 | 58.0 |
|  | permutation_bonf | 34.4 | 18.2 | 58.2 |
| exp early discrete | asymptotic | 44.0 | 26.0 | 66.6 |
|  | asymptotic_bonf | 43.1 | 25.0 | 65.4 |
|  | permutation_bonf | 39.6 | 22.3 | 64.2 |
| exp late continuous | asymptotic | 44.7 | 25.4 | 68.7 |
|  | asymptotic_bonf | 43.5 | 24.7 | 68.0 |
|  | permutation_bonf | 40.9 | 21.9 | 67.6 |
| exp late discrete | asymptotic | 51.7 | 30.6 | 75.9 |
|  | asymptotic_bonf | 51.3 | 29.6 | 75.3 |
|  | permutation_bonf | 47.7 | 25.8 | 73.7 |
| exp prop continuous | asymptotic | 43.3 | 25.0 | 61.9 |
|  | asymptotic_bonf | 42.3 | 24.2 | 61.2 |
|  | permutation_bonf | 40.9 | 22.2 | 59.7 |
| exp prop discrete | asymptotic | 48.6 | 30.9 | 68.7 |
|  | asymptotic_bonf | 48.0 | 29.7 | 68.1 |
|  | permutation_bonf | 46.2 | 27.0 | 65.9 |
| logn continuous | asymptotic | 90.7 | 73.1 | 97.2 |
|  | asymptotic_bonf | 90.4 | 72.2 | 97.2 |
|  | permutation_bonf | 88.7 | 66.7 | 96.2 |
| logn discrete | asymptotic | 94.5 | 81.9 | 99.0 |
|  | asymptotic_bonf | 94.4 | 81.7 | 98.8 |
|  | permutation_bonf | 92.7 | 75.8 | 98.5 |
| pwExp continuous | asymptotic | 34.9 | 20.2 | 58.6 |
|  | asymptotic_bonf | 34.2 | 19.1 | 57.9 |
|  | permutation_bonf | 33.5 | 17.5 | 56.0 |
| pwExp discrete | asymptotic | 41.2 | 24.2 | 66.5 |
|  | asymptotic_bonf | 40.8 | 23.9 | 66.2 |
|  | permutation_bonf | 38.3 | 21.4 | 64.5 |
| Weib late continuous | asymptotic | 91.3 | 73.3 | 98.6 |
|  | asymptotic_bonf | 91.2 | 72.7 | 98.5 |
|  | permutation_bonf | 89.7 | 68.1 | 98.1 |
| Weib late discrete | asymptotic | 95.4 | 81.4 | 99.3 |
|  | asymptotic_bonf | 95.2 | 80.4 | 99.3 |
|  | permutation_bonf | 93.5 | 76.1 | 99.0 |
| Weib prop continuous | asymptotic | 88.7 | 68.8 | 97.7 |
|  | asymptotic_bonf | 88.5 | 67.8 | 97.6 |
|  | permutation_bonf | 86.1 | 64.1 | 96.7 |
| Weib prop discrete | asymptotic | 94.2 | 78.5 | 99.1 |
|  | asymptotic_bonf | 93.9 | 77.8 | 99.1 |
|  | permutation_bonf | 91.8 | 72.8 | 98.4 |
| Weib scale continuous | asymptotic | 77.9 | 55.2 | 91.5 |
|  | asymptotic_bonf | 77.4 | 54.6 | 91.2 |
|  | permutation_bonf | 74.6 | 48.3 | 89.1 |
| Weib scale discrete | asymptotic | 85.4 | 64.6 | 95.3 |
|  | asymptotic_bonf | 85.1 | 63.8 | 95.2 |
|  | permutation_bonf | 81.6 | 58.4 | 94.2 |
| Weib shape continuous | asymptotic | 59.8 | 35.5 | 77.1 |
|  | asymptotic_bonf | 59.5 | 35.0 | 76.7 |
|  | permutation_bonf | 54.6 | 28.0 | 74.5 |
| Weib shape discrete | asymptotic | 69.7 | 48.0 | 86.4 |
|  | asymptotic_bonf | 69.1 | 47.1 | 86.1 |
|  | permutation_bonf | 65.5 | 38.2 | 83.2 |

Table S71: Rejection rates in percent for the Dunnett-type contrast matrix with $\delta = 1.5$ and unbalanced medium sample sizes under unequal, low censoring.

| Distribution | Method | $\mathcal{H}_{0,7}$ | $\mathcal{H}_{0,8}$ | $\mathcal{H}_{0,9}$ |
|---|---|---|---|---|
| exp early continuous | asymptotic | 3.5 | 2.4 | 5.2 |
| | asymptotic_bonf | 3.1 | 2.3 | 5.1 |
| | permutation_bonf | 1.4 | 0.5 | 3.9 |
| exp early discrete | asymptotic | 4.1 | 2.7 | 6.9 |
| | asymptotic_bonf | 3.6 | 2.6 | 6.4 |
| | permutation_bonf | 1.6 | 0.8 | 4.7 |
| exp late continuous | asymptotic | 4.2 | 2.4 | 6.6 |
| | asymptotic_bonf | 4.2 | 2.4 | 6.2 |
| | permutation_bonf | 2.1 | 0.8 | 4.2 |
| exp late discrete | asymptotic | 5.4 | 2.5 | 7.9 |
| | asymptotic_bonf | 4.9 | 2.5 | 7.4 |
| | permutation_bonf | 2.5 | 1.1 | 5.1 |
| exp prop continuous | asymptotic | 4.2 | 2.3 | 4.9 |
| | asymptotic_bonf | 4.0 | 2.3 | 4.3 |
| | permutation_bonf | 2.3 | 0.9 | 3.1 |
| exp prop discrete | asymptotic | 5.4 | 2.8 | 6.0 |
| | asymptotic_bonf | 4.7 | 2.7 | 5.5 |
| | permutation_bonf | 2.3 | 0.8 | 3.5 |
| logn continuous | asymptotic | 11.3 | 6.9 | 20.9 |
| | asymptotic_bonf | 10.9 | 6.4 | 20.5 |
| | permutation_bonf | 3.4 | 0.9 | 10.4 |
| logn discrete | asymptotic | 13.4 | 8.2 | 23.9 |
| | asymptotic_bonf | 13.1 | 7.8 | 23.8 |
| | permutation_bonf | 4.2 | 0.7 | 12.7 |
| pwExp continuous | asymptotic | 3.3 | 2.1 | 5.2 |
| | asymptotic_bonf | 3.0 | 1.9 | 4.9 |
| | permutation_bonf | 1.4 | 0.7 | 3.0 |
| pwExp discrete | asymptotic | 3.9 | 2.3 | 5.8 |
| | asymptotic_bonf | 3.8 | 2.1 | 5.4 |
| | permutation_bonf | 2.2 | 0.6 | 3.2 |
| Weib late continuous | asymptotic | 13.6 | 6.7 | 25.9 |
| | asymptotic_bonf | 13.3 | 6.2 | 25.1 |
| | permutation_bonf | 5.5 | 0.7 | 14.7 |
| Weib late discrete | asymptotic | 16.7 | 8.5 | 30.1 |
| | asymptotic_bonf | 16.1 | 8.3 | 29.5 |
| | permutation_bonf | 6.3 | 0.8 | 17.3 |
| Weib prop continuous | asymptotic | 12.3 | 6.0 | 21.7 |
| | asymptotic_bonf | 11.8 | 5.7 | 21.4 |
| | permutation_bonf | 4.4 | 0.3 | 11.1 |
| Weib prop discrete | asymptotic | 15.0 | 7.0 | 26.7 |
| | asymptotic_bonf | 14.4 | 7.0 | 25.9 |
| | permutation_bonf | 4.6 | 0.4 | 12.9 |
| Weib scale continuous | asymptotic | 7.6 | 3.2 | 12.0 |
| | asymptotic_bonf | 7.5 | 2.8 | 11.8 |
| | permutation_bonf | 1.9 | 0.0 | 5.4 |
| Weib scale discrete | asymptotic | 9.3 | 4.1 | 14.2 |
| | asymptotic_bonf | 9.0 | 3.6 | 13.9 |
| | permutation_bonf | 1.8 | 0.1 | 6.4 |
| Weib shape continuous | asymptotic | 3.7 | 1.1 | 5.4 |
| | asymptotic_bonf | 3.6 | 1.1 | 5.2 |
| | permutation_bonf | 0.7 | 0.0 | 2.6 |
| Weib shape discrete | asymptotic | 4.7 | 1.8 | 7.9 |
| | asymptotic_bonf | 4.5 | 1.6 | 7.7 |
| | permutation_bonf | 0.6 | 0.0 | 3.0 |

Table S72: Rejection rates in percent for the Dunnett-type contrast matrix with $\delta = 1.5$ and unbalanced small sample sizes under unequal, low censoring.

| Distribution | Method | $\mathcal{H}_{0,7}$ | $\mathcal{H}_{0,8}$ | $\mathcal{H}_{0,9}$ |
|---|---|---|---|---|
| exp early continuous | asymptotic | 1.1 | 3.3 | 2.2 |
| | asymptotic_bonf | 1.1 | 3.3 | 2.2 |
| | permutation_bonf | 0.1 | 0.0 | 0.1 |
| exp early discrete | asymptotic | 1.2 | 3.5 | 2.2 |
| | asymptotic_bonf | 1.0 | 3.3 | 2.2 |
| | permutation_bonf | 0.1 | 0.0 | 0.1 |
| exp late continuous | asymptotic | 1.3 | 2.3 | 2.7 |
| | asymptotic_bonf | 1.2 | 2.3 | 2.7 |
| | permutation_bonf | 0.1 | 0.0 | 0.2 |
| exp late discrete | asymptotic | 1.3 | 2.5 | 3.0 |
| | asymptotic_bonf | 1.2 | 2.3 | 2.7 |
| | permutation_bonf | 0.0 | 0.0 | 0.1 |
| exp prop continuous | asymptotic | 1.8 | 2.3 | 2.5 |
| | asymptotic_bonf | 1.6 | 2.3 | 2.4 |
| | permutation_bonf | 0.1 | 0.3 | 0.4 |
| exp prop discrete | asymptotic | 1.8 | 2.3 | 2.4 |
| | asymptotic_bonf | 1.8 | 2.2 | 2.3 |
| | permutation_bonf | 0.0 | 0.3 | 0.4 |
| logn continuous | asymptotic | 2.2 | 3.5 | 2.9 |
| | asymptotic_bonf | 2.2 | 3.4 | 2.8 |
| | permutation_bonf | 0.2 | 0.1 | 0.1 |
| logn discrete | asymptotic | 2.2 | 3.7 | 3.0 |
| | asymptotic_bonf | 2.2 | 3.7 | 3.0 |
| | permutation_bonf | 0.2 | 0.1 | 0.1 |
| pwExp continuous | asymptotic | 1.6 | 2.3 | 1.9 |
| | asymptotic_bonf | 1.5 | 2.3 | 1.9 |
| | permutation_bonf | 0.0 | 0.0 | 0.2 |
| pwExp discrete | asymptotic | 1.8 | 2.5 | 2.0 |
| | asymptotic_bonf | 1.5 | 2.5 | 1.8 |
| | permutation_bonf | 0.0 | 0.0 | 0.2 |
| Weib late continuous | asymptotic | 2.0 | 2.2 | 3.9 |
| | asymptotic_bonf | 2.0 | 2.1 | 3.7 |
| | permutation_bonf | 0.0 | 0.0 | 0.0 |
| Weib late discrete | asymptotic | 2.4 | 2.4 | 4.0 |
| | asymptotic_bonf | 2.4 | 2.4 | 3.9 |
| | permutation_bonf | 0.0 | 0.0 | 0.0 |
| Weib prop continuous | asymptotic | 2.0 | 2.4 | 2.9 |
| | asymptotic_bonf | 1.9 | 2.4 | 2.8 |
| | permutation_bonf | 0.0 | 0.0 | 0.0 |
| Weib prop discrete | asymptotic | 1.7 | 2.9 | 3.6 |
| | asymptotic_bonf | 1.7 | 2.9 | 3.4 |
| | permutation_bonf | 0.0 | 0.0 | 0.0 |
| Weib scale continuous | asymptotic | 2.9 | 4.0 | 2.8 |
| | asymptotic_bonf | 2.9 | 3.9 | 2.8 |
| | permutation_bonf | 0.2 | 0.1 | 0.1 |
| Weib scale discrete | asymptotic | 3.2 | 4.1 | 2.7 |
| | asymptotic_bonf | 3.0 | 4.1 | 2.6 |
| | permutation_bonf | 0.2 | 0.0 | 0.1 |
| Weib shape continuous | asymptotic | 6.2 | 9.1 | 3.5 |
| | asymptotic_bonf | 6.2 | 9.1 | 3.5 |
| | permutation_bonf | 0.7 | 0.8 | 0.8 |
| Weib shape discrete | asymptotic | 6.1 | 9.2 | 3.5 |
| | asymptotic_bonf | 6.1 | 9.2 | 3.5 |
| | permutation_bonf | 0.8 | 0.5 | 0.7 |

Table S73: Rejection rates in percent for the Tukey-type contrast matrix with $\delta = 1.5$ and balanced large sample sizes under equal censoring.

| Distribution | Method | $\mathcal{H}_{0,7}$ | $\mathcal{H}_{0,8}$ | $\mathcal{H}_{0,9}$ | $\mathcal{H}_{0,13}$ | $\mathcal{H}_{0,14}$ | $\mathcal{H}_{0,15}$ | $\mathcal{H}_{0,16}$ | $\mathcal{H}_{0,17}$ | $\mathcal{H}_{0,18}$ |
|---|---|---|---|---|---|---|---|---|---|---|
| exp early continuous | asymptotic | 79.3 | 57.0 | 92.4 | 78.5 | 57.8 | 92.5 | 77.7 | 56.0 | 95.3 |
| | asymptotic_bonf | 77.5 | 54.6 | 91.4 | 76.8 | 54.8 | 91.8 | 76.1 | 54.9 | 94.4 |
| | permutation_bonf | 75.6 | 53.5 | 90.8 | 75.3 | 53.2 | 91.2 | 74.6 | 52.6 | 93.6 |
| exp early discrete | asymptotic | 86.3 | 65.3 | 95.4 | 84.4 | 64.7 | 95.3 | 85.0 | 64.1 | 97.2 |
| | asymptotic_bonf | 84.8 | 62.7 | 95.3 | 83.1 | 63.1 | 94.8 | 83.4 | 61.2 | 97.0 |
| | permutation_bonf | 83.2 | 60.5 | 94.0 | 82.3 | 61.0 | 94.0 | 81.5 | 59.8 | 96.3 |
| exp late continuous | asymptotic | 84.2 | 61.6 | 95.1 | 81.9 | 61.6 | 94.8 | 82.6 | 61.0 | 96.8 |
| | asymptotic_bonf | 82.9 | 59.0 | 94.8 | 80.1 | 59.6 | 94.3 | 80.3 | 58.5 | 96.4 |
| | permutation_bonf | 80.7 | 57.5 | 94.2 | 79.3 | 58.1 | 93.4 | 79.2 | 57.0 | 95.8 |
| exp late discrete | asymptotic | 89.2 | 67.6 | 96.9 | 89.5 | 68.7 | 97.2 | 88.8 | 68.9 | 98.6 |
| | asymptotic_bonf | 87.6 | 66.1 | 96.4 | 87.4 | 66.9 | 96.7 | 87.3 | 67.2 | 98.1 |
| | permutation_bonf | 87.2 | 64.5 | 96.6 | 86.5 | 65.8 | 96.8 | 85.7 | 65.2 | 98.0 |
| exp prop continuous | asymptotic | 80.8 | 62.1 | 94.6 | 81.7 | 62.1 | 95.0 | 80.3 | 63.0 | 95.6 |
| | asymptotic_bonf | 78.3 | 59.2 | 94.1 | 79.7 | 60.6 | 94.6 | 78.6 | 60.2 | 95.0 |
| | permutation_bonf | 77.9 | 58.4 | 93.2 | 78.0 | 58.4 | 93.8 | 77.6 | 59.2 | 94.5 |
| exp prop discrete | asymptotic | 86.9 | 70.2 | 96.7 | 88.2 | 69.0 | 97.4 | 86.3 | 68.5 | 98.2 |
| | asymptotic_bonf | 86.0 | 68.3 | 96.2 | 86.8 | 66.8 | 96.8 | 84.5 | 67.0 | 97.9 |
| | permutation_bonf | 85.0 | 67.4 | 96.4 | 85.3 | 66.1 | 96.4 | 83.9 | 66.9 | 96.8 |
| logn continuous | asymptotic | 99.9 | 98.6 | 99.9 | 99.7 | 98.6 | 100.0 | 99.8 | 98.1 | 100.0 |
| | asymptotic_bonf | 99.9 | 98.3 | 99.9 | 99.6 | 97.5 | 100.0 | 99.8 | 97.9 | 100.0 |
| | permutation_bonf | 100.0 | 97.0 | 99.9 | 99.6 | 97.3 | 100.0 | 99.8 | 97.4 | 100.0 |
| logn discrete | asymptotic | 100.0 | 99.2 | 100.0 | 99.9 | 99.4 | 100.0 | 99.9 | 99.4 | 100.0 |
| | asymptotic_bonf | 100.0 | 99.2 | 100.0 | 99.9 | 99.4 | 100.0 | 99.9 | 99.4 | 100.0 |
| | permutation_bonf | 100.0 | 99.2 | 100.0 | 99.9 | 99.0 | 100.0 | 99.9 | 99.4 | 100.0 |
| pwExp continuous | asymptotic | 78.8 | 55.3 | 92.3 | 76.9 | 56.3 | 91.9 | 76.7 | 55.1 | 94.6 |
| | asymptotic_bonf | 76.5 | 52.7 | 91.6 | 75.8 | 54.0 | 90.8 | 75.1 | 53.2 | 93.6 |
| | permutation_bonf | 74.8 | 51.5 | 90.9 | 73.9 | 53.9 | 89.8 | 74.0 | 52.7 | 93.4 |
| pwExp discrete | asymptotic | 85.5 | 62.0 | 95.9 | 84.4 | 64.0 | 95.4 | 84.1 | 64.4 | 97.3 |
| | asymptotic_bonf | 83.3 | 60.0 | 94.8 | 83.3 | 61.9 | 94.9 | 82.2 | 62.7 | 96.6 |
| | permutation_bonf | 82.6 | 58.8 | 94.3 | 81.9 | 60.4 | 94.7 | 81.1 | 61.8 | 96.5 |
| Weib late continuous | asymptotic | 99.9 | 98.3 | 100.0 | 99.7 | 98.4 | 100.0 | 99.8 | 97.6 | 100.0 |
| | asymptotic_bonf | 99.9 | 98.2 | 100.0 | 99.7 | 98.2 | 100.0 | 99.8 | 97.2 | 100.0 |
| | permutation_bonf | 99.8 | 97.4 | 100.0 | 99.4 | 98.2 | 100.0 | 99.8 | 96.6 | 100.0 |
| Weib late discrete | asymptotic | 100.0 | 99.3 | 100.0 | 100.0 | 99.3 | 100.0 | 100.0 | 99.1 | 100.0 |
| | asymptotic_bonf | 99.9 | 99.2 | 100.0 | 99.9 | 99.3 | 100.0 | 100.0 | 98.8 | 100.0 |
| | permutation_bonf | 99.9 | 99.3 | 100.0 | 100.0 | 99.3 | 100.0 | 100.0 | 98.9 | 100.0 |
| Weib prop continuous | asymptotic | 99.7 | 98.2 | 100.0 | 99.5 | 97.9 | 100.0 | 99.8 | 97.0 | 100.0 |
| | asymptotic_bonf | 99.7 | 98.0 | 100.0 | 99.4 | 97.4 | 100.0 | 99.8 | 96.7 | 100.0 |
| | permutation_bonf | 99.7 | 97.2 | 100.0 | 99.4 | 97.6 | 100.0 | 99.8 | 96.3 | 100.0 |
| Weib prop discrete | asymptotic | 100.0 | 99.0 | 100.0 | 99.9 | 99.2 | 100.0 | 99.9 | 98.8 | 100.0 |
| | asymptotic_bonf | 100.0 | 99.0 | 100.0 | 99.9 | 99.1 | 100.0 | 99.9 | 98.6 | 100.0 |
| | permutation_bonf | 100.0 | 98.9 | 100.0 | 99.9 | 99.2 | 100.0 | 99.9 | 98.3 | 100.0 |
| Weib scale continuous | asymptotic | 99.2 | 96.4 | 100.0 | 98.9 | 95.4 | 100.0 | 99.4 | 95.3 | 100.0 |
| | asymptotic_bonf | 99.0 | 95.7 | 100.0 | 98.8 | 95.2 | 99.9 | 99.3 | 94.6 | 100.0 |
| | permutation_bonf | 99.0 | 95.4 | 100.0 | 98.9 | 94.8 | 100.0 | 98.9 | 94.2 | 100.0 |
| Weib scale discrete | asymptotic | 99.9 | 98.5 | 100.0 | 99.6 | 98.1 | 100.0 | 99.9 | 97.3 | 100.0 |
| | asymptotic_bonf | 99.9 | 98.5 | 100.0 | 99.4 | 98.0 | 100.0 | 99.9 | 97.2 | 100.0 |
| | permutation_bonf | 99.7 | 98.3 | 100.0 | 99.3 | 97.5 | 100.0 | 99.8 | 96.9 | 100.0 |
| Weib shape continuous | asymptotic | 98.4 | 90.2 | 99.7 | 97.7 | 90.6 | 99.5 | 98.0 | 90.0 | 99.7 |
| | asymptotic_bonf | 98.2 | 89.5 | 99.7 | 97.2 | 89.3 | 99.4 | 97.8 | 89.2 | 99.7 |
| | permutation_bonf | 98.0 | 88.5 | 99.5 | 97.1 | 88.9 | 99.5 | 97.2 | 88.1 | 99.6 |
| Weib shape discrete | asymptotic | 99.3 | 95.5 | 100.0 | 99.3 | 95.9 | 99.8 | 99.4 | 95.0 | 100.0 |
| | asymptotic_bonf | 99.2 | 94.8 | 100.0 | 99.1 | 95.3 | 99.8 | 99.2 | 94.7 | 100.0 |
| | permutation_bonf | 99.3 | 94.3 | 99.9 | 98.8 | 94.5 | 99.7 | 99.0 | 93.6 | 99.9 |

Table S74: Rejection rates in percent for the Tukey-type contrast matrix with $\delta = 1.5$ and balanced medium sample sizes under equal censoring.

| Distribution | Method | $\mathcal{H}_{0,7}$ | $\mathcal{H}_{0,8}$ | $\mathcal{H}_{0,9}$ | $\mathcal{H}_{0,13}$ | $\mathcal{H}_{0,14}$ | $\mathcal{H}_{0,15}$ | $\mathcal{H}_{0,16}$ | $\mathcal{H}_{0,17}$ | $\mathcal{H}_{0,18}$ |
|---|---|---|---|---|---|---|---|---|---|---|
| exp early continuous | asymptotic | 11.1 | 6.9 | 16.2 | 11.7 | 6.4 | 18.6 | 11.2 | 4.4 | 18.4 |
| | asymptotic_bonf | 10.1 | 5.8 | 14.5 | 10.6 | 6.0 | 17.1 | 10.2 | 3.6 | 16.5 |
| | permutation_bonf | 10.0 | 6.1 | 15.1 | 10.3 | 5.5 | 16.7 | 9.6 | 3.9 | 16.6 |
| exp early discrete | asymptotic | 13.3 | 8.1 | 19.5 | 13.3 | 7.9 | 22.1 | 14.1 | 5.7 | 22.9 |
| | asymptotic_bonf | 12.2 | 7.1 | 17.6 | 12.4 | 7.2 | 20.1 | 12.9 | 5.0 | 21.0 |
| | permutation_bonf | 12.8 | 6.9 | 17.1 | 11.8 | 7.1 | 19.1 | 11.8 | 5.1 | 19.3 |
| exp late continuous | asymptotic | 12.0 | 7.3 | 17.8 | 12.8 | 6.9 | 21.7 | 12.3 | 5.4 | 22.2 |
| | asymptotic_bonf | 10.5 | 6.0 | 16.8 | 11.7 | 5.9 | 20.6 | 11.5 | 4.7 | 20.2 |
| | permutation_bonf | 10.2 | 6.1 | 16.6 | 11.2 | 6.4 | 19.4 | 10.4 | 4.9 | 20.6 |
| exp late discrete | asymptotic | 14.4 | 9.1 | 21.9 | 14.2 | 8.2 | 25.5 | 14.6 | 6.3 | 26.4 |
| | asymptotic_bonf | 13.1 | 7.7 | 19.7 | 13.4 | 7.8 | 23.5 | 13.5 | 5.8 | 24.2 |
| | permutation_bonf | 13.2 | 8.1 | 19.2 | 13.1 | 8.1 | 22.6 | 13.6 | 6.0 | 23.2 |
| exp prop continuous | asymptotic | 13.1 | 7.9 | 15.8 | 12.1 | 6.6 | 16.6 | 11.4 | 6.4 | 18.0 |
| | asymptotic_bonf | 11.8 | 6.5 | 14.1 | 11.1 | 6.3 | 15.5 | 10.2 | 5.9 | 17.0 |
| | permutation_bonf | 11.6 | 7.6 | 14.8 | 10.2 | 5.8 | 14.9 | 10.1 | 6.0 | 15.0 |
| exp prop discrete | asymptotic | 15.2 | 9.6 | 18.6 | 13.7 | 7.8 | 19.6 | 13.6 | 8.0 | 21.1 |
| | asymptotic_bonf | 13.7 | 8.4 | 17.5 | 12.7 | 7.0 | 18.4 | 12.9 | 6.9 | 19.2 |
| | permutation_bonf | 12.1 | 9.1 | 17.9 | 12.3 | 6.1 | 16.6 | 12.1 | 6.7 | 17.7 |
| logn continuous | asymptotic | 37.6 | 23.7 | 58.4 | 37.0 | 25.0 | 58.0 | 38.2 | 24.3 | 56.9 |
| | asymptotic_bonf | 35.6 | 21.9 | 55.9 | 35.6 | 22.9 | 55.5 | 36.9 | 21.9 | 53.9 |
| | permutation_bonf | 34.2 | 21.5 | 54.7 | 33.9 | 22.1 | 52.4 | 35.5 | 19.8 | 51.9 |
| logn discrete | asymptotic | 45.5 | 29.6 | 66.7 | 44.2 | 30.4 | 65.0 | 45.6 | 29.1 | 67.5 |
| | asymptotic_bonf | 43.4 | 26.8 | 64.6 | 42.0 | 28.7 | 63.3 | 44.1 | 26.8 | 65.1 |
| | permutation_bonf | 41.7 | 26.8 | 63.7 | 41.3 | 28.4 | 62.3 | 41.9 | 26.3 | 62.3 |
| pwExp continuous | asymptotic | 10.9 | 6.7 | 16.5 | 11.0 | 5.7 | 19.2 | 11.5 | 4.9 | 18.8 |
| | asymptotic_bonf | 10.0 | 5.6 | 15.2 | 10.1 | 5.1 | 17.0 | 9.8 | 4.3 | 16.5 |
| | permutation_bonf | 9.5 | 5.5 | 14.3 | 9.6 | 5.3 | 16.0 | 9.7 | 4.7 | 15.5 |
| pwExp discrete | asymptotic | 13.2 | 8.2 | 20.2 | 13.2 | 6.7 | 22.3 | 12.7 | 6.1 | 22.4 |
| | asymptotic_bonf | 12.1 | 6.7 | 18.3 | 11.8 | 6.2 | 21.1 | 11.8 | 5.2 | 20.6 |
| | permutation_bonf | 11.7 | 7.6 | 18.1 | 11.7 | 6.1 | 20.0 | 11.6 | 5.6 | 18.8 |
| Weib late continuous | asymptotic | 37.1 | 24.4 | 57.4 | 36.3 | 23.2 | 60.2 | 41.7 | 23.6 | 56.7 |
| | asymptotic_bonf | 34.9 | 22.8 | 55.4 | 34.8 | 22.0 | 58.5 | 39.6 | 22.1 | 55.3 |
| | permutation_bonf | 34.7 | 22.1 | 53.3 | 35.0 | 20.4 | 56.9 | 36.9 | 21.1 | 52.6 |
| Weib late discrete | asymptotic | 44.0 | 28.5 | 64.3 | 44.6 | 27.4 | 65.8 | 47.6 | 29.6 | 63.2 |
| | asymptotic_bonf | 41.8 | 26.9 | 61.3 | 42.0 | 25.6 | 63.5 | 46.1 | 26.3 | 60.8 |
| | permutation_bonf | 40.5 | 25.1 | 59.7 | 42.2 | 24.4 | 62.0 | 44.4 | 26.5 | 58.5 |
| Weib prop continuous | asymptotic | 34.4 | 23.4 | 55.1 | 35.4 | 21.1 | 57.2 | 38.9 | 22.5 | 54.3 |
| | asymptotic_bonf | 33.0 | 21.2 | 52.9 | 33.5 | 20.1 | 54.8 | 36.9 | 20.1 | 52.0 |
| | permutation_bonf | 31.3 | 20.7 | 51.8 | 33.7 | 20.0 | 53.9 | 35.0 | 20.5 | 50.6 |
| Weib prop discrete | asymptotic | 43.9 | 28.7 | 63.6 | 42.9 | 26.8 | 65.1 | 47.6 | 28.3 | 61.7 |
| | asymptotic_bonf | 41.5 | 26.8 | 61.8 | 41.0 | 24.3 | 63.7 | 45.6 | 26.2 | 60.0 |
| | permutation_bonf | 39.7 | 25.5 | 59.5 | 39.2 | 24.5 | 61.3 | 44.5 | 26.2 | 57.9 |
| Weib scale continuous | asymptotic | 28.5 | 18.7 | 46.1 | 29.7 | 16.4 | 47.1 | 32.0 | 18.1 | 44.8 |
| | asymptotic_bonf | 26.4 | 16.9 | 44.5 | 27.9 | 15.4 | 45.0 | 29.1 | 16.5 | 42.6 |
| | permutation_bonf | 25.7 | 16.9 | 43.1 | 25.7 | 15.0 | 43.4 | 27.8 | 16.3 | 40.1 |
| Weib scale discrete | asymptotic | 36.2 | 23.9 | 54.1 | 36.3 | 21.7 | 55.7 | 40.1 | 22.8 | 52.6 |
| | asymptotic_bonf | 33.8 | 22.1 | 52.3 | 33.4 | 19.9 | 53.8 | 37.9 | 21.2 | 50.1 |
| | permutation_bonf | 31.9 | 21.9 | 50.2 | 32.6 | 19.5 | 52.2 | 36.4 | 21.2 | 48.1 |
| Weib shape continuous | asymptotic | 19.5 | 12.1 | 33.5 | 20.7 | 12.9 | 36.2 | 22.8 | 12.3 | 33.0 |
| | asymptotic_bonf | 17.5 | 11.7 | 31.9 | 18.9 | 11.2 | 34.4 | 20.5 | 10.9 | 30.8 |
| | permutation_bonf | 17.7 | 11.8 | 30.4 | 18.2 | 10.8 | 32.7 | 19.8 | 10.5 | 29.2 |
| Weib shape discrete | asymptotic | 26.7 | 16.3 | 41.3 | 28.6 | 16.6 | 44.5 | 30.3 | 16.5 | 42.2 |
| | asymptotic_bonf | 25.0 | 14.7 | 39.6 | 25.7 | 15.2 | 42.1 | 28.3 | 15.1 | 39.5 |
| | permutation_bonf | 23.1 | 15.5 | 37.7 | 24.0 | 14.8 | 41.5 | 26.5 | 15.1 | 38.0 |

Table S75: Rejection rates in percent for the Tukey-type contrast matrix with $\delta = 1.5$ and balanced small sample sizes under equal censoring.

| Distribution | Method | $\mathcal{H}_{0,7}$ | $\mathcal{H}_{0,8}$ | $\mathcal{H}_{0,9}$ | $\mathcal{H}_{0,13}$ | $\mathcal{H}_{0,14}$ | $\mathcal{H}_{0,15}$ | $\mathcal{H}_{0,16}$ | $\mathcal{H}_{0,17}$ | $\mathcal{H}_{0,18}$ |
|---|---|---|---|---|---|---|---|---|---|---|
| exp early continuous | asymptotic | 2.2 | 0.9 | 2.1 | 1.5 | 1.2 | 2.2 | 1.7 | 0.7 | 2.4 |
| | asymptotic_bonf | 2.0 | 0.5 | 1.9 | 1.2 | 1.0 | 1.7 | 1.4 | 0.6 | 2.3 |
| | permutation_bonf | 1.6 | 0.6 | 1.0 | 0.8 | 0.6 | 1.4 | 1.1 | 0.6 | 2.0 |
| exp early discrete | asymptotic | 2.2 | 1.5 | 1.7 | 1.9 | 0.9 | 2.3 | 1.7 | 0.9 | 1.9 |
| | asymptotic_bonf | 1.9 | 1.3 | 1.4 | 1.6 | 0.9 | 2.0 | 1.5 | 0.7 | 1.5 |
| | permutation_bonf | 1.5 | 1.1 | 1.1 | 1.5 | 0.7 | 1.6 | 1.1 | 0.5 | 1.3 |
| exp late continuous | asymptotic | 2.1 | 1.0 | 2.3 | 1.9 | 1.6 | 2.6 | 2.5 | 1.2 | 3.1 |
| | asymptotic_bonf | 2.0 | 1.0 | 1.9 | 1.7 | 1.5 | 2.3 | 2.2 | 1.2 | 2.7 |
| | permutation_bonf | 1.9 | 0.8 | 1.6 | 1.2 | 0.8 | 2.0 | 1.6 | 1.0 | 2.2 |
| exp late discrete | asymptotic | 2.4 | 0.9 | 2.8 | 1.8 | 1.7 | 3.0 | 2.8 | 1.3 | 2.9 |
| | asymptotic_bonf | 2.3 | 0.8 | 2.3 | 1.7 | 1.5 | 2.7 | 2.2 | 1.2 | 2.5 |
| | permutation_bonf | 2.0 | 0.9 | 1.6 | 1.4 | 1.4 | 2.1 | 2.4 | 0.7 | 2.4 |
| exp prop continuous | asymptotic | 1.1 | 0.9 | 2.0 | 1.8 | 1.0 | 2.4 | 1.2 | 0.9 | 2.4 |
| | asymptotic_bonf | 1.1 | 0.7 | 1.9 | 1.7 | 0.9 | 1.9 | 0.7 | 0.9 | 2.2 |
| | permutation_bonf | 0.8 | 0.6 | 1.8 | 1.6 | 0.9 | 1.6 | 0.6 | 0.9 | 1.5 |
| exp prop discrete | asymptotic | 1.6 | 1.1 | 3.4 | 1.9 | 0.5 | 3.2 | 1.7 | 1.0 | 3.0 |
| | asymptotic_bonf | 1.6 | 0.9 | 3.0 | 1.6 | 0.3 | 2.7 | 1.5 | 0.9 | 2.7 |
| | permutation_bonf | 1.2 | 0.9 | 2.2 | 1.2 | 0.5 | 2.3 | 1.2 | 0.7 | 2.4 |
| logn continuous | asymptotic | 4.9 | 2.6 | 7.6 | 4.5 | 2.2 | 5.7 | 4.4 | 2.7 | 7.6 |
| | asymptotic_bonf | 4.3 | 2.3 | 7.0 | 3.6 | 1.6 | 5.1 | 3.9 | 2.6 | 6.9 |
| | permutation_bonf | 3.6 | 2.0 | 5.6 | 3.4 | 1.6 | 4.6 | 3.8 | 2.3 | 5.8 |
| logn discrete | asymptotic | 5.7 | 3.4 | 9.4 | 5.6 | 2.7 | 7.1 | 5.4 | 3.2 | 9.1 |
| | asymptotic_bonf | 4.7 | 2.9 | 8.3 | 5.1 | 2.3 | 6.2 | 4.8 | 2.5 | 8.1 |
| | permutation_bonf | 4.3 | 2.3 | 7.3 | 4.2 | 1.9 | 5.1 | 4.9 | 2.3 | 7.0 |
| pwExp continuous | asymptotic | 1.6 | 1.0 | 1.4 | 1.6 | 1.3 | 2.0 | 1.8 | 0.8 | 2.1 |
| | asymptotic_bonf | 1.5 | 0.8 | 1.2 | 1.2 | 1.1 | 1.9 | 1.5 | 0.7 | 1.9 |
| | permutation_bonf | 1.2 | 0.5 | 1.0 | 0.9 | 0.4 | 1.6 | 1.4 | 0.6 | 1.6 |
| pwExp discrete | asymptotic | 1.8 | 1.1 | 2.0 | 1.7 | 1.4 | 2.5 | 2.1 | 0.9 | 2.5 |
| | asymptotic_bonf | 1.7 | 0.7 | 1.4 | 1.5 | 1.1 | 2.0 | 1.8 | 0.7 | 2.2 |
| | permutation_bonf | 1.3 | 0.5 | 1.2 | 1.2 | 0.6 | 1.8 | 1.5 | 0.4 | 1.8 |
| Weib late continuous | asymptotic | 4.3 | 4.3 | 6.1 | 4.9 | 2.9 | 7.4 | 4.4 | 3.5 | 7.2 |
| | asymptotic_bonf | 4.1 | 3.5 | 5.6 | 4.4 | 2.6 | 6.9 | 3.9 | 2.8 | 6.3 |
| | permutation_bonf | 3.0 | 3.2 | 4.6 | 4.0 | 2.5 | 5.2 | 2.7 | 2.9 | 4.6 |
| Weib late discrete | asymptotic | 5.1 | 4.4 | 7.4 | 5.7 | 3.5 | 8.1 | 5.5 | 3.9 | 7.9 |
| | asymptotic_bonf | 4.5 | 4.1 | 6.5 | 5.2 | 3.0 | 7.2 | 5.0 | 3.4 | 7.1 |
| | permutation_bonf | 3.9 | 3.7 | 5.0 | 4.2 | 2.9 | 6.6 | 3.6 | 3.1 | 5.3 |
| Weib prop continuous | asymptotic | 4.0 | 3.5 | 5.2 | 4.5 | 2.6 | 6.6 | 4.0 | 3.1 | 6.1 |
| | asymptotic_bonf | 3.7 | 3.2 | 4.5 | 4.0 | 2.4 | 6.0 | 3.7 | 2.7 | 5.3 |
| | permutation_bonf | 2.5 | 3.0 | 4.1 | 3.6 | 2.1 | 5.0 | 2.4 | 2.7 | 4.0 |
| Weib prop discrete | asymptotic | 4.6 | 3.9 | 5.8 | 5.2 | 3.0 | 8.1 | 5.2 | 3.3 | 6.8 |
| | asymptotic_bonf | 3.9 | 3.4 | 4.9 | 4.7 | 2.6 | 7.5 | 4.1 | 2.9 | 6.2 |
| | permutation_bonf | 3.6 | 3.8 | 4.3 | 4.4 | 2.8 | 6.2 | 3.4 | 2.9 | 5.1 |
| Weib scale continuous | asymptotic | 2.4 | 2.4 | 3.0 | 2.6 | 2.0 | 5.0 | 1.8 | 2.1 | 4.1 |
| | asymptotic_bonf | 1.9 | 2.2 | 2.9 | 2.1 | 1.9 | 4.5 | 1.4 | 1.7 | 3.2 |
| | permutation_bonf | 1.4 | 2.5 | 2.5 | 1.8 | 1.6 | 3.5 | 1.6 | 1.4 | 2.3 |
| Weib scale discrete | asymptotic | 3.1 | 2.9 | 3.8 | 3.9 | 2.1 | 5.5 | 1.9 | 2.1 | 5.3 |
| | asymptotic_bonf | 2.4 | 2.4 | 3.3 | 2.9 | 2.0 | 5.3 | 1.3 | 1.8 | 4.4 |
| | permutation_bonf | 1.8 | 2.1 | 3.0 | 2.4 | 1.5 | 4.3 | 1.6 | 1.7 | 3.4 |
| Weib shape continuous | asymptotic | 1.3 | 1.3 | 1.7 | 1.3 | 1.0 | 3.0 | 1.0 | 1.0 | 2.2 |
| | asymptotic_bonf | 1.0 | 1.3 | 1.7 | 1.0 | 0.6 | 2.4 | 0.8 | 0.8 | 1.6 |
| | permutation_bonf | 0.7 | 1.4 | 2.0 | 1.0 | 1.0 | 2.2 | 0.7 | 1.0 | 1.5 |
| Weib shape discrete | asymptotic | 1.5 | 1.7 | 2.6 | 1.6 | 1.0 | 3.3 | 1.0 | 1.0 | 2.8 |
| | asymptotic_bonf | 1.1 | 1.3 | 2.4 | 1.5 | 1.0 | 2.9 | 0.8 | 0.9 | 2.2 |
| | permutation_bonf | 1.0 | 1.6 | 2.0 | 1.3 | 1.0 | 2.8 | 0.8 | 0.8 | 2.0 |

Table S76: Rejection rates in percent for the Tukey-type contrast matrix with $\delta = 1.5$ and unbalanced large sample sizes under equal censoring.

| Distribution | Method | $\mathcal{H}_{0,7}$ | $\mathcal{H}_{0,8}$ | $\mathcal{H}_{0,9}$ | $\mathcal{H}_{0,13}$ | $\mathcal{H}_{0,14}$ | $\mathcal{H}_{0,15}$ | $\mathcal{H}_{0,16}$ | $\mathcal{H}_{0,17}$ | $\mathcal{H}_{0,18}$ |
|---|---|---|---|---|---|---|---|---|---|---|
| exp early continuous | asymptotic | 31.3 | 16.0 | 53.4 | 27.3 | 12.3 | 42.7 | 27.1 | 13.2 | 45.8 |
| | asymptotic_bonf | 27.3 | 13.7 | 50.6 | 23.9 | 11.2 | 39.4 | 23.3 | 11.7 | 41.3 |
| | permutation_bonf | 24.4 | 11.3 | 47.7 | 21.7 | 9.9 | 36.8 | 22.4 | 11.0 | 38.8 |
| exp early discrete | asymptotic | 36.9 | 19.3 | 59.6 | 30.7 | 15.7 | 48.3 | 32.2 | 17.1 | 51.7 |
| | asymptotic_bonf | 33.0 | 17.0 | 57.0 | 28.4 | 13.7 | 45.2 | 29.1 | 14.4 | 47.7 |
| | permutation_bonf | 29.1 | 13.4 | 53.4 | 24.8 | 12.6 | 45.3 | 26.7 | 13.4 | 46.1 |
| exp late continuous | asymptotic | 36.4 | 20.4 | 62.6 | 30.4 | 15.0 | 50.2 | 31.8 | 15.7 | 53.2 |
| | asymptotic_bonf | 33.0 | 17.4 | 59.6 | 26.8 | 13.0 | 46.5 | 28.9 | 13.7 | 49.4 |
| | permutation_bonf | 29.0 | 14.4 | 57.1 | 25.1 | 11.7 | 45.2 | 26.3 | 13.1 | 46.6 |
| exp late discrete | asymptotic | 44.3 | 23.9 | 69.1 | 35.5 | 19.0 | 57.5 | 37.8 | 18.9 | 59.1 |
| | asymptotic_bonf | 39.4 | 21.4 | 65.7 | 31.5 | 16.4 | 54.3 | 34.3 | 16.4 | 56.2 |
| | permutation_bonf | 35.3 | 16.6 | 63.2 | 28.9 | 14.4 | 52.5 | 33.0 | 15.7 | 54.4 |
| exp prop continuous | asymptotic | 35.9 | 20.0 | 54.2 | 29.8 | 16.9 | 42.7 | 28.0 | 16.1 | 46.7 |
| | asymptotic_bonf | 32.5 | 17.7 | 50.4 | 26.5 | 14.3 | 38.8 | 25.7 | 14.2 | 42.9 |
| | permutation_bonf | 30.0 | 14.6 | 47.4 | 25.1 | 12.7 | 37.2 | 23.8 | 12.6 | 41.3 |
| exp prop discrete | asymptotic | 41.7 | 24.8 | 62.3 | 34.3 | 20.3 | 49.3 | 35.4 | 19.6 | 54.9 |
| | asymptotic_bonf | 37.8 | 21.9 | 57.8 | 31.1 | 17.7 | 46.0 | 31.0 | 17.3 | 51.4 |
| | permutation_bonf | 35.1 | 17.6 | 55.2 | 29.8 | 15.3 | 42.8 | 28.6 | 14.7 | 48.5 |
| logn continuous | asymptotic | 84.8 | 63.0 | 94.1 | 74.7 | 51.8 | 88.9 | 75.6 | 56.4 | 90.3 |
| | asymptotic_bonf | 82.1 | 59.7 | 93.1 | 72.5 | 48.0 | 86.5 | 73.4 | 52.8 | 88.6 |
| | permutation_bonf | 75.4 | 50.1 | 89.9 | 69.7 | 43.1 | 83.8 | 68.9 | 48.1 | 85.9 |
| logn discrete | asymptotic | 91.0 | 71.0 | 97.3 | 81.3 | 62.2 | 93.4 | 84.0 | 65.5 | 94.1 |
| | asymptotic_bonf | 89.0 | 67.4 | 96.5 | 79.4 | 57.7 | 92.2 | 82.4 | 62.2 | 93.5 |
| | permutation_bonf | 82.6 | 59.6 | 94.3 | 76.0 | 52.2 | 90.4 | 77.4 | 56.0 | 91.3 |
| pwExp continuous | asymptotic | 29.2 | 15.3 | 52.1 | 24.1 | 12.5 | 42.5 | 25.7 | 13.6 | 43.1 |
| | asymptotic_bonf | 26.7 | 13.1 | 47.7 | 21.7 | 10.5 | 39.1 | 22.8 | 11.5 | 39.9 |
| | permutation_bonf | 24.2 | 10.4 | 45.0 | 19.7 | 9.1 | 37.2 | 21.1 | 11.8 | 37.6 |
| pwExp discrete | asymptotic | 34.8 | 19.2 | 60.8 | 29.3 | 15.3 | 49.2 | 30.8 | 16.9 | 50.7 |
| | asymptotic_bonf | 32.0 | 15.9 | 57.0 | 25.8 | 13.3 | 45.7 | 27.6 | 14.2 | 47.3 |
| | permutation_bonf | 28.9 | 12.3 | 51.8 | 25.5 | 11.9 | 44.6 | 25.6 | 13.6 | 44.7 |
| Weib late continuous | asymptotic | 84.8 | 62.0 | 96.4 | 78.6 | 52.7 | 91.7 | 78.2 | 55.2 | 94.3 |
| | asymptotic_bonf | 83.2 | 58.2 | 95.6 | 75.2 | 49.8 | 90.7 | 75.4 | 51.1 | 93.3 |
| | permutation_bonf | 80.2 | 49.2 | 94.2 | 73.2 | 44.7 | 88.3 | 71.9 | 46.0 | 91.2 |
| Weib late discrete | asymptotic | 87.8 | 69.5 | 95.9 | 81.1 | 59.8 | 93.3 | 80.8 | 60.5 | 96.0 |
| | asymptotic_bonf | 86.2 | 66.0 | 95.8 | 79.3 | 56.1 | 92.1 | 78.6 | 58.8 | 95.0 |
| | permutation_bonf | 81.6 | 57.1 | 94.8 | 77.2 | 52.6 | 91.3 | 75.8 | 52.7 | 93.4 |
| Weib prop continuous | asymptotic | 82.1 | 58.0 | 95.1 | 75.5 | 50.6 | 90.1 | 75.2 | 52.2 | 91.9 |
| | asymptotic_bonf | 80.3 | 55.1 | 94.4 | 72.6 | 46.3 | 88.0 | 73.1 | 47.7 | 90.1 |
| | permutation_bonf | 75.4 | 46.1 | 92.6 | 69.7 | 42.0 | 86.7 | 69.3 | 44.1 | 87.8 |
| Weib prop discrete | asymptotic | 87.9 | 68.2 | 97.4 | 82.4 | 59.4 | 94.7 | 82.4 | 60.6 | 95.9 |
| | asymptotic_bonf | 85.8 | 65.0 | 96.7 | 80.3 | 55.5 | 93.8 | 80.8 | 57.4 | 95.0 |
| | permutation_bonf | 82.8 | 54.9 | 95.9 | 77.6 | 51.4 | 92.5 | 77.9 | 52.5 | 93.3 |
| Weib scale continuous | asymptotic | 69.7 | 46.9 | 86.6 | 63.0 | 39.7 | 80.3 | 63.4 | 42.3 | 82.1 |
| | asymptotic_bonf | 67.4 | 42.7 | 84.9 | 59.3 | 35.8 | 77.8 | 59.9 | 38.8 | 79.8 |
| | permutation_bonf | 61.5 | 35.1 | 81.3 | 55.0 | 31.5 | 74.8 | 56.4 | 33.5 | 76.8 |
| Weib scale discrete | asymptotic | 77.2 | 54.7 | 92.3 | 72.0 | 48.7 | 88.5 | 72.5 | 50.0 | 88.5 |
| | asymptotic_bonf | 74.4 | 50.8 | 90.7 | 69.1 | 44.6 | 86.4 | 68.9 | 46.7 | 87.6 |
| | permutation_bonf | 69.1 | 42.6 | 86.8 | 65.2 | 39.7 | 83.7 | 64.9 | 41.1 | 84.6 |
| Weib shape continuous | asymptotic | 52.9 | 29.0 | 71.2 | 45.8 | 27.3 | 65.3 | 47.9 | 28.5 | 67.8 |
| | asymptotic_bonf | 48.2 | 24.3 | 67.8 | 42.6 | 23.3 | 61.1 | 43.6 | 24.5 | 63.7 |
| | permutation_bonf | 41.2 | 18.8 | 62.9 | 37.8 | 20.4 | 58.3 | 39.6 | 20.6 | 59.6 |
| Weib shape discrete | asymptotic | 64.7 | 40.1 | 81.5 | 56.2 | 35.0 | 75.2 | 57.1 | 35.9 | 77.6 |
| | asymptotic_bonf | 59.9 | 35.0 | 79.1 | 52.1 | 32.4 | 71.9 | 54.6 | 32.7 | 74.6 |
| | permutation_bonf | 51.7 | 25.6 | 73.9 | 48.8 | 27.2 | 70.3 | 50.7 | 28.0 | 70.4 |

Table S77: Rejection rates in percent for the Tukey-type contrast matrix with $\delta = 1.5$ and unbalanced medium sample sizes under equal censoring.

| Distribution | Method | $\mathcal{H}_{0,7}$ | $\mathcal{H}_{0,8}$ | $\mathcal{H}_{0,9}$ | $\mathcal{H}_{0,13}$ | $\mathcal{H}_{0,14}$ | $\mathcal{H}_{0,15}$ | $\mathcal{H}_{0,16}$ | $\mathcal{H}_{0,17}$ | $\mathcal{H}_{0,18}$ |
|---|---|---|---|---|---|---|---|---|---|---|
| exp early continuous | asymptotic | 1.6 | 1.5 | 3.8 | 1.6 | 1.1 | 3.5 | 2.7 | 1.1 | 4.2 |
| | asymptotic_bonf | 1.5 | 0.9 | 3.2 | 1.2 | 0.9 | 3.0 | 2.5 | 0.9 | 3.4 |
| | permutation_bonf | 0.5 | 0.2 | 1.6 | 0.6 | 0.3 | 2.3 | 1.8 | 0.5 | 2.5 |
| exp early discrete | asymptotic | 2.4 | 1.8 | 4.4 | 1.9 | 1.4 | 3.9 | 3.5 | 1.2 | 4.4 |
| | asymptotic_bonf | 2.0 | 1.3 | 3.9 | 1.6 | 1.1 | 3.6 | 2.8 | 0.8 | 3.9 |
| | permutation_bonf | 0.7 | 0.2 | 2.4 | 0.8 | 0.6 | 2.3 | 1.7 | 0.5 | 2.7 |
| exp late continuous | asymptotic | 2.4 | 1.7 | 5.2 | 1.8 | 1.4 | 4.1 | 3.1 | 1.4 | 4.8 |
| | asymptotic_bonf | 2.0 | 1.5 | 3.9 | 1.2 | 1.2 | 3.4 | 2.6 | 0.8 | 3.9 |
| | permutation_bonf | 0.8 | 0.2 | 1.8 | 0.8 | 0.5 | 2.7 | 2.0 | 0.4 | 3.0 |
| exp late discrete | asymptotic | 2.8 | 1.8 | 5.9 | 2.2 | 1.6 | 4.4 | 3.8 | 1.2 | 5.2 |
| | asymptotic_bonf | 2.2 | 1.7 | 5.1 | 1.8 | 1.5 | 4.0 | 3.3 | 1.1 | 4.3 |
| | permutation_bonf | 0.9 | 0.6 | 2.8 | 0.9 | 0.8 | 2.9 | 1.9 | 0.5 | 3.3 |
| exp prop continuous | asymptotic | 3.2 | 1.4 | 3.2 | 2.6 | 1.2 | 3.5 | 2.2 | 1.8 | 2.9 |
| | asymptotic_bonf | 2.2 | 1.0 | 3.0 | 2.2 | 0.8 | 2.6 | 1.9 | 1.2 | 2.3 |
| | permutation_bonf | 1.2 | 0.4 | 1.7 | 1.5 | 0.3 | 1.7 | 1.0 | 0.3 | 1.9 |
| exp prop discrete | asymptotic | 3.4 | 1.5 | 4.3 | 3.0 | 1.2 | 4.4 | 2.7 | 1.9 | 4.0 |
| | asymptotic_bonf | 2.9 | 1.0 | 3.4 | 2.5 | 0.9 | 3.3 | 2.5 | 1.7 | 3.4 |
| | permutation_bonf | 1.3 | 0.3 | 2.1 | 1.6 | 0.6 | 2.3 | 1.7 | 0.4 | 2.0 |
| logn continuous | asymptotic | 7.1 | 4.6 | 13.7 | 7.3 | 4.5 | 13.0 | 6.3 | 4.5 | 12.5 |
| | asymptotic_bonf | 5.9 | 4.0 | 12.0 | 5.7 | 3.7 | 11.5 | 5.1 | 3.3 | 10.6 |
| | permutation_bonf | 1.2 | 0.2 | 5.3 | 2.5 | 1.0 | 7.1 | 3.3 | 1.0 | 6.3 |
| logn discrete | asymptotic | 8.9 | 5.5 | 16.7 | 8.6 | 5.2 | 16.5 | 8.0 | 3.9 | 15.6 |
| | asymptotic_bonf | 7.1 | 4.4 | 15.0 | 7.4 | 4.0 | 13.3 | 6.5 | 3.6 | 13.0 |
| | permutation_bonf | 1.2 | 0.2 | 5.5 | 3.1 | 1.0 | 8.5 | 3.1 | 1.5 | 9.0 |
| pwExp continuous | asymptotic | 2.1 | 1.5 | 3.5 | 1.2 | 1.3 | 3.4 | 2.3 | 1.3 | 3.5 |
| | asymptotic_bonf | 1.5 | 1.4 | 3.3 | 0.7 | 1.2 | 2.9 | 1.8 | 1.0 | 2.7 |
| | permutation_bonf | 0.3 | 0.4 | 1.5 | 0.4 | 0.6 | 2.4 | 1.2 | 0.3 | 1.9 |
| pwExp discrete | asymptotic | 2.7 | 2.0 | 4.0 | 1.8 | 1.4 | 4.4 | 2.5 | 1.5 | 4.2 |
| | asymptotic_bonf | 2.1 | 1.5 | 3.6 | 1.4 | 1.2 | 3.1 | 2.1 | 1.1 | 3.5 |
| | permutation_bonf | 0.5 | 0.4 | 1.5 | 0.6 | 0.7 | 2.7 | 1.3 | 0.2 | 2.4 |
| Weib late continuous | asymptotic | 8.5 | 4.0 | 17.3 | 8.5 | 4.0 | 14.1 | 8.0 | 4.1 | 16.0 |
| | asymptotic_bonf | 6.7 | 2.8 | 14.3 | 6.9 | 2.7 | 11.8 | 6.6 | 3.0 | 13.9 |
| | permutation_bonf | 2.1 | 0.2 | 7.3 | 4.1 | 1.0 | 8.0 | 2.7 | 0.5 | 8.6 |
| Weib late discrete | asymptotic | 10.4 | 4.6 | 21.4 | 9.7 | 4.9 | 17.5 | 10.2 | 4.0 | 19.9 |
| | asymptotic_bonf | 9.2 | 3.6 | 18.7 | 7.5 | 3.5 | 15.8 | 8.6 | 3.2 | 17.1 |
| | permutation_bonf | 2.0 | 0.1 | 8.4 | 4.7 | 1.2 | 10.8 | 3.3 | 0.7 | 11.1 |
| Weib prop continuous | asymptotic | 7.6 | 3.4 | 14.7 | 6.7 | 3.3 | 11.8 | 6.6 | 3.5 | 12.8 |
| | asymptotic_bonf | 5.9 | 2.6 | 12.4 | 5.8 | 2.5 | 10.1 | 5.3 | 2.3 | 11.4 |
| | permutation_bonf | 1.5 | 0.1 | 5.4 | 3.9 | 0.9 | 6.8 | 2.7 | 0.4 | 6.7 |
| Weib prop discrete | asymptotic | 9.4 | 3.8 | 17.2 | 8.0 | 3.8 | 14.8 | 8.0 | 4.5 | 16.6 |
| | asymptotic_bonf | 7.9 | 2.8 | 14.6 | 6.5 | 2.7 | 12.9 | 6.8 | 2.9 | 14.0 |
| | permutation_bonf | 1.7 | 0.1 | 5.7 | 4.3 | 0.6 | 8.3 | 2.6 | 0.0 | 7.9 |
| Weib scale continuous | asymptotic | 4.6 | 1.6 | 9.4 | 4.8 | 1.7 | 8.0 | 4.4 | 1.8 | 8.6 |
| | asymptotic_bonf | 3.4 | 1.3 | 7.9 | 4.1 | 1.1 | 6.8 | 3.1 | 1.6 | 6.9 |
| | permutation_bonf | 0.6 | 0.0 | 2.3 | 2.3 | 0.6 | 3.6 | 1.3 | 0.1 | 3.8 |
| Weib scale discrete | asymptotic | 5.8 | 2.0 | 10.1 | 5.5 | 2.3 | 9.6 | 5.6 | 2.5 | 10.1 |
| | asymptotic_bonf | 4.3 | 1.4 | 9.2 | 4.8 | 1.8 | 8.4 | 4.3 | 1.8 | 8.4 |
| | permutation_bonf | 0.9 | 0.0 | 3.4 | 3.0 | 0.5 | 4.7 | 1.3 | 0.1 | 4.6 |
| Weib shape continuous | asymptotic | 2.3 | 1.1 | 4.3 | 2.8 | 1.1 | 4.0 | 2.5 | 1.0 | 3.9 |
| | asymptotic_bonf | 1.7 | 0.7 | 3.2 | 2.5 | 0.5 | 3.2 | 1.8 | 0.7 | 2.6 |
| | permutation_bonf | 0.2 | 0.0 | 1.1 | 1.1 | 0.2 | 1.9 | 0.8 | 0.1 | 1.3 |
| Weib shape discrete | asymptotic | 3.3 | 1.1 | 5.2 | 3.4 | 1.0 | 5.4 | 3.2 | 1.2 | 5.7 |
| | asymptotic_bonf | 2.3 | 1.0 | 4.6 | 2.9 | 0.9 | 4.0 | 2.6 | 0.9 | 4.1 |
| | permutation_bonf | 0.1 | 0.0 | 1.1 | 1.0 | 0.2 | 2.8 | 0.5 | 0.0 | 2.2 |

Table S78: Rejection rates in percent for the Tukey-type contrast matrix with $\delta = 1.5$ and unbalanced small sample sizes under equal censoring.

| Distribution | Method | $\mathcal{H}_{0,7}$ | $\mathcal{H}_{0,8}$ | $\mathcal{H}_{0,9}$ | $\mathcal{H}_{0,13}$ | $\mathcal{H}_{0,14}$ | $\mathcal{H}_{0,15}$ | $\mathcal{H}_{0,16}$ | $\mathcal{H}_{0,17}$ | $\mathcal{H}_{0,18}$ |
|---|---|---|---|---|---|---|---|---|---|---|
| exp early continuous | asymptotic | 0.8 | 3.0 | 1.7 | 0.8 | 1.7 | 1.5 | 0.6 | 2.2 | 1.7 |
|  | asymptotic_bonf | 0.7 | 3.0 | 1.6 | 0.6 | 1.7 | 1.3 | 0.6 | 2.0 | 1.5 |
|  | permutation_bonf | 0.1 | 0.1 | 0.2 | 0.1 | 0.1 | 0.1 | 0.0 | 0.4 | 0.4 |
| exp early discrete | asymptotic | 0.9 | 2.9 | 2.0 | 0.9 | 1.5 | 1.7 | 0.7 | 2.1 | 1.7 |
|  | asymptotic_bonf | 0.6 | 2.7 | 1.7 | 0.7 | 1.5 | 1.6 | 0.7 | 1.7 | 1.4 |
|  | permutation_bonf | 0.1 | 0.2 | 0.1 | 0.1 | 0.3 | 0.3 | 0.0 | 0.3 | 0.2 |
| exp late continuous | asymptotic | 1.0 | 1.9 | 2.2 | 0.6 | 1.0 | 1.9 | 1.0 | 1.7 | 1.7 |
|  | asymptotic_bonf | 0.7 | 1.9 | 1.7 | 0.5 | 1.0 | 1.6 | 0.6 | 1.5 | 1.5 |
|  | permutation_bonf | 0.0 | 0.1 | 0.0 | 0.0 | 0.1 | 0.0 | 0.0 | 0.3 | 0.2 |
| exp late discrete | asymptotic | 1.1 | 2.0 | 2.6 | 1.0 | 0.9 | 2.2 | 0.9 | 1.9 | 1.9 |
|  | asymptotic_bonf | 0.9 | 2.0 | 2.0 | 0.6 | 0.8 | 1.5 | 0.8 | 1.6 | 1.6 |
|  | permutation_bonf | 0.0 | 0.1 | 0.0 | 0.0 | 0.2 | 0.2 | 0.0 | 0.2 | 0.1 |
| exp prop continuous | asymptotic | 1.0 | 2.0 | 1.9 | 1.5 | 1.6 | 1.7 | 1.1 | 1.4 | 2.1 |
|  | asymptotic_bonf | 1.0 | 1.8 | 1.5 | 1.1 | 1.2 | 1.3 | 0.9 | 1.3 | 1.7 |
|  | permutation_bonf | 0.2 | 0.3 | 0.3 | 0.0 | 0.1 | 0.2 | 0.2 | 0.0 | 0.3 |
| exp prop discrete | asymptotic | 1.2 | 2.3 | 2.1 | 1.3 | 1.6 | 2.2 | 1.0 | 1.5 | 2.0 |
|  | asymptotic_bonf | 1.0 | 2.2 | 1.9 | 1.1 | 1.2 | 1.6 | 0.6 | 1.4 | 1.6 |
|  | permutation_bonf | 0.1 | 0.1 | 0.3 | 0.0 | 0.1 | 0.3 | 0.2 | 0.0 | 0.3 |
| logn continuous | asymptotic | 2.7 | 4.6 | 2.3 | 2.0 | 1.4 | 2.4 | 1.7 | 2.1 | 2.2 |
|  | asymptotic_bonf | 2.2 | 4.5 | 2.1 | 1.6 | 1.0 | 2.3 | 1.1 | 1.7 | 2.0 |
|  | permutation_bonf | 0.3 | 0.1 | 0.1 | 0.2 | 0.0 | 0.4 | 0.1 | 0.1 | 0.1 |
| logn discrete | asymptotic | 2.9 | 5.1 | 2.9 | 2.6 | 1.1 | 2.5 | 2.2 | 1.7 | 2.7 |
|  | asymptotic_bonf | 2.7 | 5.0 | 2.4 | 2.1 | 1.0 | 2.2 | 1.7 | 1.3 | 2.2 |
|  | permutation_bonf | 0.1 | 0.1 | 0.1 | 0.3 | 0.1 | 0.3 | 0.1 | 0.0 | 0.0 |
| pwExp continuous | asymptotic | 1.1 | 2.4 | 1.8 | 0.7 | 1.4 | 1.5 | 0.7 | 1.7 | 1.7 |
|  | asymptotic_bonf | 0.9 | 2.4 | 1.6 | 0.5 | 1.3 | 1.1 | 0.4 | 1.4 | 1.4 |
|  | permutation_bonf | 0.1 | 0.1 | 0.2 | 0.0 | 0.1 | 0.3 | 0.0 | 0.3 | 0.3 |
| pwExp discrete | asymptotic | 1.1 | 2.5 | 1.6 | 0.7 | 1.2 | 1.9 | 0.8 | 1.7 | 1.9 |
|  | asymptotic_bonf | 1.0 | 2.4 | 1.5 | 0.6 | 1.0 | 1.3 | 0.3 | 1.2 | 1.8 |
|  | permutation_bonf | 0.1 | 0.0 | 0.2 | 0.0 | 0.1 | 0.4 | 0.0 | 0.2 | 0.2 |
| Weib late continuous | asymptotic | 1.6 | 2.1 | 2.6 | 1.3 | 0.7 | 2.1 | 1.7 | 1.0 | 2.0 |
|  | asymptotic_bonf | 1.5 | 1.9 | 2.3 | 1.2 | 0.6 | 1.8 | 1.6 | 0.8 | 1.6 |
|  | permutation_bonf | 0.1 | 0.0 | 0.2 | 0.2 | 0.1 | 0.3 | 0.1 | 0.0 | 0.1 |
| Weib late discrete | asymptotic | 2.4 | 2.9 | 2.8 | 1.9 | 1.1 | 2.5 | 1.7 | 1.0 | 2.3 |
|  | asymptotic_bonf | 2.0 | 2.7 | 2.2 | 1.5 | 1.0 | 2.2 | 1.6 | 0.7 | 1.9 |
|  | permutation_bonf | 0.0 | 0.0 | 0.1 | 0.2 | 0.2 | 0.1 | 0.3 | 0.0 | 0.1 |
| Weib prop continuous | asymptotic | 1.9 | 3.0 | 2.4 | 1.1 | 0.8 | 1.9 | 1.6 | 1.7 | 1.9 |
|  | asymptotic_bonf | 1.7 | 2.6 | 1.9 | 1.0 | 0.7 | 1.7 | 1.5 | 1.3 | 1.8 |
|  | permutation_bonf | 0.1 | 0.1 | 0.2 | 0.2 | 0.1 | 0.0 | 0.1 | 0.0 | 0.0 |
| Weib prop discrete | asymptotic | 2.2 | 3.4 | 2.4 | 1.6 | 0.9 | 2.2 | 1.6 | 1.3 | 1.7 |
|  | asymptotic_bonf | 2.1 | 3.3 | 2.1 | 1.1 | 0.7 | 2.0 | 1.3 | 1.1 | 1.5 |
|  | permutation_bonf | 0.0 | 0.1 | 0.2 | 0.1 | 0.0 | 0.0 | 0.2 | 0.0 | 0.0 |
| Weib scale continuous | asymptotic | 2.9 | 4.3 | 1.9 | 2.3 | 1.1 | 1.7 | 2.3 | 2.2 | 1.9 |
|  | asymptotic_bonf | 2.9 | 4.2 | 1.9 | 1.9 | 1.0 | 1.4 | 1.8 | 1.8 | 1.7 |
|  | permutation_bonf | 0.1 | 0.1 | 0.3 | 0.2 | 0.2 | 0.0 | 0.1 | 0.0 | 0.0 |
| Weib scale discrete | asymptotic | 3.4 | 5.5 | 2.5 | 1.9 | 1.8 | 1.8 | 2.1 | 1.9 | 2.1 |
|  | asymptotic_bonf | 3.2 | 5.5 | 2.4 | 1.5 | 1.2 | 1.5 | 1.9 | 1.7 | 1.6 |
|  | permutation_bonf | 0.0 | 0.1 | 0.3 | 0.2 | 0.2 | 0.0 | 0.3 | 0.1 | 0.0 |
| Weib shape continuous | asymptotic | 6.1 | 9.3 | 3.0 | 4.6 | 3.3 | 2.7 | 5.2 | 5.1 | 2.9 |
|  | asymptotic_bonf | 5.9 | 9.0 | 2.6 | 4.1 | 3.0 | 2.5 | 4.6 | 4.6 | 2.5 |
|  | permutation_bonf | 0.5 | 0.5 | 0.7 | 0.8 | 0.6 | 0.3 | 0.5 | 0.3 | 0.6 |
| Weib shape discrete | asymptotic | 6.3 | 9.4 | 3.3 | 3.9 | 2.6 | 2.3 | 4.5 | 4.1 | 3.0 |
|  | asymptotic_bonf | 6.1 | 9.1 | 3.0 | 3.4 | 2.2 | 2.2 | 4.2 | 3.7 | 2.5 |
|  | permutation_bonf | 0.1 | 0.3 | 0.6 | 0.8 | 0.4 | 0.1 | 0.3 | 0.2 | 0.3 |

Table S79: Rejection rates in percent for the Tukey-type contrast matrix with $\delta = 1.5$ and balanced large sample sizes under unequal, high censoring.

| Distribution | Method | $\mathcal{H}_{0,7}$ | $\mathcal{H}_{0,8}$ | $\mathcal{H}_{0,9}$ | $\mathcal{H}_{0,13}$ | $\mathcal{H}_{0,14}$ | $\mathcal{H}_{0,15}$ | $\mathcal{H}_{0,16}$ | $\mathcal{H}_{0,17}$ | $\mathcal{H}_{0,18}$ |
|---|---|---|---|---|---|---|---|---|---|---|
| exp early continuous | asymptotic | 67.1 | 46.9 | 85.6 | 65.1 | 43.0 | 85.2 | 68.7 | 45.8 | 90.2 |
|  | asymptotic_bonf | 65.3 | 44.7 | 84.3 | 63.2 | 40.9 | 83.9 | 67.2 | 43.6 | 88.9 |
|  | permutation_bonf | 62.6 | 43.0 | 83.7 | 60.7 | 39.3 | 82.5 | 64.8 | 43.2 | 86.4 |
| exp early discrete | asymptotic | 71.6 | 50.9 | 88.1 | 69.7 | 46.3 | 87.9 | 72.9 | 51.5 | 92.4 |
|  | asymptotic_bonf | 69.9 | 49.0 | 87.0 | 68.1 | 44.4 | 86.3 | 70.7 | 49.4 | 90.9 |
|  | permutation_bonf | 68.3 | 48.1 | 86.5 | 66.6 | 41.6 | 85.2 | 69.5 | 47.8 | 89.6 |
| exp late continuous | asymptotic | 68.7 | 48.2 | 86.9 | 66.2 | 44.0 | 86.7 | 70.9 | 49.0 | 91.4 |
|  | asymptotic_bonf | 66.9 | 45.8 | 86.0 | 64.0 | 41.2 | 85.2 | 69.1 | 46.1 | 90.6 |
|  | permutation_bonf | 66.0 | 44.9 | 85.2 | 63.2 | 41.2 | 84.2 | 67.3 | 45.1 | 88.9 |
| exp late discrete | asymptotic | 73.9 | 53.2 | 89.4 | 71.9 | 47.2 | 89.1 | 74.9 | 52.1 | 93.2 |
|  | asymptotic_bonf | 72.3 | 50.5 | 88.9 | 69.6 | 45.4 | 87.4 | 73.3 | 50.2 | 92.7 |
|  | permutation_bonf | 71.2 | 48.5 | 87.0 | 68.9 | 44.0 | 87.2 | 72.7 | 48.4 | 91.1 |
| exp prop continuous | asymptotic | 69.1 | 49.0 | 87.8 | 67.0 | 47.7 | 86.7 | 69.3 | 52.0 | 89.1 |
|  | asymptotic_bonf | 67.5 | 47.5 | 86.4 | 65.1 | 46.1 | 85.8 | 67.9 | 49.8 | 88.2 |
|  | permutation_bonf | 66.2 | 46.0 | 84.7 | 63.1 | 44.7 | 84.9 | 66.7 | 48.6 | 87.5 |
| exp prop discrete | asymptotic | 74.0 | 52.9 | 89.3 | 72.0 | 51.7 | 89.4 | 75.0 | 55.6 | 91.0 |
|  | asymptotic_bonf | 72.3 | 50.9 | 88.4 | 70.4 | 49.8 | 88.4 | 72.4 | 53.5 | 90.0 |
|  | permutation_bonf | 70.3 | 49.6 | 87.3 | 68.4 | 47.0 | 86.1 | 70.8 | 52.5 | 90.4 |
| logn continuous | asymptotic | 98.0 | 89.6 | 99.8 | 96.7 | 87.5 | 99.3 | 98.0 | 88.4 | 99.8 |
|  | asymptotic_bonf | 97.6 | 88.1 | 99.7 | 96.4 | 86.8 | 99.3 | 97.9 | 87.1 | 99.8 |
|  | permutation_bonf | 97.7 | 86.9 | 99.6 | 96.2 | 84.2 | 99.1 | 97.2 | 87.4 | 99.9 |
| logn discrete | asymptotic | 99.1 | 94.8 | 100.0 | 98.4 | 92.6 | 99.7 | 99.2 | 93.5 | 100.0 |
|  | asymptotic_bonf | 99.0 | 93.9 | 100.0 | 98.2 | 92.0 | 99.7 | 99.1 | 92.6 | 100.0 |
|  | permutation_bonf | 99.0 | 93.6 | 99.9 | 98.0 | 90.8 | 99.5 | 98.4 | 92.9 | 99.9 |
| pwExp continuous | asymptotic | 65.5 | 44.9 | 84.6 | 65.1 | 41.7 | 85.8 | 66.7 | 45.3 | 89.1 |
|  | asymptotic_bonf | 63.5 | 42.3 | 83.2 | 62.4 | 39.7 | 84.7 | 65.4 | 43.2 | 87.8 |
|  | permutation_bonf | 60.9 | 41.9 | 83.7 | 61.9 | 39.4 | 82.9 | 65.2 | 43.1 | 86.3 |
| pwExp discrete | asymptotic | 70.7 | 49.6 | 87.8 | 69.9 | 44.8 | 88.0 | 70.9 | 50.2 | 91.6 |
|  | asymptotic_bonf | 68.9 | 47.2 | 87.1 | 68.2 | 43.1 | 86.8 | 69.5 | 48.2 | 91.0 |
|  | permutation_bonf | 66.7 | 46.5 | 86.8 | 66.3 | 42.4 | 86.6 | 68.4 | 47.1 | 89.2 |
| Weib late continuous | asymptotic | 98.1 | 90.1 | 100.0 | 96.5 | 89.2 | 99.7 | 97.0 | 89.4 | 99.8 |
|  | asymptotic_bonf | 98.0 | 89.1 | 100.0 | 96.3 | 87.7 | 99.7 | 96.9 | 88.6 | 99.7 |
|  | permutation_bonf | 97.5 | 88.3 | 99.6 | 95.4 | 86.9 | 99.6 | 96.0 | 87.3 | 99.2 |
| Weib late discrete | asymptotic | 99.2 | 94.8 | 99.9 | 98.2 | 92.8 | 99.9 | 98.4 | 92.9 | 99.8 |
|  | asymptotic_bonf | 99.0 | 94.1 | 99.9 | 98.0 | 91.7 | 99.9 | 98.2 | 92.2 | 99.8 |
|  | permutation_bonf | 98.9 | 93.5 | 99.8 | 97.6 | 91.0 | 99.9 | 98.1 | 91.6 | 99.6 |
| Weib prop continuous | asymptotic | 97.7 | 88.9 | 100.0 | 96.5 | 88.7 | 99.7 | 96.7 | 88.9 | 99.5 |
|  | asymptotic_bonf | 97.3 | 87.4 | 100.0 | 95.9 | 87.2 | 99.6 | 96.4 | 87.7 | 99.3 |
|  | permutation_bonf | 97.0 | 87.7 | 99.7 | 95.2 | 86.5 | 99.7 | 96.0 | 87.3 | 99.3 |
| Weib prop discrete | asymptotic | 99.0 | 94.0 | 100.0 | 97.9 | 92.2 | 99.9 | 98.5 | 93.0 | 99.9 |
|  | asymptotic_bonf | 98.9 | 93.4 | 100.0 | 97.7 | 91.6 | 99.9 | 98.2 | 92.3 | 99.8 |
|  | permutation_bonf | 99.0 | 92.9 | 99.9 | 97.5 | 90.8 | 99.9 | 97.9 | 91.7 | 99.6 |
| Weib scale continuous | asymptotic | 96.4 | 86.0 | 99.8 | 95.2 | 84.0 | 99.3 | 95.9 | 85.2 | 99.0 |
|  | asymptotic_bonf | 96.0 | 84.6 | 99.7 | 94.4 | 82.1 | 99.2 | 95.0 | 84.2 | 98.9 |
|  | permutation_bonf | 95.2 | 83.1 | 99.3 | 93.2 | 79.8 | 98.9 | 93.7 | 82.2 | 98.7 |
| Weib scale discrete | asymptotic | 98.3 | 91.0 | 100.0 | 97.0 | 89.3 | 99.7 | 98.1 | 90.4 | 99.4 |
|  | asymptotic_bonf | 98.2 | 90.3 | 99.8 | 96.9 | 88.4 | 99.7 | 97.8 | 89.1 | 99.4 |
|  | permutation_bonf | 97.4 | 90.6 | 99.7 | 96.7 | 88.0 | 99.4 | 97.2 | 88.1 | 99.2 |
| Weib shape continuous | asymptotic | 94.6 | 81.3 | 98.9 | 92.8 | 78.5 | 98.2 | 93.4 | 80.8 | 97.9 |
|  | asymptotic_bonf | 94.0 | 79.7 | 98.8 | 92.1 | 77.0 | 97.7 | 92.8 | 79.2 | 97.7 |
|  | permutation_bonf | 93.3 | 77.8 | 98.3 | 91.0 | 76.0 | 97.3 | 92.6 | 77.4 | 97.3 |
| Weib shape discrete | asymptotic | 97.1 | 87.6 | 99.7 | 95.9 | 85.4 | 99.5 | 96.0 | 85.5 | 99.0 |
|  | asymptotic_bonf | 96.5 | 86.4 | 99.7 | 95.4 | 83.7 | 99.2 | 95.9 | 84.9 | 98.9 |
|  | permutation_bonf | 96.5 | 86.2 | 99.1 | 94.3 | 82.5 | 99.2 | 95.2 | 84.2 | 98.5 |

Table S80: Rejection rates in percent for the Tukey-type contrast matrix with $\delta = 1.5$ and balanced medium sample sizes under unequal, high censoring.

| Distribution | Method | $\mathcal{H}_{0,7}$ | $\mathcal{H}_{0,8}$ | $\mathcal{H}_{0,9}$ | $\mathcal{H}_{0,13}$ | $\mathcal{H}_{0,14}$ | $\mathcal{H}_{0,15}$ | $\mathcal{H}_{0,16}$ | $\mathcal{H}_{0,17}$ | $\mathcal{H}_{0,18}$ |
|---|---|---|---|---|---|---|---|---|---|---|
| exp early continuous | asymptotic | 9.4 | 5.5 | 11.4 | 9.1 | 5.5 | 12.0 | 9.0 | 3.8 | 13.5 |
| | asymptotic_bonf | 8.6 | 4.7 | 10.5 | 8.1 | 4.7 | 10.4 | 8.4 | 3.6 | 12.3 |
| | permutation_bonf | 8.3 | 4.4 | 11.0 | 8.2 | 3.7 | 10.0 | 7.7 | 3.9 | 11.3 |
| exp early discrete | asymptotic | 11.1 | 5.1 | 12.9 | 10.4 | 4.4 | 14.8 | 9.9 | 5.0 | 15.4 |
| | asymptotic_bonf | 10.2 | 4.6 | 12.1 | 9.4 | 4.0 | 13.1 | 8.9 | 4.2 | 13.7 |
| | permutation_bonf | 9.6 | 4.4 | 12.4 | 9.1 | 3.9 | 12.9 | 8.6 | 3.8 | 13.0 |
| exp late continuous | asymptotic | 9.7 | 5.5 | 13.6 | 9.9 | 5.1 | 13.6 | 10.5 | 4.4 | 14.3 |
| | asymptotic_bonf | 8.6 | 4.7 | 12.6 | 9.3 | 4.5 | 12.7 | 9.8 | 3.7 | 12.8 |
| | permutation_bonf | 8.0 | 4.7 | 11.9 | 8.6 | 4.0 | 11.0 | 8.8 | 3.3 | 13.1 |
| exp late discrete | asymptotic | 11.0 | 5.4 | 14.5 | 11.0 | 5.5 | 15.0 | 10.8 | 4.9 | 17.6 |
| | asymptotic_bonf | 10.1 | 4.9 | 13.3 | 9.8 | 4.9 | 13.9 | 9.8 | 4.4 | 15.4 |
| | permutation_bonf | 9.8 | 4.2 | 12.4 | 9.5 | 4.8 | 12.8 | 9.8 | 4.4 | 15.1 |
| exp prop continuous | asymptotic | 10.0 | 6.4 | 11.5 | 8.3 | 6.4 | 13.1 | 9.1 | 5.8 | 13.8 |
| | asymptotic_bonf | 8.6 | 5.9 | 10.0 | 7.3 | 5.8 | 12.0 | 8.3 | 4.8 | 12.6 |
| | permutation_bonf | 8.0 | 5.6 | 10.6 | 6.6 | 6.2 | 11.1 | 7.6 | 4.7 | 11.6 |
| exp prop discrete | asymptotic | 10.8 | 7.4 | 13.2 | 9.2 | 6.7 | 14.4 | 10.7 | 6.2 | 15.1 |
| | asymptotic_bonf | 10.0 | 6.3 | 11.9 | 8.2 | 5.6 | 13.3 | 9.8 | 5.3 | 14.2 |
| | permutation_bonf | 9.7 | 5.2 | 11.5 | 7.6 | 5.6 | 12.5 | 9.1 | 4.9 | 13.4 |
| logn continuous | asymptotic | 21.7 | 13.4 | 39.2 | 20.6 | 13.2 | 36.2 | 23.8 | 12.3 | 35.7 |
| | asymptotic_bonf | 20.5 | 12.4 | 37.5 | 18.9 | 12.5 | 34.4 | 22.3 | 10.5 | 33.5 |
| | permutation_bonf | 19.1 | 11.8 | 35.3 | 18.3 | 11.4 | 32.0 | 21.5 | 10.5 | 31.8 |
| logn discrete | asymptotic | 27.1 | 17.4 | 46.2 | 24.2 | 17.2 | 42.3 | 28.9 | 15.5 | 41.9 |
| | asymptotic_bonf | 24.7 | 16.1 | 44.0 | 22.9 | 16.0 | 40.6 | 27.2 | 13.7 | 39.6 |
| | permutation_bonf | 24.2 | 15.6 | 42.7 | 22.3 | 16.3 | 40.3 | 25.9 | 14.2 | 37.2 |
| pwExp continuous | asymptotic | 8.1 | 5.5 | 12.6 | 8.3 | 5.0 | 12.0 | 8.4 | 3.6 | 13.2 |
| | asymptotic_bonf | 7.7 | 4.8 | 12.2 | 7.9 | 4.3 | 11.0 | 7.9 | 3.1 | 12.0 |
| | permutation_bonf | 6.8 | 4.5 | 11.3 | 8.4 | 3.9 | 10.9 | 7.0 | 3.3 | 12.0 |
| pwExp discrete | asymptotic | 10.1 | 5.5 | 14.9 | 9.7 | 4.4 | 14.5 | 9.3 | 4.5 | 14.8 |
| | asymptotic_bonf | 8.9 | 5.2 | 13.8 | 8.9 | 4.0 | 13.5 | 8.5 | 4.2 | 14.0 |
| | permutation_bonf | 8.8 | 4.4 | 12.5 | 9.0 | 4.2 | 12.8 | 7.6 | 4.2 | 12.8 |
| Weib late continuous | asymptotic | 24.1 | 13.8 | 39.0 | 21.2 | 12.5 | 38.1 | 24.5 | 15.0 | 37.8 |
| | asymptotic_bonf | 21.8 | 12.8 | 37.0 | 20.1 | 11.4 | 35.8 | 23.0 | 13.5 | 36.0 |
| | permutation_bonf | 20.1 | 12.0 | 35.8 | 17.9 | 10.2 | 33.3 | 20.8 | 12.7 | 33.3 |
| Weib late discrete | asymptotic | 26.8 | 16.3 | 43.9 | 25.0 | 14.5 | 40.8 | 28.4 | 16.6 | 41.8 |
| | asymptotic_bonf | 25.2 | 14.5 | 42.4 | 22.5 | 13.5 | 39.2 | 26.4 | 15.3 | 39.9 |
| | permutation_bonf | 23.5 | 14.5 | 39.8 | 21.3 | 12.6 | 37.4 | 25.3 | 15.5 | 38.0 |
| Weib prop continuous | asymptotic | 21.7 | 13.1 | 36.7 | 20.6 | 12.4 | 35.8 | 23.6 | 15.0 | 35.6 |
| | asymptotic_bonf | 19.9 | 12.2 | 35.3 | 19.1 | 11.7 | 33.2 | 22.7 | 13.7 | 34.3 |
| | permutation_bonf | 19.7 | 11.2 | 33.7 | 17.8 | 10.4 | 31.1 | 20.1 | 12.8 | 32.4 |
| Weib prop discrete | asymptotic | 28.4 | 16.1 | 44.2 | 24.5 | 15.1 | 41.4 | 27.8 | 16.5 | 40.8 |
| | asymptotic_bonf | 26.7 | 14.7 | 42.5 | 23.7 | 14.3 | 39.9 | 26.3 | 14.9 | 39.1 |
| | permutation_bonf | 24.9 | 13.9 | 39.5 | 22.5 | 13.9 | 39.1 | 24.9 | 14.9 | 38.8 |
| Weib scale continuous | asymptotic | 18.6 | 11.9 | 32.0 | 17.6 | 10.4 | 30.5 | 20.0 | 11.8 | 28.9 |
| | asymptotic_bonf | 16.9 | 10.9 | 30.1 | 16.2 | 9.6 | 29.0 | 18.0 | 11.1 | 27.3 |
| | permutation_bonf | 16.3 | 10.1 | 28.1 | 15.8 | 9.2 | 27.4 | 17.1 | 10.7 | 26.2 |
| Weib scale discrete | asymptotic | 22.9 | 13.4 | 39.2 | 21.2 | 13.3 | 36.4 | 24.3 | 14.8 | 34.8 |
| | asymptotic_bonf | 21.4 | 12.3 | 37.8 | 19.4 | 11.8 | 34.3 | 22.8 | 13.4 | 33.3 |
| | permutation_bonf | 19.5 | 12.2 | 35.7 | 19.4 | 12.4 | 32.8 | 21.4 | 12.6 | 32.3 |
| Weib shape continuous | asymptotic | 15.6 | 9.7 | 26.6 | 15.8 | 9.4 | 26.1 | 16.8 | 10.8 | 24.5 |
| | asymptotic_bonf | 14.4 | 8.6 | 24.7 | 14.0 | 8.4 | 25.0 | 14.6 | 9.1 | 23.0 |
| | permutation_bonf | 13.5 | 9.3 | 23.2 | 13.4 | 8.3 | 23.3 | 15.7 | 9.7 | 21.3 |
| Weib shape discrete | asymptotic | 18.1 | 10.9 | 32.9 | 18.9 | 11.9 | 32.1 | 20.4 | 12.4 | 28.1 |
| | asymptotic_bonf | 16.7 | 10.0 | 30.0 | 17.3 | 10.4 | 29.8 | 19.7 | 11.3 | 26.6 |
| | permutation_bonf | 15.6 | 10.5 | 27.8 | 15.8 | 10.0 | 28.3 | 18.0 | 10.5 | 24.6 |

Table S81: Rejection rates in percent for the Tukey-type contrast matrix with $\delta = 1.5$ and balanced small sample sizes under unequal, high censoring.

| Distribution | Method | $\mathcal{H}_{0,7}$ | $\mathcal{H}_{0,8}$ | $\mathcal{H}_{0,9}$ | $\mathcal{H}_{0,13}$ | $\mathcal{H}_{0,14}$ | $\mathcal{H}_{0,15}$ | $\mathcal{H}_{0,16}$ | $\mathcal{H}_{0,17}$ | $\mathcal{H}_{0,18}$ |
|---|---|---|---|---|---|---|---|---|---|---|
| exp early continuous | asymptotic | 2.1 | 1.2 | 2.2 | 1.7 | 1.4 | 2.4 | 2.1 | 0.7 | 2.1 |
| | asymptotic_bonf | 1.9 | 0.9 | 1.8 | 1.3 | 1.1 | 2.2 | 1.8 | 0.4 | 1.6 |
| | permutation_bonf | 1.3 | 0.8 | 1.5 | 1.2 | 1.3 | 1.9 | 1.1 | 0.5 | 1.5 |
| exp early discrete | asymptotic | 1.5 | 1.3 | 1.2 | 1.6 | 1.2 | 2.7 | 1.6 | 1.0 | 2.6 |
| | asymptotic_bonf | 1.3 | 1.2 | 1.0 | 1.3 | 1.1 | 2.6 | 1.3 | 0.6 | 2.0 |
| | permutation_bonf | 1.2 | 1.0 | 1.0 | 0.9 | 0.9 | 1.9 | 1.0 | 0.7 | 1.6 |
| exp late continuous | asymptotic | 2.0 | 1.3 | 2.6 | 2.1 | 1.2 | 2.5 | 2.2 | 0.8 | 2.4 |
| | asymptotic_bonf | 1.4 | 0.9 | 2.0 | 1.9 | 1.0 | 2.2 | 2.1 | 0.6 | 2.3 |
| | permutation_bonf | 1.1 | 0.7 | 1.5 | 1.3 | 1.1 | 1.7 | 1.3 | 0.5 | 1.7 |
| exp late discrete | asymptotic | 2.3 | 1.1 | 2.4 | 2.2 | 1.1 | 2.5 | 2.7 | 1.0 | 2.4 |
| | asymptotic_bonf | 2.0 | 0.9 | 2.3 | 2.1 | 1.1 | 2.2 | 2.1 | 0.8 | 2.2 |
| | permutation_bonf | 1.4 | 0.4 | 1.5 | 1.6 | 1.0 | 1.4 | 1.4 | 0.5 | 1.5 |
| exp prop continuous | asymptotic | 0.8 | 1.3 | 2.6 | 1.9 | 0.6 | 1.6 | 1.0 | 1.3 | 2.1 |
| | asymptotic_bonf | 0.6 | 1.3 | 2.2 | 1.8 | 0.6 | 1.4 | 0.9 | 1.1 | 1.9 |
| | permutation_bonf | 0.5 | 1.0 | 1.5 | 1.1 | 0.5 | 1.3 | 0.5 | 0.8 | 1.6 |
| exp prop discrete | asymptotic | 1.9 | 0.8 | 2.4 | 1.4 | 1.5 | 2.8 | 1.1 | 0.8 | 2.9 |
| | asymptotic_bonf | 1.5 | 0.7 | 2.2 | 1.4 | 1.2 | 2.6 | 0.9 | 0.7 | 2.2 |
| | permutation_bonf | 0.9 | 0.6 | 1.4 | 0.7 | 1.1 | 1.6 | 0.7 | 0.5 | 1.6 |
| logn continuous | asymptotic | 3.4 | 1.9 | 6.0 | 3.6 | 1.7 | 4.7 | 3.0 | 1.9 | 5.2 |
| | asymptotic_bonf | 3.0 | 1.5 | 5.3 | 3.2 | 1.5 | 4.3 | 2.3 | 1.7 | 4.6 |
| | permutation_bonf | 2.5 | 1.6 | 4.3 | 2.4 | 1.1 | 3.9 | 1.8 | 1.7 | 3.2 |
| logn discrete | asymptotic | 3.9 | 2.3 | 6.3 | 4.1 | 2.5 | 5.2 | 4.0 | 2.8 | 6.1 |
| | asymptotic_bonf | 3.6 | 1.8 | 5.8 | 3.6 | 2.3 | 4.7 | 3.6 | 2.4 | 5.2 |
| | permutation_bonf | 3.0 | 1.8 | 4.5 | 2.8 | 1.9 | 3.6 | 2.6 | 2.4 | 3.8 |
| pwExp continuous | asymptotic | 1.5 | 1.3 | 2.1 | 1.4 | 1.4 | 2.4 | 2.0 | 0.6 | 1.8 |
| | asymptotic_bonf | 1.1 | 1.3 | 1.8 | 1.2 | 1.3 | 2.0 | 1.7 | 0.6 | 1.5 |
| | permutation_bonf | 0.9 | 0.6 | 1.0 | 0.7 | 1.1 | 1.3 | 1.3 | 0.4 | 1.4 |
| pwExp discrete | asymptotic | 1.8 | 1.5 | 2.4 | 2.0 | 1.0 | 3.0 | 2.6 | 0.6 | 2.4 |
| | asymptotic_bonf | 1.6 | 1.1 | 2.1 | 1.7 | 0.9 | 2.3 | 2.3 | 0.5 | 1.9 |
| | permutation_bonf | 1.1 | 0.8 | 2.1 | 1.2 | 0.5 | 1.9 | 1.7 | 0.5 | 1.6 |
| Weib late continuous | asymptotic | 2.1 | 3.6 | 5.4 | 2.7 | 2.9 | 5.9 | 3.3 | 2.9 | 5.2 |
| | asymptotic_bonf | 2.0 | 3.2 | 4.4 | 2.3 | 2.6 | 5.2 | 3.0 | 2.6 | 4.5 |
| | permutation_bonf | 1.3 | 2.7 | 3.1 | 1.8 | 2.0 | 3.7 | 2.0 | 1.6 | 3.1 |
| Weib late discrete | asymptotic | 2.7 | 4.3 | 6.4 | 3.2 | 3.0 | 6.3 | 3.9 | 3.2 | 5.4 |
| | asymptotic_bonf | 2.4 | 3.8 | 5.6 | 2.8 | 2.7 | 5.8 | 3.6 | 3.2 | 4.8 |
| | permutation_bonf | 1.4 | 3.2 | 4.1 | 2.2 | 2.3 | 3.8 | 2.7 | 2.7 | 3.9 |
| Weib prop continuous | asymptotic | 2.0 | 3.3 | 5.0 | 2.5 | 2.7 | 5.5 | 2.8 | 2.8 | 4.6 |
| | asymptotic_bonf | 1.8 | 2.8 | 3.8 | 2.0 | 2.5 | 4.8 | 2.7 | 2.0 | 4.4 |
| | permutation_bonf | 1.2 | 2.2 | 3.3 | 1.8 | 1.6 | 3.5 | 1.9 | 1.6 | 2.7 |
| Weib prop discrete | asymptotic | 2.3 | 3.4 | 5.7 | 2.6 | 2.8 | 6.0 | 3.3 | 3.1 | 4.9 |
| | asymptotic_bonf | 1.8 | 3.2 | 5.2 | 2.3 | 2.5 | 5.8 | 2.8 | 2.8 | 4.6 |
| | permutation_bonf | 1.9 | 2.5 | 3.5 | 1.8 | 1.9 | 3.9 | 2.1 | 2.0 | 3.5 |
| Weib scale continuous | asymptotic | 1.2 | 2.3 | 3.1 | 1.4 | 1.9 | 4.2 | 1.8 | 2.0 | 3.6 |
| | asymptotic_bonf | 0.8 | 2.2 | 2.9 | 1.2 | 1.4 | 2.8 | 1.5 | 1.8 | 2.8 |
| | permutation_bonf | 0.7 | 1.7 | 2.7 | 1.0 | 1.0 | 2.0 | 1.0 | 1.4 | 2.2 |
| Weib scale discrete | asymptotic | 1.4 | 2.4 | 3.8 | 1.7 | 1.7 | 4.6 | 2.2 | 2.0 | 3.6 |
| | asymptotic_bonf | 1.1 | 2.1 | 3.4 | 1.5 | 1.2 | 3.8 | 1.8 | 1.4 | 3.0 |
| | permutation_bonf | 0.8 | 1.9 | 2.7 | 1.4 | 1.0 | 3.1 | 1.2 | 1.6 | 2.0 |
| Weib shape continuous | asymptotic | 1.1 | 1.5 | 2.4 | 1.2 | 1.1 | 2.6 | 1.2 | 1.5 | 2.2 |
| | asymptotic_bonf | 0.7 | 1.3 | 2.4 | 1.0 | 0.9 | 2.5 | 0.9 | 1.1 | 1.9 |
| | permutation_bonf | 0.6 | 1.1 | 1.8 | 0.6 | 0.9 | 1.8 | 0.8 | 1.0 | 1.8 |
| Weib shape discrete | asymptotic | 0.8 | 1.7 | 2.9 | 1.4 | 1.3 | 3.4 | 1.4 | 1.7 | 2.5 |
| | asymptotic_bonf | 0.8 | 1.5 | 2.6 | 1.2 | 0.8 | 3.1 | 1.2 | 1.6 | 2.1 |
| | permutation_bonf | 0.7 | 1.2 | 1.9 | 1.2 | 1.1 | 2.9 | 0.8 | 1.3 | 1.8 |

Table S82: Rejection rates in percent for the Tukey-type contrast matrix with $\delta = 1.5$ and unbalanced large sample sizes under unequal, high censoring.

| Distribution | Method | $\mathcal{H}_{0,7}$ | $\mathcal{H}_{0,8}$ | $\mathcal{H}_{0,9}$ | $\mathcal{H}_{0,13}$ | $\mathcal{H}_{0,14}$ | $\mathcal{H}_{0,15}$ | $\mathcal{H}_{0,16}$ | $\mathcal{H}_{0,17}$ | $\mathcal{H}_{0,18}$ |
|---|---|---|---|---|---|---|---|---|---|---|
| exp early continuous | asymptotic | 24.4 | 12.2 | 46.6 | 20.1 | 11.2 | 34.4 | 21.2 | 10.6 | 38.4 |
|  | asymptotic_bonf | 22.2 | 10.6 | 43.6 | 17.4 | 9.4 | 30.8 | 18.9 | 8.7 | 35.3 |
|  | permutation_bonf | 18.9 | 8.4 | 38.9 | 16.0 | 7.9 | 29.0 | 17.7 | 7.1 | 32.3 |
| exp early discrete | asymptotic | 28.3 | 14.5 | 50.3 | 22.3 | 11.4 | 37.0 | 23.8 | 12.1 | 41.0 |
|  | asymptotic_bonf | 24.9 | 12.7 | 47.0 | 20.1 | 9.7 | 34.0 | 21.1 | 10.2 | 37.7 |
|  | permutation_bonf | 22.4 | 9.2 | 41.4 | 19.3 | 8.5 | 32.3 | 19.1 | 8.2 | 35.5 |
| exp late continuous | asymptotic | 27.8 | 13.8 | 50.6 | 20.8 | 12.0 | 36.4 | 22.9 | 10.7 | 41.1 |
|  | asymptotic_bonf | 24.2 | 12.2 | 47.6 | 18.7 | 10.1 | 33.4 | 20.2 | 9.0 | 38.1 |
|  | permutation_bonf | 21.1 | 9.7 | 42.0 | 17.5 | 8.4 | 31.6 | 18.9 | 7.7 | 35.2 |
| exp late discrete | asymptotic | 31.6 | 16.0 | 55.1 | 23.1 | 12.0 | 39.8 | 25.3 | 13.2 | 45.0 |
|  | asymptotic_bonf | 26.7 | 14.2 | 52.0 | 20.9 | 9.7 | 36.5 | 22.9 | 11.4 | 40.8 |
|  | permutation_bonf | 23.6 | 11.1 | 46.5 | 19.8 | 9.3 | 34.9 | 21.6 | 9.4 | 39.4 |
| exp prop continuous | asymptotic | 26.7 | 15.2 | 43.6 | 21.5 | 12.8 | 30.8 | 20.7 | 12.0 | 38.0 |
|  | asymptotic_bonf | 23.8 | 13.9 | 40.5 | 19.0 | 10.8 | 28.1 | 18.7 | 10.6 | 34.5 |
|  | permutation_bonf | 20.1 | 10.1 | 36.5 | 18.2 | 9.4 | 26.2 | 16.1 | 9.7 | 32.9 |
| exp prop discrete | asymptotic | 28.7 | 18.1 | 48.6 | 23.3 | 14.5 | 35.3 | 23.4 | 14.9 | 41.4 |
|  | asymptotic_bonf | 25.5 | 15.9 | 44.0 | 20.9 | 12.7 | 32.7 | 20.8 | 13.0 | 38.2 |
|  | permutation_bonf | 22.7 | 12.3 | 41.1 | 18.9 | 11.3 | 31.0 | 18.1 | 10.8 | 36.6 |
| logn continuous | asymptotic | 64.7 | 40.9 | 80.7 | 54.4 | 32.2 | 68.2 | 56.4 | 35.3 | 73.4 |
|  | asymptotic_bonf | 61.6 | 37.0 | 77.3 | 50.0 | 28.8 | 64.3 | 52.6 | 32.2 | 70.1 |
|  | permutation_bonf | 52.6 | 24.1 | 70.9 | 46.4 | 24.6 | 61.4 | 45.4 | 27.1 | 65.9 |
| logn discrete | asymptotic | 72.7 | 48.4 | 86.9 | 61.9 | 39.3 | 77.1 | 63.4 | 41.6 | 81.1 |
|  | asymptotic_bonf | 69.6 | 43.6 | 84.3 | 57.5 | 35.4 | 73.5 | 60.3 | 38.6 | 78.4 |
|  | permutation_bonf | 59.2 | 29.3 | 77.9 | 54.8 | 30.6 | 69.7 | 53.8 | 32.2 | 75.0 |
| pwExp continuous | asymptotic | 24.0 | 11.9 | 46.1 | 18.3 | 9.9 | 35.1 | 21.8 | 11.4 | 36.6 |
|  | asymptotic_bonf | 22.6 | 9.9 | 42.9 | 16.1 | 8.8 | 31.5 | 19.5 | 9.4 | 33.7 |
|  | permutation_bonf | 19.1 | 8.2 | 38.1 | 15.3 | 7.3 | 29.7 | 17.1 | 7.4 | 31.0 |
| pwExp discrete | asymptotic | 27.5 | 14.0 | 50.7 | 20.7 | 11.1 | 37.3 | 23.6 | 11.7 | 40.8 |
|  | asymptotic_bonf | 24.8 | 11.9 | 47.5 | 18.8 | 9.4 | 34.2 | 21.3 | 10.3 | 37.3 |
|  | permutation_bonf | 21.0 | 9.6 | 42.3 | 18.5 | 8.3 | 33.0 | 19.8 | 8.3 | 33.4 |
| Weib late continuous | asymptotic | 65.2 | 40.7 | 84.4 | 53.6 | 32.9 | 74.6 | 55.2 | 34.8 | 78.2 |
|  | asymptotic_bonf | 61.2 | 36.7 | 83.2 | 49.4 | 29.9 | 71.2 | 51.9 | 31.0 | 75.2 |
|  | permutation_bonf | 51.6 | 26.2 | 77.9 | 44.9 | 25.8 | 68.3 | 47.7 | 25.3 | 70.6 |
| Weib late discrete | asymptotic | 70.0 | 45.0 | 86.6 | 60.9 | 37.2 | 77.3 | 62.2 | 39.0 | 81.6 |
|  | asymptotic_bonf | 66.6 | 42.9 | 84.6 | 58.0 | 34.0 | 75.2 | 59.7 | 35.7 | 78.9 |
|  | permutation_bonf | 57.4 | 30.2 | 80.4 | 52.3 | 29.5 | 71.6 | 53.0 | 30.6 | 75.0 |
| Weib prop continuous | asymptotic | 62.9 | 38.8 | 82.9 | 51.0 | 29.8 | 71.4 | 54.4 | 32.3 | 74.5 |
|  | asymptotic_bonf | 58.9 | 34.2 | 80.5 | 47.1 | 28.2 | 68.0 | 50.6 | 29.7 | 71.8 |
|  | permutation_bonf | 50.7 | 24.1 | 75.0 | 43.2 | 23.4 | 64.9 | 46.0 | 24.1 | 67.0 |
| Weib prop discrete | asymptotic | 69.9 | 45.7 | 87.2 | 60.7 | 35.7 | 79.6 | 62.4 | 38.6 | 81.9 |
|  | asymptotic_bonf | 66.7 | 42.3 | 85.5 | 56.2 | 32.8 | 76.7 | 58.1 | 34.9 | 79.9 |
|  | permutation_bonf | 56.8 | 28.6 | 80.8 | 51.7 | 28.6 | 73.2 | 52.0 | 30.1 | 75.3 |
| Weib scale continuous | asymptotic | 54.2 | 32.8 | 72.4 | 42.9 | 24.8 | 62.4 | 45.0 | 26.4 | 67.5 |
|  | asymptotic_bonf | 49.7 | 29.3 | 70.0 | 39.4 | 21.9 | 59.5 | 42.1 | 24.9 | 63.6 |
|  | permutation_bonf | 42.8 | 21.0 | 63.9 | 35.7 | 19.1 | 55.4 | 37.6 | 20.1 | 58.9 |
| Weib scale discrete | asymptotic | 61.5 | 37.5 | 79.2 | 49.3 | 29.5 | 69.7 | 52.7 | 31.6 | 74.5 |
|  | asymptotic_bonf | 57.2 | 34.1 | 76.4 | 46.0 | 26.7 | 66.4 | 48.0 | 28.9 | 71.1 |
|  | permutation_bonf | 48.3 | 23.6 | 70.9 | 41.6 | 21.7 | 62.8 | 44.0 | 24.0 | 67.0 |
| Weib shape continuous | asymptotic | 44.2 | 23.9 | 62.6 | 35.5 | 22.3 | 53.7 | 39.6 | 22.1 | 58.7 |
|  | asymptotic_bonf | 41.2 | 20.8 | 58.8 | 31.3 | 18.7 | 50.2 | 35.1 | 19.0 | 54.7 |
|  | permutation_bonf | 32.6 | 13.0 | 51.4 | 28.8 | 15.8 | 46.4 | 29.6 | 14.6 | 48.3 |
| Weib shape discrete | asymptotic | 51.6 | 30.0 | 70.0 | 42.7 | 25.5 | 61.1 | 45.5 | 26.8 | 64.3 |
|  | asymptotic_bonf | 46.6 | 26.2 | 66.3 | 38.9 | 22.4 | 57.6 | 40.7 | 22.8 | 61.4 |
|  | permutation_bonf | 36.1 | 15.2 | 57.8 | 34.5 | 18.1 | 52.5 | 35.1 | 17.7 | 55.5 |

Table S83: Rejection rates in percent for the Tukey-type contrast matrix with $\delta = 1.5$ and unbalanced medium sample sizes under unequal, high censoring.

| Distribution | Method | $\mathcal{H}_{0,7}$ | $\mathcal{H}_{0,8}$ | $\mathcal{H}_{0,9}$ | $\mathcal{H}_{0,13}$ | $\mathcal{H}_{0,14}$ | $\mathcal{H}_{0,15}$ | $\mathcal{H}_{0,16}$ | $\mathcal{H}_{0,17}$ | $\mathcal{H}_{0,18}$ |
|---|---|---|---|---|---|---|---|---|---|---|
| exp early continuous | asymptotic | 1.6 | 1.5 | 3.4 | 1.0 | 0.9 | 2.8 | 2.1 | 0.8 | 3.4 |
|  | asymptotic_bonf | 1.0 | 1.0 | 2.9 | 0.8 | 0.8 | 2.5 | 1.6 | 0.6 | 2.8 |
|  | permutation_bonf | 0.3 | 0.1 | 1.3 | 0.4 | 0.3 | 1.5 | 1.0 | 0.1 | 1.4 |
| exp early discrete | asymptotic | 1.9 | 1.2 | 3.6 | 1.7 | 0.9 | 3.4 | 1.6 | 0.7 | 3.5 |
|  | asymptotic_bonf | 1.2 | 0.9 | 3.0 | 1.3 | 0.7 | 2.9 | 1.4 | 0.3 | 2.8 |
|  | permutation_bonf | 0.5 | 0.0 | 1.0 | 0.2 | 0.5 | 1.7 | 0.6 | 0.1 | 1.7 |
| exp late continuous | asymptotic | 2.4 | 1.3 | 3.6 | 1.6 | 1.1 | 3.3 | 2.3 | 0.8 | 3.9 |
|  | asymptotic_bonf | 1.7 | 1.0 | 3.1 | 1.1 | 0.9 | 2.6 | 1.9 | 0.5 | 2.8 |
|  | permutation_bonf | 0.4 | 0.1 | 1.5 | 0.4 | 0.3 | 1.7 | 0.8 | 0.1 | 1.6 |
| exp late discrete | asymptotic | 2.3 | 1.2 | 3.7 | 2.1 | 1.0 | 3.7 | 2.4 | 1.2 | 4.0 |
|  | asymptotic_bonf | 1.6 | 1.0 | 2.9 | 1.8 | 1.0 | 2.9 | 2.1 | 0.7 | 3.4 |
|  | permutation_bonf | 0.3 | 0.1 | 1.6 | 0.4 | 0.3 | 1.9 | 1.0 | 0.2 | 1.9 |
| exp prop continuous | asymptotic | 2.4 | 1.2 | 3.0 | 2.1 | 1.2 | 2.8 | 1.9 | 1.5 | 2.6 |
|  | asymptotic_bonf | 1.9 | 0.8 | 2.5 | 1.2 | 0.9 | 2.3 | 1.6 | 1.1 | 2.1 |
|  | permutation_bonf | 0.5 | 0.1 | 1.2 | 1.1 | 0.5 | 1.8 | 0.8 | 0.2 | 1.7 |
| exp prop discrete | asymptotic | 2.3 | 1.1 | 3.2 | 1.8 | 1.2 | 3.1 | 2.4 | 1.6 | 2.8 |
|  | asymptotic_bonf | 2.0 | 0.7 | 2.4 | 1.2 | 1.2 | 3.0 | 1.8 | 1.5 | 2.6 |
|  | permutation_bonf | 0.8 | 0.1 | 1.2 | 0.8 | 0.3 | 2.0 | 0.8 | 0.4 | 1.8 |
| logn continuous | asymptotic | 4.2 | 2.3 | 8.1 | 4.6 | 3.0 | 6.9 | 4.0 | 2.3 | 7.5 |
|  | asymptotic_bonf | 3.4 | 1.9 | 6.8 | 3.6 | 2.7 | 5.7 | 3.0 | 2.3 | 6.7 |
|  | permutation_bonf | 0.4 | 0.0 | 1.3 | 0.8 | 0.4 | 2.2 | 1.0 | 0.1 | 2.8 |
| logn discrete | asymptotic | 4.8 | 2.8 | 9.0 | 4.8 | 2.8 | 8.9 | 4.2 | 2.6 | 8.0 |
|  | asymptotic_bonf | 4.0 | 2.3 | 7.8 | 3.8 | 2.8 | 7.4 | 3.6 | 2.2 | 7.2 |
|  | permutation_bonf | 0.2 | 0.0 | 0.7 | 1.0 | 0.3 | 2.4 | 1.0 | 0.1 | 3.5 |
| pwExp continuous | asymptotic | 1.8 | 1.3 | 3.7 | 1.0 | 1.4 | 3.1 | 1.8 | 1.1 | 2.7 |
|  | asymptotic_bonf | 1.2 | 1.1 | 2.7 | 1.0 | 1.2 | 2.6 | 1.6 | 1.1 | 2.3 |
|  | permutation_bonf | 0.1 | 0.1 | 1.0 | 0.4 | 0.3 | 1.5 | 0.7 | 0.1 | 1.5 |
| pwExp discrete | asymptotic | 1.9 | 1.0 | 3.4 | 1.2 | 1.2 | 3.1 | 1.8 | 1.2 | 3.2 |
|  | asymptotic_bonf | 1.4 | 1.0 | 3.0 | 0.8 | 1.2 | 2.6 | 1.3 | 1.1 | 2.5 |
|  | permutation_bonf | 0.3 | 0.0 | 1.3 | 0.3 | 0.7 | 1.8 | 0.6 | 0.2 | 1.2 |
| Weib late continuous | asymptotic | 4.6 | 1.5 | 9.5 | 4.5 | 2.2 | 6.8 | 3.8 | 2.5 | 8.0 |
|  | asymptotic_bonf | 3.9 | 1.1 | 8.1 | 3.3 | 1.8 | 5.6 | 3.0 | 1.8 | 7.1 |
|  | permutation_bonf | 0.3 | 0.0 | 1.7 | 1.5 | 0.3 | 2.9 | 0.9 | 0.0 | 3.1 |
| Weib late discrete | asymptotic | 5.4 | 1.6 | 12.3 | 5.0 | 2.3 | 10.0 | 5.2 | 2.8 | 10.9 |
|  | asymptotic_bonf | 3.9 | 1.3 | 10.4 | 4.2 | 1.9 | 8.3 | 3.9 | 2.1 | 9.4 |
|  | permutation_bonf | 0.0 | 0.1 | 2.4 | 1.4 | 0.2 | 3.5 | 0.5 | 0.2 | 3.5 |
| Weib prop continuous | asymptotic | 4.7 | 1.3 | 7.8 | 3.7 | 1.9 | 6.9 | 3.7 | 1.9 | 8.0 |
|  | asymptotic_bonf | 3.5 | 0.9 | 6.5 | 3.1 | 1.7 | 5.9 | 2.9 | 1.4 | 7.0 |
|  | permutation_bonf | 0.3 | 0.0 | 1.5 | 1.6 | 0.4 | 2.7 | 0.7 | 0.1 | 2.7 |
| Weib prop discrete | asymptotic | 5.0 | 1.4 | 10.1 | 4.7 | 2.1 | 9.4 | 4.4 | 2.5 | 9.5 |
|  | asymptotic_bonf | 3.5 | 0.8 | 8.8 | 3.7 | 1.5 | 7.5 | 3.7 | 1.7 | 7.5 |
|  | permutation_bonf | 0.2 | 0.0 | 1.5 | 1.4 | 0.1 | 2.8 | 0.7 | 0.3 | 2.8 |
| Weib scale continuous | asymptotic | 2.9 | 1.0 | 4.9 | 2.9 | 1.2 | 4.9 | 2.8 | 1.2 | 5.5 |
|  | asymptotic_bonf | 1.9 | 0.8 | 4.1 | 2.4 | 1.0 | 4.1 | 1.9 | 0.6 | 4.3 |
|  | permutation_bonf | 0.3 | 0.0 | 0.7 | 0.7 | 0.5 | 1.5 | 0.3 | 0.0 | 1.6 |
| Weib scale discrete | asymptotic | 3.4 | 1.0 | 6.1 | 3.4 | 1.5 | 5.6 | 3.0 | 1.3 | 7.1 |
|  | asymptotic_bonf | 3.0 | 0.6 | 4.6 | 2.8 | 1.0 | 4.7 | 2.1 | 0.9 | 5.4 |
|  | permutation_bonf | 0.2 | 0.0 | 0.4 | 0.8 | 0.3 | 1.3 | 0.4 | 0.0 | 2.0 |
| Weib shape continuous | asymptotic | 2.3 | 1.0 | 3.3 | 2.8 | 0.7 | 3.2 | 2.0 | 0.8 | 3.3 |
|  | asymptotic_bonf | 1.6 | 0.6 | 2.8 | 2.1 | 0.7 | 2.5 | 1.5 | 0.6 | 2.9 |
|  | permutation_bonf | 0.2 | 0.0 | 0.5 | 0.2 | 0.3 | 1.3 | 0.3 | 0.0 | 1.0 |
| Weib shape discrete | asymptotic | 2.2 | 0.9 | 3.8 | 2.9 | 0.8 | 3.6 | 2.2 | 1.0 | 3.8 |
|  | asymptotic_bonf | 1.8 | 0.6 | 3.2 | 2.1 | 0.7 | 3.3 | 1.8 | 0.5 | 2.9 |
|  | permutation_bonf | 0.1 | 0.0 | 0.4 | 0.4 | 0.2 | 1.0 | 0.2 | 0.0 | 0.7 |

Table S84: Rejection rates in percent for the Tukey-type contrast matrix with $\delta = 1.5$ and unbalanced small sample sizes under unequal, high censoring.

| Distribution | Method | $\mathcal{H}_{0,7}$ | $\mathcal{H}_{0,8}$ | $\mathcal{H}_{0,9}$ | $\mathcal{H}_{0,13}$ | $\mathcal{H}_{0,14}$ | $\mathcal{H}_{0,15}$ | $\mathcal{H}_{0,16}$ | $\mathcal{H}_{0,17}$ | $\mathcal{H}_{0,18}$ |
|---|---|---|---|---|---|---|---|---|---|---|
| exp early continuous | asymptotic | 0.9 | 4.2 | 1.3 | 1.0 | 1.7 | 1.2 | 0.9 | 2.4 | 1.4 |
| | asymptotic_bonf | 0.9 | 4.1 | 1.2 | 0.8 | 1.4 | 0.9 | 0.7 | 2.1 | 1.2 |
| | permutation_bonf | 0.2 | 0.1 | 0.1 | 0.0 | 0.0 | 0.1 | 0.0 | 0.3 | 0.0 |
| exp early discrete | asymptotic | 1.9 | 5.1 | 1.7 | 1.4 | 1.6 | 1.2 | 1.4 | 2.9 | 1.8 |
| | asymptotic_bonf | 1.9 | 4.9 | 1.5 | 1.1 | 1.4 | 1.1 | 1.3 | 2.4 | 1.3 |
| | permutation_bonf | 0.2 | 0.1 | 0.0 | 0.1 | 0.1 | 0.0 | 0.1 | 0.5 | 0.0 |
| exp late continuous | asymptotic | 1.1 | 3.7 | 1.7 | 1.0 | 1.6 | 1.1 | 0.8 | 1.8 | 1.5 |
| | asymptotic_bonf | 1.0 | 3.5 | 1.1 | 1.0 | 1.4 | 0.8 | 0.5 | 1.5 | 1.4 |
| | permutation_bonf | 0.2 | 0.1 | 0.1 | 0.0 | 0.0 | 0.0 | 0.0 | 0.2 | 0.0 |
| exp late discrete | asymptotic | 1.6 | 4.1 | 2.1 | 1.6 | 1.4 | 1.7 | 0.9 | 1.7 | 1.8 |
| | asymptotic_bonf | 1.6 | 3.9 | 1.9 | 1.3 | 1.3 | 1.4 | 0.8 | 1.6 | 1.3 |
| | permutation_bonf | 0.2 | 0.0 | 0.0 | 0.0 | 0.0 | 0.0 | 0.1 | 0.3 | 0.1 |
| exp prop continuous | asymptotic | 2.0 | 4.5 | 1.9 | 1.3 | 1.9 | 1.8 | 1.7 | 2.1 | 1.6 |
| | asymptotic_bonf | 1.7 | 4.4 | 1.7 | 0.7 | 1.3 | 1.5 | 1.5 | 1.8 | 1.2 |
| | permutation_bonf | 0.3 | 0.1 | 0.0 | 0.0 | 0.2 | 0.3 | 0.2 | 0.1 | 0.2 |
| exp prop discrete | asymptotic | 2.6 | 5.0 | 2.2 | 1.6 | 1.5 | 2.1 | 2.1 | 2.1 | 1.5 |
| | asymptotic_bonf | 2.5 | 5.0 | 1.8 | 1.4 | 1.3 | 1.6 | 1.8 | 1.9 | 1.3 |
| | permutation_bonf | 0.3 | 0.1 | 0.0 | 0.1 | 0.1 | 0.1 | 0.1 | 0.0 | 0.2 |
| logn continuous | asymptotic | 5.3 | 10.5 | 3.6 | 1.9 | 0.8 | 2.1 | 2.9 | 1.1 | 3.0 |
| | asymptotic_bonf | 5.1 | 10.1 | 3.5 | 1.0 | 0.7 | 1.8 | 2.5 | 1.0 | 2.6 |
| | permutation_bonf | 0.2 | 0.3 | 0.2 | 0.2 | 0.5 | 0.2 | 0.1 | 0.2 | 0.3 |
| logn discrete | asymptotic | 5.6 | 11.3 | 4.2 | 1.6 | 0.8 | 2.3 | 2.6 | 0.9 | 2.9 |
| | asymptotic_bonf | 5.3 | 9.9 | 3.8 | 1.0 | 0.7 | 1.8 | 2.2 | 0.7 | 2.5 |
| | permutation_bonf | 0.2 | 0.5 | 0.0 | 0.4 | 0.4 | 0.3 | 0.1 | 0.3 | 0.4 |
| pwExp continuous | asymptotic | 1.2 | 3.9 | 1.5 | 0.9 | 1.4 | 1.4 | 0.8 | 1.9 | 1.3 |
| | asymptotic_bonf | 1.2 | 3.9 | 1.1 | 0.7 | 1.1 | 1.2 | 0.8 | 1.7 | 1.0 |
| | permutation_bonf | 0.0 | 0.2 | 0.0 | 0.0 | 0.0 | 0.0 | 0.0 | 0.1 | 0.2 |
| pwExp discrete | asymptotic | 1.3 | 4.7 | 1.4 | 1.1 | 1.3 | 1.5 | 1.3 | 2.2 | 1.8 |
| | asymptotic_bonf | 1.2 | 4.5 | 1.2 | 0.9 | 1.0 | 1.2 | 1.0 | 1.8 | 1.5 |
| | permutation_bonf | 0.0 | 0.1 | 0.0 | 0.0 | 0.1 | 0.1 | 0.1 | 0.2 | 0.2 |
| Weib late continuous | asymptotic | 5.0 | 6.5 | 3.1 | 1.8 | 0.7 | 2.6 | 2.3 | 1.1 | 2.4 |
| | asymptotic_bonf | 5.0 | 6.2 | 3.0 | 1.5 | 0.6 | 2.1 | 2.2 | 0.9 | 2.3 |
| | permutation_bonf | 0.2 | 0.1 | 0.4 | 0.2 | 0.0 | 0.5 | 0.1 | 0.1 | 0.4 |
| Weib late discrete | asymptotic | 6.5 | 6.7 | 4.1 | 2.1 | 0.5 | 2.8 | 2.1 | 1.4 | 3.3 |
| | asymptotic_bonf | 6.2 | 6.4 | 3.8 | 1.8 | 0.5 | 2.6 | 1.8 | 1.1 | 2.9 |
| | permutation_bonf | 0.0 | 0.3 | 0.3 | 0.2 | 0.4 | 0.2 | 0.2 | 0.3 | 0.2 |
| Weib prop continuous | asymptotic | 5.2 | 6.9 | 3.1 | 1.8 | 0.7 | 2.2 | 2.3 | 1.3 | 2.3 |
| | asymptotic_bonf | 5.0 | 6.4 | 2.7 | 1.4 | 0.7 | 1.8 | 2.1 | 1.0 | 1.9 |
| | permutation_bonf | 0.2 | 0.0 | 0.4 | 0.2 | 0.0 | 0.2 | 0.1 | 0.1 | 0.2 |
| Weib prop discrete | asymptotic | 6.6 | 7.5 | 3.8 | 1.8 | 0.5 | 2.6 | 2.3 | 0.7 | 2.1 |
| | asymptotic_bonf | 6.3 | 7.1 | 3.3 | 1.3 | 0.5 | 2.0 | 1.9 | 0.5 | 1.8 |
| | permutation_bonf | 0.1 | 0.1 | 0.3 | 0.2 | 0.1 | 0.1 | 0.2 | 0.0 | 0.3 |
| Weib scale continuous | asymptotic | 6.4 | 10.0 | 3.9 | 2.1 | 1.8 | 2.2 | 2.6 | 2.2 | 2.5 |
| | asymptotic_bonf | 6.2 | 9.3 | 3.7 | 1.8 | 1.6 | 2.2 | 2.3 | 1.9 | 2.4 |
| | permutation_bonf | 0.2 | 0.1 | 0.3 | 0.2 | 0.0 | 0.1 | 0.3 | 0.2 | 0.2 |
| Weib scale discrete | asymptotic | 7.6 | 10.1 | 4.1 | 2.0 | 1.0 | 2.0 | 2.1 | 1.8 | 2.2 |
| | asymptotic_bonf | 7.5 | 9.8 | 3.9 | 1.6 | 0.8 | 1.5 | 2.0 | 1.2 | 2.0 |
| | permutation_bonf | 0.1 | 0.1 | 0.1 | 0.3 | 0.1 | 0.0 | 0.4 | 0.0 | 0.2 |
| Weib shape continuous | asymptotic | 7.4 | 12.2 | 3.9 | 2.4 | 1.3 | 2.0 | 3.1 | 2.5 | 2.4 |
| | asymptotic_bonf | 7.3 | 11.4 | 3.9 | 2.1 | 1.0 | 2.0 | 2.3 | 1.7 | 2.3 |
| | permutation_bonf | 0.6 | 0.1 | 0.3 | 0.2 | 0.0 | 0.0 | 0.4 | 0.3 | 0.0 |
| Weib shape discrete | asymptotic | 9.1 | 13.0 | 4.8 | 2.3 | 0.8 | 1.9 | 2.6 | 1.8 | 2.5 |
| | asymptotic_bonf | 8.9 | 12.5 | 4.6 | 1.6 | 0.6 | 1.6 | 2.1 | 1.1 | 2.4 |
| | permutation_bonf | 0.3 | 0.2 | 0.5 | 0.3 | 0.1 | 0.1 | 0.2 | 0.2 | 0.0 |

Table S85: Rejection rates in percent for the Tukey-type contrast matrix with $\delta = 1.5$ and balanced large sample sizes under unequal, low censoring.

| Distribution | Method | $\mathcal{H}_{0,7}$ | $\mathcal{H}_{0,8}$ | $\mathcal{H}_{0,9}$ | $\mathcal{H}_{0,13}$ | $\mathcal{H}_{0,14}$ | $\mathcal{H}_{0,15}$ | $\mathcal{H}_{0,16}$ | $\mathcal{H}_{0,17}$ | $\mathcal{H}_{0,18}$ |
|---|---|---|---|---|---|---|---|---|---|---|
| exp early continuous | asymptotic | 79.6 | 55.9 | 92.2 | 78.6 | 57.6 | 92.6 | 76.7 | 54.9 | 93.5 |
| | asymptotic_bonf | 77.0 | 53.9 | 91.7 | 76.7 | 55.3 | 92.1 | 75.1 | 52.8 | 93.0 |
| | permutation_bonf | 74.9 | 52.0 | 90.9 | 76.0 | 54.1 | 91.6 | 72.8 | 51.9 | 92.0 |
| exp early discrete | asymptotic | 84.4 | 63.6 | 95.1 | 84.9 | 64.9 | 95.6 | 82.9 | 61.6 | 96.1 |
| | asymptotic_bonf | 83.6 | 61.5 | 94.8 | 83.0 | 63.5 | 95.0 | 81.3 | 59.8 | 95.8 |
| | permutation_bonf | 80.9 | 58.8 | 94.0 | 82.8 | 62.7 | 94.4 | 79.9 | 58.0 | 94.5 |
| exp late continuous | asymptotic | 82.9 | 61.0 | 94.7 | 82.6 | 62.3 | 94.6 | 80.7 | 59.5 | 95.6 |
| | asymptotic_bonf | 81.7 | 58.2 | 94.0 | 80.9 | 59.7 | 93.9 | 78.6 | 57.2 | 95.0 |
| | permutation_bonf | 79.6 | 56.7 | 93.3 | 79.6 | 60.3 | 93.5 | 77.0 | 54.7 | 94.5 |
| exp late discrete | asymptotic | 87.4 | 68.1 | 96.5 | 89.3 | 70.7 | 97.2 | 87.3 | 65.7 | 98.0 |
| | asymptotic_bonf | 86.2 | 66.5 | 96.2 | 88.2 | 68.4 | 96.8 | 85.8 | 63.5 | 97.9 |
| | permutation_bonf | 85.9 | 64.7 | 95.8 | 87.0 | 66.7 | 96.4 | 83.9 | 61.2 | 97.1 |
| exp prop continuous | asymptotic | 79.0 | 60.0 | 93.5 | 82.4 | 62.0 | 95.0 | 79.5 | 60.6 | 94.1 |
| | asymptotic_bonf | 77.8 | 57.8 | 93.1 | 80.2 | 60.2 | 94.3 | 77.7 | 58.8 | 93.4 |
| | permutation_bonf | 75.7 | 56.1 | 92.6 | 79.6 | 58.5 | 94.4 | 76.6 | 57.9 | 92.6 |
| exp prop discrete | asymptotic | 85.1 | 69.5 | 96.2 | 88.6 | 69.9 | 98.2 | 84.4 | 67.6 | 96.8 |
| | asymptotic_bonf | 83.6 | 67.6 | 95.6 | 87.6 | 67.6 | 97.6 | 83.3 | 65.0 | 96.3 |
| | permutation_bonf | 82.6 | 65.3 | 95.7 | 85.9 | 66.7 | 96.5 | 81.5 | 64.7 | 95.3 |
| logn continuous | asymptotic | 100.0 | 97.9 | 100.0 | 99.8 | 98.4 | 100.0 | 99.9 | 97.8 | 100.0 |
| | asymptotic_bonf | 100.0 | 97.5 | 100.0 | 99.6 | 98.4 | 100.0 | 99.9 | 97.6 | 100.0 |
| | permutation_bonf | 100.0 | 97.3 | 100.0 | 99.6 | 97.9 | 100.0 | 99.9 | 96.9 | 100.0 |
| logn discrete | asymptotic | 100.0 | 99.5 | 100.0 | 99.9 | 99.5 | 100.0 | 99.9 | 99.3 | 100.0 |
| | asymptotic_bonf | 100.0 | 99.3 | 100.0 | 99.9 | 99.5 | 100.0 | 99.9 | 99.2 | 100.0 |
| | permutation_bonf | 100.0 | 99.0 | 100.0 | 99.9 | 99.1 | 100.0 | 99.9 | 99.4 | 100.0 |
| pwExp continuous | asymptotic | 76.6 | 54.3 | 92.3 | 78.0 | 56.6 | 92.1 | 74.4 | 53.3 | 92.6 |
| | asymptotic_bonf | 74.1 | 51.6 | 91.5 | 76.0 | 54.5 | 91.0 | 72.7 | 50.9 | 91.6 |
| | permutation_bonf | 73.7 | 51.1 | 89.9 | 75.2 | 53.2 | 90.2 | 72.0 | 49.6 | 91.5 |
| pwExp discrete | asymptotic | 83.6 | 62.2 | 95.0 | 84.7 | 64.2 | 95.3 | 82.1 | 60.2 | 95.3 |
| | asymptotic_bonf | 82.1 | 59.8 | 94.4 | 83.4 | 62.5 | 94.9 | 80.1 | 57.6 | 94.9 |
| | permutation_bonf | 80.9 | 57.7 | 94.3 | 83.2 | 61.2 | 94.4 | 79.3 | 56.6 | 94.2 |
| Weib late continuous | asymptotic | 99.9 | 98.6 | 100.0 | 99.8 | 99.1 | 100.0 | 99.9 | 97.2 | 100.0 |
| | asymptotic_bonf | 99.8 | 98.4 | 100.0 | 99.8 | 98.9 | 100.0 | 99.9 | 96.8 | 100.0 |
| | permutation_bonf | 99.9 | 98.1 | 100.0 | 99.7 | 98.8 | 100.0 | 99.4 | 96.5 | 100.0 |
| Weib late discrete | asymptotic | 100.0 | 99.8 | 100.0 | 100.0 | 99.8 | 100.0 | 100.0 | 98.7 | 100.0 |
| | asymptotic_bonf | 100.0 | 99.7 | 100.0 | 100.0 | 99.8 | 100.0 | 100.0 | 98.6 | 100.0 |
| | permutation_bonf | 100.0 | 99.6 | 100.0 | 100.0 | 99.6 | 100.0 | 100.0 | 98.7 | 100.0 |
| Weib prop continuous | asymptotic | 99.8 | 98.0 | 100.0 | 99.7 | 98.3 | 100.0 | 99.8 | 96.2 | 100.0 |
| | asymptotic_bonf | 99.7 | 97.5 | 100.0 | 99.7 | 97.9 | 100.0 | 99.8 | 96.0 | 100.0 |
| | permutation_bonf | 99.8 | 97.5 | 100.0 | 99.7 | 98.3 | 100.0 | 99.4 | 96.2 | 100.0 |
| Weib prop discrete | asymptotic | 100.0 | 99.6 | 100.0 | 100.0 | 99.7 | 100.0 | 99.9 | 98.4 | 100.0 |
| | asymptotic_bonf | 100.0 | 99.5 | 100.0 | 100.0 | 99.7 | 100.0 | 99.9 | 98.3 | 100.0 |
| | permutation_bonf | 100.0 | 99.5 | 100.0 | 100.0 | 99.3 | 100.0 | 100.0 | 98.3 | 100.0 |
| Weib scale continuous | asymptotic | 99.5 | 96.0 | 100.0 | 99.3 | 96.1 | 100.0 | 99.2 | 94.2 | 100.0 |
| | asymptotic_bonf | 99.3 | 94.8 | 100.0 | 99.3 | 95.5 | 100.0 | 98.9 | 93.5 | 100.0 |
| | permutation_bonf | 99.2 | 95.3 | 100.0 | 99.1 | 94.8 | 99.9 | 98.7 | 92.9 | 100.0 |
| Weib scale discrete | asymptotic | 99.8 | 98.8 | 100.0 | 99.8 | 99.0 | 100.0 | 99.9 | 97.5 | 100.0 |
| | asymptotic_bonf | 99.8 | 98.5 | 100.0 | 99.7 | 98.9 | 100.0 | 99.9 | 97.0 | 100.0 |
| | permutation_bonf | 99.8 | 98.6 | 100.0 | 99.8 | 98.3 | 100.0 | 99.9 | 96.3 | 100.0 |
| Weib shape continuous | asymptotic | 98.0 | 89.8 | 99.6 | 97.6 | 90.7 | 99.5 | 97.6 | 87.8 | 99.5 |
| | asymptotic_bonf | 97.7 | 89.2 | 99.6 | 97.2 | 89.7 | 99.5 | 97.2 | 87.0 | 99.4 |
| | permutation_bonf | 97.4 | 88.0 | 99.6 | 96.9 | 89.5 | 99.4 | 96.4 | 85.3 | 99.3 |
| Weib shape discrete | asymptotic | 99.2 | 95.1 | 99.9 | 99.3 | 96.3 | 99.8 | 98.9 | 94.1 | 99.9 |
| | asymptotic_bonf | 99.1 | 94.1 | 99.8 | 99.2 | 95.8 | 99.8 | 98.8 | 93.4 | 99.9 |
| | permutation_bonf | 99.0 | 93.8 | 99.9 | 98.9 | 95.3 | 99.9 | 98.9 | 92.8 | 99.9 |

Table S86: Rejection rates in percent for the Tukey-type contrast matrix with $\delta = 1.5$ and balanced medium sample sizes under unequal, low censoring.

| Distribution | Method | $\mathcal{H}_{0,7}$ | $\mathcal{H}_{0,8}$ | $\mathcal{H}_{0,9}$ | $\mathcal{H}_{0,13}$ | $\mathcal{H}_{0,14}$ | $\mathcal{H}_{0,15}$ | $\mathcal{H}_{0,16}$ | $\mathcal{H}_{0,17}$ | $\mathcal{H}_{0,18}$ |
|---|---|---|---|---|---|---|---|---|---|---|
| exp early continuous | asymptotic | 11.1 | 7.0 | 15.1 | 11.7 | 6.7 | 18.7 | 11.5 | 4.2 | 17.7 |
| | asymptotic_bonf | 10.5 | 6.3 | 14.1 | 11.1 | 5.9 | 17.3 | 10.0 | 3.9 | 16.4 |
| | permutation_bonf | 9.9 | 6.2 | 14.2 | 10.2 | 5.8 | 16.3 | 9.4 | 3.6 | 15.4 |
| exp early discrete | asymptotic | 13.2 | 8.4 | 18.6 | 14.0 | 8.0 | 22.9 | 11.9 | 6.2 | 21.2 |
| | asymptotic_bonf | 12.0 | 6.9 | 17.1 | 13.0 | 7.2 | 21.2 | 11.2 | 5.1 | 19.3 |
| | permutation_bonf | 12.8 | 7.1 | 17.4 | 11.7 | 7.0 | 19.7 | 10.6 | 5.1 | 17.4 |
| exp late continuous | asymptotic | 11.4 | 8.2 | 17.7 | 13.2 | 7.1 | 22.8 | 12.7 | 6.0 | 20.6 |
| | asymptotic_bonf | 10.5 | 7.5 | 16.5 | 12.2 | 6.0 | 20.6 | 11.5 | 4.8 | 19.4 |
| | permutation_bonf | 10.3 | 7.5 | 15.3 | 11.2 | 6.7 | 19.2 | 11.2 | 5.1 | 17.4 |
| exp late discrete | asymptotic | 14.1 | 9.7 | 21.0 | 15.3 | 8.5 | 25.5 | 14.6 | 6.2 | 24.8 |
| | asymptotic_bonf | 12.8 | 8.9 | 19.5 | 13.9 | 7.7 | 24.0 | 13.4 | 5.4 | 22.2 |
| | permutation_bonf | 13.7 | 8.4 | 18.5 | 13.4 | 7.8 | 22.7 | 12.4 | 6.2 | 20.7 |
| exp prop continuous | asymptotic | 13.2 | 8.3 | 16.3 | 12.0 | 6.1 | 16.3 | 11.4 | 6.2 | 17.1 |
| | asymptotic_bonf | 12.3 | 7.7 | 15.1 | 11.3 | 5.6 | 15.2 | 9.8 | 5.7 | 15.2 |
| | permutation_bonf | 10.5 | 7.2 | 14.8 | 10.6 | 5.8 | 14.3 | 9.8 | 5.6 | 14.4 |
| exp prop discrete | asymptotic | 14.7 | 9.4 | 19.1 | 13.6 | 7.5 | 20.9 | 13.6 | 7.6 | 19.7 |
| | asymptotic_bonf | 13.3 | 8.5 | 17.7 | 13.1 | 6.6 | 18.9 | 12.5 | 6.7 | 17.9 |
| | permutation_bonf | 11.8 | 8.3 | 16.0 | 11.8 | 6.5 | 17.1 | 11.5 | 6.4 | 17.2 |
| logn continuous | asymptotic | 37.8 | 25.7 | 59.4 | 40.1 | 26.5 | 60.4 | 38.4 | 21.9 | 54.9 |
| | asymptotic_bonf | 36.1 | 24.6 | 56.9 | 38.5 | 24.5 | 58.0 | 36.2 | 21.0 | 52.7 |
| | permutation_bonf | 34.5 | 23.6 | 54.6 | 36.3 | 24.7 | 56.7 | 34.7 | 20.3 | 51.1 |
| logn discrete | asymptotic | 45.3 | 32.5 | 66.8 | 47.5 | 32.2 | 69.1 | 47.7 | 28.9 | 64.3 |
| | asymptotic_bonf | 43.4 | 30.3 | 64.9 | 45.2 | 30.6 | 67.1 | 45.0 | 26.7 | 62.5 |
| | permutation_bonf | 43.5 | 30.2 | 63.0 | 45.3 | 30.7 | 64.9 | 44.0 | 25.3 | 58.8 |
| pwExp continuous | asymptotic | 10.1 | 7.4 | 16.4 | 11.3 | 6.0 | 19.0 | 11.1 | 4.8 | 16.8 |
| | asymptotic_bonf | 9.0 | 6.2 | 15.2 | 10.2 | 5.1 | 16.8 | 9.7 | 4.3 | 15.5 |
| | permutation_bonf | 8.4 | 6.7 | 14.8 | 9.3 | 5.1 | 16.0 | 9.0 | 4.4 | 14.4 |
| pwExp discrete | asymptotic | 12.1 | 8.3 | 19.5 | 13.1 | 6.6 | 22.5 | 12.3 | 5.8 | 20.9 |
| | asymptotic_bonf | 11.2 | 7.5 | 17.9 | 12.0 | 6.2 | 21.3 | 10.9 | 4.9 | 19.2 |
| | permutation_bonf | 10.3 | 7.5 | 17.6 | 11.4 | 6.1 | 20.3 | 10.3 | 5.4 | 17.9 |
| Weib late continuous | asymptotic | 38.4 | 26.6 | 58.8 | 39.8 | 26.3 | 63.1 | 40.7 | 24.5 | 56.9 |
| | asymptotic_bonf | 36.3 | 25.6 | 57.8 | 37.7 | 24.3 | 61.1 | 37.9 | 23.4 | 55.2 |
| | permutation_bonf | 35.3 | 24.1 | 55.2 | 36.8 | 24.1 | 59.8 | 36.4 | 21.5 | 53.3 |
| Weib late discrete | asymptotic | 47.1 | 32.2 | 67.9 | 48.1 | 31.8 | 71.4 | 48.9 | 29.2 | 64.8 |
| | asymptotic_bonf | 44.9 | 30.5 | 65.8 | 46.2 | 29.6 | 69.3 | 45.9 | 27.8 | 63.1 |
| | permutation_bonf | 43.7 | 29.1 | 62.9 | 44.9 | 28.6 | 66.9 | 45.3 | 27.0 | 60.3 |
| Weib prop continuous | asymptotic | 36.7 | 25.3 | 56.6 | 37.6 | 24.0 | 60.7 | 38.3 | 22.9 | 53.6 |
| | asymptotic_bonf | 35.0 | 23.3 | 54.0 | 36.1 | 22.2 | 58.9 | 35.8 | 21.6 | 51.5 |
| | permutation_bonf | 33.0 | 22.1 | 53.0 | 35.1 | 21.9 | 55.8 | 33.6 | 21.2 | 49.1 |
| Weib prop discrete | asymptotic | 45.3 | 30.7 | 65.3 | 46.5 | 30.0 | 68.4 | 47.9 | 28.2 | 63.1 |
| | asymptotic_bonf | 43.2 | 29.1 | 62.9 | 44.2 | 28.3 | 66.8 | 45.9 | 27.2 | 61.2 |
| | permutation_bonf | 41.9 | 28.1 | 60.1 | 42.3 | 27.4 | 65.6 | 43.1 | 27.7 | 59.7 |
| Weib scale continuous | asymptotic | 29.4 | 19.6 | 46.2 | 30.8 | 18.1 | 49.3 | 30.2 | 18.5 | 42.8 |
| | asymptotic_bonf | 27.4 | 18.0 | 43.9 | 29.0 | 16.9 | 47.1 | 27.8 | 16.8 | 41.2 |
| | permutation_bonf | 25.4 | 18.0 | 43.3 | 28.2 | 16.3 | 45.5 | 25.7 | 15.6 | 39.0 |
| Weib scale discrete | asymptotic | 37.4 | 24.4 | 55.5 | 37.7 | 23.3 | 57.8 | 38.8 | 22.3 | 51.8 |
| | asymptotic_bonf | 35.1 | 22.8 | 53.8 | 36.3 | 21.6 | 55.5 | 36.6 | 20.6 | 49.4 |
| | permutation_bonf | 32.9 | 22.0 | 52.1 | 35.2 | 21.5 | 53.6 | 34.2 | 21.0 | 47.6 |
| Weib shape continuous | asymptotic | 19.5 | 12.8 | 33.6 | 21.4 | 12.9 | 37.1 | 21.5 | 11.8 | 30.6 |
| | asymptotic_bonf | 18.3 | 11.9 | 31.7 | 19.6 | 11.3 | 34.7 | 19.9 | 10.4 | 29.1 |
| | permutation_bonf | 18.0 | 11.5 | 30.3 | 19.0 | 11.1 | 32.3 | 19.2 | 10.2 | 27.6 |
| Weib shape discrete | asymptotic | 26.5 | 17.4 | 41.1 | 28.7 | 17.2 | 44.6 | 28.4 | 16.2 | 40.4 |
| | asymptotic_bonf | 23.5 | 15.2 | 39.7 | 26.8 | 15.2 | 42.5 | 25.9 | 14.7 | 38.0 |
| | permutation_bonf | 22.5 | 16.0 | 39.4 | 26.1 | 14.9 | 40.8 | 25.9 | 14.0 | 35.9 |

Table S87: Rejection rates in percent for the Tukey-type contrast matrix with $\delta = 1.5$ and balanced small sample sizes under unequal, low censoring.

| Distribution | Method | $\mathcal{H}_{0,7}$ | $\mathcal{H}_{0,8}$ | $\mathcal{H}_{0,9}$ | $\mathcal{H}_{0,13}$ | $\mathcal{H}_{0,14}$ | $\mathcal{H}_{0,15}$ | $\mathcal{H}_{0,16}$ | $\mathcal{H}_{0,17}$ | $\mathcal{H}_{0,18}$ |
|---|---|---|---|---|---|---|---|---|---|---|
| exp early continuous | asymptotic | 2.3 | 0.8 | 2.1 | 1.7 | 1.5 | 2.1 | 2.4 | 0.6 | 2.8 |
| | asymptotic_bonf | 2.1 | 0.8 | 1.9 | 1.5 | 1.1 | 1.8 | 2.1 | 0.6 | 2.6 |
| | permutation_bonf | 1.9 | 0.6 | 1.2 | 1.1 | 0.6 | 1.1 | 1.8 | 0.5 | 1.5 |
| exp early discrete | asymptotic | 2.0 | 1.4 | 1.7 | 2.2 | 1.2 | 2.4 | 2.1 | 1.1 | 1.7 |
| | asymptotic_bonf | 1.8 | 1.2 | 1.6 | 1.7 | 1.0 | 1.9 | 1.6 | 1.1 | 1.3 |
| | permutation_bonf | 1.6 | 0.8 | 1.2 | 1.3 | 0.8 | 1.8 | 1.1 | 1.0 | 1.3 |
| exp late continuous | asymptotic | 2.3 | 1.2 | 2.2 | 1.9 | 1.8 | 2.7 | 2.7 | 1.3 | 2.7 |
| | asymptotic_bonf | 2.0 | 1.0 | 2.0 | 1.9 | 1.6 | 2.2 | 2.6 | 1.2 | 2.6 |
| | permutation_bonf | 1.6 | 0.8 | 1.4 | 1.2 | 1.0 | 1.6 | 1.9 | 0.9 | 2.3 |
| exp late discrete | asymptotic | 2.4 | 1.4 | 2.8 | 2.1 | 1.7 | 2.9 | 2.9 | 1.6 | 3.3 |
| | asymptotic_bonf | 2.2 | 1.2 | 2.6 | 2.0 | 1.7 | 2.6 | 2.7 | 1.1 | 3.1 |
| | permutation_bonf | 1.5 | 0.9 | 2.0 | 1.6 | 1.2 | 1.7 | 2.2 | 1.1 | 2.4 |
| exp prop continuous | asymptotic | 1.3 | 0.9 | 2.4 | 1.7 | 1.2 | 1.9 | 1.1 | 1.1 | 2.4 |
| | asymptotic_bonf | 1.0 | 0.9 | 2.2 | 1.6 | 0.8 | 1.4 | 0.9 | 0.8 | 1.9 |
| | permutation_bonf | 0.7 | 0.9 | 1.5 | 1.2 | 0.9 | 1.3 | 0.8 | 0.7 | 1.6 |
| exp prop discrete | asymptotic | 2.3 | 1.2 | 3.1 | 1.6 | 0.5 | 3.5 | 1.9 | 1.2 | 3.1 |
| | asymptotic_bonf | 2.2 | 0.9 | 2.9 | 1.4 | 0.5 | 3.1 | 1.8 | 1.0 | 2.7 |
| | permutation_bonf | 1.7 | 0.5 | 2.4 | 1.0 | 0.3 | 2.3 | 1.4 | 1.0 | 2.2 |
| logn continuous | asymptotic | 5.4 | 3.4 | 7.9 | 4.7 | 2.8 | 5.7 | 5.0 | 3.2 | 7.5 |
| | asymptotic_bonf | 4.7 | 2.8 | 7.2 | 4.0 | 1.9 | 5.3 | 4.6 | 2.9 | 7.1 |
| | permutation_bonf | 4.1 | 2.6 | 5.5 | 3.2 | 1.8 | 4.4 | 4.2 | 2.5 | 5.8 |
| logn discrete | asymptotic | 6.8 | 4.1 | 10.3 | 6.5 | 3.4 | 7.7 | 5.7 | 3.4 | 9.2 |
| | asymptotic_bonf | 6.0 | 3.4 | 9.0 | 5.3 | 2.8 | 6.8 | 5.2 | 3.0 | 8.9 |
| | permutation_bonf | 4.8 | 3.5 | 7.9 | 4.5 | 2.9 | 6.5 | 5.1 | 2.8 | 7.5 |
| pwExp continuous | asymptotic | 1.4 | 1.1 | 1.6 | 1.4 | 1.2 | 2.0 | 2.2 | 0.9 | 2.6 |
| | asymptotic_bonf | 1.4 | 1.0 | 1.4 | 1.0 | 1.0 | 1.9 | 2.2 | 0.8 | 2.4 |
| | permutation_bonf | 1.2 | 0.5 | 1.0 | 1.0 | 0.3 | 1.2 | 1.8 | 0.7 | 1.9 |
| pwExp discrete | asymptotic | 1.7 | 1.0 | 2.4 | 1.6 | 1.5 | 2.5 | 2.4 | 1.1 | 3.1 |
| | asymptotic_bonf | 1.4 | 1.0 | 1.6 | 1.5 | 1.1 | 1.9 | 2.4 | 1.0 | 2.7 |
| | permutation_bonf | 1.1 | 0.7 | 1.1 | 0.9 | 0.5 | 1.6 | 2.0 | 1.0 | 2.0 |
| Weib late continuous | asymptotic | 4.9 | 4.3 | 8.1 | 5.9 | 3.6 | 8.4 | 4.3 | 4.1 | 6.8 |
| | asymptotic_bonf | 4.1 | 4.1 | 7.1 | 5.3 | 3.2 | 7.4 | 3.9 | 3.9 | 6.4 |
| | permutation_bonf | 3.8 | 3.6 | 6.0 | 3.8 | 2.6 | 5.5 | 3.0 | 3.3 | 5.2 |
| Weib late discrete | asymptotic | 5.9 | 4.6 | 9.7 | 6.9 | 4.1 | 9.7 | 5.7 | 4.6 | 8.6 |
| | asymptotic_bonf | 5.4 | 4.2 | 8.3 | 6.4 | 3.6 | 8.9 | 5.1 | 4.3 | 8.2 |
| | permutation_bonf | 4.7 | 4.5 | 6.6 | 5.2 | 3.2 | 7.3 | 3.5 | 4.2 | 7.4 |
| Weib prop continuous | asymptotic | 4.6 | 4.0 | 6.8 | 5.4 | 3.2 | 7.0 | 3.8 | 3.9 | 6.2 |
| | asymptotic_bonf | 3.6 | 3.5 | 6.3 | 4.5 | 2.7 | 6.0 | 3.3 | 3.6 | 5.1 |
| | permutation_bonf | 3.0 | 3.2 | 5.2 | 3.6 | 2.4 | 4.9 | 2.7 | 3.0 | 4.8 |
| Weib prop discrete | asymptotic | 5.4 | 4.8 | 8.4 | 6.4 | 3.6 | 8.8 | 4.7 | 4.8 | 7.4 |
| | asymptotic_bonf | 5.0 | 4.0 | 7.6 | 6.0 | 2.9 | 8.1 | 4.0 | 4.1 | 7.2 |
| | permutation_bonf | 4.2 | 3.6 | 5.6 | 4.5 | 2.6 | 6.8 | 3.3 | 3.9 | 5.9 |
| Weib scale continuous | asymptotic | 2.9 | 2.6 | 4.7 | 2.5 | 1.6 | 4.9 | 1.9 | 1.9 | 3.8 |
| | asymptotic_bonf | 2.7 | 2.5 | 4.3 | 2.1 | 1.5 | 4.5 | 1.5 | 1.7 | 3.4 |
| | permutation_bonf | 1.7 | 2.3 | 3.0 | 1.9 | 1.3 | 3.9 | 1.6 | 1.6 | 3.0 |
| Weib scale discrete | asymptotic | 3.5 | 3.2 | 5.3 | 3.5 | 2.0 | 5.7 | 2.2 | 2.5 | 5.4 |
| | asymptotic_bonf | 2.6 | 3.2 | 4.5 | 3.0 | 1.9 | 5.3 | 1.8 | 2.0 | 4.7 |
| | permutation_bonf | 2.0 | 2.4 | 3.8 | 2.5 | 1.6 | 4.8 | 1.8 | 1.8 | 4.3 |
| Weib shape continuous | asymptotic | 1.3 | 1.9 | 2.0 | 1.3 | 1.2 | 3.0 | 1.0 | 1.0 | 2.4 |
| | asymptotic_bonf | 1.0 | 1.6 | 1.9 | 1.3 | 0.9 | 2.4 | 0.9 | 0.9 | 2.1 |
| | permutation_bonf | 0.9 | 1.4 | 2.0 | 0.7 | 0.9 | 2.4 | 0.9 | 0.7 | 2.3 |
| Weib shape discrete | asymptotic | 1.5 | 2.3 | 3.0 | 1.7 | 1.3 | 3.8 | 1.3 | 1.4 | 3.2 |
| | asymptotic_bonf | 1.3 | 2.2 | 2.4 | 1.5 | 1.2 | 3.0 | 1.1 | 1.1 | 3.0 |
| | permutation_bonf | 1.1 | 2.0 | 2.2 | 1.4 | 1.1 | 2.7 | 0.9 | 0.8 | 2.9 |

Table S88: Rejection rates in percent for the Tukey-type contrast matrix with $\delta = 1.5$ and unbalanced large sample sizes under unequal, low censoring.

| Distribution | Method | $\mathcal{H}_{0,7}$ | $\mathcal{H}_{0,8}$ | $\mathcal{H}_{0,9}$ | $\mathcal{H}_{0,13}$ | $\mathcal{H}_{0,14}$ | $\mathcal{H}_{0,15}$ | $\mathcal{H}_{0,16}$ | $\mathcal{H}_{0,17}$ | $\mathcal{H}_{0,18}$ |
|---|---|---|---|---|---|---|---|---|---|---|
| exp early continuous | asymptotic | 31.0 | 16.3 | 53.5 | 26.6 | 12.9 | 42.8 | 26.9 | 12.7 | 44.1 |
| | asymptotic_bonf | 28.1 | 14.4 | 51.2 | 23.6 | 11.1 | 39.9 | 23.9 | 11.2 | 40.3 |
| | permutation_bonf | 24.3 | 11.7 | 48.3 | 21.1 | 10.8 | 37.7 | 22.4 | 10.3 | 38.2 |
| exp early discrete | asymptotic | 36.6 | 20.6 | 60.6 | 31.1 | 15.4 | 49.1 | 31.9 | 16.4 | 52.0 |
| | asymptotic_bonf | 33.3 | 17.6 | 57.6 | 27.6 | 13.8 | 45.3 | 28.5 | 14.0 | 47.7 |
| | permutation_bonf | 28.8 | 14.0 | 54.7 | 26.4 | 13.1 | 44.5 | 26.3 | 12.7 | 44.2 |
| exp late continuous | asymptotic | 37.8 | 19.9 | 63.5 | 31.1 | 15.1 | 51.5 | 31.4 | 15.7 | 52.0 |
| | asymptotic_bonf | 33.8 | 17.8 | 60.1 | 27.7 | 13.1 | 48.1 | 28.4 | 13.6 | 47.9 |
| | permutation_bonf | 30.2 | 14.6 | 58.4 | 25.6 | 11.8 | 46.2 | 26.8 | 12.9 | 45.3 |
| exp late discrete | asymptotic | 44.5 | 24.4 | 70.0 | 37.5 | 19.1 | 59.4 | 38.1 | 19.0 | 59.8 |
| | asymptotic_bonf | 41.1 | 22.0 | 67.4 | 33.4 | 16.7 | 55.4 | 34.4 | 17.0 | 57.5 |
| | permutation_bonf | 36.7 | 17.2 | 64.5 | 31.2 | 15.8 | 53.5 | 32.2 | 15.2 | 53.9 |
| exp prop continuous | asymptotic | 36.7 | 20.0 | 54.5 | 30.0 | 17.0 | 43.1 | 28.6 | 16.2 | 45.8 |
| | asymptotic_bonf | 33.1 | 17.8 | 50.3 | 27.3 | 14.6 | 39.6 | 26.1 | 14.3 | 43.2 |
| | permutation_bonf | 30.5 | 16.2 | 47.1 | 26.1 | 12.1 | 37.3 | 23.1 | 12.6 | 41.8 |
| exp prop discrete | asymptotic | 42.9 | 24.4 | 62.8 | 35.0 | 20.5 | 49.7 | 33.9 | 19.7 | 53.9 |
| | asymptotic_bonf | 40.0 | 21.9 | 59.4 | 31.6 | 18.0 | 46.8 | 30.3 | 16.9 | 50.2 |
| | permutation_bonf | 36.1 | 18.0 | 56.2 | 29.8 | 15.8 | 45.2 | 27.4 | 14.8 | 49.4 |
| logn continuous | asymptotic | 87.8 | 66.7 | 95.6 | 77.5 | 57.1 | 91.3 | 77.3 | 58.8 | 90.9 |
| | asymptotic_bonf | 85.6 | 63.8 | 94.4 | 76.4 | 54.3 | 89.9 | 74.3 | 55.5 | 89.2 |
| | permutation_bonf | 81.0 | 54.7 | 93.0 | 72.9 | 48.3 | 87.3 | 71.5 | 49.7 | 86.9 |
| logn discrete | asymptotic | 92.9 | 76.3 | 98.3 | 86.4 | 66.4 | 95.4 | 86.2 | 68.6 | 95.2 |
| | asymptotic_bonf | 92.0 | 73.5 | 98.0 | 84.1 | 63.3 | 94.6 | 84.1 | 66.3 | 94.1 |
| | permutation_bonf | 87.1 | 62.7 | 96.8 | 81.0 | 57.7 | 92.6 | 80.8 | 60.0 | 92.5 |
| pwExp continuous | asymptotic | 29.9 | 15.4 | 52.1 | 24.2 | 13.0 | 43.0 | 24.9 | 13.4 | 41.4 |
| | asymptotic_bonf | 26.8 | 12.9 | 48.4 | 21.6 | 11.1 | 38.7 | 23.2 | 11.9 | 38.4 |
| | permutation_bonf | 24.8 | 9.8 | 46.3 | 20.1 | 9.8 | 38.0 | 21.4 | 10.4 | 36.0 |
| pwExp discrete | asymptotic | 34.9 | 18.9 | 60.6 | 29.5 | 15.9 | 50.0 | 29.4 | 16.6 | 50.3 |
| | asymptotic_bonf | 32.7 | 16.3 | 57.1 | 26.0 | 13.6 | 47.0 | 26.6 | 15.0 | 46.5 |
| | permutation_bonf | 28.7 | 12.5 | 53.3 | 25.4 | 12.4 | 43.8 | 25.1 | 13.1 | 44.1 |
| Weib late continuous | asymptotic | 89.1 | 66.8 | 97.6 | 82.9 | 57.9 | 94.4 | 80.8 | 56.9 | 94.9 |
| | asymptotic_bonf | 88.1 | 64.3 | 97.2 | 81.4 | 54.7 | 93.4 | 79.8 | 54.0 | 93.5 |
| | permutation_bonf | 83.9 | 56.3 | 96.1 | 78.4 | 51.4 | 92.3 | 76.7 | 48.8 | 92.5 |
| Weib late discrete | asymptotic | 93.3 | 75.9 | 98.9 | 88.5 | 67.6 | 97.3 | 87.8 | 66.7 | 97.9 |
| | asymptotic_bonf | 92.6 | 73.3 | 98.7 | 87.1 | 64.1 | 96.6 | 86.8 | 64.4 | 97.4 |
| | permutation_bonf | 89.3 | 65.1 | 97.8 | 85.0 | 59.9 | 95.7 | 83.8 | 59.1 | 96.1 |
| Weib prop continuous | asymptotic | 85.4 | 63.8 | 96.4 | 79.4 | 54.5 | 92.4 | 78.6 | 53.6 | 93.0 |
| | asymptotic_bonf | 83.9 | 60.5 | 95.4 | 77.4 | 51.6 | 91.3 | 76.4 | 49.7 | 92.0 |
| | permutation_bonf | 79.9 | 51.7 | 94.2 | 74.8 | 47.5 | 89.3 | 73.5 | 45.5 | 89.9 |
| Weib prop discrete | asymptotic | 92.1 | 73.5 | 98.7 | 86.9 | 64.4 | 96.4 | 86.2 | 64.7 | 96.9 |
| | asymptotic_bonf | 90.4 | 70.1 | 98.2 | 85.1 | 60.6 | 96.0 | 84.3 | 61.4 | 95.9 |
| | permutation_bonf | 86.6 | 61.0 | 97.6 | 82.8 | 56.4 | 95.0 | 80.7 | 55.1 | 95.0 |
| Weib scale continuous | asymptotic | 72.7 | 48.3 | 88.5 | 66.2 | 42.2 | 83.3 | 65.6 | 41.6 | 83.7 |
| | asymptotic_bonf | 69.7 | 44.7 | 86.7 | 62.4 | 38.7 | 80.7 | 63.1 | 38.8 | 81.6 |
| | permutation_bonf | 64.8 | 35.9 | 83.7 | 57.8 | 33.5 | 77.3 | 58.2 | 32.1 | 79.7 |
| Weib scale discrete | asymptotic | 81.6 | 58.5 | 93.4 | 74.8 | 51.2 | 90.2 | 74.7 | 49.5 | 90.1 |
| | asymptotic_bonf | 78.9 | 54.4 | 92.6 | 73.0 | 48.2 | 88.2 | 72.3 | 46.2 | 89.0 |
| | permutation_bonf | 72.7 | 43.8 | 89.2 | 68.8 | 41.4 | 86.6 | 67.2 | 40.8 | 87.3 |
| Weib shape continuous | asymptotic | 53.6 | 30.1 | 72.3 | 47.0 | 27.8 | 66.4 | 46.9 | 28.5 | 67.6 |
| | asymptotic_bonf | 48.5 | 25.9 | 68.7 | 43.2 | 24.5 | 62.6 | 43.1 | 24.2 | 63.3 |
| | permutation_bonf | 40.3 | 19.2 | 62.5 | 38.9 | 21.4 | 59.4 | 38.2 | 20.3 | 59.6 |
| Weib shape discrete | asymptotic | 65.3 | 40.0 | 82.6 | 57.4 | 35.8 | 76.0 | 58.1 | 36.1 | 77.8 |
| | asymptotic_bonf | 61.5 | 35.3 | 79.7 | 54.2 | 32.7 | 72.6 | 54.3 | 33.2 | 74.4 |
| | permutation_bonf | 51.7 | 24.8 | 74.2 | 49.2 | 27.4 | 70.0 | 49.8 | 26.5 | 71.1 |

Table S89: Rejection rates in percent for the Tukey-type contrast matrix with $\delta = 1.5$ and unbalanced medium sample sizes under unequal, low censoring.

| Distribution | Method | $\mathcal{H}_{0,7}$ | $\mathcal{H}_{0,8}$ | $\mathcal{H}_{0,9}$ | $\mathcal{H}_{0,13}$ | $\mathcal{H}_{0,14}$ | $\mathcal{H}_{0,15}$ | $\mathcal{H}_{0,16}$ | $\mathcal{H}_{0,17}$ | $\mathcal{H}_{0,18}$ |
|---|---|---|---|---|---|---|---|---|---|---|
| exp early continuous | asymptotic | 1.8 | 1.3 | 3.9 | 1.5 | 1.1 | 3.5 | 2.3 | 1.3 | 3.9 |
|  | asymptotic_bonf | 1.3 | 0.8 | 3.3 | 1.1 | 1.0 | 3.2 | 2.0 | 0.9 | 3.0 |
|  | permutation_bonf | 0.6 | 0.2 | 1.6 | 0.5 | 0.4 | 2.3 | 1.6 | 0.5 | 2.2 |
| exp early discrete | asymptotic | 2.7 | 1.4 | 4.6 | 1.9 | 1.3 | 4.2 | 3.0 | 1.3 | 4.7 |
|  | asymptotic_bonf | 1.8 | 1.1 | 4.0 | 1.2 | 1.1 | 3.6 | 2.5 | 0.9 | 3.8 |
|  | permutation_bonf | 0.6 | 0.3 | 2.3 | 0.9 | 0.5 | 2.9 | 1.3 | 0.6 | 2.6 |
| exp late continuous | asymptotic | 2.6 | 1.7 | 4.7 | 1.9 | 1.4 | 3.8 | 2.9 | 1.6 | 4.3 |
|  | asymptotic_bonf | 2.1 | 1.5 | 4.5 | 1.2 | 1.2 | 3.3 | 2.4 | 1.2 | 3.6 |
|  | permutation_bonf | 0.8 | 0.2 | 1.6 | 0.8 | 0.6 | 2.8 | 1.7 | 0.3 | 2.4 |
| exp late discrete | asymptotic | 3.3 | 2.0 | 5.9 | 2.6 | 1.6 | 4.3 | 3.5 | 2.0 | 5.3 |
|  | asymptotic_bonf | 2.4 | 1.6 | 5.0 | 1.7 | 1.3 | 4.0 | 2.5 | 1.2 | 4.2 |
|  | permutation_bonf | 1.0 | 0.5 | 2.7 | 1.2 | 1.1 | 2.8 | 1.7 | 0.3 | 3.2 |
| exp prop continuous | asymptotic | 2.8 | 1.3 | 3.4 | 2.8 | 1.2 | 3.8 | 2.2 | 1.9 | 3.2 |
|  | asymptotic_bonf | 2.3 | 1.0 | 2.7 | 2.4 | 1.1 | 2.7 | 1.9 | 1.4 | 2.7 |
|  | permutation_bonf | 1.3 | 0.5 | 1.5 | 1.9 | 0.4 | 1.7 | 0.9 | 0.3 | 2.2 |
| exp prop discrete | asymptotic | 3.5 | 1.7 | 3.9 | 3.2 | 1.2 | 4.3 | 3.0 | 2.0 | 3.7 |
|  | asymptotic_bonf | 2.8 | 1.2 | 3.5 | 2.6 | 1.1 | 3.5 | 2.2 | 1.5 | 3.3 |
|  | permutation_bonf | 1.5 | 0.3 | 1.6 | 1.7 | 0.2 | 2.2 | 1.5 | 0.5 | 1.9 |
| logn continuous | asymptotic | 8.1 | 5.2 | 16.2 | 7.2 | 5.3 | 13.6 | 7.0 | 4.4 | 13.7 |
|  | asymptotic_bonf | 6.7 | 4.4 | 14.1 | 5.9 | 4.7 | 12.2 | 5.6 | 3.7 | 12.4 |
|  | permutation_bonf | 1.3 | 0.3 | 4.7 | 2.3 | 0.8 | 7.8 | 2.2 | 1.3 | 6.7 |
| logn discrete | asymptotic | 10.6 | 6.3 | 19.8 | 9.1 | 6.4 | 16.7 | 8.4 | 5.2 | 16.4 |
|  | asymptotic_bonf | 8.7 | 5.0 | 17.7 | 7.9 | 5.5 | 14.8 | 7.1 | 4.3 | 14.9 |
|  | permutation_bonf | 1.2 | 0.2 | 5.5 | 3.5 | 1.1 | 10.4 | 3.1 | 1.0 | 8.4 |
| pwExp continuous | asymptotic | 2.0 | 1.5 | 3.6 | 1.1 | 1.3 | 3.5 | 1.8 | 1.1 | 2.9 |
|  | asymptotic_bonf | 1.3 | 1.3 | 3.1 | 0.6 | 1.0 | 3.0 | 1.8 | 0.9 | 2.6 |
|  | permutation_bonf | 0.7 | 0.4 | 1.6 | 0.4 | 0.7 | 2.2 | 1.1 | 0.3 | 2.0 |
| pwExp discrete | asymptotic | 2.5 | 1.7 | 4.3 | 1.8 | 1.4 | 3.6 | 2.3 | 1.5 | 4.0 |
|  | asymptotic_bonf | 1.7 | 1.4 | 3.7 | 0.8 | 1.0 | 3.2 | 1.9 | 1.1 | 3.1 |
|  | permutation_bonf | 0.5 | 0.4 | 1.7 | 0.4 | 0.6 | 2.6 | 1.3 | 0.1 | 2.6 |
| Weib late continuous | asymptotic | 10.3 | 4.2 | 20.4 | 9.1 | 4.8 | 16.5 | 9.7 | 4.5 | 18.2 |
|  | asymptotic_bonf | 8.8 | 3.4 | 18.0 | 7.8 | 3.7 | 14.1 | 8.3 | 3.6 | 15.5 |
|  | permutation_bonf | 2.4 | 0.0 | 8.0 | 4.3 | 1.3 | 8.7 | 3.2 | 0.6 | 8.1 |
| Weib late discrete | asymptotic | 12.8 | 5.3 | 25.0 | 10.6 | 5.4 | 20.2 | 12.1 | 5.6 | 21.4 |
|  | asymptotic_bonf | 11.5 | 4.2 | 22.8 | 9.5 | 4.6 | 18.0 | 9.8 | 4.3 | 18.8 |
|  | permutation_bonf | 2.4 | 0.3 | 8.4 | 5.8 | 1.6 | 11.2 | 4.1 | 0.7 | 11.4 |
| Weib prop continuous | asymptotic | 8.2 | 2.9 | 16.9 | 7.9 | 3.6 | 14.2 | 8.9 | 3.5 | 14.4 |
|  | asymptotic_bonf | 7.0 | 2.4 | 13.9 | 6.6 | 2.6 | 11.6 | 7.0 | 2.9 | 11.8 |
|  | permutation_bonf | 1.6 | 0.0 | 5.9 | 3.6 | 1.1 | 7.0 | 2.8 | 0.5 | 6.4 |
| Weib prop discrete | asymptotic | 11.6 | 4.4 | 20.8 | 9.5 | 4.6 | 17.4 | 10.6 | 4.4 | 17.8 |
|  | asymptotic_bonf | 9.6 | 2.9 | 17.7 | 8.2 | 3.5 | 15.5 | 8.6 | 3.3 | 15.2 |
|  | permutation_bonf | 1.9 | 0.0 | 6.3 | 4.7 | 0.9 | 8.7 | 3.6 | 0.4 | 8.3 |
| Weib scale continuous | asymptotic | 5.3 | 1.4 | 8.7 | 4.6 | 1.7 | 7.9 | 4.7 | 1.7 | 6.9 |
|  | asymptotic_bonf | 4.1 | 0.8 | 6.5 | 3.9 | 1.2 | 6.4 | 3.4 | 1.2 | 5.7 |
|  | permutation_bonf | 0.5 | 0.0 | 2.6 | 2.0 | 0.5 | 4.0 | 1.4 | 0.1 | 3.0 |
| Weib scale discrete | asymptotic | 5.9 | 2.4 | 10.9 | 6.2 | 2.4 | 10.4 | 6.2 | 2.8 | 10.2 |
|  | asymptotic_bonf | 5.1 | 1.2 | 9.1 | 5.2 | 1.5 | 8.1 | 4.6 | 1.8 | 7.4 |
|  | permutation_bonf | 0.7 | 0.0 | 2.7 | 2.1 | 0.5 | 4.3 | 1.4 | 0.1 | 3.6 |
| Weib shape continuous | asymptotic | 2.2 | 0.9 | 4.2 | 3.1 | 0.8 | 3.6 | 2.8 | 0.9 | 3.9 |
|  | asymptotic_bonf | 1.8 | 0.5 | 3.4 | 2.4 | 0.6 | 3.2 | 1.5 | 0.4 | 2.7 |
|  | permutation_bonf | 0.2 | 0.0 | 1.1 | 1.1 | 0.2 | 1.9 | 0.4 | 0.0 | 1.4 |
| Weib shape discrete | asymptotic | 3.4 | 1.1 | 5.3 | 3.3 | 1.0 | 5.1 | 3.9 | 1.2 | 5.2 |
|  | asymptotic_bonf | 1.9 | 0.9 | 4.4 | 2.8 | 0.7 | 4.2 | 2.7 | 0.9 | 3.9 |
|  | permutation_bonf | 0.2 | 0.0 | 1.0 | 1.2 | 0.2 | 2.6 | 0.5 | 0.0 | 1.7 |

Table S90: Rejection rates in percent for the Tukey-type contrast matrix with $\delta = 1.5$ and unbalanced small sample sizes under unequal, low censoring.

| Distribution | Method | $\mathcal{H}_{0,7}$ | $\mathcal{H}_{0,8}$ | $\mathcal{H}_{0,9}$ | $\mathcal{H}_{0,13}$ | $\mathcal{H}_{0,14}$ | $\mathcal{H}_{0,15}$ | $\mathcal{H}_{0,16}$ | $\mathcal{H}_{0,17}$ | $\mathcal{H}_{0,18}$ |
|---|---|---|---|---|---|---|---|---|---|---|
| exp early continuous | asymptotic | 0.8 | 3.0 | 2.0 | 0.8 | 1.5 | 1.5 | 0.8 | 2.2 | 1.3 |
|  | asymptotic_bonf | 0.6 | 2.9 | 1.6 | 0.7 | 1.5 | 1.3 | 0.6 | 1.9 | 1.0 |
|  | permutation_bonf | 0.1 | 0.0 | 0.1 | 0.1 | 0.1 | 0.1 | 0.0 | 0.3 | 0.3 |
| exp early discrete | asymptotic | 0.7 | 3.0 | 1.9 | 0.8 | 1.4 | 1.5 | 0.9 | 1.9 | 1.4 |
|  | asymptotic_bonf | 0.7 | 2.9 | 1.8 | 0.7 | 1.4 | 1.5 | 0.6 | 1.7 | 1.0 |
|  | permutation_bonf | 0.1 | 0.0 | 0.1 | 0.1 | 0.2 | 0.2 | 0.0 | 0.3 | 0.2 |
| exp late continuous | asymptotic | 1.0 | 1.9 | 1.9 | 0.7 | 1.0 | 1.8 | 0.7 | 1.6 | 1.4 |
|  | asymptotic_bonf | 0.7 | 1.7 | 1.7 | 0.6 | 1.0 | 1.5 | 0.5 | 1.5 | 1.1 |
|  | permutation_bonf | 0.0 | 0.0 | 0.0 | 0.0 | 0.1 | 0.0 | 0.0 | 0.2 | 0.0 |
| exp late discrete | asymptotic | 0.8 | 1.8 | 2.0 | 1.0 | 1.0 | 2.3 | 0.9 | 1.6 | 1.8 |
|  | asymptotic_bonf | 0.7 | 1.7 | 1.7 | 0.6 | 0.9 | 1.8 | 0.7 | 1.4 | 1.3 |
|  | permutation_bonf | 0.0 | 0.0 | 0.0 | 0.0 | 0.3 | 0.1 | 0.0 | 0.1 | 0.1 |
| exp prop continuous | asymptotic | 1.1 | 2.0 | 1.7 | 1.3 | 1.1 | 1.5 | 0.8 | 0.9 | 1.8 |
|  | asymptotic_bonf | 1.1 | 1.9 | 1.7 | 0.9 | 0.9 | 1.4 | 0.6 | 0.9 | 1.6 |
|  | permutation_bonf | 0.0 | 0.3 | 0.2 | 0.0 | 0.1 | 0.2 | 0.2 | 0.0 | 0.2 |
| exp prop discrete | asymptotic | 1.2 | 2.1 | 1.9 | 1.2 | 1.5 | 1.8 | 0.8 | 1.1 | 1.7 |
|  | asymptotic_bonf | 1.2 | 2.0 | 1.9 | 1.0 | 1.1 | 1.5 | 0.6 | 0.8 | 1.6 |
|  | permutation_bonf | 0.0 | 0.2 | 0.1 | 0.0 | 0.1 | 0.2 | 0.0 | 0.0 | 0.2 |
| logn continuous | asymptotic | 1.7 | 3.1 | 2.2 | 1.3 | 0.7 | 1.8 | 0.9 | 1.0 | 1.9 |
|  | asymptotic_bonf | 1.5 | 3.0 | 1.8 | 0.8 | 0.6 | 1.6 | 0.8 | 0.8 | 1.5 |
|  | permutation_bonf | 0.2 | 0.1 | 0.1 | 0.0 | 0.2 | 0.0 | 0.0 | 0.0 | 0.0 |
| logn discrete | asymptotic | 1.9 | 3.2 | 2.3 | 1.3 | 0.8 | 1.6 | 1.1 | 0.9 | 1.8 |
|  | asymptotic_bonf | 1.5 | 3.1 | 2.0 | 1.2 | 0.7 | 1.5 | 0.8 | 0.7 | 1.6 |
|  | permutation_bonf | 0.2 | 0.1 | 0.1 | 0.0 | 0.1 | 0.0 | 0.1 | 0.0 | 0.0 |
| pwExp continuous | asymptotic | 1.0 | 2.3 | 1.6 | 0.7 | 1.3 | 1.5 | 0.6 | 1.8 | 1.5 |
|  | asymptotic_bonf | 0.9 | 2.3 | 1.3 | 0.6 | 1.2 | 1.1 | 0.5 | 1.6 | 1.1 |
|  | permutation_bonf | 0.0 | 0.0 | 0.1 | 0.0 | 0.1 | 0.3 | 0.0 | 0.2 | 0.2 |
| pwExp discrete | asymptotic | 1.2 | 2.4 | 1.6 | 0.7 | 1.2 | 1.8 | 0.5 | 1.4 | 1.5 |
|  | asymptotic_bonf | 1.1 | 2.4 | 1.5 | 0.6 | 1.0 | 1.4 | 0.5 | 1.2 | 1.2 |
|  | permutation_bonf | 0.0 | 0.0 | 0.1 | 0.0 | 0.2 | 0.3 | 0.0 | 0.2 | 0.2 |
| Weib late continuous | asymptotic | 1.7 | 1.6 | 2.7 | 1.3 | 0.6 | 2.5 | 1.3 | 0.8 | 2.1 |
|  | asymptotic_bonf | 1.3 | 1.5 | 2.0 | 1.0 | 0.4 | 2.1 | 1.2 | 0.6 | 2.0 |
|  | permutation_bonf | 0.0 | 0.0 | 0.0 | 0.1 | 0.0 | 0.4 | 0.1 | 0.0 | 0.0 |
| Weib late discrete | asymptotic | 1.5 | 2.0 | 3.2 | 1.3 | 0.6 | 2.9 | 1.5 | 0.9 | 2.1 |
|  | asymptotic_bonf | 1.4 | 1.6 | 2.4 | 1.0 | 0.5 | 2.3 | 1.2 | 0.7 | 1.9 |
|  | permutation_bonf | 0.0 | 0.0 | 0.0 | 0.1 | 0.0 | 0.3 | 0.1 | 0.0 | 0.1 |
| Weib prop continuous | asymptotic | 1.4 | 2.2 | 2.0 | 1.2 | 0.6 | 2.4 | 1.1 | 0.7 | 2.0 |
|  | asymptotic_bonf | 1.2 | 2.1 | 1.7 | 1.0 | 0.6 | 2.0 | 1.0 | 0.6 | 1.7 |
|  | permutation_bonf | 0.0 | 0.0 | 0.0 | 0.1 | 0.0 | 0.1 | 0.1 | 0.0 | 0.0 |
| Weib prop discrete | asymptotic | 1.2 | 2.3 | 2.4 | 0.8 | 0.6 | 2.6 | 1.2 | 0.9 | 1.9 |
|  | asymptotic_bonf | 1.1 | 2.3 | 2.1 | 0.6 | 0.5 | 2.1 | 0.9 | 0.7 | 1.7 |
|  | permutation_bonf | 0.0 | 0.0 | 0.0 | 0.1 | 0.0 | 0.1 | 0.1 | 0.0 | 0.0 |
| Weib scale continuous | asymptotic | 2.7 | 3.5 | 2.5 | 1.6 | 0.9 | 1.1 | 1.2 | 1.2 | 1.6 |
|  | asymptotic_bonf | 2.6 | 3.4 | 2.1 | 1.3 | 0.9 | 1.0 | 0.8 | 0.7 | 1.1 |
|  | permutation_bonf | 0.0 | 0.0 | 0.1 | 0.1 | 0.3 | 0.0 | 0.0 | 0.0 | 0.0 |
| Weib scale discrete | asymptotic | 2.9 | 3.8 | 2.5 | 1.5 | 0.8 | 1.6 | 0.9 | 1.2 | 1.7 |
|  | asymptotic_bonf | 2.7 | 3.7 | 2.3 | 0.9 | 0.8 | 1.0 | 0.6 | 0.7 | 1.4 |
|  | permutation_bonf | 0.0 | 0.0 | 0.0 | 0.1 | 0.2 | 0.0 | 0.0 | 0.0 | 0.0 |
| Weib shape continuous | asymptotic | 6.0 | 8.7 | 3.2 | 4.7 | 3.0 | 2.7 | 4.0 | 3.2 | 2.7 |
|  | asymptotic_bonf | 5.9 | 8.5 | 2.7 | 4.0 | 2.6 | 2.3 | 3.4 | 2.7 | 2.0 |
|  | permutation_bonf | 0.6 | 0.2 | 0.4 | 0.8 | 0.8 | 0.2 | 0.0 | 0.1 | 0.3 |
| Weib shape discrete | asymptotic | 6.1 | 9.0 | 3.2 | 4.0 | 2.5 | 2.4 | 3.2 | 2.6 | 2.4 |
|  | asymptotic_bonf | 6.0 | 8.7 | 2.9 | 3.3 | 2.1 | 2.1 | 2.8 | 2.0 | 2.0 |
|  | permutation_bonf | 0.3 | 0.2 | 0.4 | 1.0 | 0.5 | 0.2 | 0.0 | 0.2 | 0.3 |



\end{document}